\documentclass[a4paper,12pt]{report}
\usepackage{Stylefile}

\onehalfspace

\title
 {
\vspace*{1.0cm}
\LARGE{\bf Pore-Scale Modeling of Non-Newtonian \vspace{0.3cm} \\
Flow in Porous Media} \vspace*{2.0cm} \\
\Large{\bf Taha Sochi} \vspace*{4.0cm} \\
\large{\bf  A dissertation submitted to the \\
            Department of Earth Science and Engineering \\
            Imperial College London \\
            in fulfillment of the requirements for the degree of \\
            Doctor of Philosophy}%\vspace*{2.0cm} \\
 }

\date{October 2007}

\ifpdf
\fi

\makeindex

\begin{document}

\maketitle %
\pagenumbering{roman}

%\linespread{2}

\phantomsection \addcontentsline{toc}{section}{Abstract}
\noindent
{\LARGE \bf \vspace{1.0cm} \\ Abstract} \vspace{0.5cm}\\
\begin{spacing}{1.5}
\noindent %
The thesis investigates the flow of \nNEW\ fluids in porous media
using network modeling. \NNEW\ fluids occur in diverse natural and
synthetic forms and have many important applications including in
the oil industry. They show very complex time and strain dependent
behavior and may have initial yield stress. Their common feature is
that they do not obey the simple \NEW\ relation of proportionality
between stress and rate of deformation. They are generally
classified into three main categories: \timeind\ in which strain
rate solely depends on the instantaneous stress, \timedep\ in which
strain rate is a function of both magnitude and duration of the
applied stress and \vc\ which shows partial elastic recovery on
removal of the deforming stress and usually demonstrates both time
and strain dependency.
\vspace{0.2cm}

The methodology followed in this investigation is pore-scale network
modeling. Two three-dimensional topologically-disordered networks
representing a \sandp\ and \Berea\ sandstone were used. The networks
are built from topologically-equivalent three-dimensional voxel
images of the pore space with the pore sizes, shapes and
connectivity reflecting the real medium. Pores and throats are
modeled as having triangular, square or circular cross-section by
assigning a shape factor, which is the ratio of the area to the
perimeter squared and is obtained from the pore space description.
An iterative numerical technique is used to solve the pressure field
and obtain the total volumetric flow rate and apparent viscosity. In
some cases, analytical expressions for the volumetric flow rate in a
single tube are derived and implemented in each throat to simulate
the flow in the pore space.
\vspace{0.2cm}

The \timeind\ category of the \nNEW\ fluids is investigated using
two \timeind\ fluid models: \ELLIS\ and \HB. Thorough comparison
between the two random networks and the uniform bundle-of-tubes
model is presented. The analysis confirmed the reliability of the
\nNEW\ network model used in this study. Good results are obtained,
especially for the \ELLIS\ model, when comparing the network model
results to experimental data sets found in the literature. The
\yields\ phenomenon is also investigated and several numerical
algorithms were developed and implemented to predict threshold yield
pressure of the network.
\vspace{0.2cm}

An extensive literature survey and investigation were carried out to
understand the phenomenon of \vy\ and clearly identify its
characteristic features, with special attention paid to flow in
porous media. The extensional flow and viscosity and \convdiv\
geometry were thoroughly examined as the basis of the peculiar \vc\
behavior in porous media. The modified \BauMan\ model, which
successfully describes \shThin, elasticity and \thixotropic\
time-dependency, was identified as a promising candidate for
modeling the flow of \vc\ materials which also show \thixotropic\
attributes. An algorithm that employs this model to simulate
\steadys\ \timedep\ \vc\ flow was implemented in the \nNEW\ code and
the initial results were analyzed. The findings are encouraging for
further future development.
\vspace{0.2cm}

The \timedep\ category of the \nNEW\ fluids was examined and several
problems in modeling and simulating the flow of these fluids were
identified. Several suggestions were presented to overcome these
difficulties.
\end{spacing}

\phantomsection \addcontentsline{toc}{section}{Acknowledgements}
\noindent
{\LARGE \bf \vspace{3.0cm} \\ Acknowledgements} \vspace{0.5cm}\\
\begin{spacing}{1.5}
\noindent %
I would like to thank
\begin{itemize}

    \item My supervisor Prof. Martin Blunt for his guidance and advice and for
    offering this opportunity to study \nNEW\ flow in porous media at
    Imperial College London funded by the Pore-Scale Modeling Consortium.

    \item Our sponsors in the Pore-Scale Modeling Consortium (BHP, DTI, ENI, JOGMEC,
    Saudi Aramco, Schlumberger, Shell, Statoil and Total),
    for their financial support, with special thanks to Schlumberger for
    funding this research.

    \item Schlumberger Cambridge Research Centre for hosting a number of
    productive meetings in which many important aspects of the work presented in
    this thesis were discussed and assessed.

    \item Dr Valerie Anderson and Dr John Crawshaw for their
    help and advice with regards to the \Tardy\ algorithm.

    \item The Internal Examiner Prof. Geoffrey Maitland from the
    Department of Chemical Engineering Imperial College London, and the
    External Examiner Prof. William Rossen from the Faculty of Civil
    Engineering and Geosciences Delft University for their helpful
    remarks and  corrections which improved the quality of this
    work.

    \item Staff and students in Imperial College London for their
    kindness and support.

    \item Family and friends for support and encouragement, with special
    thanks to my wife.

\end{itemize}
\end{spacing}

\phantomsection \addcontentsline{toc}{section}{Contents} %
\tableofcontents

\newpage
\phantomsection \addcontentsline{toc}{section}{List of Figures} %
\listoffigures

\newpage
\phantomsection \addcontentsline{toc}{section}{List of Tables} %
\listoftables

\newpage
\phantomsection \addcontentsline{toc}{section}{Nomenclature}
\noindent \vspace{1.0cm} \\
{\LARGE \bf Nomenclature}
\vspace{0.7cm} %\noindent

\begin{supertabular}{ll}
  {\bf \em Symbol}      & {\bf \em Meaning and units} \vspace{0.3cm}                       \\
  $\eAlpha$             & parameter in \ELLIS\ model (---)                                 \\
%  $\alpha$              & scale factor (---)                                               \\
%  $\sR$                 & shear rate (s$^{-1}$)                                             \\
  $\sR$                 & strain rate (s$^{-1}$)                                            \\
  $\sR_{c}$             & characteristic strain rate (s$^{-1}$)                             \\
  $\crsR$               & critical shear rate (s$^{-1}$)                                    \\
  $\rsTen$              & rate-of-strain tensor                                             \\
  $\epsilon$            & porosity (---)                                                   \\
  $\elong$              & elongation (---)                                                 \\
  $\elongR$             & elongation rate (s$^{-1}$)                                        \\
  $\rxTimF$             & structural relaxation time in \FRED\ model (s)                    \\
  $\rxTim$              & relaxation time (s)                                               \\
  $\rdTim$              & retardation time (s)                                              \\
  $\fTim_{c}$           & characteristic time of fluid (s)                                  \\
  $\fgTim$              & first time constant in \GODF\ model (s)                          \\
  $\sgTim$              & second time constant in \GODF\ model (s)                         \\
  $\seTim$              & time constant in \SEM\ (s)                  \\
  $\Vis$                & viscosity (Pa.s)                                                  \\
  $\aVis$               & apparent viscosity (Pa.s)                                         \\
  $\eVis$               & effective viscosity (Pa.s)                                        \\
  $\iVis$               & initial-time viscosity (Pa.s)                                     \\
  $\inVis$              & infinite-time viscosity (Pa.s)                                    \\
  $\lVis$               & zero-shear viscosity (Pa.s)                                        \\
  $\sVis$               & shear viscosity (Pa.s)                                            \\
  $\exVis$              & extensional (elongational) viscosity (Pa.s)                       \\
  $\hVis$               & infinite-shear viscosity (Pa.s)                                   \\
  $\fdVis$              & viscosity deficit associated with $\fgTim$ in \GODF\ model (Pa.s) \\
  $\sdVis$              & viscosity deficit associated with $\sgTim$ in \GODF\ model (Pa.s) \\
  $\rho$                & fluid mass density (kg.m$^{-3}$)                                  \\
  $\sS$                 & stress (Pa)                                                 \\
  $\sTen$               & stress tensor                                                     \\
  $\hsS$                & stress when $\Vis = \lVis / 2$ in \ELLIS\ model (Pa)                         \\
  $\ysS$                & \yields\ (Pa)                                                     \\
  $\wsS$                & stress at tube wall ($=\Delta PR/2L$) (Pa)                        \\
  $\delta \sS$          & small change in stress (Pa)                                       \\
  $\phi$                & porosity (---)                                                    \\
  $\verb|       |$
%  $\flui$               & fluidity (Pa$^{-1}$.s$^{-1}$)                                     \\
%  $\fluiZ$              & fluidity  at zero shear rate (Pa$^{-1}$.s$^{-1}$)                 \\
%  $\fluiH$              & fluidity  at high shear rate (Pa$^{-1}$.s$^{-1}$)                 \\
%
 \\
%
% XXXXXXXXXXXXXXXXXXXXXXXXXXXXXXXXXXXX  Latin  XXXXXXXXXXXXXXXXXXXXXXXXXXXXXXXXXXXX
%
  $\ve a$               & acceleration vector                                               \\
  $\textrm{a}$          & magnitude of acceleration (m.s$^{-2}$)                            \\
  $\WPo$                & exponent in truncated \Weibull\ distribution (---)                            \\
  $\WPt$                & exponent in truncated \Weibull\ distribution (---)                            \\
  $c$                   & dimensionless constant in \SEM\ (---)  \\
  $C$                   & consistency factor in \HB\ model (Pa.s$^{n}$)                                   \\
  $C^{'}$               & tortuosity factor (---)                                      \\
  $C^{''}$              & packed bed parameter (---)                                   \\
  $De$                  & \Deborah\ number (---)                                         \\
  $D_{p}$               & particle diameter (m)                                             \\
  $E$                   & elastic modulus (Pa)                                              \\
  ${\ve {e}}_{z}$       & unit vector in $z$-direction                                      \\
  $\fe$                 & scale factor for the entry of corrugated tube (---)               \\
  $\fm$                 & scale factor for the middle of corrugated tube (---)               \\
  $\ve {F}$             & force vector                                                      \\
  $G$                   & geometric conductance (m$^4$)                                     \\
  $G'$                  & flow conductance (m$^3$.Pa$^{-1}$.s$^{-1}$)                       \\
  $Go$                  & elastic modulus (Pa)                                               \\
  $\kF$                 & parameter in \FRED\ model (Pa$^{-1}$)                             \\
  $K$                   & absolute permeability (m$^{2}$)                                   \\
  $L$                   & tube length (m)                                                   \\
  $l_{c}$               & characteristic length of the flow system (m)                      \\
  $m$                   & mass (kg)                                                         \\
  $n$                   & flow behavior index (---)                                    \\
  $\overline{n}$        & average \powlaw\ behavior index inside porous medium (---)  \\
  $\fNSD$               & first normal stress difference (Pa)                               \\
  $\sNSD$               & second normal stress difference (Pa)                              \\
  $P$                   & pressure (Pa)                                                     \\
  $P_{y}$               & yield pressure (Pa)                                               \\
  $\Delta P$            & pressure drop (Pa)                                                \\
  $\Delta P_{th}$       & threshold pressure drop (Pa)                                      \\
  $\nabla P$            & pressure gradient (Pa.m$^{-1}$)                                   \\
  $\nabla P_{th}$       & threshold pressure gradient (Pa.m$^{-1}$)                         \\
  $q$                   & Darcy velocity (m.s$^{-1}$)                                       \\
  $Q$                   & volumetric flow rate (m$^{3}$.s$^{-1}$)                           \\
  $r$                   & radius (m)                                                        \\
  $R$                   & tube radius (m)                                                   \\
  $Re$                  & \Rey\ number (---)                                                \\
  $R_{eq}$              & equivalent radius (m)                                             \\
  $\Rmax$               & maximum radius of corrugated capillary                            \\
  $\Rmin$               & minimum radius of corrugated capillary                            \\
  $\mathrm{d}r$         & infinitesimal change in radius (m)                                \\
  $\delta r$            & small change in radius (m)                                        \\
  $t$                   & time (s)                                                          \\
  $t_{c}$               & characteristic time of flow system (s)                            \\
  T                     & temperature (K, $^{\circ}$C)                                      \\
  $\Tr$                  & \TR\ ratio (---)                                                 \\
  $u_{z}$               & flow speed in $z$-direction (m.s$^{-1}$)                          \\
  $\fVel$               & fluid velocity vector                                             \\
  $\WV$                 & variable in truncated \Weibull\ distribution                      \\
  $v_{c}$               & characteristic speed of flow (m.s$^{-1}$)                         \\
  $We$                  & \Weissenberg\ number (---)                                                 \\
  $\Rand$               & random number between 0 and 1 (---)                               \\
  $\verb|       |$ &                                                                        \\
\end{supertabular}

%%%%%%%%%%%%%%%%%%%%%%%%%%%%%%%%%%%%%%%%%%%%%%%%%%%%%%%%%%%%%%
\vspace{0.2cm}

{\bf \em \noindent Abbreviations and Notations}: %
\vspace{-0.3cm} \noindent

\begin{supertabular}{ll}
  $\verb|       |$ &                                                                 \\
  (.)$_{a}$              & apparent                                                  \\
  AMG                    & Algebraic Multi-Grid                                      \\
  ATP                    & Actual Threshold Pressure                                 \\
%  Ave.                   & Average                                                   \\
  (.)$_{e}$              & effective                                                 \\
  (.)$_{eq}$             & equivalent                                                \\
%  Eq.                    & Equation                                                  \\
%  Exp                    & Experimental                                              \\
(.)$_{exp}$             & experimental                                              \\
  {\em iff}              & if and only if                                            \\
  IPM                    & \InvPM\                          \\
  k                    & kilo                          \\
%  Max.                   & Maximum                                                   \\
 (.)$_{max}$             & maximum                                                   \\
%  Min.                   & Minimum                                                   \\
 (.)$_{min}$             & minimum                                                   \\
  mm                     & millimeter                                                \\
  MTP                    & Minimum Threshold Path                                    \\
%  Net                    & Network                                                   \\
(.)$_{net}$             & network                                              \\
  No.                    & Number                                                    \\
  PMP                    & \PathMP\                                  \\
%  St. Dev.               & Standard Deviation                                        \\
  (.)$_{th}$             & threshold                                                 \\
  $x_{_{l}}$             & network lower boundary in the \nNEW\ code                 \\
  $x_{_{u}}$             & network upper boundary in the \nNEW\ code                 \\
  $\mu$m                 & micrometer                                                \\
%  {\te \cdot}            & tensor    (notice bf)                                     \\
  {$\ucd \cdot$}         & upper convected time derivative                           \\
  $(\cdot)^{T}$          & matrix transpose                                          \\
  vs.          & versus                                          \\
  $|\verb| ||$           & modulus                                                   \\
  $\verb|       |$       &                                                           \\
\end{supertabular}

\vspace{0.5cm}

\noindent %
{\bf Note}: units, when relevant, are given in the SI system.
Vectors and tensors are marked with boldface. Some symbols may rely
on the context for unambiguous identification.

%

%\end{document} %XXXXXXXXXXXXXXXXXXXXXXXXXXXXXXXXXXXXXXXXXXXXXXXXXXXXXXXXXXXXXXX

{\setlength{\parskip}{6pt plus 2pt minus 1pt}

%%%%%%%%%%%%%%%%%%%%%%%%%%%%%%%%%%%  Head style  %%%%%%%%%%%%%%%%%%%%%%%%%%%%%%%%%%%
\pagestyle{headings} %
\addtolength{\headheight}{+1.6pt}
\lhead[{Chapter \thechapter \thepage}]%
      {{\bfseries\rightmark}}
\rhead[{\bfseries\leftmark}]%
     {{\bfseries\thepage}} %tell it to put page number at rhead
\headsep = 1.0cm               % Added 07 Sep 2006
%%%%%%%%%%%%%%%%%%%%%%%%%%%%%%%%%%%%%%%%%%%%%%%%%%%%%%%%%%%%%%%%%%%%%%%%%%%%%%%%%%%%

\pagenumbering{arabic}

\def\baselinestretch{1}
\chapter{Introduction} \label{Introduction}
\def\baselinestretch{1.66}
\NEW\ fluids are defined to be those fluids exhibiting a direct
proportionality between stress $\sS$ and strain rate $\sR$ in
laminar flow, that is
\begin{equation}\label{}
    \sS = \Vis \sR
\end{equation}
where the viscosity $\Vis$ is independent of the strain rate
although it might be affected by other physical parameters, such as
temperature and pressure, for a given fluid system. A stress versus
strain rate graph will be a straight line through the origin
\cite{skellandbook, ChhabraR1999}. In more precise technical terms,
\NEW\ fluids are characterized by the assumption that the extra
stress tensor, which is the part of the total stress tensor that
represents the shear and extensional stresses caused by the flow
excluding hydrostatic pressure, is a linear isotropic function of
the components of the velocity gradient, and therefore exhibits a
linear relationship between stress and the rate of strain
\cite{owensbook2002, OsP2004}. In tensor form, which takes into
account both shear and extension flow components, this linear
relationship is expressed by
\begin{equation}\label{NewtonianTensForm}
    \sTen = \Vis \rsTen
\end{equation}
where $\sTen$ is the extra stress tensor and $\rsTen$ is the
rate-of-strain tensor which describes the rate at which neighboring
particles move with respect to each other independent of superposed
rigid rotations. \NEW\ fluids are generally featured by having
shear- and \timeind\ viscosity, zero normal stress differences in
simple shear flow and simple proportionality between the viscosities
in different types of deformation \cite{barnesbookHW1993, birdbook}.

\vspace{0.2cm}

All those fluids for which the proportionality between stress and
strain rate is not satisfied, due to nonlinearity or initial
\yields, are said to be \nNEW. Some of the most characteristic
features of \nNEW\ behavior are: strain-dependent viscosity where
the viscosity depends on the type and rate of deformation, \timedep\
viscosity where the viscosity depends on duration of deformation,
\yields\ where a certain amount of stress should be reached before
the flow starts, and stress relaxation where the resistance force on
stretching the fluid element will first rise sharply then decay with
a characteristic relaxation time.

\NNEW\ fluids are commonly divided into three broad groups, although
in reality these classifications are often by no means distinct or
sharply defined \cite{skellandbook, ChhabraR1999}:
\begin{enumerate}
    \item \Timeind\ fluids are those for which the strain rate at a given point is solely
    dependent upon the instantaneous stress at that point.

    \item \Vc\ fluids are those that show partial elastic recovery upon the removal of a
    deforming stress. Such materials possess properties of both fluids and elastic solids.

    \item \Timedep\ fluids are those for which the strain rate is a function of both the
    magnitude and the duration of stress and possibly of the time lapse between consecutive
    applications of stress. These fluids are classified as \thixotropic\ (work softening) or
    \rheopectic\ (work hardening or anti-\thixotropic) depending upon whether the stress decreases or
    increases with time at a given strain rate and constant temperature.

\end{enumerate}
Those fluids that exhibit a combination of properties from more than
one of the above groups may be described as complex fluids
\cite{Collyer1974}, though this term may be used for \nNEW\ fluids
in general.

\vspace{0.2cm}

The generic rheological behavior of the three groups of \nNEW\
fluids is graphically presented in Figures
(\ref{TimeIndependent}-\ref{TimeDependent}). Figure
(\ref{TimeIndependent}) demonstrates the six principal rheological
classes of the \timeind\ fluids in shear flow. These represent
\shThin, \shThik\ and shear-independent fluids each with and without
\yields. It should be emphasized that these rheological classes are
idealizations as the rheology of the actual fluids is usually more
complex where the fluid may behave differently under various
deformation and environmental conditions. However, these basic
rheological trends can describe the actual behavior under specific
conditions and the overall behavior consists of a combination of
stages each modeled with one of these basic classes.

\vspace{0.2cm}

Figures (\ref{VERheology1}-\ref{VERheology3}) display several
aspects of the rheology of the \vc\ fluids in bulk and \insitu. In
Figure (\ref{VERheology1}) a stress versus time graph reveals a
distinctive feature of time-dependency largely observed in \vc\
fluids. As seen, the overshoot observed on applying a sudden
deformation cycle relaxes eventually to the equilibrium steady
state. This \timedep\ behavior has an impact not only on the flow
development in time, but also on the dilatancy behavior observed in
porous media flow under \steadys\ conditions where the \convdiv\
geometry contributes to the observed increase in viscosity when the
relaxation time characterizing the fluid becomes comparable in size
to the characteristic time of the flow.

\vspace{0.2cm}

In Figure (\ref{VERheology2}) a rheogram reveals another
characteristic \vc\ feature observed in porous media flow. The
intermediate plateau may be attributed to the time-dependent nature
of the \vc\ fluid when the relaxation time of the fluid and the
characteristic time of the flow become comparable in size. The
\convdiv\ nature of the pore structure accentuates this phenomenon
where the overshoot in stress interacts with the tightening of the
throats to produce this behavior. This behavior was also attributed
to build-up and break-down due to sudden change in radius and hence
rate of strain on passing through the \convdiv\ pores
\cite{TardyA2005}. This may suggest a \thixotropic\ origin for this
feature.

\vspace{0.2cm}

In Figure (\ref{VERheology3}) a rheogram of a typical \vc\ fluid is
presented. In addition to the low-deformation \NEW\ plateau and the
\shThin\ region which are widely observed in many \timeind\ fluids
and modeled by various \timeind\ rheological models such as
\CARREAU\ and \ELLIS, there is a thickening region which is believed
to be originating from the dominance of extension over shear at high
flow rates. This behavior is mainly observed in porous media flow
and the \convdiv\ geometry is usually given as an explanation to the
shift from shear flow to extension flow at high flow rates. However,
this behavior may also be observed in bulk at high strain rates.

\vspace{0.2cm}

In Figure (\ref{TimeDependent}) the two basic classes of \timedep\
fluids are presented and compared to the \timeind\ fluid in a graph
of stress against time of deformation under constant strain rate
condition. As seen, \thixotropy\ is the equivalent in time of
\shThin, while \rheopexy\ is the equivalent in time of \shThik.

\vspace{0.2cm}

A large number of models have been proposed in the literature to
model all types of \nNEW\ fluids under various flow conditions.
However, it should be emphasized that most these models are
basically empirical in nature and arising from curve-fitting
exercises \cite{barnesbookHW1993}. In this thesis, we investigate a
few models of the \timeind, \timedep\ and \vc\ fluids.

\vspace{1cm}

%%%%%%%%%%%%%%%%%%%%%%%%% Time-independent
\begin{figure}[!h]
  \centering{}
  \includegraphics
  [scale=0.6]
  {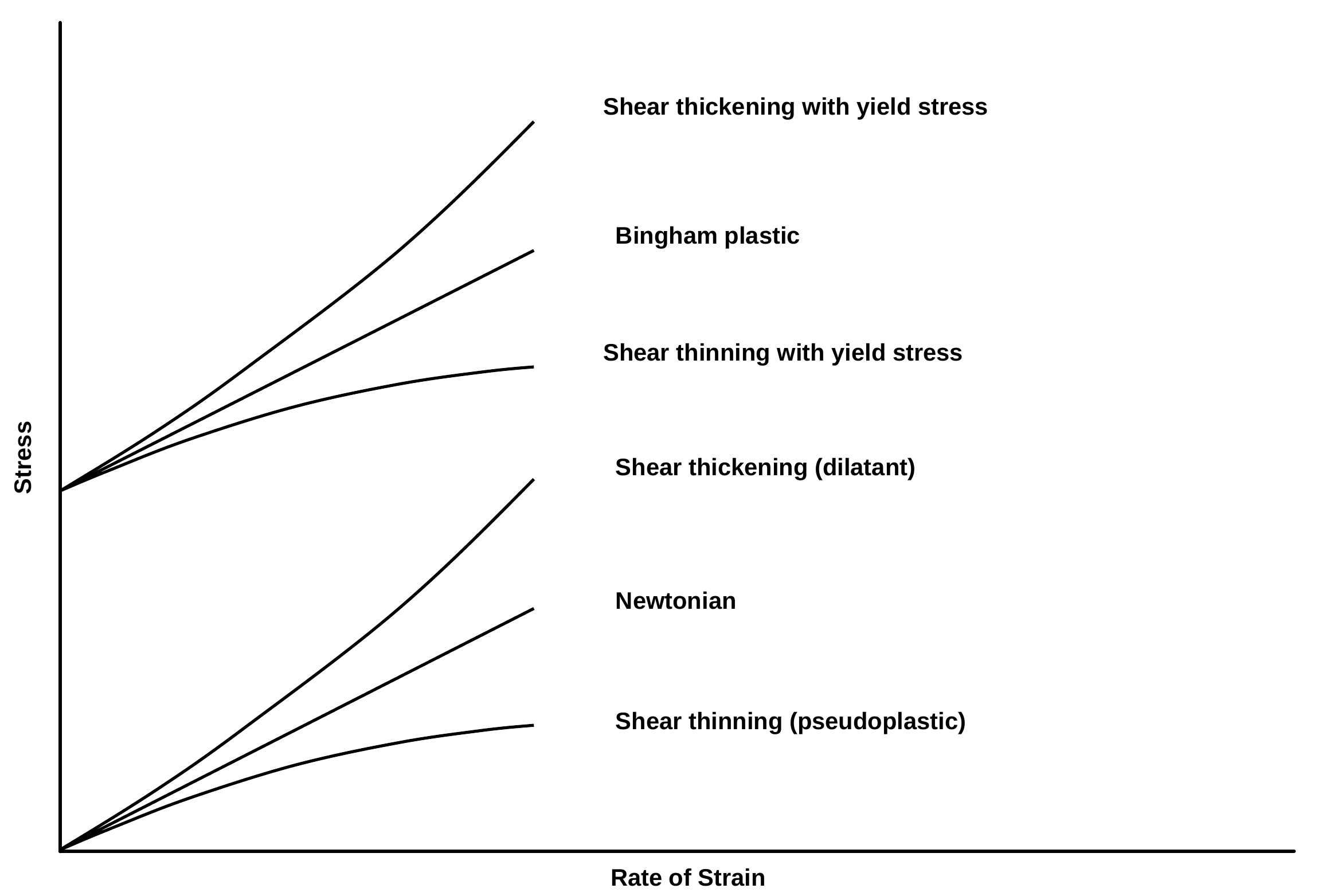}
  \caption[The six main classes of the \timeind\ fluids presented in
  a generic graph of stress against strain rate in shear flow]
  {The six main classes of the \timeind\ fluids presented in
  a generic graph of stress against strain rate in shear flow.}
  \label{TimeIndependent}
\end{figure}

%\vspace{2.0cm}

%%%%%%%%%%%%%%%%%%%%%%%%% Viscoelasticity
%%%%%%%%%%%%%%% Graph 1
\begin{figure}[!h]
  \centering{}
  \includegraphics
  [scale=0.57]
  {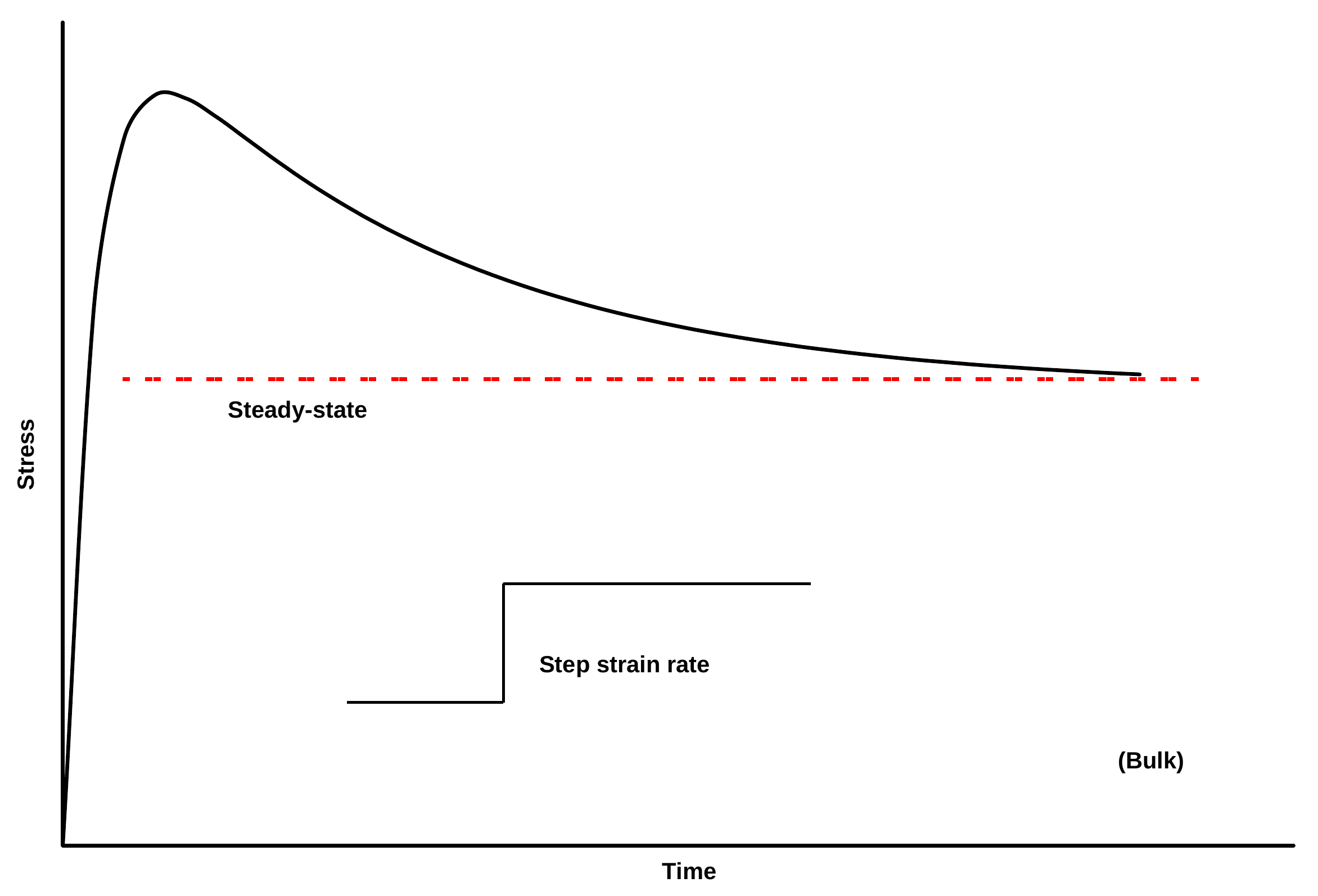}
  \caption[Typical time-dependence behavior of \vc\ fluids due to delayed response and relaxation
           following a step increase in strain rate]
  {Typical time-dependence behavior of \vc\ fluids due to delayed response and relaxation
   following a step increase in strain rate.}
  \label{VERheology1}
\end{figure}

\vspace{2.0cm}

%%%%%%%%%%%%%%% Graph 2

\begin{figure}[!h]
  \centering{}
  \includegraphics
  [scale=0.57]
  {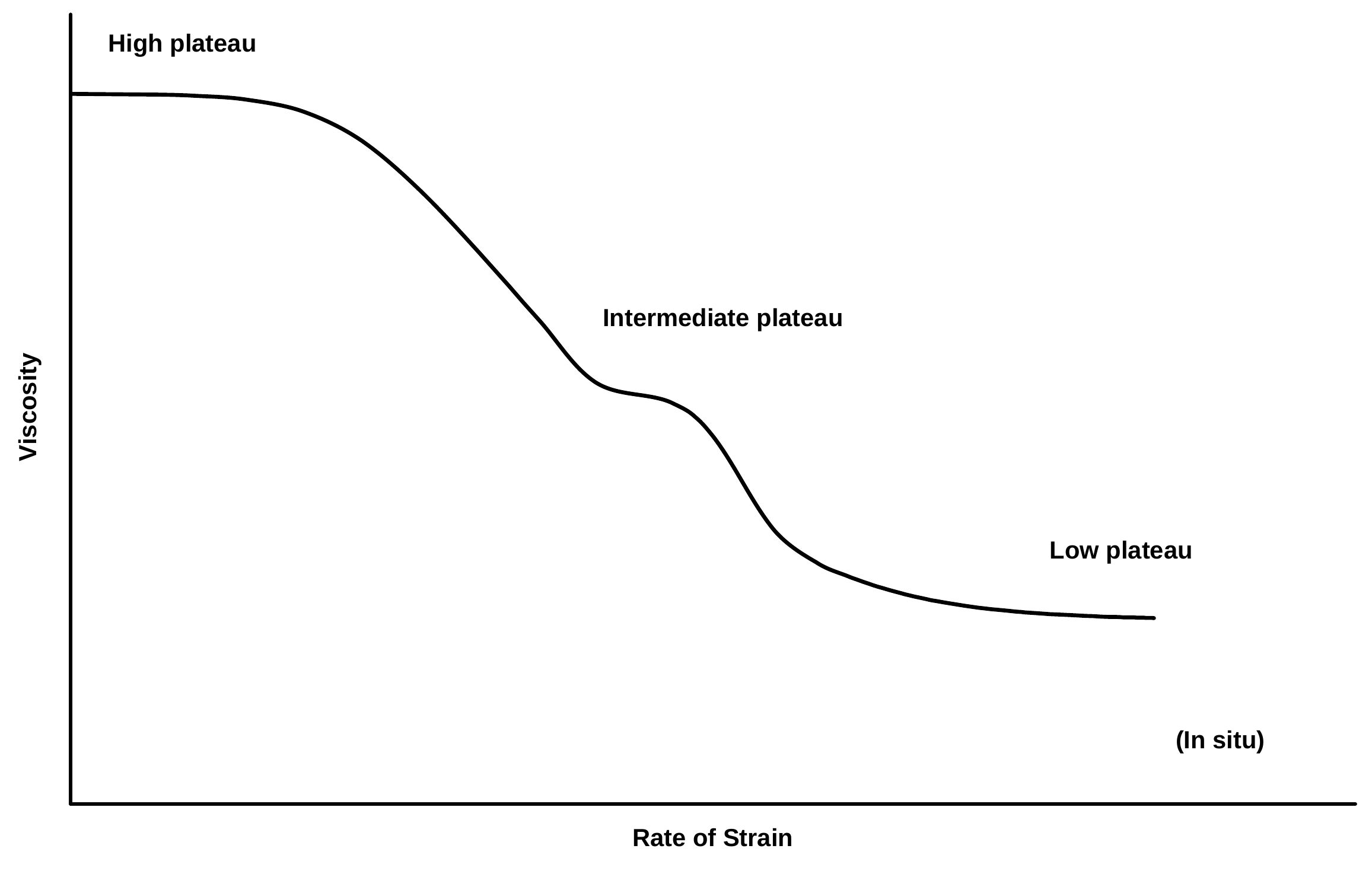}
  \caption[Intermediate plateau typical of \insitu\ \vc\ behavior due to
  \convdiv\ geometry with the characteristic time of fluid being comparable to the time of flow]
  {Intermediate plateau typical of \insitu\ \vc\ behavior due to \convdiv\
  geometry with the characteristic time of fluid being comparable to the time of flow.}
  \label{VERheology2}
\end{figure}

%%%%%%%%%%%%%%% Graph 3
\begin{figure}[!h]
  \centering{}
  \includegraphics
  [scale=0.57]
  {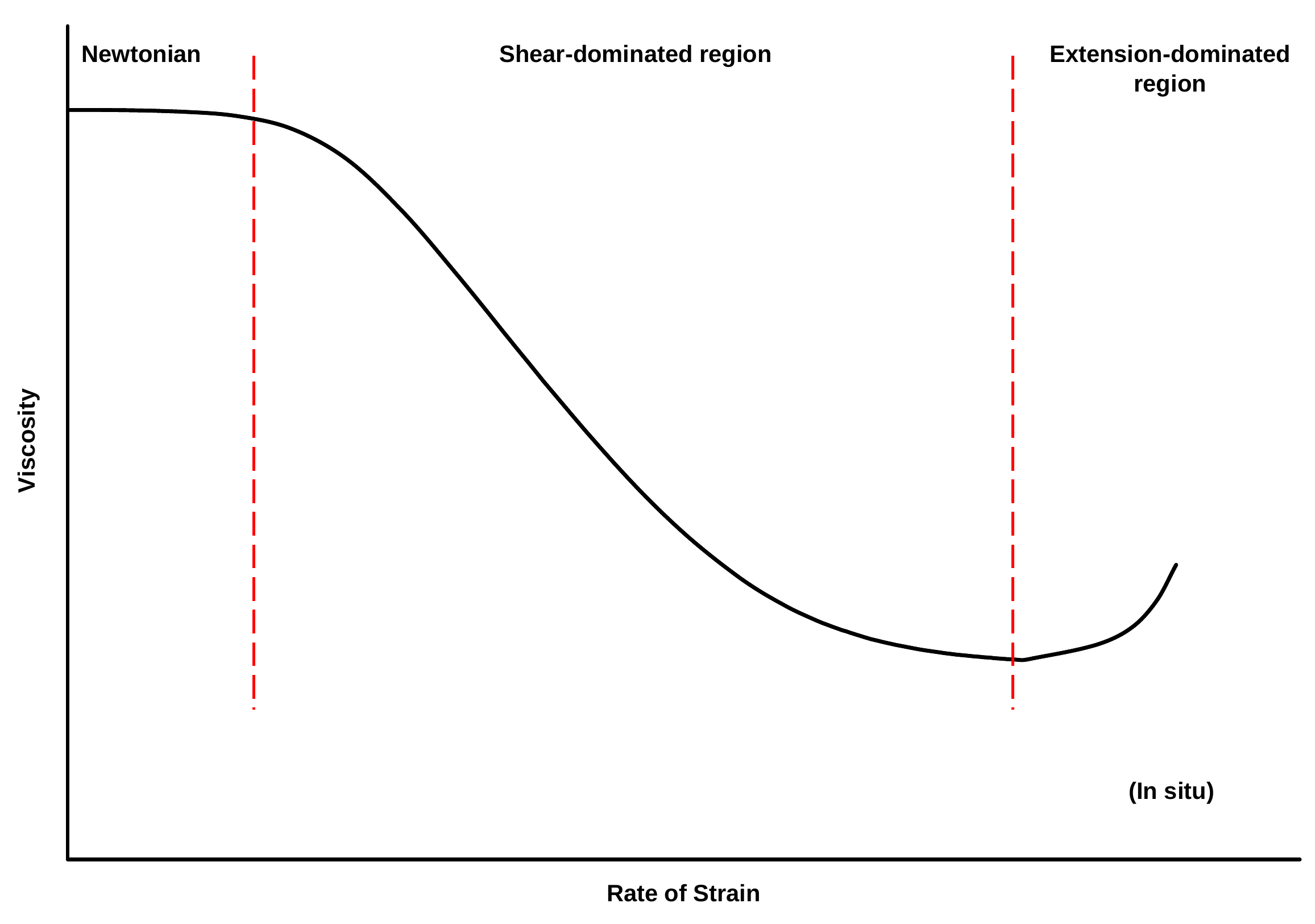}
  \caption[Strain hardening at high strain rates characteristic of \vc\ fluid mainly observed
  \insitu\ due to the dominance of extension over shear at high flow rates]
  {Strain hardening at high strain rates characteristic of \vc\ fluid mainly observed
  \insitu\ due to the dominance of extension over shear at high flow rates.}
  \label{VERheology3}
\end{figure}

\vspace{0.5cm}

%%%%%%%%%%%%%%%%%%%%%%%%% Time-dependent
\begin{figure}[!h]
  \centering{}
  \includegraphics
  [scale=0.57]
  {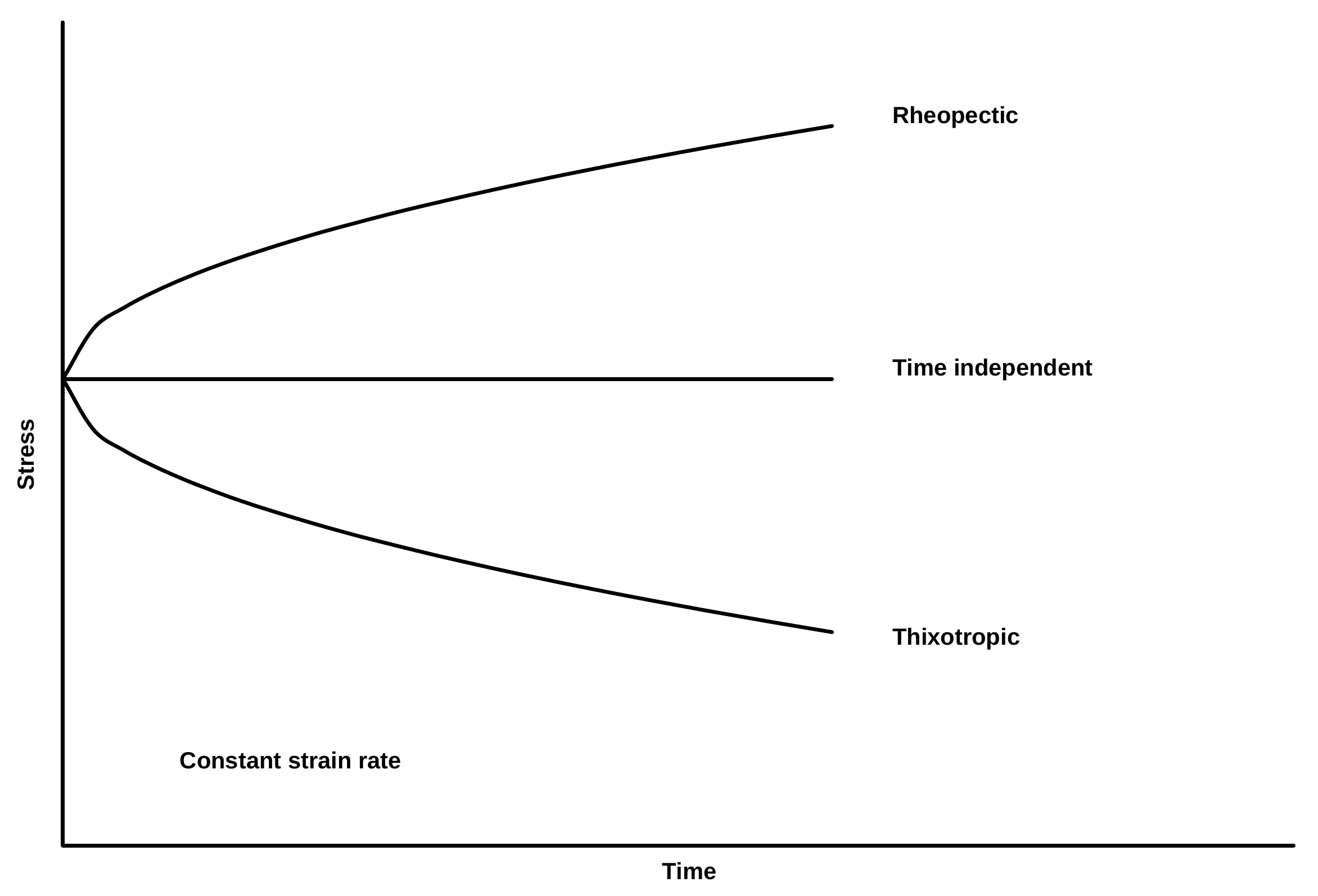}
  \caption[The two classes of \timedep\ fluids compared to the \timeind\
           presented in a generic graph of stress against time]
  {The two classes of \timedep\ fluids compared to the \timeind\
  presented in a generic graph of stress against time.}
  \label{TimeDependent}
\end{figure}

\newpage

%XXXXXXXXXXXXXXXXXXXXXXXXXXXXXXXXXXXXXXXXXXXXXXXXXXXXXXXXXXXXXXXXXXXXXX
\section{\TimeInd\ Fluids} \label{}
Shear-rate dependence is one of the most important and defining
characteristics of \nNEW\ fluids in general and \timeind\ fluids in
particular. When a typical \nNEW\ fluid experiences a shear flow the
viscosity appears to be \NEW\ at low-shear rates. After this initial
\NEW\ plateau the viscosity is found to vary with increasing
shear-rate. The fluid is described as \shThin\ or pseudoplastic if
the viscosity decreases, and \shThik\ or dilatant if the viscosity
increases on increasing shear-rate. After this shear-dependent
regime, the viscosity reaches a limiting constant value at high
shear-rate. This region is described as the upper \NEW\ plateau. If
the fluid sustains initial stress without flowing, it is called a
\yields\ fluid.

\vspace{0.2cm}

Almost all polymer solutions that exhibit a shear-rate dependent
viscosity are \shThin, with relatively few polymer solutions
demonstrating dilatant behavior. Moreover, in most known cases of
shear-thickening there is a region of \shThin\ at lower shear rates
\cite{birdbook, owensbook2002, barnesbookHW1993}.

\vspace{0.2cm}

In this thesis, two fluid models of the \timeind\ group are
investigated: \ELLIS\ and \HB.

%SSSSSSSSSSSSSSSSSSSSSSSSSSSSSSSSSSSSS
\subsection{\ELLIS\ Model} \label{}
This is a three-parameter model which describes \timeind\ \shThin\
yield-free \nNEW\ fluids. It is used as a substitute for the
\powlaw\ and is appreciably better than the \powlaw\ model in
matching experimental measurements. Its distinctive feature is the
low-shear \NEW\ plateau without a high-shear plateau. According to
this model, the fluid viscosity $\Vis$ is given by \cite
{SadowskiB1965, Savins1969, birdbook, carreaubook}
\begin{equation}
    \Vis = \frac{\lVis}{1+ \left(\frac{\sS}{\hsS} \right)^{\eAlpha - 1}}
\end{equation}
where $\lVis$ is the low-shear viscosity, $\sS$ is the shear stress,
$\hsS$ is the shear stress at which $\Vis=\lVis/2$ and $\eAlpha$ is
an indicial parameter. A generic graph demonstrating the bulk
rheology, that is viscosity versus shear rate on logarithmic scale,
is shown in Figure (\ref{BulkRheoEllis}).

\vspace{0.2cm}

For \ELLIS\ fluids, the volumetric flow rate in circular cylindrical
tube is given by \cite {SadowskiB1965, Savins1969, birdbook,
carreaubook}:
\begin{equation}\label{QEllis}
    Q = \frac{\pi R^{4} \Delta P}{8 L \lVis}
    \left[ 1 + \frac{4}{\eAlpha + 3} \left( \frac{R \Delta P}{2 L \hsS}\right)^{\eAlpha-1} \right]
\end{equation}
where $\lVis$, $\hsS$ and $\eAlpha$ are the \ELLIS\ parameters, $R$
is the tube radius, $\Delta P$ is the pressure drop across the tube
and $L$ is the tube length. The derivation of this expression is
given in Appendix \ref{AppEllisQ}.

\vspace{0.2cm}

%%%%%%%%%%%%%%%%%%%%%%%%%%%%%%%%%%% Ellis bulk rheology %%%%%%%%%%%%%%%%%%%%%%%%
\begin{figure} [!h]
  \centering{}
  \includegraphics
  [scale=0.6]
  {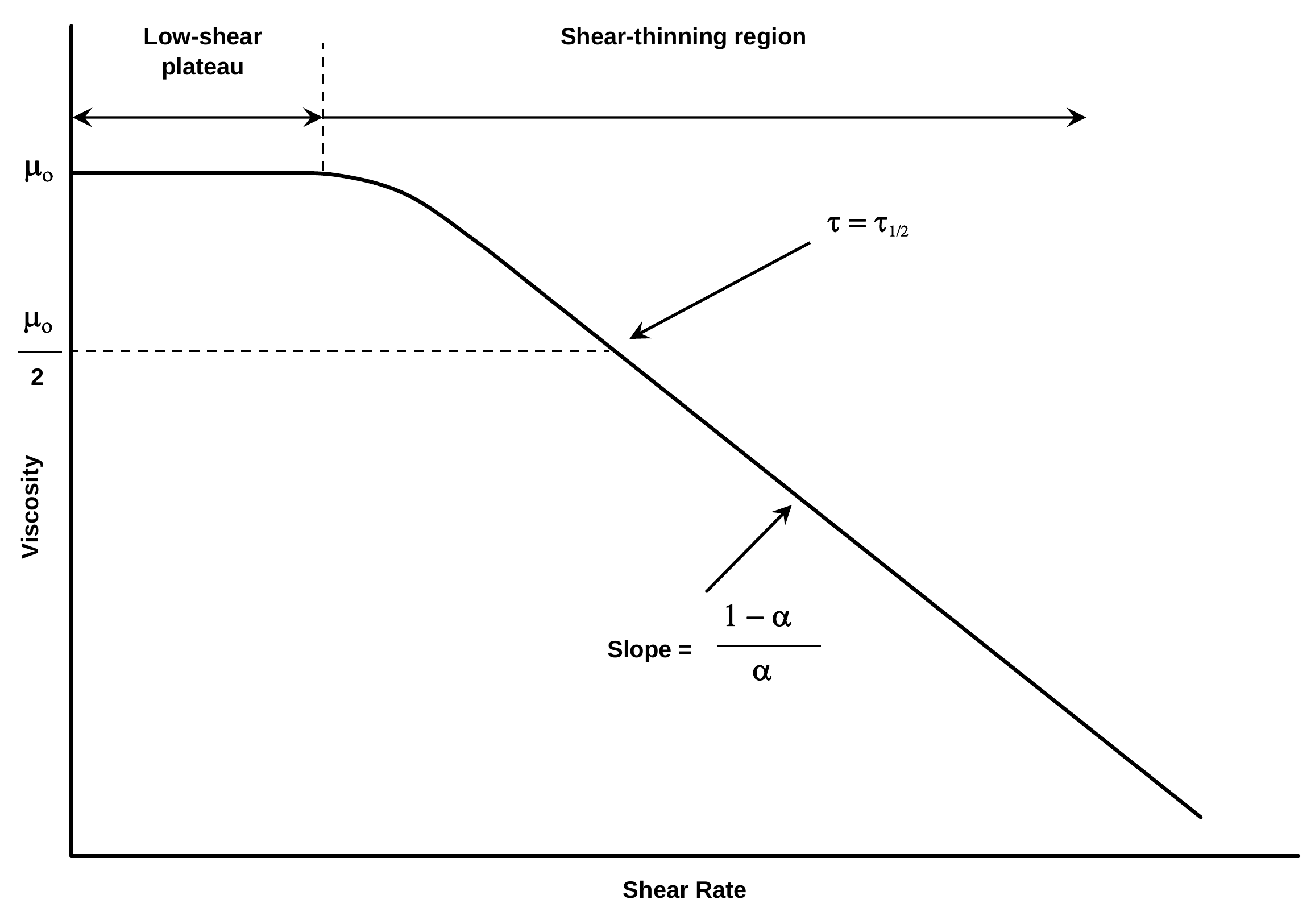}
  \caption[The bulk rheology of an \ELLIS\ fluid on logarithmic scale]
  {The bulk rheology of an \ELLIS\ fluid on logarithmic scale.}
  \label{BulkRheoEllis}
\end{figure}

%SSSSSSSSSSSSSSSSSSSSSSSSSSSSSSSSSSSSS
\subsection{\HB\ Model} \label{}
The \HB\ model has three parameters and can describe \NEW\ and a
large group of \timeind\ \nNEW\ fluids. It is given by \cite
{skellandbook}
\begin{equation}\label{}
    \sS = \ysS + C \sR^{n}
    \verb|       | (\sS > \ysS)
\end{equation}
where $\sS$ is the shear stress, $\ysS$ is the \yields\ above which
the substance starts flowing, $C$ is the consistency factor, $\sR$
is the shear rate and $n$ is the flow behavior index. The \HB\ model
reduces to the \powlaw, or \OstWae\ model, when the \yields\ is
zero, to the \BING\ plastic model when the flow behavior index is
unity, and to the \Newton's law for viscous fluids when both the
\yields\ is zero and the flow behavior index is unity
\cite{LiuM1998}.

There are six main classes to this model:
\begin{enumerate}

    \item \ShThin\ (pseudoplastic)\hspace* {1.5cm} [$\ysS = 0, n < 1.0$]

    \item \NEW\ \hspace* {4.9cm} [$\ysS = 0, n = 1.0$]

    \item \ShThik\ (dilatant) \hspace* {2.0cm} [$\ysS = 0, n > 1.0$]

    \item \ShThin\ with \yields\ \hspace* {1.15cm} [$\ysS > 0, n < 1.0$]

    \item \BING\ plastic \hspace* {3.9cm} [$\ysS > 0, n = 1.0$]

    \item \ShThik\ with \yields\ \hspace* {0.8cm} [$\ysS > 0, n > 1.0$]

\end{enumerate}
These classes are graphically illustrated in Figure
(\ref{TimeIndependent}). We would like to remark that dubbing the
sixth class as ``\shThik'' may look awkward because the viscosity of
this fluid actually decreases on yield. However, describing this
fluid as ``\shThik'' is accurate as thickening takes place on
shearing the fluid after yield with no indication to the sudden
viscosity drop on yield.

\vspace{0.2cm}

For \HB\ fluids, the volumetric flow rate in circular cylindrical
tube, assuming that the forthcoming yield condition is satisfied, is
given by \cite {skellandbook}:
\begin{eqnarray} \label{QHerschel}
    Q = \frac{8\pi}{C^\frac{1}{n}}\left(\frac{L}{\Delta P}\right)^{3}
    \left(\wsS - \ysS\right)^{1+\frac{1}{n}}\left[\frac{\left(\wsS - \ysS\right)^{2}}{3+1/n}+
    \frac{2\ysS \left(\wsS - \ysS\right)}{2+1/n} + \frac{\ysS^{2}}{1+1/n}\right]
    \nonumber \\
    (\wsS > \ysS) \hspace{1.0cm}
\end{eqnarray}
where $\ysS$, $C$ and $n$ are the \HB\ parameters, $L$ is the tube
length, $\Delta P$ is the pressure drop across the tube and $\wsS$
is the shear stress at the tube wall ($=\Delta PR/2L$). The
derivation of this expression can be found in Appendix
\ref{AppHerQ}.
\vspace{0.2cm}

For \yields\ fluids, the threshold pressure drop above which the
flow in a single tube starts is given by
\begin{equation}\label{yieldCondition}
    \Delta P_{th} = {\frac{2 L \ysS}{R}}
\end{equation}
where $\Delta P_{th}$ is the threshold pressure, $\ysS$ is the
\yields\ and $R$ and $L$ are the tube radius and length
respectively. The derivation of this expression is presented in
Appendix \ref{AppTubeY}.

%XXXXXXXXXXXXXXXXXXXXXXXXXXXXXXXXXXXXXXXXXXXXXXXXXXXXXXXXXXXXXXXXXXXXXX
\section{\Vc\ Fluids} \label{}
Polymeric fluids often show strong \vc\ effects, which can include
\shThin, extension thickening, \vc\ normal stresses, and \timedep\
rheology phenomena. The equations describing the flow of \vc\ fluids
consist of the basic laws of continuum mechanics and the rheological
equation of state, or constitutive equation, describing a particular
fluid and relates the \vc\ stress to the deformation history. The
quest is to derive a model that is as simple as possible, involving
the minimum number of variables and parameters, and yet having the
capability to predict the \vc\ behavior in complex flows
\cite{larsonbook1999}.

\vspace{0.2cm}

No theory is yet available that can adequately describe all of the
observed \vc\ phenomena in a variety of flows. However, many
differential and integral \vc\ constitutive models have been
proposed in the literature. What is common to all these is the
presence of at least one characteristic time parameter to account
for the fluid memory, that is the stress at the present time depends
upon the strain or rate-of-strain for all past times, but with an
exponentially fading memory \cite{Hulsen1996-2, Keunings2004,
owensbook2002, Denn1990, deiberthesis}.

\vspace{0.2cm}

Broadly speaking, \vy\ is divided into two major fields: linear and
nonlinear.

%SSSSSSSSSSSSSSSSSSSSSSSSSSSSSSSSSSSSS
\subsection{Linear \Vy} \label{}
Linear \vy\ is the field of rheology devoted to the study of \vc\
materials under very small strain or deformation where the
displacement gradients are very small and the flow regime can be
described by a linear relationship between stress and rate of
strain. In principle, the strain has to be small enough so that the
structure of the material remains unperturbed by the flow history.
If the strain rate is small enough, deviation from linear \vy\ may
not occur at all. The equations of linear \vy\ cannot be valid for
deformations of arbitrary magnitude and rate because the equations
violate the principle of frame invariance. The validity of the
linear \vy\ when the small-deformation condition is satisfied with a
large magnitude of the rate of strain is still an open question,
though it is generally accepted that the linear \vc\ constitutive
equations are valid in general for any strain rate as long as the
total strain remains small. However, the higher the strain rate the
shorter the time at which the critical strain for departure from
linear regime is reached \cite{birdbook, carreaubook,
larsonbook1999}.

\vspace{0.2cm}

The linear \vc\ models have several limitations. For example, they
cannot describe strain rate dependence of viscosity in general, and
are unable to describe normal stress phenomena since they are
nonlinear effects. Due to the restriction to infinitesimal
deformations, the linear models may be more appropriate to the
description of \vc\ solids rather than \vc\ fluids
\cite{carreaubook, birdbook, larsonbook1988, kronjagerthesis}.

\vspace{0.2cm}

Despite the limitations of the linear \vc\ models and despite the
fact that they are not of primary interest to the study of flow
where the material is usually subject to large deformation, they are
very important in the study of \vy\ for several reasons
\cite{birdbook, carreaubook, kronjagerthesis}:
\begin{enumerate}

    \item They are used to characterize the behavior of \vc\ materials
    at small deformations.

    \item They serve as a motivation and starting point for developing nonlinear models since the
    latter are generally extensions of the linear.

    \item They are used for analyzing experimental data obtained in small-deformation
    experiments and for interpreting important \vc\ phenomena, at least
    qualitatively.

\end{enumerate}
Here, we present two of the most widely used linear \vc\ models in
differential form.

%SSSSSSSSSSSSSSSSSS
\subsubsection{The \Maxwell\ Model} \label{}
This is the first known attempt to obtain a \vc\ constitutive
equation. This simple model, with only one elastic parameter,
combines the ideas of viscosity of fluids and elasticity of solids
to arrive at an equation for \vc\ materials \cite{birdbook,
MenaML1987}. \Maxwell\ \cite{maxwell1} proposed that fluids with
both viscosity and elasticity could be described, in modern
notation, by the relation:
\begin{equation}\label{maxwell}
    {\sTen} + \rxTim \frac{\partial {\sTen}}{\partial t} =
    \lVis {\rsTen}
\end{equation}
where $\sTen$ is the extra stress tensor, $\rxTim$ is the fluid
relaxation time, $t$ is time, $\lVis$ is the low-shear viscosity and
$\rsTen$ is the rate-of-strain tensor.

%SSSSSSSSSSSSSSSSSS
\subsubsection{The \JEF\ Model} \label{}
This is an extension to the \Maxwell\ model by including a time
derivative of the strain rate, that is \cite{birdbook,
Jeffreysbook}:
\begin{equation} \label{Jeffreys}
    {\sTen} + \rxTim \frac{\partial {\sTen}}{\partial t} =
    \lVis \left( {\rsTen} + \rdTim \frac{\partial {\rsTen}}{\partial t} \right)
\end{equation}
where $\rdTim$ is the retardation time that accounts for the
corrections of this model.

\vspace{0.2cm}

The \JEF\ model has three constants: a viscous parameter $\lVis$,
and two elastic parameters,  $\rxTim$ and $\rdTim$. The model
reduces to the linear \Maxwell\ when $\rdTim = 0$, and to the \NEW\
when $\rxTim = \rdTim = 0$. As observed by several authors, the
\JEF\ model is one of the most suitable linear models to compare
with experiment \cite{MardonesG1990}.

%SSSSSSSSSSSSSSSSSSSSSSSSSSSSSSSSSSSSS
\subsection{Nonlinear \Vy} \label{}
This is the field of rheology devoted to the study of \vc\ materials
under large deformation, and hence it is the subject to investigate
for the purpose of studying the flow of \vc\ fluids. It should be
remarked that the nonlinear \vc\ constitutive equations are
sufficiently complex that very few flow problems can be solved
analytically. Moreover, there appears to be no differential or
integral constitutive equation general enough to explain the
observed behavior of polymeric systems undergoing large deformations
but still simple enough to provide a basis for engineering design
procedures \cite{birdbook, skellandbook, white1}.

\vspace{0.2cm}

As the equations of linear \vy\ cannot be valid for deformations of
large magnitude because they do not satisfy the principle of frame
invariance, \OLD\ and others developed a set of frame-invariant
differential constitutive equations by defining time derivatives in
frames that deform with the material elements. Examples of these
equations include rotational, upper and lower convected time
derivative models \cite{larsonbook1988}.

\vspace{0.2cm}

There is a large number of proposed constitutive equations and
rheological models for the nonlinear \vy, as a quick survey to the
literature reveals. However, many of these models are extensions or
modifications to others. The two most popular nonlinear \vc\ models
in differential form are the \UpCoMa\ and the \OldB\ models.

%SSSSSSSSSSSSSSSSSS
\subsubsection{The \UpCoMa\ (UCM) Model} \label{}
To extend the linear \Maxwell\ model to the nonlinear regime,
several time derivatives (e.g. upper convected, lower convected and
corotational) are proposed to replace the ordinary time derivative
in the original model. The idea of these derivatives is to express
the constitutive equation in real space coordinates rather than
local coordinates and hence fulfilling the \OLD's admissibility
criteria for constitutive equations. These admissibility criteria
ensures that the equations are invariant under a change of
coordinate system, value invariant under a change of translational
or rotational motion of the fluid element as it goes through space,
and value invariant under a change of rheological history of
neighboring fluid elements. The most commonly used of these
derivatives in conjunction with the \Maxwell\ model is the upper
convected. On purely continuum mechanical grounds there is no reason
to prefer one of these \Maxwell\ equations to the others as they all
satisfy frame invariance. The popularity of the upper convected is
due to its more realistic features \cite{birdbook, larsonbook1988,
carreaubook, owensbook2002, kronjagerthesis}.

\vspace{0.2cm}

The \UpCoMa\ (UCM) model is the simplest nonlinear \vc\ model which
parallels the \Maxwell\ linear model accounting for frame invariance
in the nonlinear flow regime, and is one of the most popular models
in numerical modeling and simulation of \vc\ flow. It is a simple
combination of the \Newton's law for viscous fluids and the
derivative of the \Hook's law for elastic solids, and therefore does
not fit the rich variety of \vc\ effects that can be observed in
complex rheological materials \cite{MardonesG1990}. Despite its
simplicity, it is largely used as the basis for other more
sophisticated \vc\ models. It represents, like its linear equivalent
\Maxwell, purely elastic fluids with shear-independent viscosity.
The UCM model is obtained by replacing the partial time derivative
in the differential form of the linear \Maxwell\ model with the
upper convected time derivative
\begin{equation}\label{UCM1}
    {\sTen} + \rxTim {\ucd \sTen} = \lVis {\rsTen}
\end{equation}
where $\sTen$ is the extra stress tensor, $\rxTim$ is the relaxation
time, $\lVis$ is the low-shear viscosity, $\rsTen$ is the
rate-of-strain tensor, and {$\ucd \sTen$} is the upper convected
time derivative of the stress tensor:

\begin{equation}\label{UCM2}
    \ucd \sTen =
    \frac{\partial {\sTen}}{\partial t} +
    \fVel \cdot \nabla \sTen -
    \left( \nabla \fVel \right)^{T} \cdot \sTen -
    \sTen \cdot \nabla \fVel
\end{equation}
where $t$ is time, $\fVel$ is the fluid velocity, $\left( \cdot
\right)^{T}$ is the transpose of the tensor and $\nabla \fVel$ is
the fluid velocity gradient tensor defined by Equation
(\ref{fVelGradTen}) in Appendix \ref{AppVE}. The convected
derivative expresses the rate of change as a fluid element moves and
deforms. The first two terms in Equation (\ref{UCM2}) comprise the
material or substantial derivative of the extra stress tensor. This
is the time derivative following a material element and describes
time changes taking place at a particular element of the
``material'' or ``substance''. The two other terms in (\ref{UCM2})
are the deformation terms. The presence of these terms, which
account for convection, rotation and stretching of the fluid motion,
ensures that the principle of frame invariance holds, that is the
relationship between the stress tensor and the deformation history
does not depend on the particular coordinate system used for the
description \cite{OsP2004, birdbook, carreaubook}.

\vspace{0.2cm}

The three main material functions predicted by the UCM model are
\cite{tannerbook2000}
\begin{equation}\label{UCMN1}
    \fNSD = \frac{3 \lVis \elongR} {(1 - 2 \rxTim \elongR) (1 + \rxTim \elongR)}
    \verb|   |
    \sNSD = 0
    \verb|   | \& \verb|   | \sVis = \lVis
\end{equation}
where $\fNSD$ and $\sNSD$ are the first and second normal stress
difference respectively, $\sVis$ is the shear viscosity, $\lVis$ is
the low-shear viscosity, $\rxTim$ is the relaxation time and
$\elongR$ is the rate of elongation. As seen, the viscosity is
constant and therefore the model represents \Boger\ fluids
\cite{tannerbook2000}. UCM also predicts a \NEW\ elongation
viscosity that is three times the \NEW\ shear viscosity, i.e.
$\exVis = 3 \lVis$.

\vspace{0.2cm}

Despite the simplicity of this model, it predicts important
properties of \vc\ fluids such as first normal stress difference in
shear and strain hardening in elongation. It also predicts the
existence of stress relaxation after cessation of flow and elastic
recoil. However, it predicts that both the shear viscosity and the
first normal stress difference are independent of shear rate and
hence fails to describe the behavior of most complex fluids.
Furthermore, it predicts that the \steadys\ elongational viscosity
is infinite at a finite elongation rate, which is far from physical
reality \cite{larsonbook1988}.

%SSSSSSSSSSSSSSSSSS
\subsubsection{The \OldB\ Model} \label{OldroydB}
The \OldB\ model is a simplification of the more elaborate and
rarely used \OLD\ 8-constant model which also contains the upper
convected, the lower convected, and the corotational \Maxwell\
equations as special cases. \OldB\ is the second simplest nonlinear
\vc\ model and is apparently the most popular in \vc\ flow modeling
and simulation. It is the nonlinear equivalent of the linear \JEF\
model, and hence it takes account of frame invariance in the
nonlinear regime, as presented in the last section. Consequently, in
the linear \vc\ regime the \OldB\ model reduces to the linear \JEF\
model. The \OldB\ model can be obtained by replacing the partial
time derivatives in the differential form of the \JEF\ model with
the upper convected time derivatives \cite{birdbook}
\begin{equation}\label{OBM1}
    {\sTen} + \rxTim {\ucd \sTen} =
    \lVis \left( {\rsTen} + \rdTim {\ucd \rsTen} \right)
\end{equation}
where $\rdTim$ is the retardation time, which may be seen as a
measure of the time the material needs to respond to deformation,
and {$\ucd \rsTen$} is the upper convected time derivative of the
rate-of-strain tensor:
\begin{equation}\label{OBM2}
    \ucd \rsTen =
    \frac{\partial {\rsTen}}{\partial t} +
    \fVel \cdot \nabla \rsTen -
    \left( \nabla \fVel \right)^{T} \cdot \rsTen -
    \rsTen \cdot \nabla \fVel
\end{equation}

The \OLD\ model reduces to the UCM model when $\rdTim = 0$, and to
\NEW\ when $\rxTim = \rdTim = 0$.
\vspace{0.2cm}

Despite the simplicity of the \OldB\ model, it shows good
qualitative agreement with experiments especially for dilute
solutions of macromolecules and \Boger\ fluids. The model is able to
describe two of the main features of \vy, namely normal stress
differences and stress relaxation. It predicts a constant viscosity
and first normal stress difference, with a zero second normal stress
difference. Like UCM, the \OldB\ model predicts a \NEW\ elongation
viscosity that is three times the \NEW\ shear viscosity, i.e.
$\exVis = 3 \lVis$. An important weakness of this model is that it
predicts an infinite extensional viscosity at a finite extensional
rate \cite{larsonbook1999, owensbook2002, MardonesG1990,
kronjagerthesis, birdbook, BalmforthC2001}.

\vspace{0.2cm}

A major limitation on the UCM and \OldB\ models is that they do not
allow for strain dependency and second normal stress difference. To
account for strain dependent viscosity and non-zero second normal
stress difference in the \vc\ fluids behavior, other more
sophisticated models such as Giesekus and Phan-Thien-Tanner (PTT)
which introduce additional parameters should be considered. However,
such equations have rarely been used because of the theoretical and
experimental complications they introduce \cite{Boger1987}.

%XXXXXXXXXXXXXXXXXXXXXXXXXXXXXXXXXXXXXXXXXXXXXXXXXXXXXXXXXXXXXXXXXXXXXX
\section{\TimeDep\ Fluids} \label{}
It is generally recognized that there are two main types of
\timedep\ fluids: \thixotropic\ (work softening) and \rheopectic\
(work hardening). There is also a general consensus that the
time-dependent feature is caused by reversible structural change
during the flow process. However, there are many controversies about
the details, and the theory of the time-dependent fluids are not
well developed.

\vspace{0.2cm}

Many models have been proposed in the literature for the \timedep\
rheological behavior. Here we present two models of this category.

%SSSSSSSSSSSSSSSSSSSSSSSSSSSSSSSSSSSSS
\subsection{\GODF\ Model} \label{Godfrey}
\GODF\ \cite{Godfrey1973} suggested that at a particular shear rate
the time dependence for \thixotropic\ fluids can be described by the
relation
\begin{equation}\label{godfrey}
    \Vis(t) = \iVis - \fdVis ( 1 - e^{-t/\fgTim} )
                    - \sdVis ( 1 - e^{-t/\sgTim} )
\end{equation}
where $\Vis(t)$ is the \timedep\ viscosity, $\iVis$ is the viscosity
at the commencement of deformation, $\fdVis$ and $\sdVis$ are the
viscosity deficits associated with the decay time constants $\fgTim$
and $\sgTim$ respectively, and $t$ is the time of shearing. The
initial viscosity specifies a maximum value while the viscosity
deficits specify the reduction associated with particular time
constants. In the usual way the time constants define the time
scales of the processes under examination.

\vspace{0.2cm}

Although \GODF\ model is proposed for \thixotropic\ fluids, it can
be easily generalized to include \rheopectic\ behavior.

%SSSSSSSSSSSSSSSSSSSSSSSSSSSSSSSSSSSSS
\subsection{\SEM} \label{}
This is a general model for the \timedep\ fluids \cite{Barnes1997}
\begin{equation}\label{SEM}
    \Vis(t) = \iVis + ( \inVis - \iVis ) ( 1 - e^{-(t/\seTim)^{c}} )
\end{equation}
where $\Vis(t)$ is the \timedep\ viscosity, $\iVis$ is the viscosity
at the commencement of deformation, $\inVis$ is the equilibrium
viscosity at infinite time, $t$ is the time of deformation, $\seTim$
is a time constant and $c$ is a dimensionless constant which in the
simplest case is unity.

\def\baselinestretch{1}
\chapter{Literature Review} \label{Literature}
\def\baselinestretch{1.66}
The study of the flow of \nNEW\ fluids in porous media is of immense
importance and serves a wide variety of practical applications in
processes such as enhanced oil recovery from underground reservoirs,
filtration of polymer solutions and soil remediation through the
removal of liquid pollutants. It will therefore come as no surprise
that a huge quantity of literature on all aspects of this subject do
exist. In this Chapter we present a short literature review focusing
on those studies which are closely related to our investigation.

%XXXXXXXXXXXXXXXXXXXXXXXXXXXXXXXXXXXXXXXXXXXXXXXXXXXXXXXXXXXXXXXXXXXXXX
\section{\TimeInd\ Fluids} \label{LTimeI}
Here we present two \timeind\ models which we investigated and
implemented in our \nNEW\ computer code.

%XXXXXXXXXXXXXXXXXXXXXXXXXXXXXXXXXXXXXXXXXXXXXXXXXXXXXXXXXXXXXXXXXXXXXX
\subsection{\ELLIS\ Fluids} \label{ELLIS}
Sadowski and Bird \cite{SadowskiB1965, Sadowski1965, sadowskithesis}
applied the \ELLIS\ model to a \nNEW\ fluid flowing through a porous
medium modeled by a bundle of capillaries of varying cross-section
but of a definite length. This led to a generalized form of \Darcy's
law and of \Ergun's friction factor correlation in each of which the
\NEW\ viscosity was replaced by an effective viscosity. The
theoretical investigation was backed by extensive experimental work
on the flow of aqueous polymeric solutions through packed beds. They
recommended the \ELLIS\ model for the description of the \steadys\
\nNEW\ behavior of dilute polymer solutions, and introduced a
relaxation time term as a correction to account for \vc\ effects in
the case of polymer solutions of high molecular weight. The unsteady
and irreversible flow behavior observed in the constant pressure
runs was explained by polymer adsorption and gel formation that
occurred throughout the bed. Finally, they suggested a general
procedure to determine the three parameters of this model.

\vspace{0.2cm}

Park \etal\ \cite{parkthesis, ParkHB1973} used an \ELLIS\ model as
an alternative to a \powlaw\ form in their investigation to the flow
of various aqueous polymeric solutions in packed beds of glass
beads. They experimentally investigated the flow of aqueous
polyacrylamide solutions which they modeled by an \ELLIS\ fluid and
noticed that neither the \powlaw\ nor the \ELLIS\ model will predict
the proper shapes of the apparent viscosity versus shear rate. They
therefore concluded that neither model would be very useful for
predicting the effective viscosities for calculations of friction
factors in packed beds.

\vspace{0.2cm}

Balhoff and Thompson \cite{balhoffthesis, balhoff2} carried out a
limited amount of experimental work on the flow of guar gum
solution, which they modeled as an \ELLIS\ fluid, in packed beds of
glass beads. Their network simulation results matched the
experimental data within an adjustable constant.

%XXXXXXXXXXXXXXXXXXXXXXXXXXXXXXXXXXXXXXXXXXXXXXXXXXXXXXXXXXXXXXXXXXXXXX
\subsection{\HB\ and \YieldS\ Fluids} \label{HB}
The flow of \HB\ and \yields\ fluids in porous media has been
examined by several investigators. Park \etal\ \cite{parkthesis,
ParkHB1973} used the \Ergun\ equation
\begin{equation}
    \frac{\Delta P}{L}
    = \frac{150 \Vis q}{D_{p}^{2}}   \frac{(1 - \epsilon)^{2}}{\epsilon^{3}}
             + \frac{1.75 \rho q^{2}}{D_{p}}   \frac{(1 - \epsilon)}{\epsilon^{3}}
\end{equation}
to correlate pressure drop-flow rate for a \HB\ fluid flowing
through packed beds by using the effective viscosity calculated from
the \HB\ model. They validated their model by experimental work on
the flow of Polymethylcellulose (PMC) in packed beds of glass beads.

\vspace{0.2cm}

To describe the non-steady flow of a \yields\ fluid in porous media,
Pascal \cite{pascal1} modified \Darcy's law by introducing a
threshold pressure gradient to account for the \yields. This
threshold gradient is directly proportional to the \yields\ and
inversely proportional to the square root of the absolute
permeability. However, the constant of proportionality must be
determined experimentally.

\vspace{0.2cm}

Al-Fariss and Pinder \cite{alfariss1, alfariss2} produced a general
form of \Darcy's law by modifying the \Blake-\Kozeny\ equation
\begin{equation}
    q = \frac{\Delta P}{L \Vis}  \frac{D_{p}^{2} \epsilon^{3}}{ 72 C^{'} (1 - \epsilon)^{2}}
\end{equation}
to describe the flow in porous media of \HB\ fluids. They ended with
very similar equations to those obtained by Pascal. They also
extended their work to include experimental investigation on the
flow of waxy oils through packed beds of sand.

\vspace{0.2cm}

Wu \etal\ \cite{wu1} applied an integral analytical method to obtain
an approximate analytical solution for single-phase flow of \BING\
fluids through porous media. They also developed a Buckley-Leverett
analytical solution for one-dimensional flow in porous media to
study the displacement of a \BING\ fluid by a \NEW\ fluid.

\vspace{0.2cm}

Chaplain \etal\ \cite{chaplain1} modeled the flow of a \BING\ fluid
through porous media by generalizing Saffman \cite{saffman1}
analysis for the \NEW\ flow to describe the dispersion in a porous
medium by a random walk. The porous medium was assumed to be
statistically homogeneous and isotropic so dispersion can be defined
by lateral and longitudinal coefficients. They demonstrated that the
pore size distribution of a porous medium can be obtained from the
characteristics of the flow of a \BING\ fluid.

\vspace{0.2cm}

Vradis and Protopapas \cite{vradis1} extended the ``capillary tube''
and the ``resistance to flow'' models to describe the flow of \BING\
fluids in porous media and presented a solution in which the flow is
zero below a threshold head gradient and Darcian above it. They
analytically demonstrated that in both models the minimum head
gradient required for the initiation of flow in the porous medium is
proportional to the \yields\ and inversely proportional to the
characteristic length scale of the porous medium, i.e. the capillary
tube diameter in the first model and the grain diameter in the
second model.

\vspace{0.2cm}

Chase and Dachavijit \cite{chase1} modified the \Ergun\ equation to
describe the flow of \yields\ fluids through porous media. They
applied the bundle of capillary tubes approach similar to that of
Al-Fariss and Pinder. Their work includes experimental validation on
the flow of \BING\ aqueous solutions of Carbopol 941 through packed
beds of glass beads.

\vspace{0.2cm}

Kuzhir \etal\ \cite{kuzhir1} presented a theoretical and
experimental investigation for the flow of a magneto-rheological
(MR) fluid through different types of porous medium. They showed
that the mean \yields\ of a \BING\ MR fluid, as well as the pressure
drop, depends on the mutual orientation of the external magnetic
field and the main axis of the flow.

\vspace{0.2cm}

Recently, Balhoff and Thompson \cite{balhoffthesis, balhoff1} used
their three-dimensional network model which is based on a
computer-generated random sphere packing to investigate the flow of
\BING\ fluids in packed beds. To model \nNEW\ flow in the throats,
they used analytical expressions for a capillary tube but
empirically adjusted key parameters to more accurately represent the
throat geometry and simulate the fluid dynamics in the real throats
of the packing. The adjustments were made specifically for each
individual fluid type using numerical techniques.

%XXXXXXXXXXXXXXXXXXXXXXXXXXXXXXXXXXXXXXXXXXXXXXXXXXXXXXXXXXXXXXXXXXXXXX
\section{\Vc\ Fluids} \label{VE}
Sadowski and Bird \cite{SadowskiB1965, Sadowski1965, sadowskithesis}
were the first to include elastic effects in their model to account
for a departure of the experimental data in porous media from the
modified \Darcy's law \cite{Savins1969, sorbiebook}. In testing
their modified friction factor-\Rey\ number correlation, they found
very good agreement with the experimental data except for the high
molecular weight Natrosol at high \Rey\ numbers. They argued that
this is a \vc\ effect and introduced a modified correlation which
used a characteristic time to designate regions of behavior where
elastic effects are important, and hence found an improved agreement
between this correlation and the data.

\vspace{0.2cm}

Investigating the flow of \powlaw\ fluids through a packed tube,
Christopher and Middleman \cite{christopher1965} remarked that the
capillary model for flow in a packed bed is deceptive, because such
a flow actually involves continual acceleration and deceleration as
fluid moves through the irregular interstices between particles.
Hence, for flow of \nNEW s in porous media it might be expected to
observe \vc\ effects which do not show up in the \steadys\ spatially
homogeneous flows usually used to establish the rheological
parameters. However, their experimental work with dilute aqueous
solutions of carboxymethylcellulose through tube packed with
spherical particles failed to detect \vc\ effects. This study was
extended later by Gaintonde and Middleman \cite{gaitonde1966} by
examining a more elastic fluid, polyisobutylene, through tubes
packed with sand and glass spheres
confirming the earlier failure.%

\vspace{0.2cm}

Marshall and Metzner \cite{MarshallM1967} investigated the flow of
\vc\ fluids through porous media and concluded that the analysis of
flow in converging channels suggests that the pressure drop should
increase to values well above those expected for purely viscous
fluids at \Deborah\ number levels of the order of 0.1 to 1.0. Their
experimental results using a porous medium support this analysis and
yield a critical value of the \Deborah\ number of about 0.05 at
which \vc\ effects were first found to be measurable.%

\vspace{0.2cm}

Wissler \cite{Wissler1971} was the first to account quantitatively
for the elongational stresses developed in \vc\ flow through porous
media \cite{TalwarK1992}. In this context, he presented a
third-order perturbation analysis of flow of a \vc\ fluid through a
\convdiv\ channel, and an analysis of the flow of a visco-inelastic
\powlaw\ fluid through the same system. The latter provides a basis
for experimental study of \vc\ effects in polymer solutions in the
sense that if the measured pressure drop exceeds the value predicted
on the basis of viscometric data alone, \vc\ effects are probably
important and the fluid can be expected to have reduced mobility in
a porous medium.

\vspace{0.2cm}

Gogarty \etal\ \cite{GogartyLF1972} developed a modified form of the
\nNEW\ \Darcy\ equation to predict the viscous and elastic pressure
drop versus flow rate relationships for flow of the elastic
Carboxymethylcellulose (CMC) solutions in beds of different
geometry, and validated their model with experimental work.
According to this model, the pressure gradient across the bed,
$\nabla P$, is related to the Darcy velocity, $q$, by
\begin{equation}\label{gogarty}
    | \nabla P | = \frac{q \aVis}{K} \left[ 1+0.243q^{(1.5-m)} \right]
\end{equation}
where $\aVis$ is the apparent viscosity, $K$ is the absolute permeability and $m$ is an
elastic correction factor.%

\vspace{0.2cm}

Park \etal\ \cite{ParkHB1973} experimentally studied the flow of
various polymeric solutions through packed beds using several fluid
models to characterize the rheological behavior. In the case of one
type of solution at high \Rey\ numbers, significant deviation of the
experimental data from the modified \Ergun\ equation was observed.
Empirical corrections based upon a pseudo \vc\ \Deborah\ number were
able to greatly improve the data fit in this case. However, they
remarked it is not clear that this is a general correction procedure
or that the deviations are in fact due to \vc\ fluid
characteristics, and hence recommended further
investigation.%

\vspace{0.2cm}

In their investigation to the flow of polymers through porous media,
Hirasaki and Pope \cite{hirasaki1974} modeled the dilatant behavior
by the \vc\ properties of the polymer solution. The additional \vc\
resistance to the flow, which is a function of \Deborah\ number, was
modeled as a simple elongational flow to represent the elongation
and contraction that occurs as the fluid
flows through a pore with varying cross sectional area.%

\vspace{0.2cm}

In their theoretical and experimental investigation, Deiber and
Schowalter \cite{DeiberS1981, deiberthesis} used a circular tube
with a radius which varies sinusoidally in the axial direction as a
first step toward modeling the flow channels of a porous medium.
They concluded that such a tube exhibits similar phenomenological
aspects to those found for the flow of \vc\ fluids through packed
beds and is a successful model for the flow of these fluids through
porous media in the qualitative sense that one predicts an increase
in pressure drop due to fluid elasticity. The numerical technique of
geometric iteration which they employed to solve the nonlinear
equations of motion demonstrated qualitative agreement with the
experimental results.

\vspace{0.2cm}

Durst \etal\ \cite{DurstHI1987} pointed out that the pressure drop
of a porous media flow is only due to a small extent to the shear
force term usually employed to derive the \Kozeny-\Darcy\ law. For a
more correct derivation, additional terms have to be taken into
account since the fluid is also exposed to elongational forces when
it passes through the porous media matrix. According to this
argument, ignoring these additional terms explains why the available
theoretical derivations of the \Kozeny-\Darcy\ relationship, which
are based on one part of the shear-caused pressure drop only,
require an adjustment of the constant in the theoretically derived
equation to be applicable to experimental results. They suggested
that the straight channel is not a suitable model flow geometry to
derive theoretically pressure loss-flow rate relationships for
porous media flows. Consequently, the derivations have to be based
on more complex flow channels showing cross-sectional variations
that result in elongational straining similar to that which the
fluid experiences when it passes through the porous medium. Their
experimental work verified some aspects of their theoretical
derivation.

\vspace{0.2cm}

Chmielewski and coworkers \cite{ChmielewskiPJ1990, ChmielewskiJ1992,
ChmielewskiJ1993} conducted experimental work and used visualization
techniques to investigate the elastic behavior of the flow of
polyisobutylene solutions through arrays of cylinders with various
geometries. They recognized that the \convdiv\ geometry of the pores
in porous media causes an extensional flow component that may be
associated with the increased flow resistance
for \vc\ liquids.%

\vspace{0.2cm}

Pilitsis \etal\ \cite{PilitsisSB1991} numerically simulated the flow
of a \shThin\ \vc\ fluid through a periodically constricted tube
using a variety of constitutive rheological models, and the results
were compared against the experimental data of Deiber and Schowalter
\cite{DeiberS1981}. It was found that the presence of the elasticity
in the mathematical modeling caused an increase in the flow
resistance over the value calculated for the viscous fluid. Another
finding is that the use of more complex constitutive equations which
can represent the dynamic or \trans\ properties of the fluid can
shift the calculated data towards the correct direction. However, in
all cases the numerical results seriously underpredicted
the experimentally measured flow resistance.%

\vspace{0.2cm}

Talwar and Khomami \cite{TalwarK1992} developed a higher order
Galerkin finite element scheme for simulation of two-dimensional
\vc\ fluid flow and successfully tested the algorithm with the
problem of flow of \UpCoMa\ (UCM) and \OldB\ fluids in undulating
tube. It was subsequently used to solve the problem of transverse
steady flow of UCM and \OldB\ fluids past an infinite square
arrangement of infinitely long, rigid, motionless cylinders. While
the experimental evidence indicates a dramatic increase in the flow
resistance above a certain \Weissenberg\ number, their numerical
results revealed a steady
decline of this quantity.%

\vspace{0.2cm}

Souvaliotis and Beris \cite{souvaliotis1996} developed a domain
decomposition spectral collocation method for the solution of
\steadys, nonlinear \vc\ flow problems. It was then applied in
simulations of \vc\ flows of UCM and \OldB\ fluids through model
porous media, represented by a square array of cylinders and a
single row of cylinders. Their results suggested that \steadys\ \vc\
flows in periodic geometries cannot explain the experimentally
observed excess pressure drop and time dependency. They eventually
concluded that the experimentally observed enhanced flow resistance
for both model and actual porous media should be, possibly,
attributed to three-dimensional and/or \timedep\ effects, which may
trigger a significant extensional response. Another possibility
which they considered is the inadequacy of the available
constitutive equations to describe unsteady,
non-viscometric flow behavior.%

\vspace{0.2cm}

Hua and Schieber \cite{HuaS1998} used a combined finite element and
Brownian dynamics technique (CONNFFESSIT) to predict the \steadys\
flow field around an infinite array of squarely-arranged cylinders
using two kinetic theory models. The attempt was concluded with
numerical convergence failure and limited success.

\vspace{0.2cm}

Garrouch \cite{garrouch1999} developed a generalized \vc\ model for
polymer flow in porous media analogous to \Darcy's law, and hence
concluded a nonlinear relationship between fluid velocity and
pressure gradient. The model accounts for polymer elasticity by
using the longest relaxation time, and accounts for polymer viscous
properties by using an average porous medium \powlaw\ behavior
index. According to this model, the correlation between the Darcy
velocity $q$ and the pressure gradient across the bed $\nabla P$ is
given by
\begin{equation}\label{garrouch}
    q = \left( \frac{\sqrt{K \phi}}{\alpha \fTim} \right)^{\frac{1}{\overline{n}-1}}
    | \nabla P |^{^{\frac{\beta}{1-\overline{n}}}}
\end{equation}
where $K$ and $\phi$ are the permeability and porosity of the bed
respectively, $\alpha$ and $\beta$ are model parameters, $\fTim$ is
a relaxation time and $\overline{n}$ is the average \powlaw\
behavior index inside the porous medium.

\vspace{0.2cm}

Investigating the \vc\ flow through an undulating channel, Koshiba
\etal\ \cite{koshiba2000} remarked that the excess pressure loss
occurs at the same \Deborah\ number as that for the cylinder arrays,
and the flow of \vc\ fluids through the undulating channel is proper
to model the flow of \vc\ fluids through porous medium. They
concluded that the stress in the flow through the undulating channel
should rapidly increase with increasing flow rate because of the
stretch-thickening elongational viscosity. Moreover, the \trans\
properties of a \vc\ fluid in an elongational flow should be
considered in the analysis of the flow through the undulating
channel.

\vspace{0.2cm}

Khuzhayorov \etal\ \cite{KhuzhayorovAR2000} applied a homogenization
technique to derive a macroscopic filtration law for describing
\trans\ linear \vc\ fluid flow in porous media. The macroscopic
filtration law is expressed in Fourier space as a generalized
\Darcy's law. The results obtained in the particular case of the
flow of an \OLD\ fluid in a bundle of capillary tubes show that the
\vc\ behavior strongly differs from the \NEW\ behavior.

\vspace{0.2cm}

Huifen \etal\ \cite{HuifenXDQX2001} developed a model for the
variation of rheological parameters along the seepage flow direction
and constructed a constitutive equation for \vc\ fluids in which the
variation of the rheological parameters of polymer solutions in
porous media is taken into account. A formula of critical elastic
flow velocity was presented. Using the proposed constitutive
equation, they investigated the seepage flow behavior of \vc\ fluids
with variable rheological parameters and concluded that during the
process of \vc\ polymer solution flooding, liquid production and
corresponding water cut decrease with the increase in relaxation
time.

\vspace{0.2cm}

Mendes and Naccache \cite{MendesN2002} employed a simple theoretical
approach to obtain a constitutive relation for flows of
extension-thickening fluids through porous media. The \nNEW\
behavior of the fluid is accounted for by a generalized \NEW\ fluid
with a viscosity function that has a \powlaw\ type dependence on the
extension rate. The pore morphology is assumed to be composed of a
bundle of periodically \convdiv\ tubes. Their predictions were
compared with the experimental data of Chmielewski and Jayaraman
\cite{ChmielewskiJ1992}. The comparisons showed that the developed
constitutive equation reproduces quite well the data within a low to
moderate range of pressure drop. In this range the flow resistance
increases as the flow rate is increased, exactly as predicted by the
constitutive relation.

\vspace{0.2cm}

Dolej$\check{\rm{s}}$ \etal\ \cite{DolejsCSD2002} presented a method
for the pressure drop calculation during the \vc\ fluid flow through
fixed beds of particles. The method is based on the application of
the modified \Rabinowitsch-\Mooney\ equation together with the
corresponding relations for consistency variables. The dependence of
a dimensionless quantity coming from the momentum balance equation
and expressing the influence of elastic effects on a suitably
defined elasticity number is determined experimentally. The validity
of the suggested approach has been verified for pseudoplastic \vc\
fluids characterized by the \powlaw\ flow model.

\vspace{0.2cm}

Numerical techniques have been exploited by many researchers to
investigate the flow of \vc\ fluids in \convdiv\ geometries. As an
example, Momeni-Masuleh and Phillips \cite{MasulehP2004} used
spectral methods to investigate \vc\ flow in an undulating tube.

%XXXXXXXXXXXXXXXXXXXXXXXXXXXXXXXXXXXXXXXXXXXXXXXXXXXXXXXXXXXXXXXXXXXXXX
\section{\TimeDep\ Fluids} \label{LTimeD}
The flow of \timedep\ fluids in porous media has not been vigorously
investigated. Consequently, very few studies can be found in the
literature on this subject. One reason is that the \timedep\ effects
are usually investigated in the context of \vy. Another reason is
that there is apparently no comprehensive framework to describe the
dynamics of \timedep\ fluids in porous media \cite{ChengE1965,
PearsonT2002, PritchardP2006}.

\vspace{0.2cm}

Among the few studies found on this subject is the investigation of
Pritchard and Pearson \cite{PritchardP2006} of viscous fingering
instability during the injection of a \thixotropic\ fluid into a
porous medium or a narrow fracture. The conclusion of this
investigation is that the perturbations decay or grow exponentially
rather than algebraically in time because of the presence of an
independent timescale in the problem.

\vspace{0.2cm}

Wang \etal\ \cite{WangHC2006} also examined \thixotropic\ effects in
the context of heavy oil flow through porous media.

\def\baselinestretch{1}
\chapter{Modeling the Flow of Fluids} \label{Modeling}
\def\baselinestretch{1.66}
The basic equations describing the flow of fluids consist of the
basic laws of continuum mechanics which are the conservation
principles of mass, energy and linear and angular momentum. These
governing equations indicate how the mass, energy and momentum of
the fluid change with position and time. The basic equations have to
be supplemented by a suitable rheological equation of state, or
constitutive equation describing a particular fluid, which is a
differential or integral mathematical relationship that relates the
extra stress tensor to the rate-of-strain tensor in general flow
condition and closes the set of governing equations. One then solves
the constitutive model together with the conservation laws using a
suitable method to predict velocity and stress fields of the flows
\cite{birdbook, Hulsen1990, Hulsen1986, carreaubook, Keunings2003,
Keunings2004}.

\vspace{0.2cm}

In the case of Navier-Stokes flows the constitutive equation is the
\NEW\ stress relation \cite{OsP2004} as given in
(\ref{NewtonianTensForm}). In the case of more rheologically complex
flows other \nNEW\ constitutive equations, such as \ELLIS\ and
\OldB, should be used to bridge the gap and obtain the flow fields.
To simplify the situation, several assumptions are usually made
about the flow and the fluid. Common assumptions include laminar,
incompressible, \steadys\ and isothermal flow. The last assumption,
for instance, makes the energy conservation equation redundant.

\vspace{0.2cm}

The constitutive equation should be frame-invariant. Consequently
sophisticated mathematical techniques are usually employed to
satisfy this condition. No single choice of constitutive equation is
best for all purposes. A constitutive equation should be chosen
considering several factors such as the type of flow (shear or
extension, steady or \trans, etc.), the important phenomena to
capture, the required level of accuracy, the available computational
resources and so on. These considerations can be influenced strongly
by personal preference or bias. Ideally the rheological equation of
state is required to be as simple as possible, involving the minimum
number of variables and parameters, and yet having the capability to
predict the behavior of complex fluids in complex flows. So far, no
constitutive equation has been developed that can adequately
describe the behavior of complex fluids in general flow situation
\cite{larsonbook1988, owensbook2002}.

%XXXXXXXXXXXXXXXXXXXXXXXXXXXXXXXXXXXXXXXXXXXXXXXXXXXXXXXXXXXXXXXXXXXXXX
\section{Modeling the Flow in Porous Media} \label{ModelingFlowPorous}
In the context of fluid flow, ``porous medium'' can be defined as a
solid matrix through which small interconnected cavities occupying a
measurable fraction of its volume are distributed. These cavities
are of two types: large ones, called pores and throats, which
contribute to the bulk flow of fluid; and small ones, comparable to
the size of the molecules, which do not have an impact on the bulk
flow though they may participate in other transportation phenomena
like diffusion. The complexities of the microscopic pore structure
are usually avoided by resorting to macroscopic physical properties
to describe and characterize the porous medium. The macroscopic
properties are strongly related to the underlying microscopic
structure. The best known examples of these properties are the
porosity and the permeability. The first describes the relative
fractional volume of the void space to the total volume while the
second quantifies the capacity of the medium to transmit fluid.

\vspace{0.2cm}

Another important property is the macroscopic homogeneity which may
be defined as the absence of local variation in the relevant
macroscopic properties such as permeability on a scale comparable to
the size of the medium under consideration. Most natural and
synthetic porous media have an element of inhomogeneity as the
structure of the porous medium is typically very complex with a
degree of randomness and can seldom be completely uniform. However,
as long as the scale and magnitude of this variation have negligible
impact on the macroscopic properties under discussion, the medium
can still be treated as homogeneous.

\vspace{0.2cm}

The mathematical description of the flow in porous media is
extremely complex task and involves many approximations. So far, no
analytical fluid mechanics solution to the flow through porous media
has been found. Furthermore, such a solution is apparently out of
reach for the foreseeable future. Therefore, to investigate the flow
through porous media other methodologies have been developed, the
main ones are the macroscopic continuum approach and the pore-scale
numerical approach. In the continuum, the porous medium is treated
as a continuum and all the complexities of the microscopic pore
structure are lumped into terms such as permeability. Semi-empirical
equations such as the \Ergun\ equation, \Darcy's law or the
\Carman-\Kozeny\ equation are usually employed \cite{balhoff1}.

\vspace{0.2cm}

In the numerical approach, a detailed description of the porous
medium at pore-scale level is adopted and the relevant physics of
flow at this level is applied. To find the solution, numerical
methods, such as finite volume and finite difference, usually in
conjunction with computational implementation are used.

\vspace{0.2cm}

The advantage of the continuum method is that it is simple and easy
to implement with no computational cost. The disadvantage is that it
does not account for the detailed physics at the pore level. One
consequence of this is that in most cases it can only deal with
\steadys\ situations with no \timedep\ \trans\ effects.

\vspace{0.2cm}

The advantage of the numerical method is that it is the most direct
approach to describe the physical situation and the closest to full
analytical solution. It is also capable, in principle at least, to
deal with \timedep\ \trans\ situations. The disadvantage is that it
requires a detailed pore space description. Moreover, it is usually
very complex and hard to implement and has a huge computational cost
with serious convergence difficulties. Due to these complexities,
the flow processes and porous media that are currently within the
reach of numerical investigation are the most simple ones.

\vspace{0.2cm}

Pore-scale network modeling is a relatively novel method developed
to deal with flow through porous media. It can be seen as a
compromise between these two extreme approaches as it partly
accounts for the physics and void space description at the pore
level with reasonable and generally affordable computational cost.
Network modeling can be used to describe a wide range of properties
from capillary pressure characteristics to interfacial area and mass
transfer coefficients. The void space is described as a network of
pores connected by throats. The pores and throats are assigned some
idealized geometry, and rules which determine the transport
properties in these elements are incorporated in the network to
compute effective transport properties on a mesoscopic scale. The
appropriate pore-scale physics combined with a geologically
representative description of the pore space gives models that can
successfully predict average behavior \cite{blunt2, blunt1}.

\vspace{0.2cm}

In our investigation to the flow of \nNEW\ fluids in porous media we
use network modeling. Our model was originally developed by Valvatne
and co-workers \cite{valvatnethesis, valvatne1, lopezthesis, lopez1}
and modified and extended by the author. The main aspects introduced
are the inclusion of the \HB\ and \ELLIS\ models and implementing a
number of \yields\ and \vc\ algorithms.

\vspace{0.2cm}

In this context, Lopez \etal\ \cite{lopez1, LopezB2004, lopezthesis}
investigated single- and two-phase flow of \shThin\ fluids in porous
media using \CARREAU\ model in conjunction with network modeling.
They were able to predict several experimental datasets found in the
literature and presented motivating theoretical analysis to several
aspects of single- and multi-pahse flow of \nNEW\ fluids in porous
media. The main features of their model will be outlined below as it
is the foundation for our model. Recently, Balhoff and Thompson
\cite{balhoffthesis, balhoff1} used a three-dimensional network
model which is based on a computer-generated random sphere packing
to investigate the flow of \nNEW\ fluids in packed beds. They used
analytical expressions for a capillary tube with empirical tuning to
key parameters to more accurately represent the throat geometry and
simulate the fluid dynamics in the real throats of the packing.

\vspace{0.2cm}

Our model uses three-dimensional networks built from a
topologically-equivalent three-dimensional voxel image of the pore
space with the pore sizes, shapes and connectivity reflecting the
real medium. Pores and throats are modeled as having triangular,
square or circular cross-section by assigning a shape factor which
is the ratio of the area to the perimeter squared and obtained from
the pore space image. Most of the network elements are not circular.
To account for the non-circularity when calculating the volumetric
flow rate analytically or numerically for a cylindrical capillary,
an equivalent radius $R_{eq}$ is defined:
\begin{equation}
    \verb|       |    R_{eq} = \left( \frac{8G}{\pi} \right)^{1/4}
\end{equation}
where the geometric conductance, $G$, is obtained empirically from
numerical simulation. Two networks obtained from Statoil and
representing two different porous media have been used: a \sandp\
and a \Berea\ sandstone. These networks are constructed by {\O}ren
and coworkers \cite{OrenBA1997, OrenB2003} from voxel images
generated by simulating the geological processes by which the porous
medium was formed. The physical and statistical properties of the
networks are presented in Tables (\ref{SPPropertiesTable}) and
(\ref{BereaPropertiesTable}).

\vspace{0.2cm}

Assuming a laminar, isothermal and incompressible flow at low \Rey\
number, the only equations that need to be considered are the
constitutive equation for the particular fluid and the conservation
of volume as an expression for the conservation of mass. Because
initially the pressure drop in each network element is not known, an
iterative method is used. This starts by assigning an effective
viscosity $\eVis$ to each network element. The effective viscosity
is defined as that viscosity which makes Poiseulle's equation fit
any set of laminar flow conditions for \timeind\ fluids
\cite{skellandbook}. By invoking the conservation of volume for
incompressible fluid, the pressure field across the entire network
is solved using a numerical solver \cite{rugebook}. Knowing the
pressure drops, the effective viscosity of each element is updated
using the expression for the flow rate with a pseudo-\POIS\
definition. The pressure field is then recomputed using the updated
viscosities and the iteration continues until convergence is
achieved when a specified error tolerance in total flow rate between
two consecutive iteration cycles is reached. Finally, the total
volumetric flow rate and the apparent viscosity in porous media,
defined as the viscosity calculated from the \Darcy's law, are
obtained.

\vspace{0.2cm}

Due to nonlinearities in the case of \nNEW\ flow, the convergence
may be hindered or delayed. To overcome these difficulties, a series
of measures were taken to guarantee convergence to the correct value
in a reasonable time. These measures include:
\begin{itemize}

    \item Scanning a smooth pressure line and archiving the values when convergence
    occurs and ignoring the pressure point if convergence did not happen within
    a certain number of iterations.

    \item Imposing certain conditions on the initialization of the solver vectors.

    \item Initializing the size of these vectors appropriately.

\end{itemize}

The new code was tested extensively. All the cases that we
investigated were verified and proved to be qualitatively correct.
Quantitatively, the \NEW, as a special case of the \nNEW, and the
\BING\ asymptotic behavior at high pressures are confirmed.

\vspace{0.2cm}

For \yields\ fluids, total blocking of the elements below their
threshold yield pressure is not allowed because if the pressure have
to communicate, the substance before yield should be considered a
fluid with high viscosity. Therefore, to simulate the state of the
fluid before yield the viscosity was set to a very high but finite
value ($10^{50}$\,Pa.s) so the flow is virtually zero. As long as
the \yields\ substance is assumed a fluid, the pressure field will
be solved as for yield-free fluids since the high viscosity
assumption will not change the situation fundamentally. It is
noteworthy that the assumption of very high but finite zero stress
viscosity for \yields\ fluids is realistic and supported by
experimental evidence.

\vspace{0.2cm}

In the case of \yields\ fluids, a further condition is imposed
before any element is allowed to flow, that is the element must be
part of a non-blocked path spanning the network from the inlet to
the outlet. What necessitates this condition is that any flowing
element should have a source on one side and a sink on the other,
and this will not be satisfied if either or both sides are blocked.

\vspace{0.3cm}

With regards to modeling the flow in porous media of complex fluids
which have time dependency due to \thixotropic\ or elastic nature,
there are three major difficulties :
\begin{itemize}

    \item The difficulty of tracking the fluid elements in the pores and throats
    and identifying their deformation history, as the path followed by these elements is
    random and can have several unpredictable outcomes.

    \item The mixing of fluid elements with various deformation history in
    the individual pores and throats. As a result, the viscosity is not
    a well-defined property of the fluid in the pores and throats.

    \item The change of viscosity along the streamline since the deformation history is
    constantly changing over the path of each fluid element.

\end{itemize}
In the current work, we deal only with one case of \steadys\ \vc\
flow, and hence we did not consider these complications in depth.
Consequently, the tracking of fluid elements or flow history in the
network and other dynamic aspects are not implemented in the \nNEW\
code as the code currently has no dynamic \timedep\ capability.
However, \timedep\ effects in \steadys\ conditions are accounted for
in the \Tardy\ algorithm which is implemented in the code and will
be presented in Section (\ref{TardyAlgorithm}). Though this may be
unrealistic in many situations of complex flow in porous media where
the flow field is \timedep\ in a dynamic sense, there are situations
where this assumption is sensible and realistic. In fact even if the
possibility of reaching a \steadys\ in porous media flow is
questioned, this remains a good approximation in many situations.
Anyway, the legitimacy of this assumption as a first step in
simulating complex flow in porous media cannot be questioned.

\vspace{0.2cm}

In our network model we adopt the widely accepted assumption of
no-slip at wall condition. This means that the fluid at the boundary
is stagnant relative to the solid boundary. Some slip indicators are
the dependence of viscosity on the geometry size and the emergence
of an apparent \yields\ at low stresses in single flow curves
\cite{Barnes1995}. The effect of slip, which includes reducing
shear-related effects and influencing \yields\ behavior, is very
important in certain circumstances and cannot be ignored. However,
this simplifying assumption is not unrealistic for the cases of flow
in porous media that are of prime interest to us. Furthermore, wall
roughness, which is the norm in the real porous media, usually
prevents wall slip or reduces its effect.

\vspace{0.2cm}

Another simplification that we adopt in our modeling strategy is the
disassociation of the \nNEW\ phenomena. For example, in modeling the
flow of \yields\ fluids through porous media an implicit assumption
has been made that there is no time dependence or \vy. Though this
assumption is unrealistic in many situations of complex flows where
various \nNEW\ events take place simultaneously, it is a reasonable
assumption in modeling the dominant effect and is valid in many
practical situations where the other effects are absent or
insignificant. Furthermore, it is a legitimate pragmatic
simplification to make when dealing with extremely complex
situations.

\vspace{0.2cm}

In this thesis we use the term ``consistent'' or ``stable'' pressure
field to describe the solution which we obtain from the solver on
solving the pressure field. We would like to clarify this term and
define it with mathematical rigor as it is a key concept in our
modeling methodology. Moreover, it is used to justify the failure of
the Invasion Percolation with Memory (IPM) and the Path of Minimum
Pressure (PMP) algorithms which will be discussed in Chapter
(\ref{Yield}). In our modeling approach, to solve the pressure field
across a network of $n$ nodes we write $n$ equations in $n$ unknowns
which are the pressure values at the nodes. The essence of these
equations is the continuity of flow of incompressible fluid at each
node in the absence of source and sink. We solve this set of
equations subject to the boundary conditions which are the pressures
at the inlet and outlet. This unique solution is ``consistent'' and
``stable'' as it is the only mathematically acceptable solution to
the problem, and, assuming the modeling process and the mathematical
technicalities are correct, should mimic the unique physical reality
of the pressure field in the porous medium.

%XXXXXXXXXXXXXXXXXXXXXXXXXXXXXXXXXXXXXXXXXXXXXXXXXXXXXXXXXXXXXXXXXXXXXX
\section{Algebraic Multi-Grid (AMG) Solver} \label{AMG}
The numerical solver which we used in our \nNEW\ code to solve the
pressure field iteratively is an Algebraic Multi-Grid (AMG) solver
\cite{rugebook}. The basis of the multigrid methods is to build an
approximate solution to the problem on a coarse grid. The
approximate solution is then interpolated to a finer mesh and used
as a starting guess in the iteration. By repeated transfer between
coarse and fine meshes, and using an iterative scheme such as Jacobi
or Gauss-Seidel which reduces errors on a length scale defined by
the mesh, reduction of errors on all length scales occurs at the
same rate. This process of transfer back and forth between two
levels of discretizations is repeated recursively until convergence
is achieved. Iterative schemes are known to rapidly reduce high
frequency modes of the error, but perform poorly on the lower
frequency modes; that is they rapidly smooth the error, which is why
they are often called smoothers \cite{FattalK2005, MoresiDM2003}.

\vspace{0.2cm}

A major advantage of using multigrid solvers is a speedy and smooth
convergence. If carefully tuned, they are capable of solving
problems with $n$ unknowns in a time proportional to $n$, in
contrast to $n^{2}$ or even $n^{3}$ for direct solvers
\cite{MoresiDM2003}. This comes on the expense of extra memory
required for storing large grids. In our case, this memory cost is
affordable on a typical modern workstation for all available
networks. A typical convergence time for the \sandp\ and \Berea\
sandstone networks used in this study is a second for the \timeind\
models and a few seconds for the \vc\ model. The time requirement
for the \yields\ algorithms is discussed in Chapter (\ref{Yield}).
In all cases, the memory requirement does not exceed a few tens of
megabytes for a network with up to 12000 pores.

\def\baselinestretch{1}
\chapter{Network Model Results for \TimeInd\ Flow} \label{NetTube}
\def\baselinestretch{1.66}
In this chapter we study generic trends in behavior for the \HB\
model of the \timeind\ category. We do not do this for the \ELLIS\
model, since its behavior as a \shThin\ fluid is included within the
\HB\ model. Moreover, \shThin\ behavior has already been studied
previously by Lopez \etal\ \cite{lopez1, LopezB2004, lopezthesis}
whose work is the basis for our model. \Vc\ behavior is investigated
in Chapter (\ref{Viscoelasticity}).

%XXXXXXXXXXXXXXXXXXXXXXXXXXXXXXXXXXXXXXXXXXXXXXXXXXXXXXXXXXXXXXXXXXXXXX
\section{Random Networks vs. Bundle of Tubes Comparison} \label{}
In capillary models the pores are described as a bundle of tubes,
which are placed in parallel. The simplest form is the model with
identical tubes, which means that the tubes are straight,
cylindrical, and of equal radius. \Darcy's law combined with the
\POIS\ law gives the following relationship for the permeability
\begin{equation}\label{BundleOfTubesKR}
    K = \frac{\phi R^{2}}{8}
\end{equation}
where $K$ and $\phi$ are the permeability and porosity of the bundle
respectively, and $R$ is the radius of the tubes. The derivation of
this relation is given in Appendix \ref{AppTubeRad}.

\vspace{0.2cm}

A limitation of this model is that it neglects the topology of the
pore space and the heterogeneity of the medium. Moreover, as it is a
unidirectional model its application is limited to simple
one-dimensional flow situations. Another shortcoming of this simple
model is that the permeability is considered in the direction of
flow only, and hence may not correctly correspond to the
permeability of the porous medium. As for \yields\ fluids, this
model predicts a universal yield point at a particular pressure
drop, whereas in real porous media yield occurs gradually.
Furthermore, possible percolation effects due to the size
distribution and connectivity of the elements of the porous media
are not reflected in this model.

\vspace{0.2cm}

It should be remarked that although this simple model is adequate
for modeling some cases of slow flow of purely viscous fluids
through porous media, it does not allow the prediction of an
increase in the pressure drop when used with a \vc\ constitutive
equation. Presumably, the \convdiv\ nature of the flow field gives
rise to an additional pressure drop, in excess to that due to
shearing forces, since porous media flow involves elongational flow
components. Therefore, a corrugated capillary bundle model is a much
better candidate when trying to approach porous medium flow
conditions in general \cite{denysthesis, PilitsisB1989}.

\vspace{0.2cm}

In this thesis a comparison is made between a network representing a
porous medium and a bundle of capillary tubes of uniform radius,
having the same \NEW\ Darcy velocity and porosity. The use in this
comparison of the uniform bundle of tubes model is within the range
of its validity. The advantage of using this simple model, rather
than more sophisticated models, is its simplicity and clarity. The
derivation of the relevant expressions for this comparison is
presented in Appendix \ref{AppTubeRad}.

\vspace{0.2cm}

Two networks representing two different porous media have been
investigated: a \sandp\ and a \Berea\ sandstone. \Berea\ is a
sedimentary consolidated rock with some clay having relatively high
porosity and permeability. This makes it a good reservoir rock and
widely used by the petroleum industry as a test bed. For each
network, two model fluids were studied: a fluid with no \yields\ and
a fluid with a \yields. For each fluid, the flow behavior index,
$n$, takes the values 0.6, 0.8, 1.0, 1.2 and 1.4. In all cases the
consistency factor, $C$, is kept constant at 0.1\,Pa.s$^{n}$, as it
is considered a viscosity scale factor.

\vspace{0.2cm}

For \yields\ fluids, the continuity of flow across the network was
tested by numerical and visual inspections. The results confirmed
that the flow condition is satisfied in all cases. A sample of the
three-dimensional visualization images for the non-blocked elements
of the \sandp\ and \Berea\ networks at various yield stages is
displayed in Figures (\ref {SPVisualYieldStages}) and (\ref
{BerVisualYieldStages}).

%SSSSSSSSSSSSSSSSSSSSSSSSSSSSSSSSSSSSS
\subsection{\SandP\ Network} \label{SPBundleComparison}
The physical and statistical properties of this network are given in
Appendix \ref{AppNetProp} Table (\ref{SPPropertiesTable}). The
comparison between the \sandp\ network and the bundle of tubes model
for the case of a fluid with no \yields\ is displayed in Figure
(\ref{SPTubeNetTau0}). By definition, the results of the network and
the bundle of tubes are identical in the \NEW\ case. There is a
clear symmetry between the \shThin\ and \shThik\ cases. This has to
be expected as the \sandp\ is a homogeneous network. Therefore
\shThin\ and \shThik\ effects take place on equal footing.

\vspace{0.2cm}

The comparison between the \sandp\ network and the bundle of tubes
model for the case of a fluid with a \yields\ is displayed in Figure
(\ref{SPTubeNetTau1}) alongside a magnified view to the yield zone
in Figure (\ref{SPTubeNetTau1Mag}). The first thing to remark is
that the \sandp\ network starts flowing at a lower pressure gradient
than the bundle of tubes. Plotting a graph of the radius of the
tubes and the average radius of the non-blocked throats of the
network as a function of the magnitude of pressure gradient reveals
that the average radius of the flowing throats at yield is slightly
greater than the tubes radius, as can be seen in Figure (\ref
{SPTubeNetRadius}). This can explain the higher yield pressure
gradient of the bundle relative to the network. However, the tubes
radius is ultimately greater than the average radius. This has an
observable impact on the network-bundle of tubes relation, as will
be discussed in the next paragraphs.

\vspace{0.2cm}

For a \BING\ fluid, the flow of the \sandp\ network exceeds the
tubes flow at low pressure gradients, but the trend is reversed
eventually. The reason is that the network yields before the tubes
but because some network elements are still blocked, even at high
pressure gradients, the tubes flow will eventually exceed the
network flow.

\vspace{0.2cm}

For the \shThin\ cases, the prominent feature is the crossover
between the network and the bundle of tubes curves. This is due to
the shift in the relation between the average radius of the
non-blocked elements and the radius of the tubes as seen in Figure
(\ref {SPTubeNetRadius}). Intersection for $n = 0.6$ occurs at
higher pressure gradient than that for $n = 0.8$ because of the flow
enhancement in the network, which yields before the bundle, caused
by more \shThin\ in the $n = 0.6$ case. This enhancement delays the
catchup of the tubes to a higher pressure gradient.

\vspace{0.2cm}

For the \shThik\ cases, the situation is more complex. There are
three main factors affecting the network-bundle of tubes relation:
the partial blocking of the network, the shift in the relation
between the bundle radius and the average radius of the non-blocked
elements, and the flow hindrance caused by the \shThik\ effect.
There are two crossovers: lower and upper. The occurrence and
relative location of each crossover is determined by the overall
effect of the three factors, some of which are competing. The lower
one is caused mainly by the partial blocking of the network plus the
shift in the radius relationship. The upper one is caused by
\shThik\ effects because the bundle of tubes is subject to more
\shThik\ at high pressure gradients.

\vspace{0.2cm}

Changing the value of the \yields\ will not affect the general
pattern observed already although some features may become
disguised. This has to be expected because apart from scaling, the
fundamental properties do not change.

%\vspace{1cm}

\begin{figure}[!h]
  \centering{}
  \includegraphics
  [scale=0.53]
  {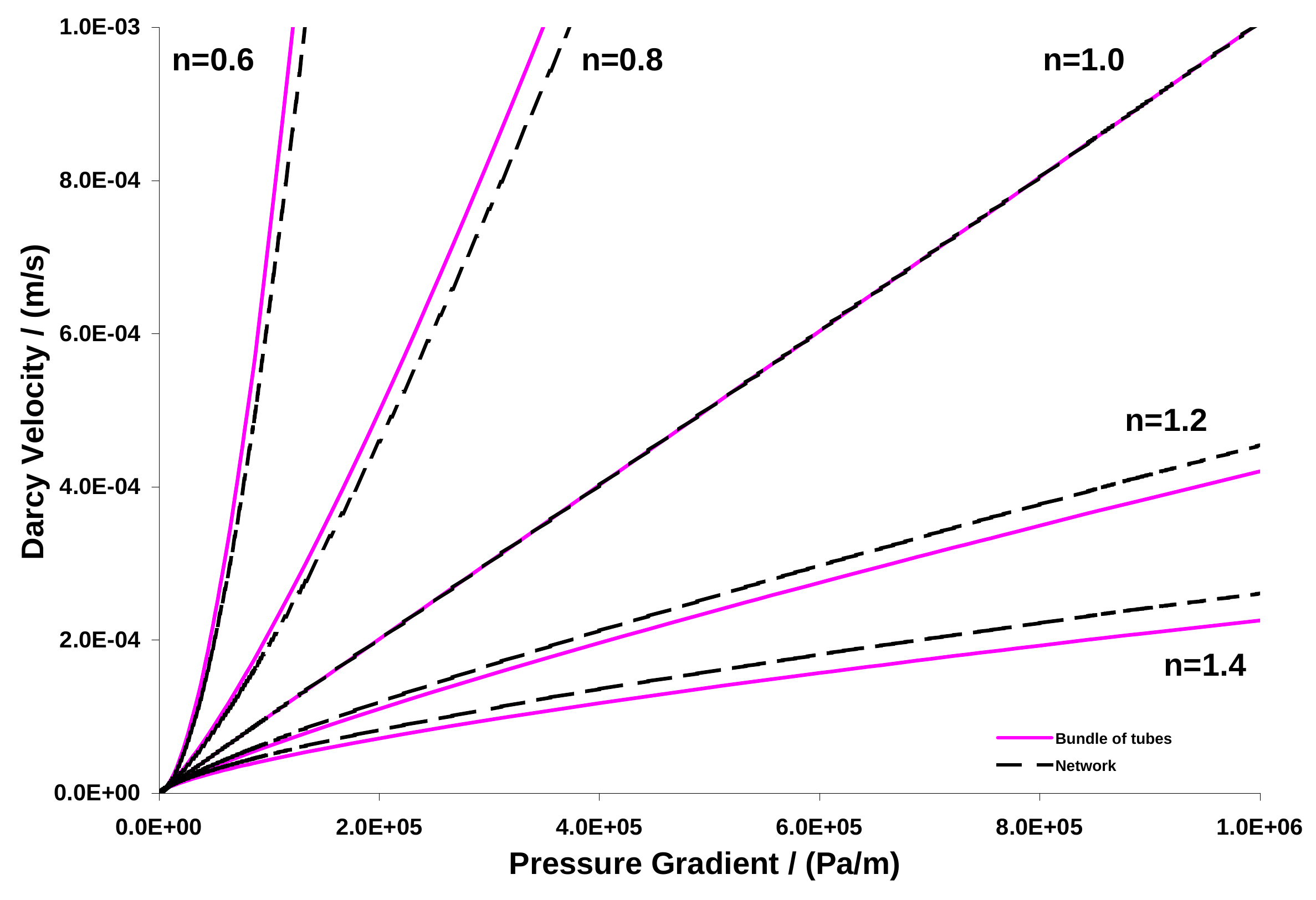}
  \caption[Comparison between the \sandp\ network ($x_{_{l}}=0.5$, $x_{_{u}}=0.95$, $K=102$\,Darcy,
  $\phi=0.35$) and a bundle of tubes ($R=48.2\mu$m) for a \HB\ fluid with $\ysS=0.0$Pa and $C=0.1$Pa.s$^n$]
  {Comparison between the \sandp\ network ($x_{_{l}}=0.5$, $x_{_{u}}=0.95$, $K=102$\,Darcy, $\phi=0.35$)
  and a bundle of tubes ($R=48.2\mu$m) for a \HB\ fluid with $\ysS=0.0$Pa and $C=0.1$Pa.s$^n$.}
  \label{SPTubeNetTau0}
\end{figure}

\newpage

\begin{figure}[!h]
  \centering{}
  \includegraphics
  [scale=0.53]
  {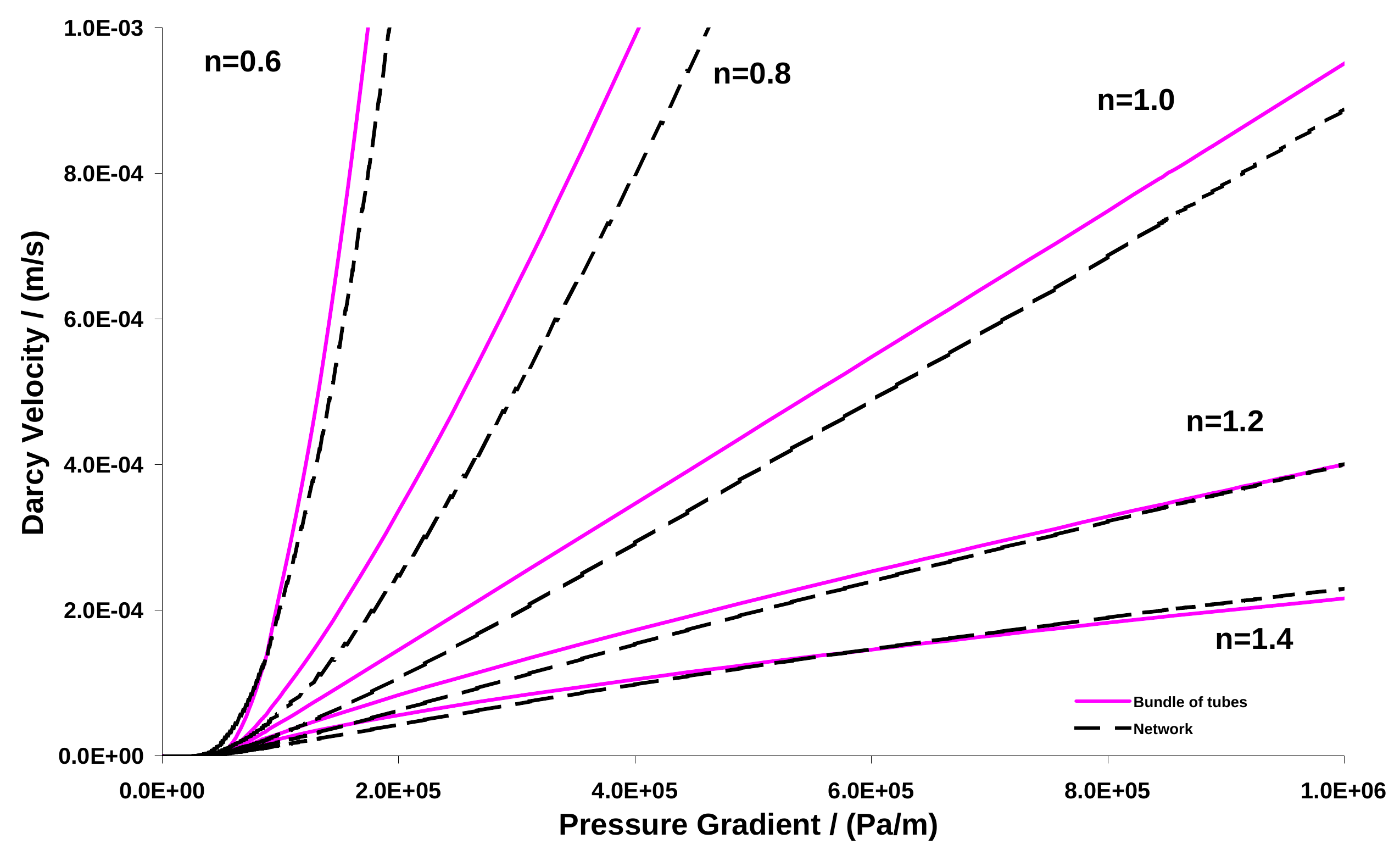}
  \caption[Comparison between the \sandp\ network ($x_{_{l}}=0.5$, $x_{_{u}}=0.95$, $K=102$\,Darcy,
  $\phi=0.35$) and a bundle of tubes ($R=48.2\mu$m) for a \HB\ fluid with $\ysS=1.0$Pa and $C=0.1$Pa.s$^n$]
  {Comparison between the \sandp\ network ($x_{_{l}}=0.5$, $x_{_{u}}=0.95$, $K=102$\,Darcy, $\phi=0.35$)
  and a bundle of tubes ($R=48.2\mu$m) for a \HB\ fluid with $\ysS=1.0$Pa and $C=0.1$Pa.s$^n$.}
  \label{SPTubeNetTau1}
\end{figure}

\begin{figure}[!b]
  \centering{}
  \includegraphics
  [scale=0.53]
  {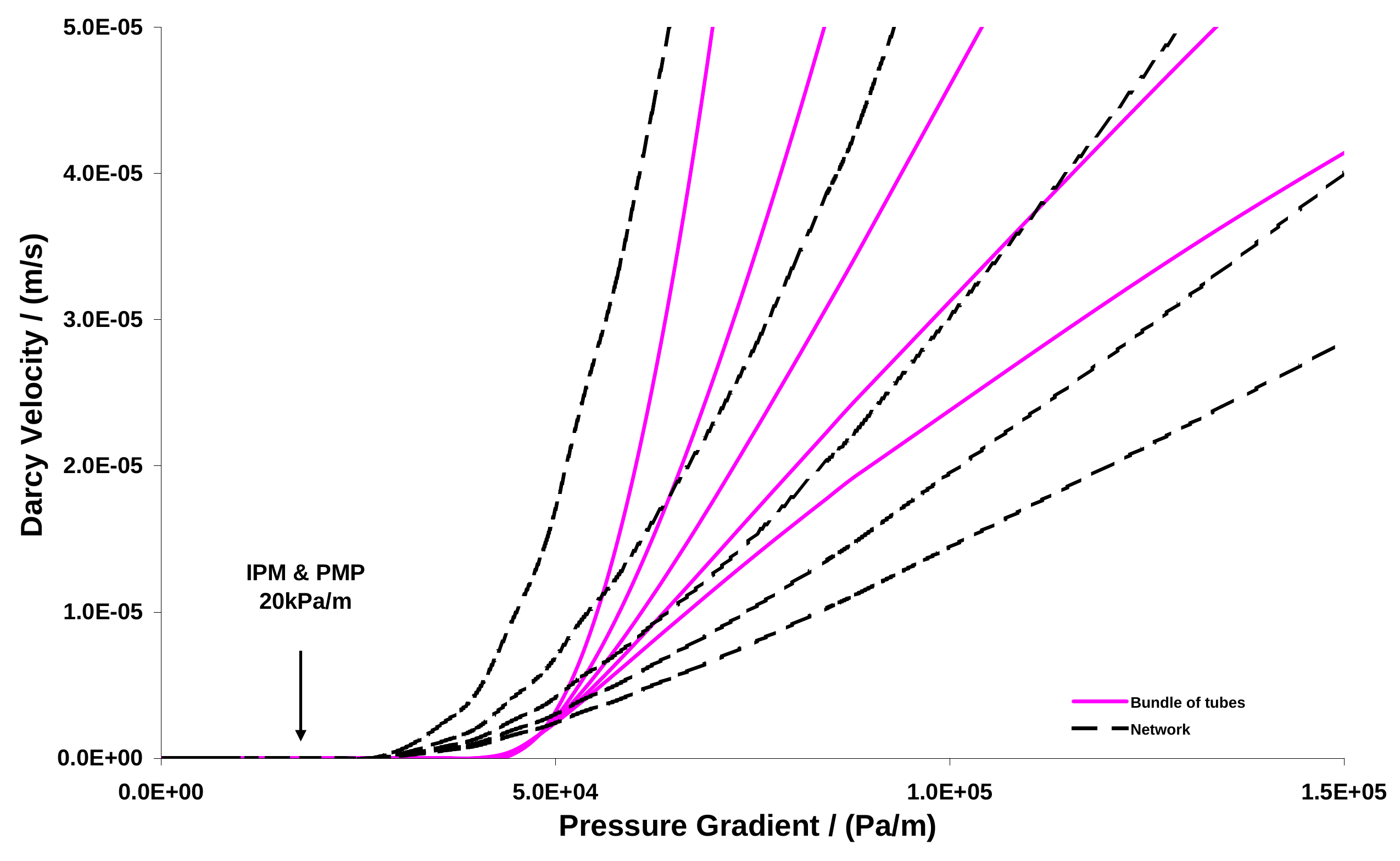}
  \caption[A magnified view of Figure (\ref{SPTubeNetTau1}) at the yield zone. The prediction of Invasion Percolation with
  Memory (IPM) and Path of Minimum Pressure (PMP) algorithms is indicated by the arrow]
  {A magnified view of Figure (\ref{SPTubeNetTau1}) at the yield zone. The prediction of Invasion Percolation with
  Memory (IPM) and Path of Minimum Pressure (PMP) algorithms is indicated by the arrow.}
  \label{SPTubeNetTau1Mag}
\end{figure}

%%%%%%%%%%%%%%%%%%%%%%%%%%%%%%%%%%%%%%%%%%%%%%%%%%%%%%%%%%

\begin{figure}[!h]
  \centering{}
  \includegraphics
  [scale=0.53]
  {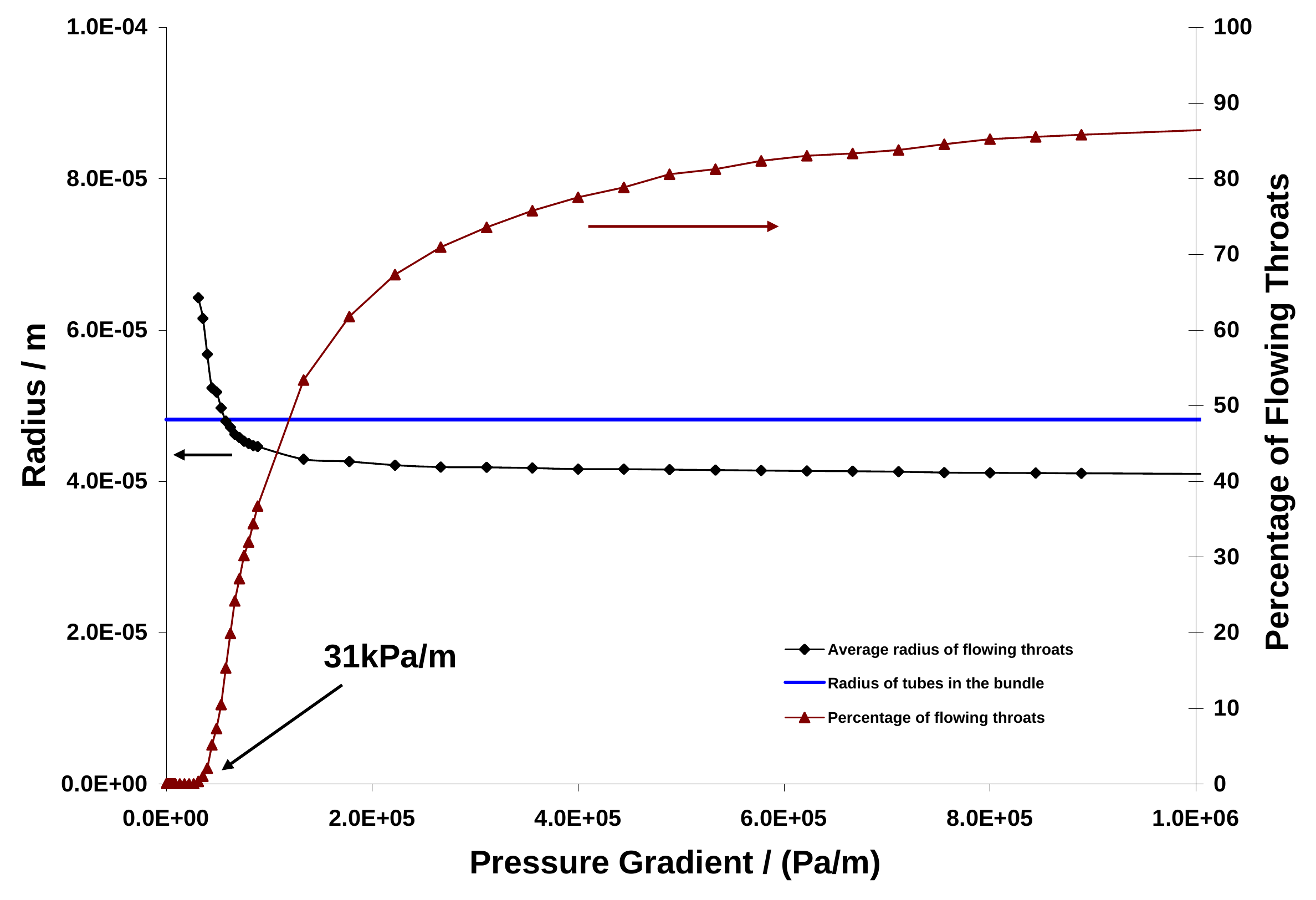}
  \caption[The radius of the bundle of tubes and the average radius of the non-blocked
  throats of the \sandp\ network, with their percentage of the total number of throats, as a
  function of pressure gradient for a \BING\ fluid ($n = 1.0$) with $\ysS=1.0$Pa and $C=0.1$Pa.s]
  {The radius of the bundle of tubes and the average radius of the non-blocked
  throats of the \sandp\ network, with their percentage of the total number of throats, as a
  function of pressure gradient for a \BING\ fluid ($n = 1.0$) with $\ysS=1.0$Pa and $C=0.1$Pa.s.}
  \label{SPTubeNetRadius}
\end{figure}

\begin{figure}[!b]
  \centering{}
  \includegraphics
  [scale=0.55]
  {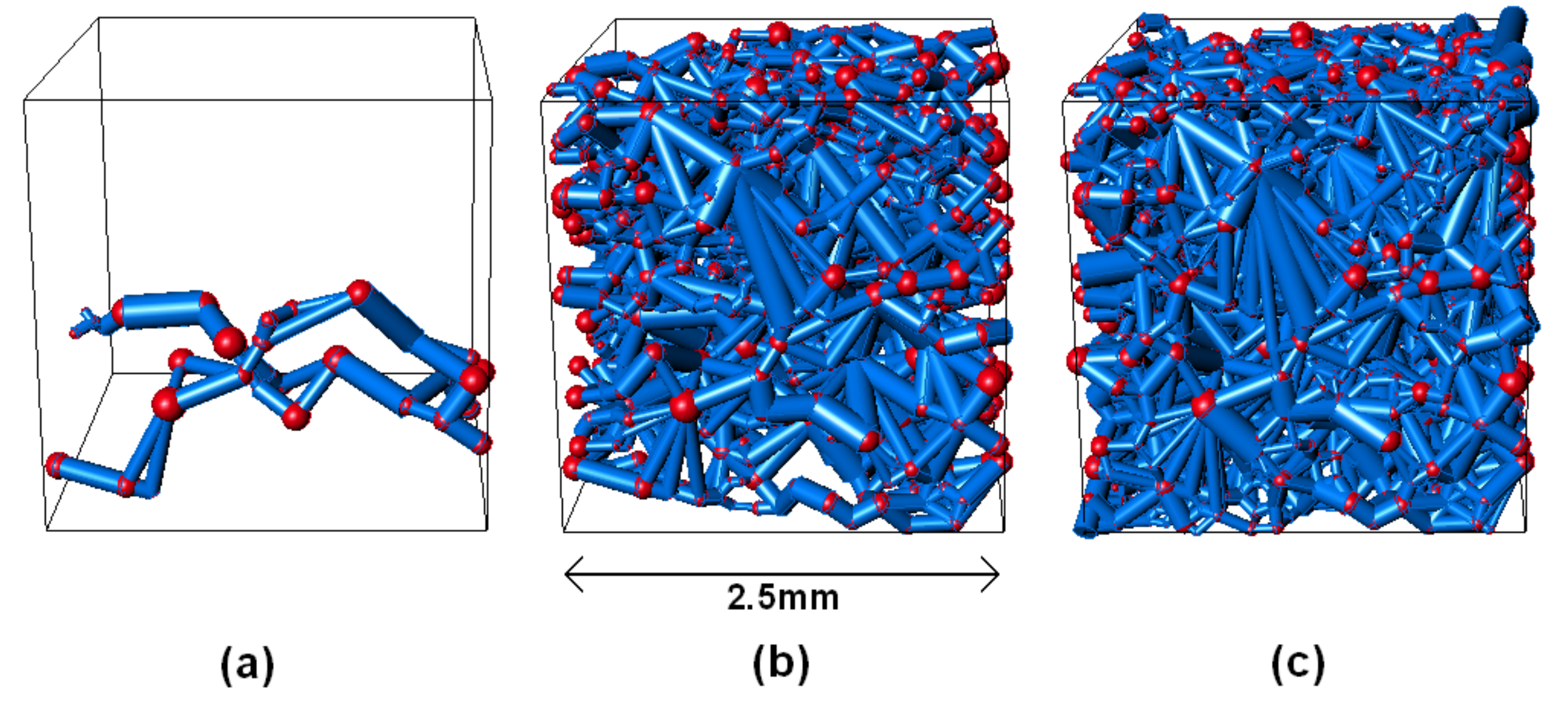}
  \caption[Visualization of the non-blocked elements of the \sandp\ network
  for a \BING\ fluid ($n=1.0$) with $\ysS=1.0$Pa and $C=0.1$Pa.s]
  {Visualization of the non-blocked elements of the \sandp\ network
  for a \BING\ fluid ($n=1.0$) with $\ysS=1.0$Pa and $C=0.1$Pa.s. The fraction of
  flowing elements is (a) 0.4\% (b) 25\% and (c) 69\%.}
  \label{SPVisualYieldStages}
\end{figure}

%SSSSSSSSSSSSSSSSSSSSSSSSSSSSSSSSSSSSS
\subsection{\Berea\ Sandstone Network} \label{}
The physical and statistical details of this network can be found in
Appendix \ref{AppNetProp} Table (\ref{BereaPropertiesTable}). This
network is more tortuous and less homogeneous than the \sandp.

\vspace{0.2cm}

The comparison between the results of the \Berea\ network simulation
and the bundle of tubes for the case of a fluid with no \yields\ is
displayed in Figure (\ref{BerTubeNetTau0}). By definition, the
results of the network and the bundle of tubes are identical in the
\NEW\ case. There is a lack of symmetry between the \shThin\ and
\shThik\ cases relative to the \NEW\ case, i.e. while the network
flow in the \shThin\ cases is considerably higher than the flow in
the tubes, the difference between the two flows in the \shThik\
cases is tiny. The reason is the inhomogeneity of the \Berea\
network coupled with the shear effects.

\vspace{0.2cm}

Another feature is that for the \shThin\ cases the average
discrepancy between the two flows is much larger for the $n=0.6$
case than for $n=0.8$. The reason is that as the fluid becomes more
\shThin\ by decreasing $n$, the disproportionality due to
inhomogeneity in the contribution of the two groups of large and
small throats is magnified resulting in the flow being carried out
largely by a relatively small number of large throats. The opposite
is observed in the two \shThik\ cases for the corresponding reason
that \shThik\ smooths out the inhomogeneity because it damps the
flow in the largest elements resulting in a more uniform flow
throughout the network.

\vspace{0.4cm}

The comparison between the \Berea\ network and the bundle of tubes
for the case of a fluid with a \yields\ is displayed in Figure
(\ref{BerTubeNetTau1}) beside a magnified view to the yield zone in
Figure (\ref{BerTubeNetTau1Mag}). As in the case of the \sandp\
network, the \Berea\ network starts flowing before the tubes for the
same reason that is the average radius of the non-blocked elements
in the network at yield is larger than the radius of the tubes in
the bundle, as can be seen in Figure (\ref {BerTubeNetRadius}).

\vspace{0.2cm}

For a \BING\ plastic, the network flow exceeds the bundle of tubes
flow at low pressure gradients, but the trend is reversed at high
pressure gradients because some elements in the network are still
blocked, as in the case of \sandp.

\vspace{0.2cm}

For the two \shThin\ cases, the general features of the
network-bundle of tubes relation are similar to those in the case of
fluid with no \yields. However, the discrepancy between the two
flows is now larger especially at low pressure gradients. There are
three factors affecting the network-bundle of tubes relation. The
first is the shift because the network yields before the bundle of
tubes. This factor dominates at low pressure gradients. The second
is the inhomogeneity coupled with \shThin, as outlined in the case
of \sandp. The third is the blocking of some network elements with
the effect of reducing the total flow. The second and the third
factors compete, especially at high pressure gradients, and they
determine the network-bundle of tubes relation which can take any
form depending on the network and fluid properties and the pressure
gradient. In the current case, the graphs suggest that for the fluid
with $n=0.6$ the second factor dominates, while for the fluid with
$n=0.8$ the two factors have almost similar impact.

\vspace{0.2cm}

For the two \shThik\ cases, the network flow exceeds the bundle of
tubes flow at the beginning as the network yields before the bundle,
but this eventually is overturned as in the case of a \BING\ fluid
with enforcement by the \shThik\ effect which impedes the network
flow and reduces it at high pressure gradients. Because the average
radius of the non-blocked throats in the \Berea\ network is greater
than the radius of the tubes in the bundle for all pressure
gradients, as seen in Figure (\ref {BerTubeNetRadius}), the fluid in
the network will be subject to more \shThik\ than the fluid in the
bundle. Unlike \shThin\ cases, the inhomogeneity coupled with shear
effects and partial blocking of the network are now enforcing each
other to deter the flow. Consequently, the discrepancy between the
network and the bundle of tubes in the two cases, i.e. $n=1.2$ and
$n=1.4$, is larger than that in the corresponding cases of a fluid
with no \yields.

\vspace{0.4cm}

Finally, it should be remarked that the superiority of the \sandp\
results in comparison to the \Berea\ can be regarded as a sign of
good behavior for our network model. The reason is that it has been
shown in previous studies that the capillary bundle model is better
suited for describing porous media that are unconsolidated and have
high permeability \cite{ZaitounK1986}.

%%%%%%%%%%%%%%%%%%%%%%%%%%%%%%%   Berea tube-net comparison   %%%%%%%%%%%%%%%%%%%%%%%%%%%%%%%%%%%%%%%%%
\newpage
\begin{figure}[!h]
  \centering{}
  \includegraphics
  [scale=0.53]
  {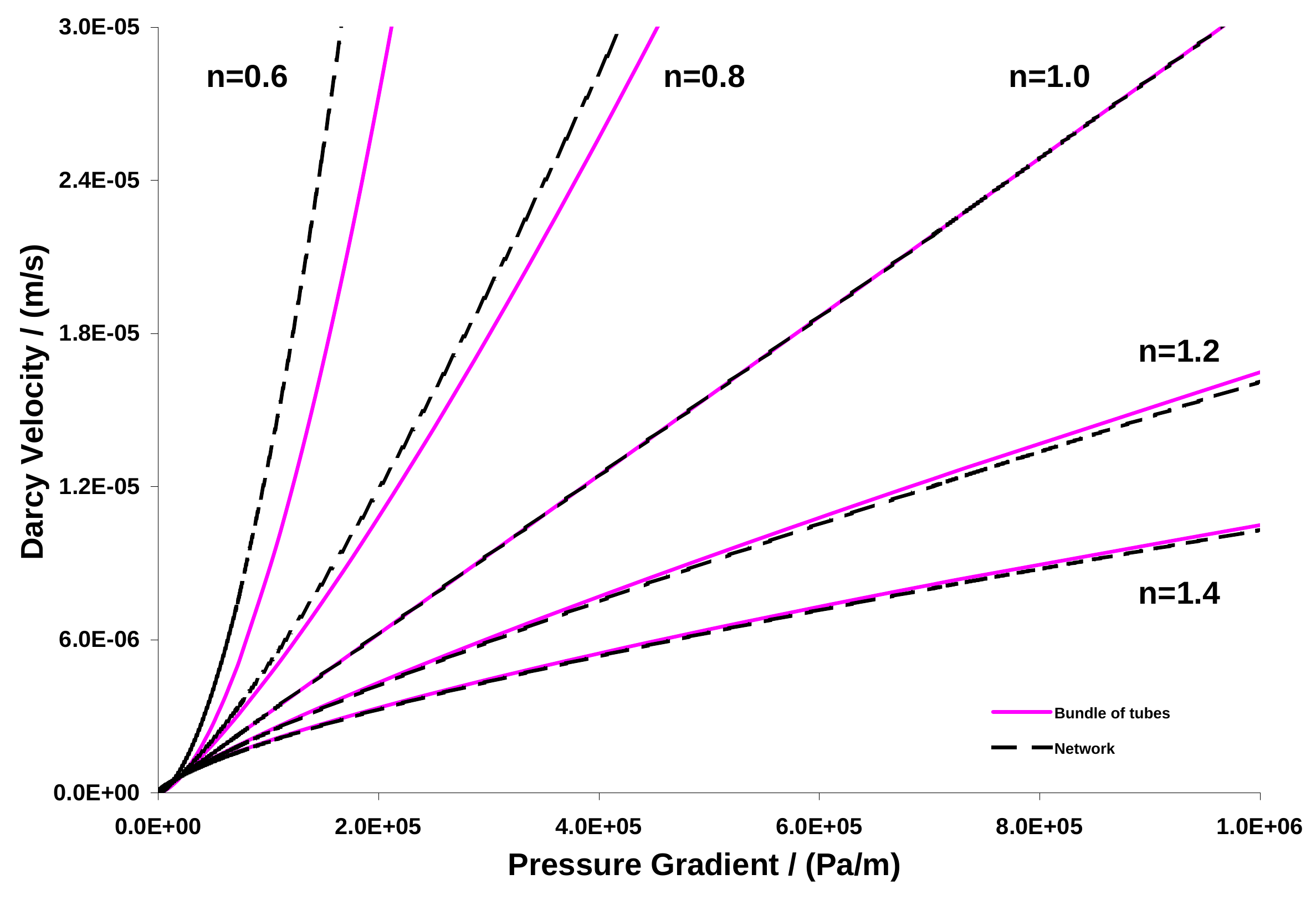}
  \caption[Comparison between \Berea\ network ($x_{_{l}}=0.5$, $x_{_{u}}=0.95$, $K=3.15$\,Darcy,
  $\phi=0.19$) and a bundle of tubes ($R=11.6\mu$m) for a \HB\ fluid with $\ysS=0.0$Pa and $C=0.1$Pa.s$^n$]
  {Comparison between \Berea\ network ($x_{_{l}}=0.5$, $x_{_{u}}=0.95$, $K=3.15$\,Darcy, $\phi=0.19$) and
  a bundle of tubes ($R=11.6\mu$m) for a \HB\ fluid with $\ysS=0.0$Pa and $C=0.1$Pa.s$^n$.}
  \label{BerTubeNetTau0}
\end{figure}

\begin{figure}[!b]
  \centering{}
  \includegraphics
  [scale=0.53]
  {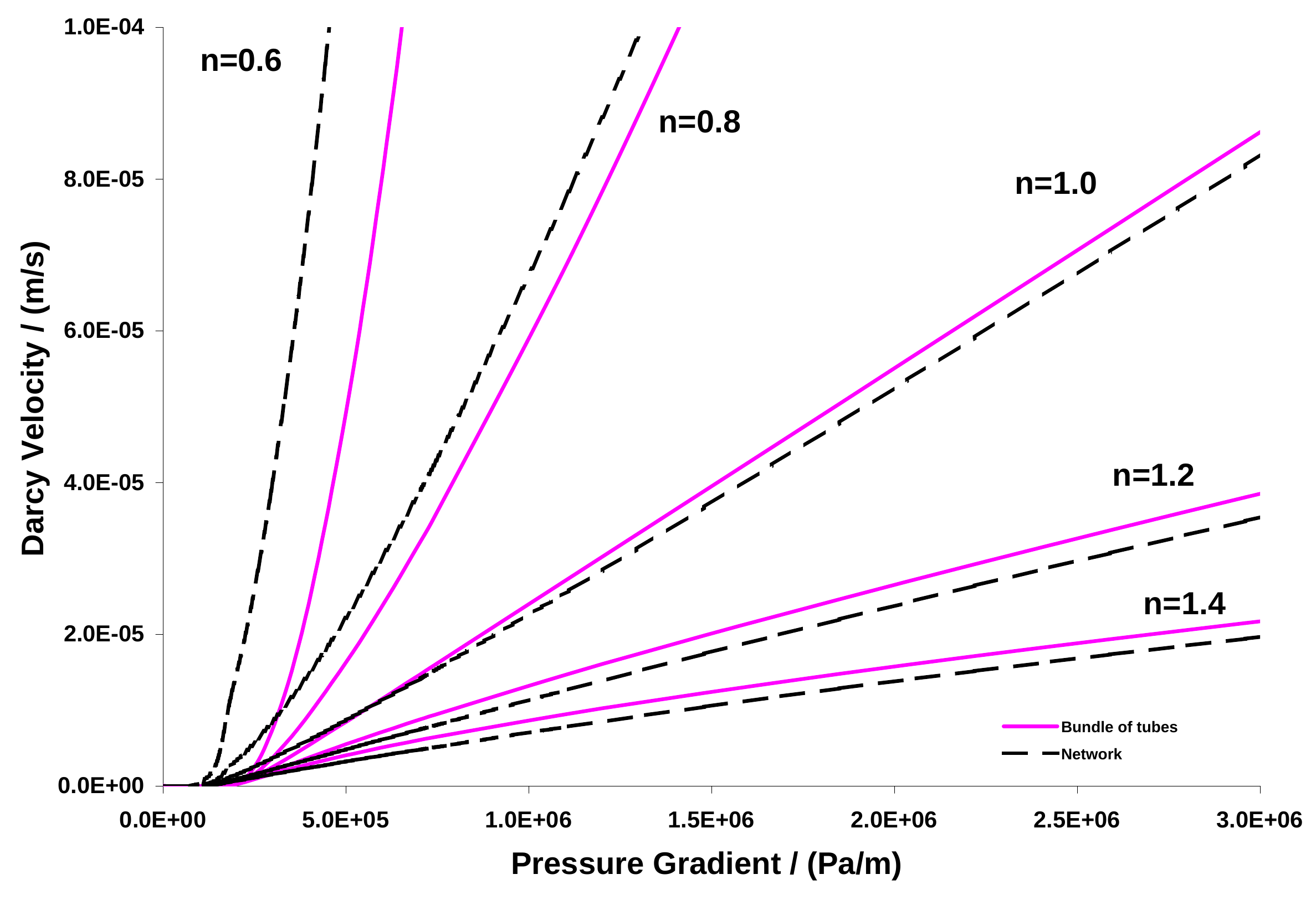}
  \caption[Comparison between \Berea\ network ($x_{_{l}}=0.5$, $x_{_{u}}=0.95$, $K=3.15$\,Darcy,
  $\phi=0.19$) and a bundle of tubes ($R=11.6\mu$m) for a \HB\ fluid with $\ysS=1.0$Pa and $C=0.1$Pa.s$^n$]
  {Comparison between \Berea\ network ($x_{_{l}}=0.5$, $x_{_{u}}=0.95$, $K=3.15$\,Darcy, $\phi=0.19$) and
  a bundle of tubes ($R=11.6\mu$m) for a \HB\ fluid with $\ysS=1.0$Pa and $C=0.1$Pa.s$^n$.}
  \label{BerTubeNetTau1}
\end{figure}

\begin{figure}[!h]
  \centering{}
  \includegraphics
  [scale=0.53]
  {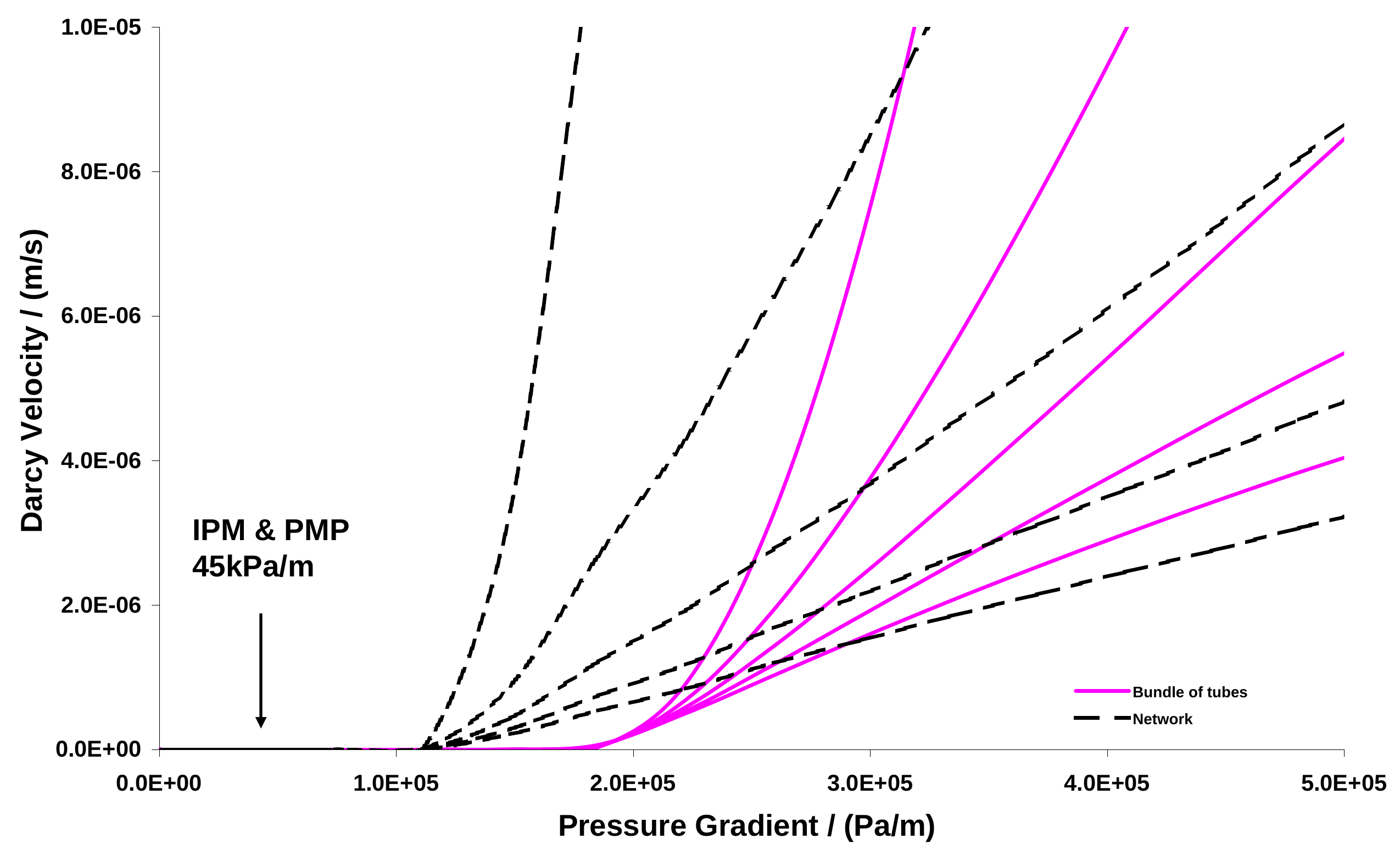}
  \caption[A magnified view of Figure (\ref{BerTubeNetTau1}) at the yield zone. The prediction of Invasion Percolation with
  Memory (IPM) and Path of Minimum Pressure (PMP) algorithms is indicated by the arrow]
  {A magnified view of Figure (\ref{BerTubeNetTau1}) at the yield zone. The prediction of Invasion Percolation with
  Memory (IPM) and Path of Minimum Pressure (PMP) algorithms is indicated by the arrow.}
  \label{BerTubeNetTau1Mag}
\end{figure}

\begin{figure}[!b]
  \centering{}
  \includegraphics
  [scale=0.53]
  {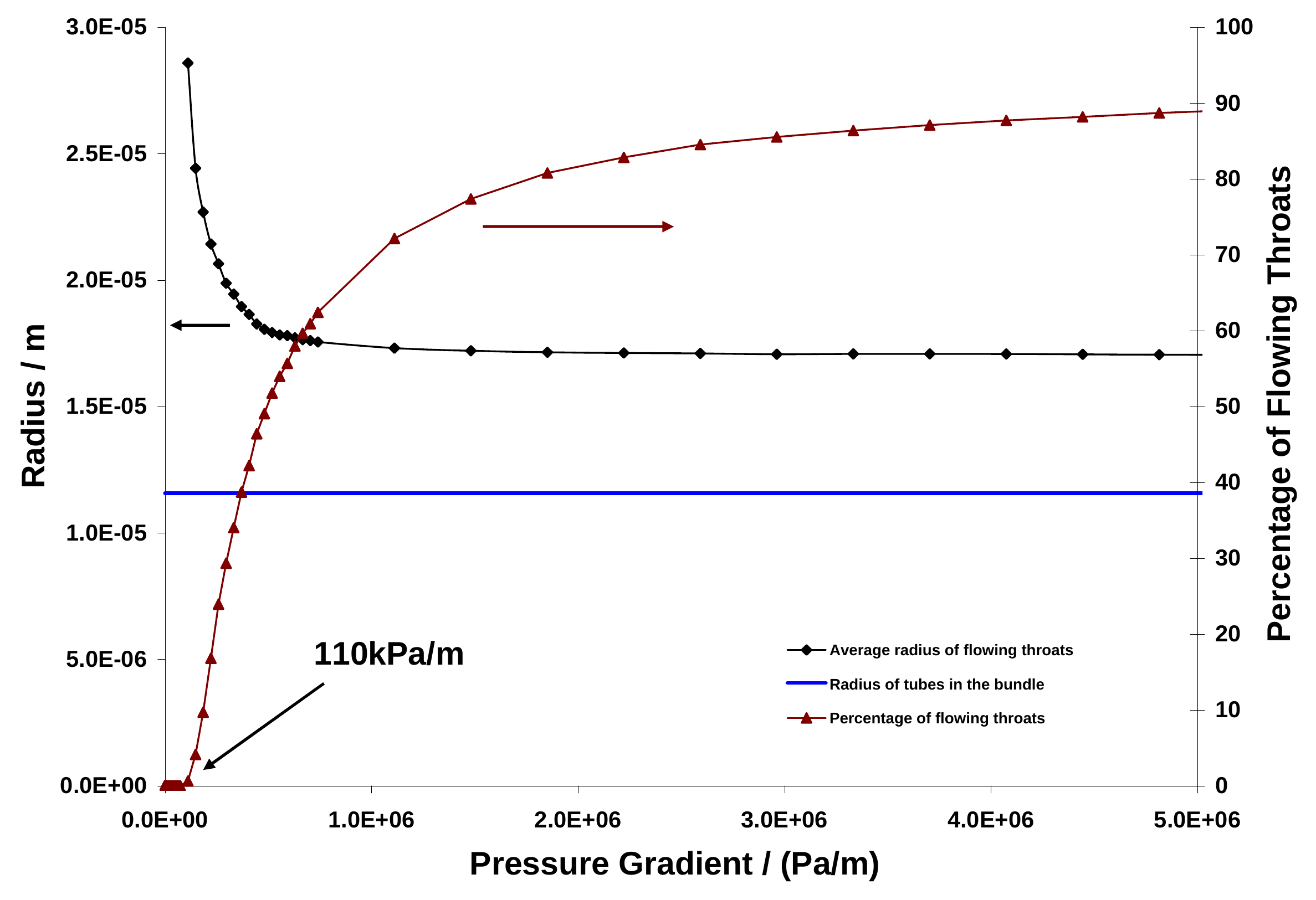}
  \caption[The radius of the bundle of tubes and the average radius of the non-blocked
  throats of the \Berea\ network, with the percentage of the total number of throats, as a
  function of pressure gradient for a \BING\ fluid ($n=1.0$) with $\ysS=1.0$Pa and $C=0.1$Pa.s]
  {The radius of the bundle of tubes and the average radius of the non-blocked
  throats of the \Berea\ network, with the percentage of the total number of throats, as a
  function of pressure gradient for a \BING\ fluid ($n=1.0$) with $\ysS=1.0$Pa and $C=0.1$Pa.s.}
  \label{BerTubeNetRadius}
\end{figure}

\begin{figure}[!h]
  \centering{}
  \includegraphics
  [scale=0.46]
  {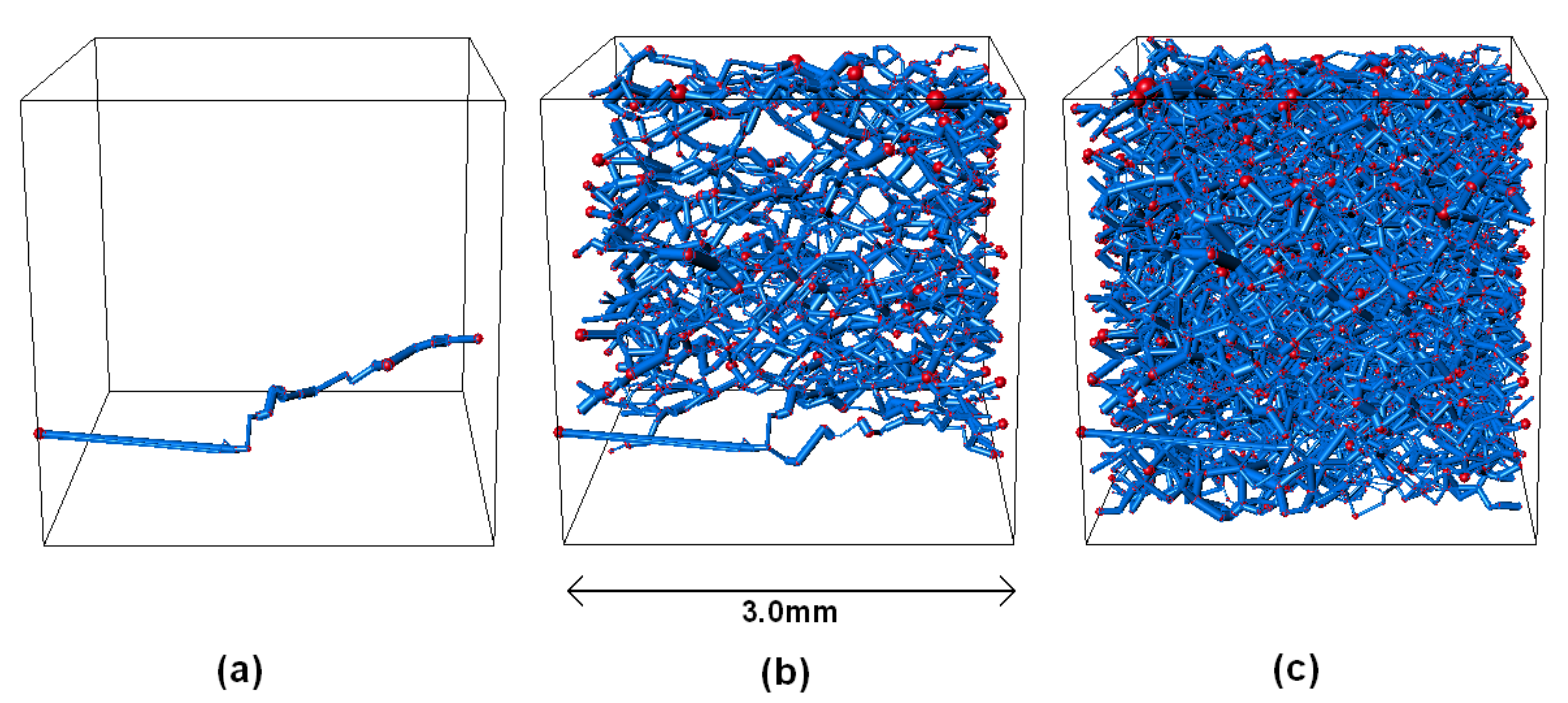}
  \caption[Visualization of the non-blocked elements of the \Berea\ network
  for a \BING\ fluid ($n=1.0$) with $\ysS=1.0$Pa and $C=0.1$Pa.s]
  {Visualization of the non-blocked elements of the \Berea\ network
  for a \BING\ fluid ($n=1.0$) with $\ysS=1.0$Pa and $C=0.1$Pa.s. The fraction of
  flowing elements is (a) 0.1\% (b) 9\% and (c) 41\%.}
  \label{BerVisualYieldStages}
\end{figure}

%XXXXXXXXXXXXXXXXXXXXXXXXXXXXXXXXXXXXXXXXXXXXXXXXXXXXXXXXXXXXXXXXXXXXXX
\section{Random vs. Cubic Networks Comparison} \label{RandomCubicComp}
In this section we present a brief comparison between a cubic
network on one hand and each one of the \sandp\ and \Berea\ networks
on the other. In both cases, the cubic network was generated by
``netgen'' of Valvatne with a truncated \Weibull\ distribution,
which is the only distribution available in this code, subject to
the two extreme limits for the throat radius and length of the
corresponding random network. The truncated \Weibull\ distribution
of a variable $\WV$ with a minimum value $\WV_{min}$ and a maximum
value $\WV_{max}$ is given, according to Valvatne
\cite{valvatnethesis}, by
\begin{equation}\label{Weibull}
    \WV = \WV_{min} + (\WV_{max} - \WV_{min}) \,
    \left[ -\WPo \ln \{ \Rand (1 - e^{-1/\WPo}) + e^{-1/\WPo} \}\right]^{1/\WPt}
\end{equation}
where $\WPo$ and $\WPt$ are parameters defining the shape of the
distribution and $\Rand$ is a random number between 0 and 1. The
values of $\WPo$ and $\WPt$ in each case were chosen to have a
reasonable match to the corresponding random network distribution.

\vspace{0.2cm}

The cubic network has also been modified to have the same
coordination number as the random network. Using ``poreflow'' of
Valvatne \cite{valvatnethesis} the cubic network was then tuned to
match the permeability and porosity of the corresponding random
network. The simulation results from our \nNEW\ code for a \HB\
fluid with $C=0.1$Pa.s$^n$ and $n=0.6, 0.8, 1.0, 1.2, 1.4$ for both
cases of yield-free fluid ($\ysS=0.0$Pa) and \yields\ fluid
($\ysS=1.0$Pa) are presented in Figures
(\ref{CubicSPcomparison0}-\ref{CubicBcomparison1}).

\vspace{0.2cm}

As can be seen, the results match very well for the \sandp\ network
with a yield-free fluid, as in Figure (\ref{CubicSPcomparison0}).
For the \sandp\ with a \yields\ fluid, as in Figure
(\ref{CubicSPcomparison1}), there is a shift between the two
networks and the cubic flow lags behind the \sandp. This can be
explained by the dependency of the network yield on the actual void
space characteristics rather than the network bulk properties such
as permeability and porosity.

\vspace{0.2cm}

For the \Berea\ network, the match between the two networks is poor
in both cases. For non-\yields\ fluids, as presented in Figure
(\ref{CubicBcomparison0}), the two networks match in the \NEW\ case
by definition. However, the match is poor for the \nNEW\ fluids. As
for \yields\ fluids, seen in Figure (\ref{CubicBcomparison1}), the
divergence between the two networks in the \shThin\ cases is large
with the cubic network underestimating the flow. Similarly for the
\BING\ fluid though the divergence is less serious. As for the two
cases of \shThik, the two networks produce reasonably close
predictions with the cubic network giving higher flow rate. Several
interacting factors, like the ones presented in the random networks
section, can explain this behavior. However, the general conclusion
is that the \nNEW\ behavior in general depends on the actual void
space characteristics rather than the network bulk properties. The
inhomogeneity of the \Berea\ network may also have aggravated the
situation and exacerbated the shift.

\vspace{0.2cm}

It should be remarked that the cubic results in general depends on
the cubic network realization as there is an element of randomness
in the cubic network distributions during its generation. The
dependency is more obvious in the case of \yields\ fluids as the
yield of a network is highly dependent on the topology and geometry
of the void space. Similar argument can be presented for the random
networks since they are generated by simulating the geological
processes by which the porous medium was formed.

%%%%%%%%%%%%%%%%%%%%%%%%%%%%%%%   cubic-SP comparison   %%%%%%%%%%%%%%%%%%%%%%%%%%%%%%%%%%%%%%%%%
\newpage
\begin{figure}[!h]
  \centering{}
  \includegraphics
  [scale=0.51]
  {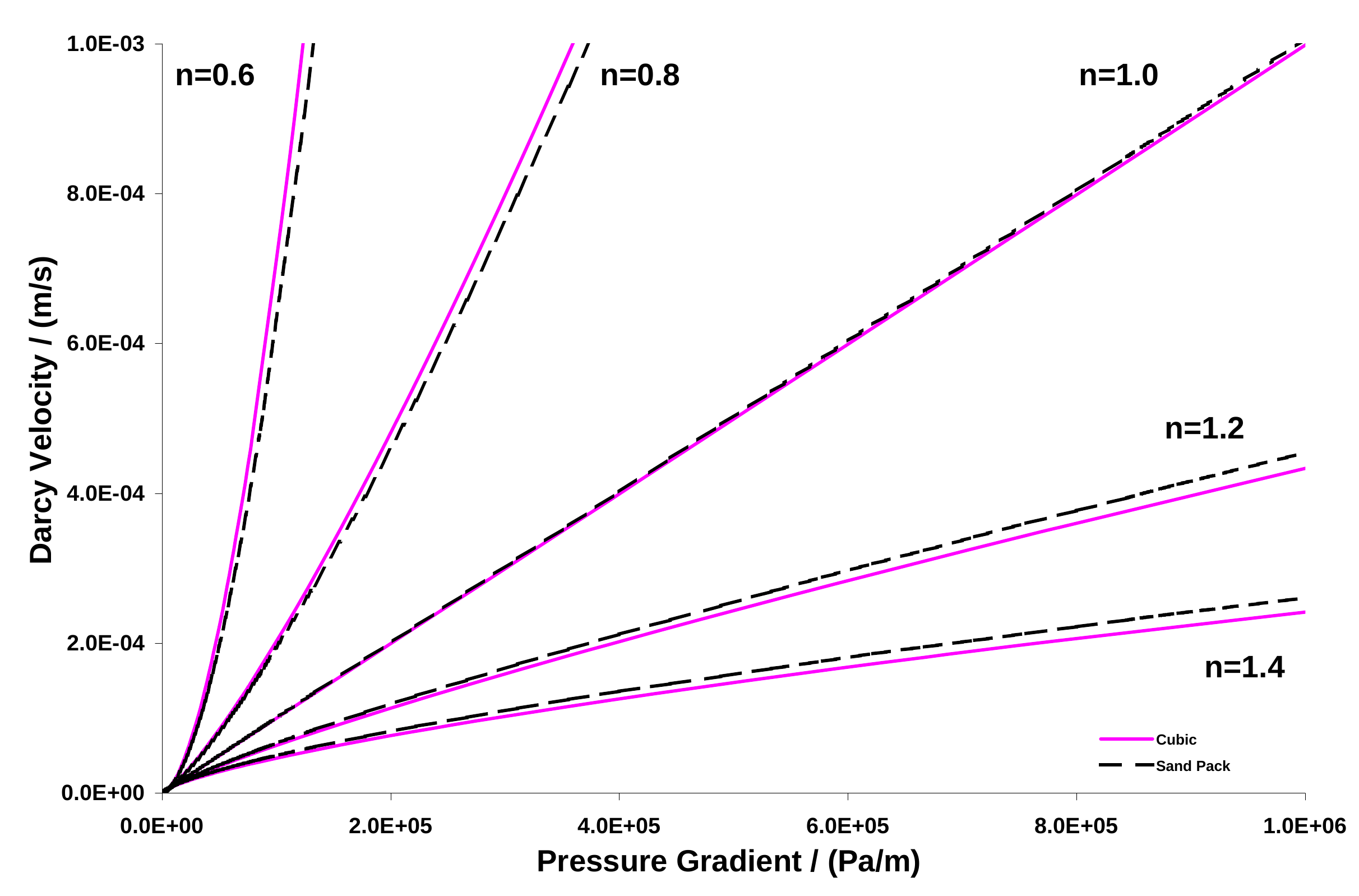}
  \caption[Comparison between the \sandp\ network ($x_{_{l}}=0.5$, $x_{_{u}}=0.95$, $K=102$\,Darcy,
  $\phi=0.35$) and a cubic network having the same throat distribution, coordination
  number, permeability and porosity for a \HB\ fluid with $\ysS=0.0$Pa and $C=0.1$Pa.s$^n$]
  {Comparison between the \sandp\ network ($x_{_{l}}=0.5$, $x_{_{u}}=0.95$, $K=102$\,Darcy,
  $\phi=0.35$) and a cubic network having the same throat distribution, coordination
  number, permeability and porosity for a \HB\ fluid with $\ysS=0.0$Pa and $C=0.1$Pa.s$^n$.}
  \label{CubicSPcomparison0}
\end{figure}
\vspace{0.5cm}
\begin{figure}[!h]
  \centering{}
  \includegraphics
  [scale=0.51]
  {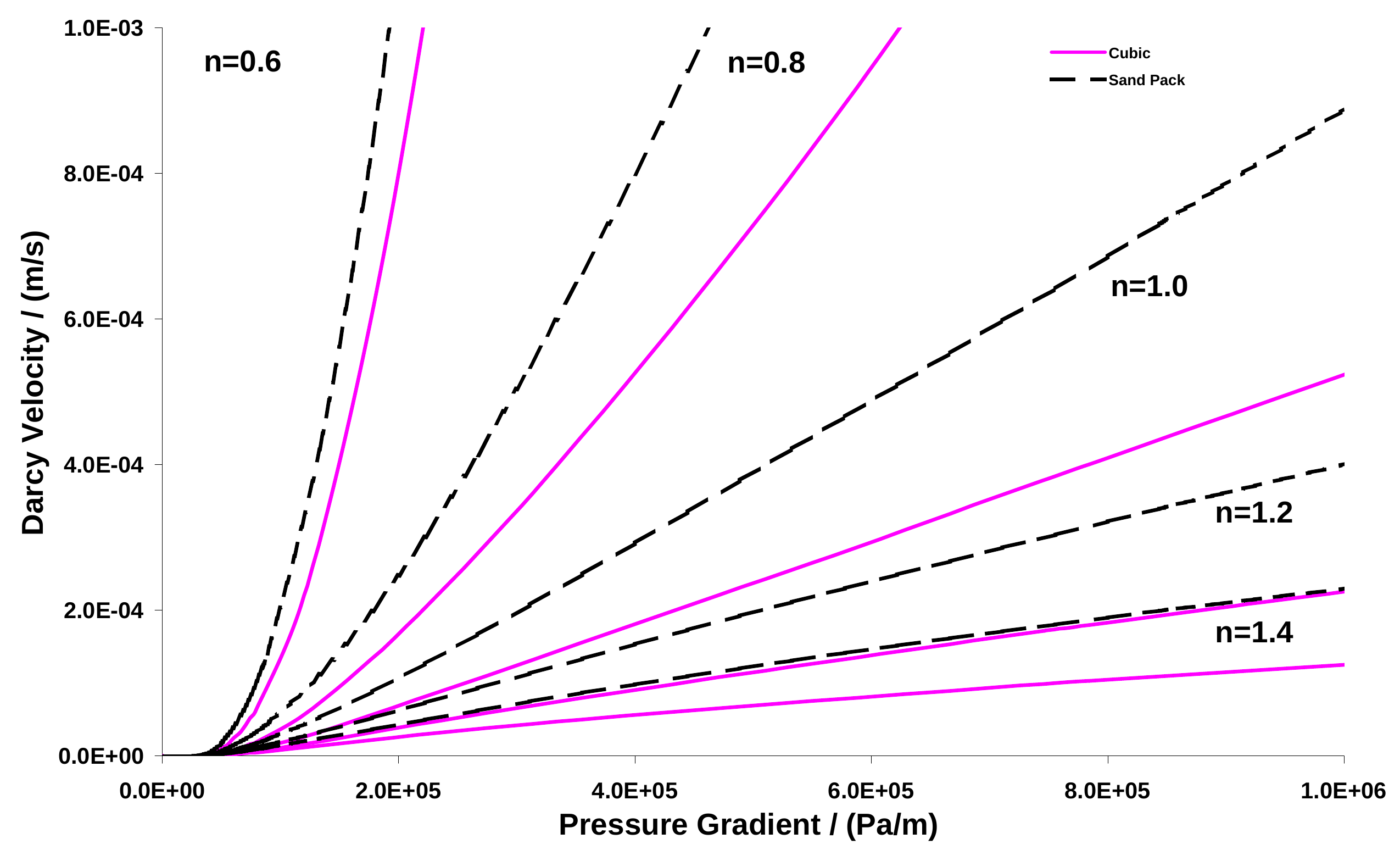}
  \caption[Comparison between the \sandp\ network ($x_{_{l}}=0.5$, $x_{_{u}}=0.95$, $K=102$\,Darcy,
  $\phi=0.35$) and a cubic network having the same throat distribution, coordination
  number, permeability and porosity for a \HB\ fluid with $\ysS=1.0$Pa and $C=0.1$Pa.s$^n$]
  {Comparison between the \sandp\ network ($x_{_{l}}=0.5$, $x_{_{u}}=0.95$, $K=102$\,Darcy,
  $\phi=0.35$) and a cubic network having the same throat distribution, coordination
  number, permeability and porosity for a \HB\ fluid with $\ysS=1.0$Pa and $C=0.1$Pa.s$^n$.}
  \label{CubicSPcomparison1}
\end{figure}

%%%%%%%%%%%%%%%%%%%%%%%%%%%%%%%   cubic-Berea comparison   %%%%%%%%%%%%%%%%%%%%%%%%%%%%%%%%%%%%%%%%%
\newpage
\begin{figure}[!h]
  \centering{}
  \includegraphics
  [scale=0.51]
  {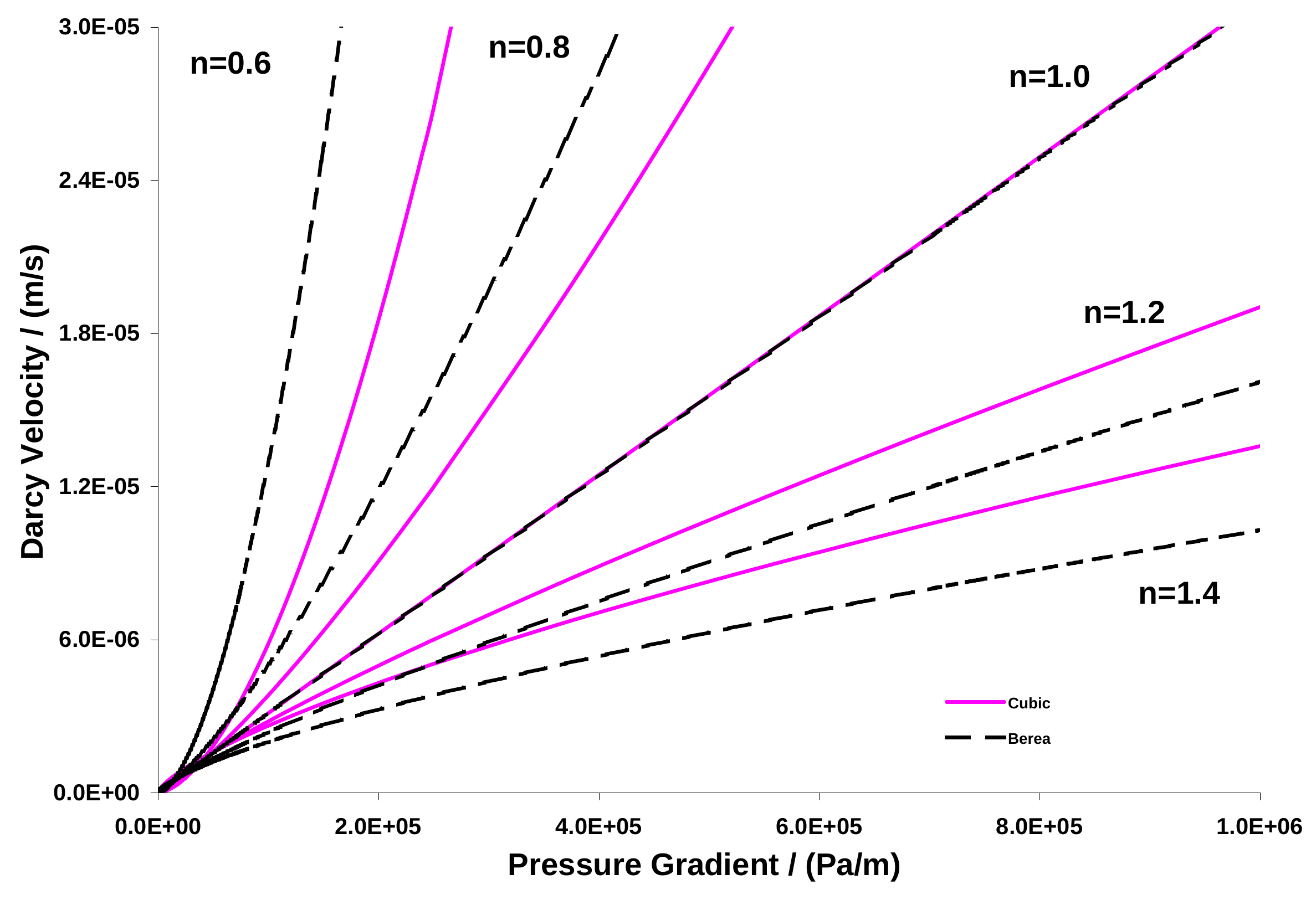}
  \caption[Comparison between the \Berea\ network ($x_{_{l}}=0.5$, $x_{_{u}}=0.95$, $K=3.15$\,Darcy,
  $\phi=0.19$) and a cubic network having the same throat distribution, coordination
  number, permeability and porosity for a \HB\ fluid with $\ysS=0.0$Pa and $C=0.1$Pa.s$^n$]
  {Comparison between the \Berea\ network ($x_{_{l}}=0.5$, $x_{_{u}}=0.95$, $K=3.15$\,Darcy,
  $\phi=0.19$) and a cubic network having the same throat distribution, coordination
  number, permeability and porosity for a \HB\ fluid with $\ysS=0.0$Pa and $C=0.1$Pa.s$^n$.}
  \label{CubicBcomparison0}
\end{figure}
\vspace{0.5cm}
\begin{figure}[!h]
  \centering{}
  \includegraphics
  [scale=0.51]
  {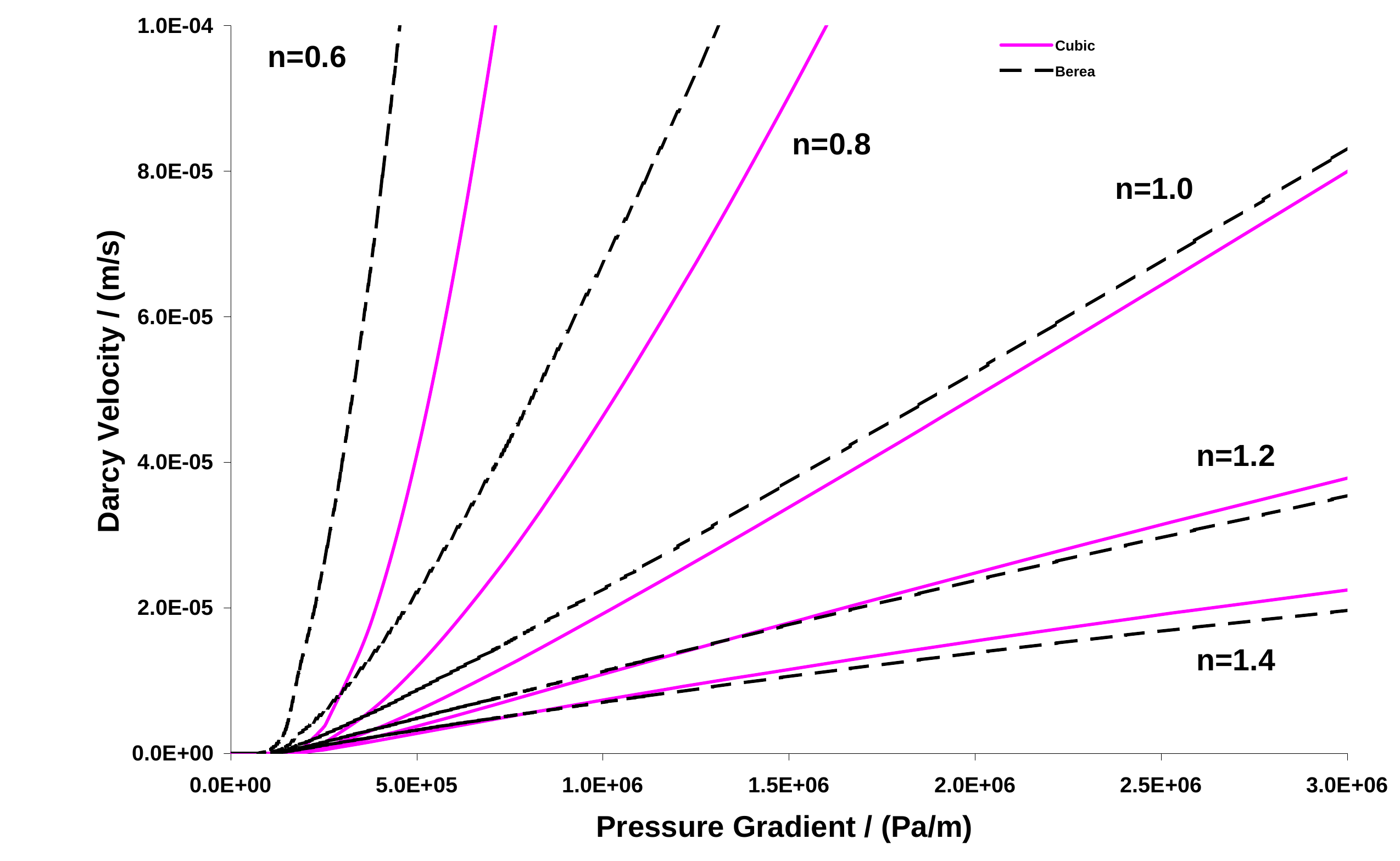}
  \caption[Comparison between the \Berea\ network ($x_{_{l}}=0.5$, $x_{_{u}}=0.95$, $K=3.15$\,Darcy,
  $\phi=0.19$) and a cubic network having the same throat distribution, coordination
  number, permeability and porosity for a \HB\ fluid with $\ysS=1.0$Pa and $C=0.1$Pa.s$^n$]
  {Comparison between the \Berea\ network ($x_{_{l}}=0.5$, $x_{_{u}}=0.95$, $K=3.15$\,Darcy,
  $\phi=0.19$) and a cubic network having the same throat distribution, coordination
  number, permeability and porosity for a \HB\ fluid with $\ysS=1.0$Pa and $C=0.1$Pa.s$^n$.}
  \label{CubicBcomparison1}
\end{figure}

%XXXXXXXXXXXXXXXXXXXXXXXXXXXXXXXXXXXXXXXXXXXXXXXXXXXXXXXXXXXXXXXXXXXXXX
\section{Bundle of Tubes vs. Experimental Data Comparison} \label{RandomCubicComp}
To gain a better insight into the bundle of tubes model, which we
used to carry a preliminary assessment to the random network model,
we present in Figures (\ref{ParkBundleCoarse}) and
(\ref{ParkBundleFine}) a comparison between the bundle of tubes
model predictions and a sample of the Park's \HB\ experimental data
that will be thoroughly discussed in Section (\ref{ParkHB}). The
bulk rheology of these data sets can be found in Table
(\ref{parkHerschelTable}).

\vspace{0.2cm}

Although, the experimental data do not match well to the bundle of
tubes model results, the agreement is good enough for such a crude
model. A remarkable feature is the close similarity between the
bundle of tubes curves seen in Figures (\ref{ParkBundleCoarse}) and
(\ref{ParkBundleFine}) and the network curves seen in Figures
(\ref{ParkHBCoarse}) and (\ref{ParkHBFine}). This should be expected
since the network results are produced with a scaled version of the
\sandp\ network, and as we observed in Section
(\ref{SPBundleComparison}) there is a good match between the \sandp\
network and the bundle of tubes model.

%%%%%%%%%%%%%%%%%%%%%%%%%%%%%%%   Park-Bundle Comparison  %%%%%%%%%%%%%%%%%%%%%%%%%%%%%%%%%%%%%%%%%
\newpage
\begin{figure}[!h]
  \centering{}
  \includegraphics
  [scale=0.5]
  {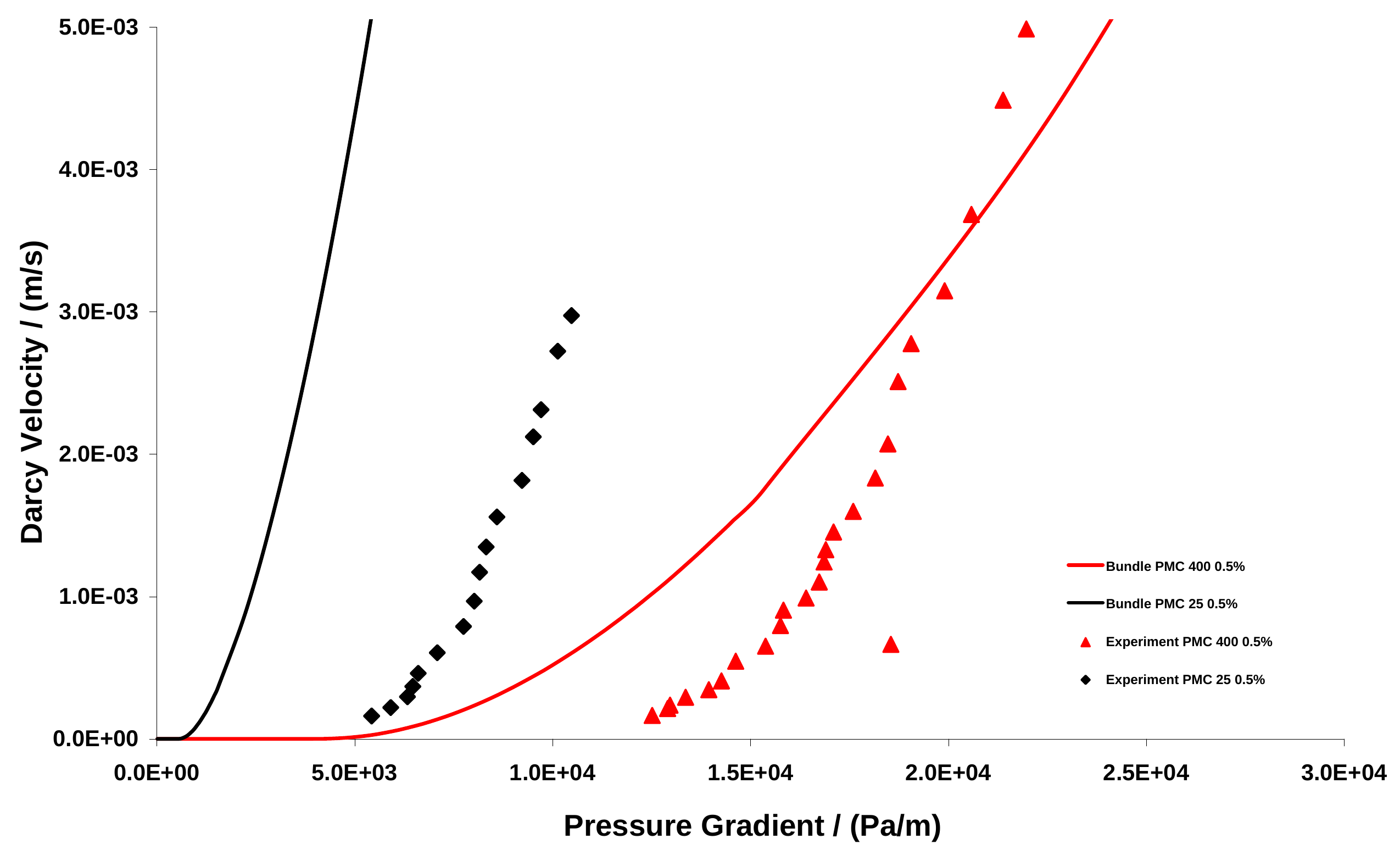}
  \caption[Sample of Park's \HB\ experimental data group for aqueous solutions of PMC 400 and PMC 25
  flowing through a coarse packed bed of glass beads having $K=3413$\,Darcy and $\phi=0.42$
  alongside the results of a bundle of tubes with $R=255\mu$m]
  {Sample of Park's \HB\ experimental data group for aqueous solutions of PMC 400 and PMC 25
  flowing through a coarse packed bed of glass beads having $K=3413$\,Darcy and $\phi=0.42$
  alongside the results of a bundle of tubes with $R=255\mu$m.}
  \label{ParkBundleCoarse}
\end{figure}
\vspace{0.5cm}
\begin{figure}[!h]
  \centering{}
  \includegraphics
  [scale=0.5]
  {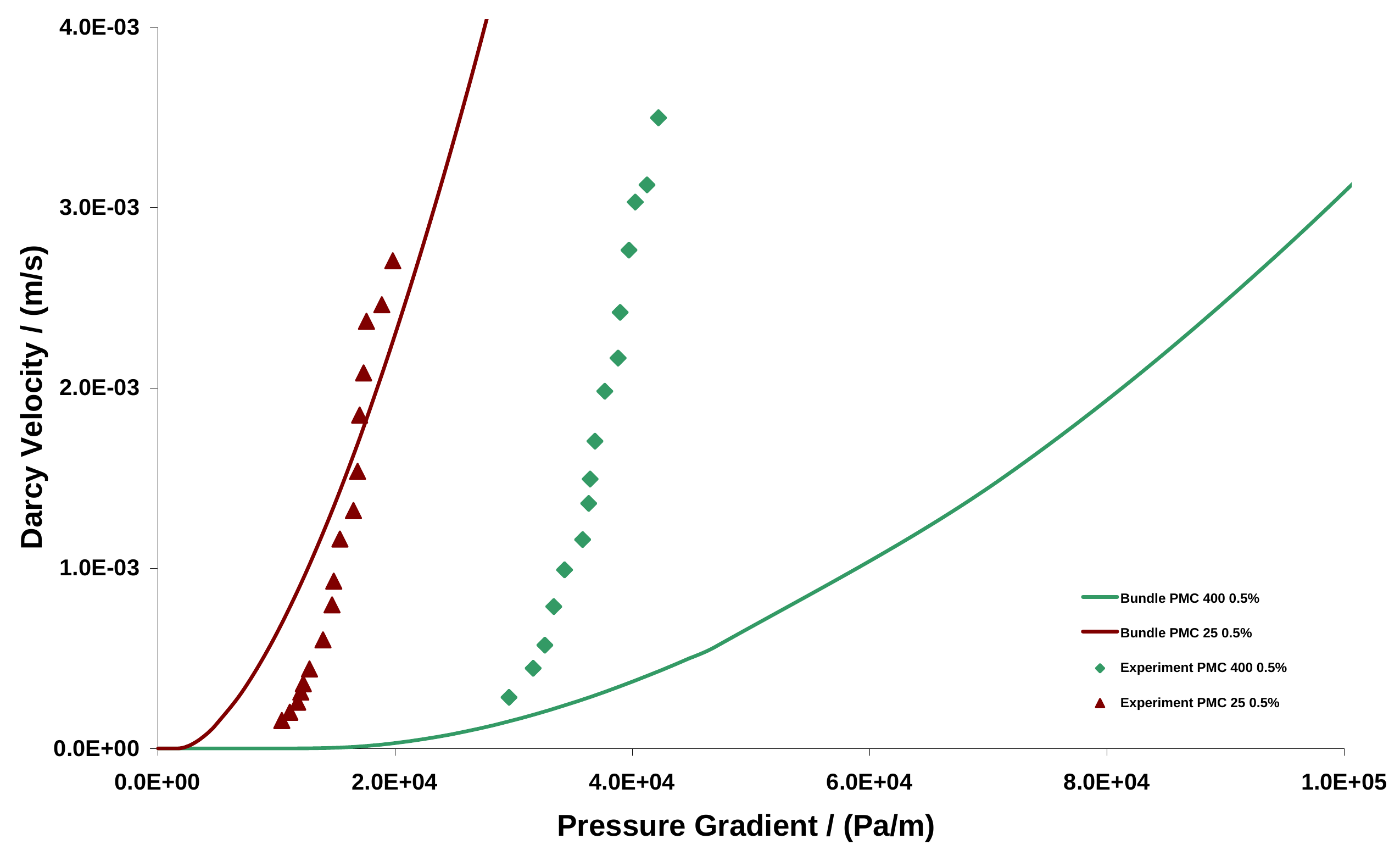}
  \caption[Sample of Park's \HB\ experimental data group for aqueous solutions of PMC 400 and PMC 25
  flowing through a fine packed bed of glass beads having $K=366$\,Darcy and $\phi=0.39$
  alongside the results of a bundle of tubes with $R=87\mu$m]
  {Sample of Park's \HB\ experimental data group for aqueous solutions of PMC 400 and PMC 25
  flowing through a fine packed bed of glass beads having $K=366$\,Darcy and $\phi=0.39$
  alongside the results of a bundle of tubes with $R=87\mu$m.}
  \label{ParkBundleFine}
\end{figure}

\def\baselinestretch{1}
\chapter{Experimental Validation for \ELLIS\ and \HB\ Models} \label{Experimental}
\def\baselinestretch{1.66}
We implemented the \ELLIS\ and \HB\ models in our \nNEW\ code using
the analytical expressions for the volumetric flow rate, i.e.
Equation (\ref{QEllis}) for \ELLIS\ and Equation (\ref{QHerschel})
for \HB. In this chapter, we will discuss the validation of our
network model by the few complete experimental data collections that
we found in the literature. In all the cases presented in this
chapter, the \sandp\ network was used after scaling to match the
permeability of the porous media used in the experiments. The reason
for using the \sandp\ instead of \Berea\ is that the \sandp\ is a
better match, though not ideal, to the packed beds used in the
experiments in terms of homogeneity and tortuosity. The length
factor which we used to scale the networks is given by
$\sqrt{\frac{K_{_{exp}}}{K_{_{net}}}}$ where $K_{_{exp}}$ is the
permeability of the experimental pack and $K_{_{net}}$ is the
permeability of the original network as obtained from \NEW\ flow
simulation.

%XXXXXXXXXXXXXXXXXXXXXXXXXXXXXXXXXXXXXXXXXXXXXXXXXXXXXXXXXXXXXXXXXXXXXX
\section{\ELLIS\ Model} \label{}
Three complete collections of experimental data found in the
literature on \ELLIS\ fluid were investigated. Good agreement with
the network model results was obtained in most cases.

%SSSSSSSSSSSSSSSSSSSSSSSSSSSSSSSSSSSSS
\subsection{Sadowski} \label{}
In this collection \cite{sadowskithesis}, twenty complete data sets
of ten aqueous polymeric solutions flowing through packed beds of
lead shot or glass beads with various properties were investigated.
The bulk rheology was given by Sadowski in his dissertation and is
shown in Table (\ref{sadowskiEllisTable}) with the corresponding bed
properties. The \insitu\ experimental data was obtained from the
relevant tables in the dissertation. The permeability of the beds,
which is needed to scale our \sandp\ network, was obtained from the
formula suggested by Sadowski, that is
\begin{equation}\label{Kformula}
    K = \frac{D_{p}^{2}}{C^{''}} \frac{\epsilon^{3}}{(1-\epsilon)^{2}}
\end{equation}
where $K$ is the absolute permeability of the bed, $D_{p}$ is the
diameter of the bed particles, $\epsilon$ is the porosity and
$C^{''}$ is a dimensionless constant assigned a value of 180 by
Sadowski.

\vspace{0.2cm}

A sample of the simulation results, with the corresponding
experimental data sets, is presented in Figures
(\ref{SadowskiEllisQP1}) and (\ref{SadowskiEllisQP2}) as Darcy
velocity versus pressure gradient, and Figures
(\ref{SadowskiEllisVQ1}) and (\ref{SadowskiEllisVQ2}) as apparent
viscosity versus Darcy velocity. As seen, the agreement between the
experimental data and the network simulation is very good in most
cases.

%%%%%%%%%%%%%%%%%%%%%%%%%%%%%%%   Sadowski Ellis q-P  %%%%%%%%%%%%%%%%%%%%%%%%%%%%%%%%%%%%%%
\newpage
\begin{figure}[!h]
  \centering{}
  \includegraphics
  [scale=0.5]
  {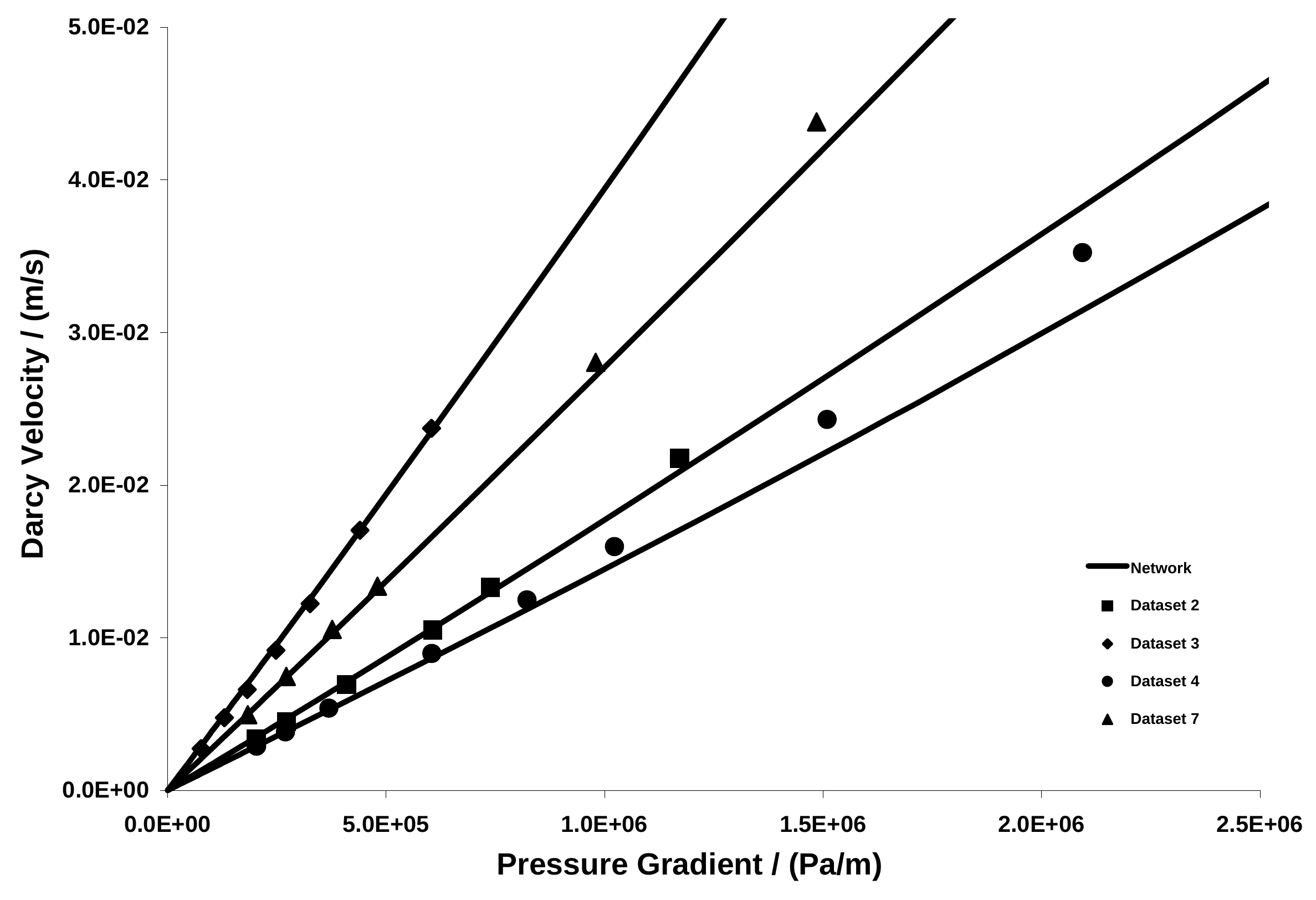}
  \caption[Sample of the Sadowski's \ELLIS\ experimental data sets (2,3,4,7) for a number of solutions with various
  concentrations and different bed properties alongside the simulation results obtained with
  scaled \sandp\ networks having the same $K$ presented as $q$ vs. $|\nabla P|$]
  {Sample of the Sadowski's \ELLIS\ experimental data sets (2,3,4,7) for a number of solutions with various
  concentrations and different bed properties alongside the simulation results obtained with
  scaled \sandp\ networks having the same $K$ presented as $q$ vs. $|\nabla P|$.}
  \label{SadowskiEllisQP1}
\end{figure}
\vspace{0.5cm}
\begin{figure}[!h]
  \centering{}
  \includegraphics
  [scale=0.5]
  {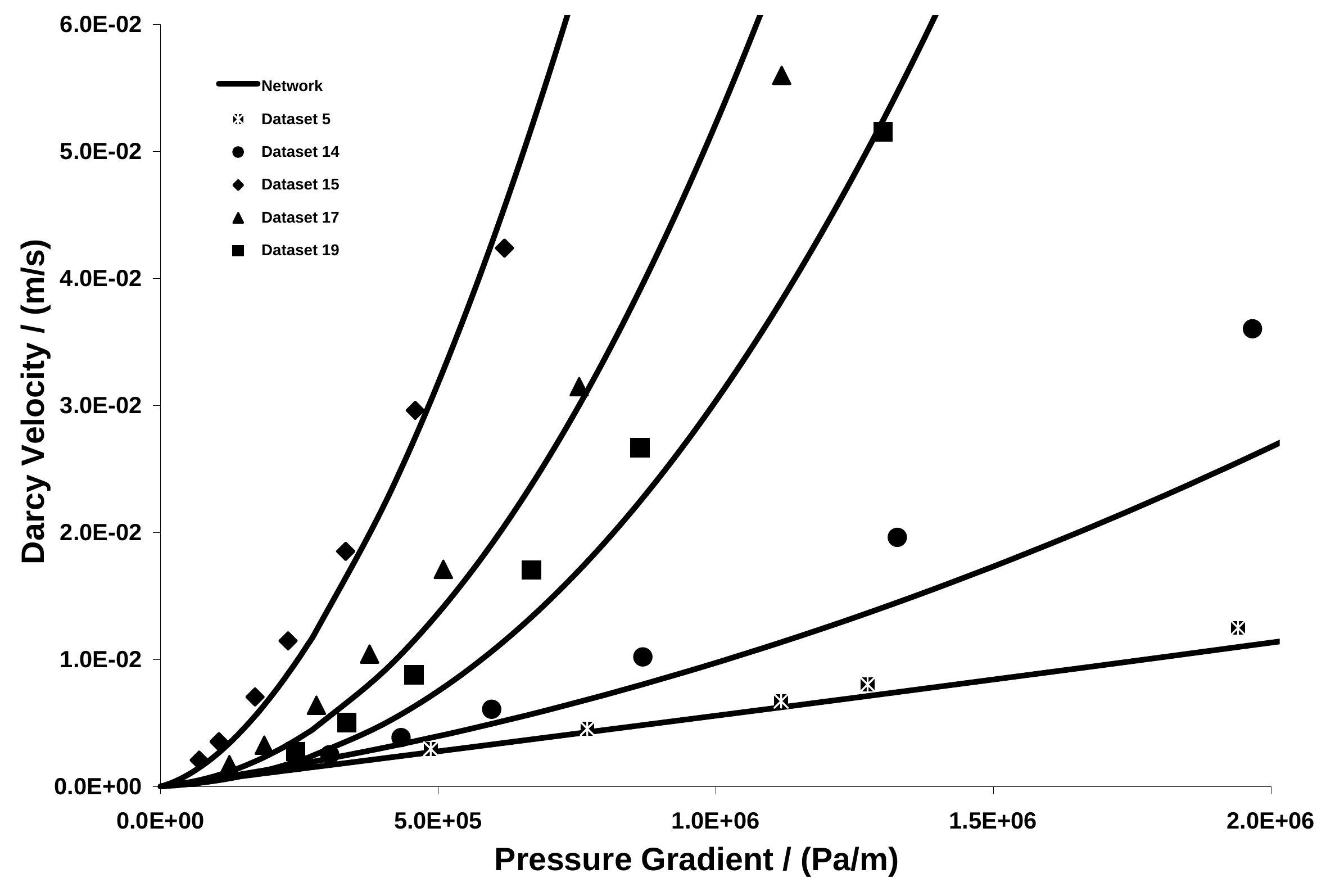}
  \caption[Sample of the Sadowski's \ELLIS\ experimental data sets (5,14,15,17,19) for a number of solutions with various
  concentrations and different bed properties alongside the simulation results obtained with
  scaled \sandp\ networks having the same $K$ presented as $q$ vs. $|\nabla P|$]
  {Sample of the Sadowski's \ELLIS\ experimental data sets (5,14,15,17,19) for a number of solutions with various
  concentrations and different bed properties alongside the simulation results obtained with
  scaled \sandp\ networks having the same $K$ presented as $q$ vs. $|\nabla P|$.}
  \label{SadowskiEllisQP2}
\end{figure}

%%%%%%%%%%%%%%%%%%%%%%%%%%%%%%%   Sadowski Ellis v-q  %%%%%%%%%%%%%%%%%%%%%%%%%%%%%%%%%%%%%%
\newpage
\begin{figure}[!h]
  \centering{}
  \includegraphics
  [scale=0.5]
  {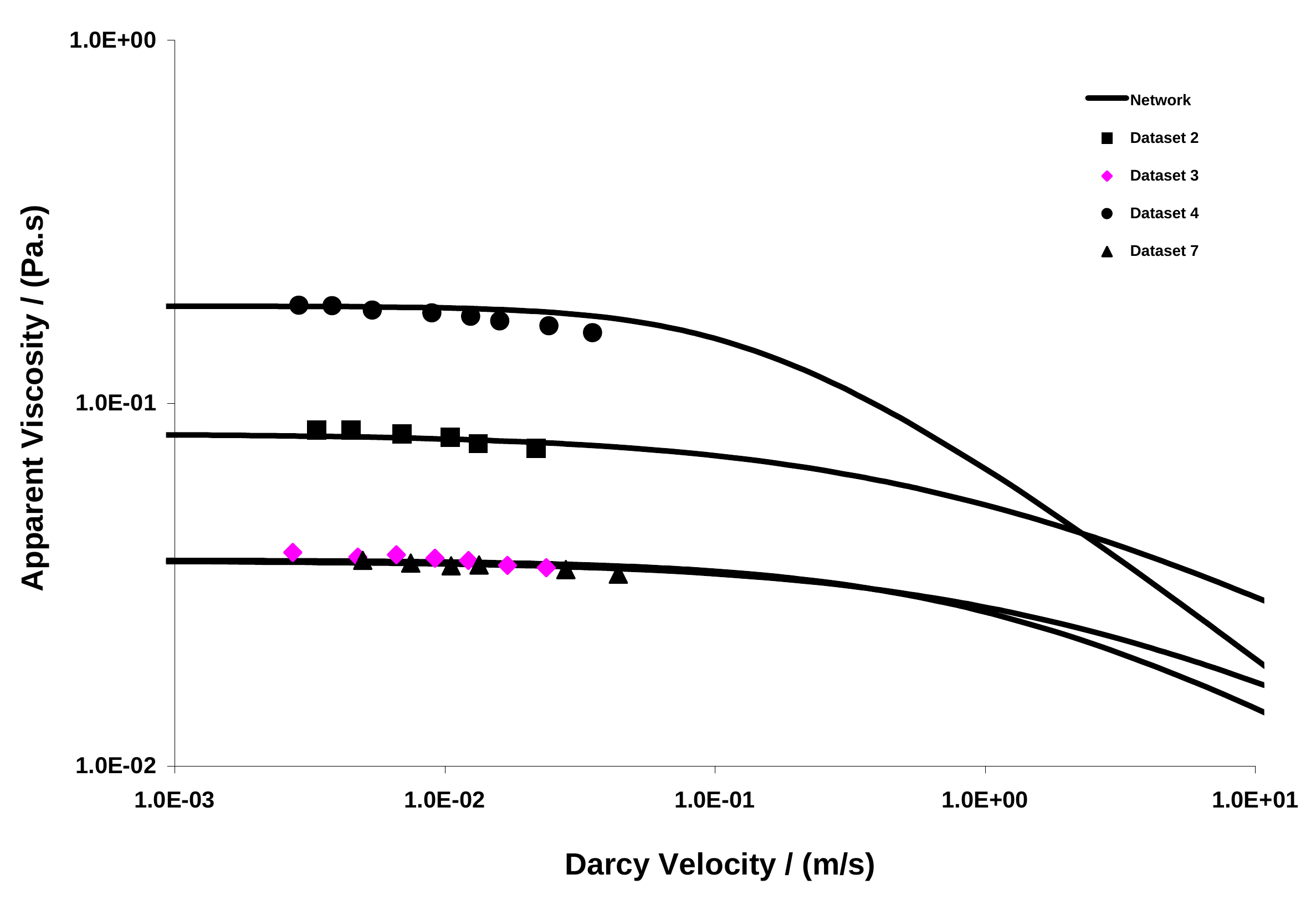}
  \caption[Sample of the Sadowski's \ELLIS\ experimental data sets (2,3,4,7) for a number of solutions with various
  concentrations and different bed properties alongside the simulation results obtained with
  scaled \sandp\ networks having the same $K$ presented as $\aVis$ vs. $q$]
  {Sample of the Sadowski's \ELLIS\ experimental data sets (2,3,4,7) for a number of solutions with various
  concentrations and different bed properties alongside the simulation results obtained with
  scaled \sandp\ networks having the same $K$ presented as $\aVis$ vs. $q$.}
  \label{SadowskiEllisVQ1}
\end{figure}
\vspace{0.5cm}
\begin{figure}[!h]
  \centering{}
  \includegraphics
  [scale=0.5]
  {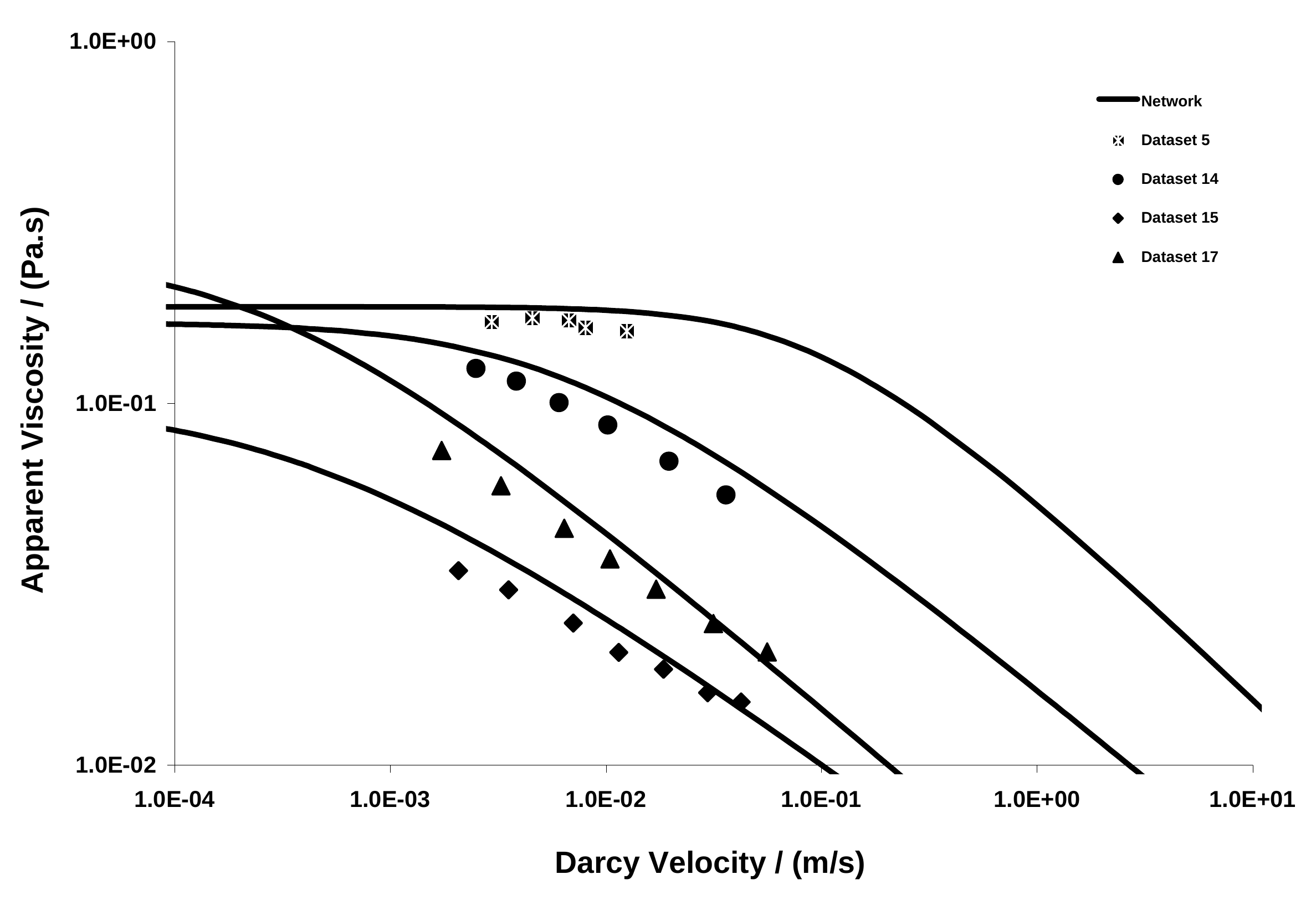}
  \caption[Sample of the Sadowski's \ELLIS\ experimental data sets (5,14,15,17) for a number of solutions with various
  concentrations and different bed properties alongside the simulation results obtained with
  scaled \sandp\ networks having the same $K$ presented as $\aVis$ vs. $q$]
  {Sample of the Sadowski's \ELLIS\ experimental data sets (5,14,15,17) for a number of solutions with various
  concentrations and different bed properties alongside the simulation results obtained with
  scaled \sandp\ networks having the same $K$ presented as $\aVis$ vs. $q$.}
  \label{SadowskiEllisVQ2}
\end{figure}

%%%%%%%%%%%%%%%%%%%%%%%%%%%%%%%%%%%%%%%%%%%%%
\begin{table} [!h]
\centering
\begin{tabular}{|c|c|c|c|c|c|c|}
\hline
           &      \multicolumn{ 4}{c}{{\bf Fluid Properties}} & \multicolumn{ 2}{|c|}{{\bf Bed Properties}} \\
\hline
{\bf Set} & {\bf Solution} & {$\lVis$ (Pa.s)} & {$\eAlpha$} & {$\hsS$ (Pa)} & {$K$ (m$^{2}$)} &  {$\phi$} \\
\hline
         1 & 18.5\% Carbowax 20-M &     0.0823 &      1.674 &     3216.0 &   3.80E-09 &     0.3690 \\
\hline
         2 & 18.5\% Carbowax 20-M &     0.0823 &      1.674 &     3216.0 &   1.39E-09 &     0.3812 \\
\hline
         3 & 14.0\% Carbowax 20-M &     0.0367 &      1.668 &     3741.0 &   1.38E-09 &     0.3807 \\
\hline
         4 & 6.0\% Elvanol 72-51 &     0.1850 &      2.400 &     1025.0 &   2.63E-09 &     0.3833 \\
\hline
         5 & 6.0\% Elvanol 72-51 &     0.1850 &      2.400 &     1025.0 &   1.02E-09 &     0.3816 \\
\hline
         6 & 6.0\% Elvanol 72-51 &     0.1850 &      2.400 &     1025.0 &   3.93E-09 &     0.3720 \\
\hline
         7 & 3.9\% Elvanol 72-51 &     0.0369 &      1.820 &     2764.0 &   9.96E-10 &     0.3795 \\
\hline
         8 & 1.4\% Natrosol - 250G &     0.0688 &      1.917 &       59.9 &   2.48E-09 &     0.3780 \\
\hline
         9 & 1.4\% Natrosol - 250G &     0.0688 &      1.917 &       59.9 &   1.01E-09 &     0.3808 \\
\hline
        10 & 1.4\% Natrosol - 250G &     0.0688 &      1.917 &       59.9 &   4.17E-09 &     0.3774 \\
\hline
        11 & 1.6\% Natrosol - 250G &     0.1064 &      1.971 &       59.1 &   2.57E-09 &     0.3814 \\
\hline
        12 & 1.6\% Natrosol - 250G &     0.1064 &      1.971 &       59.1 &   1.01E-09 &     0.3806 \\
\hline
        13 & 1.85\% Natrosol - 250G &     0.1670 &      2.006 &       60.5 &   3.91E-09 &     0.3717 \\
\hline
        14 & 1.85\% Natrosol - 250G &     0.1670 &      2.006 &       60.5 &   1.02E-09 &     0.3818 \\
\hline
        15 & 0.4\% Natrosol - 250H &     0.1000 &      1.811 &        2.2 &   1.02E-09 &     0.3818 \\
\hline
        16 & 0.4\% Natrosol - 250H &     0.1000 &      1.811 &        2.2 &   4.21E-09 &     0.3783 \\
\hline
        17 & 0.5\% Natrosol - 250H &     0.2500 &      2.055 &        3.5 &   1.03E-09 &     0.3824 \\
\hline
        18 & 0.5\% Natrosol - 250H &     0.2500 &      2.055 &        3.5 &   5.30E-09 &     0.3653 \\
\hline
        19 & 0.6\% Natrosol - 250H &     0.4000 &      2.168 &        5.2 &   1.07E-09 &     0.3862 \\
\hline
        20 & 0.6\% Natrosol - 250H &     0.4000 &      2.168 &        5.2 &   5.91E-09 &     0.3750 \\
\hline
\end{tabular}
\vspace{0.1cm} %
\caption[The bulk rheology and bed properties of Sadowski's Ellis
experimental data]
{The bulk rheology and bed properties of Sadowski's Ellis
experimental data.}
\label{sadowskiEllisTable} %
\end{table}

%SSSSSSSSSSSSSSSSSSSSSSSSSSSSSSSSSSSSS
\subsection{Park} \label{}
In this collection \cite{parkthesis}, four complete data sets on the
flow of aqueous polyacrylamide solutions with different weight
concentration in packed beds of glass beads were investigated. The
bulk rheology was given by Park and is shown in Table
(\ref{parkEllisTable}). The \insitu\ experimental data was obtained
from the relevant tables in his dissertation. The permeability of
the bed, which is needed to scale our \sandp\ network, was obtained
from Equation (\ref{Kformula}), as suggested by Park, with the
constant $C^{''}$ obtained from fitting his \NEW\ flow data, as will
be described in the Park's \HB\ experimental data section. The
simulation results compared to the experimental data points are
shown in Figure (\ref{ParkEllisQP}) as Darcy velocity versus
pressure gradient, and Figure (\ref{ParkEllisVQ}) as apparent
viscosity versus Darcy velocity. As seen, the agreement in all cases
is very good. The discrepancy observed in some cases in the high
flow rate region is due apparently to the absence of a high shear
\NEW\ plateau in \ELLIS\ model.

%%%%%%%%%%%%%%%%%%%%%%%%%%%%%%%%%%%%   Park Ellis q-P  %%%%%%%%%%%%%%%%%%%%%%%%%%%%%%%%%%%%%%%%
\newpage
\begin{figure} [!h]
  \centering{}
  \includegraphics
  [scale=0.47]
  {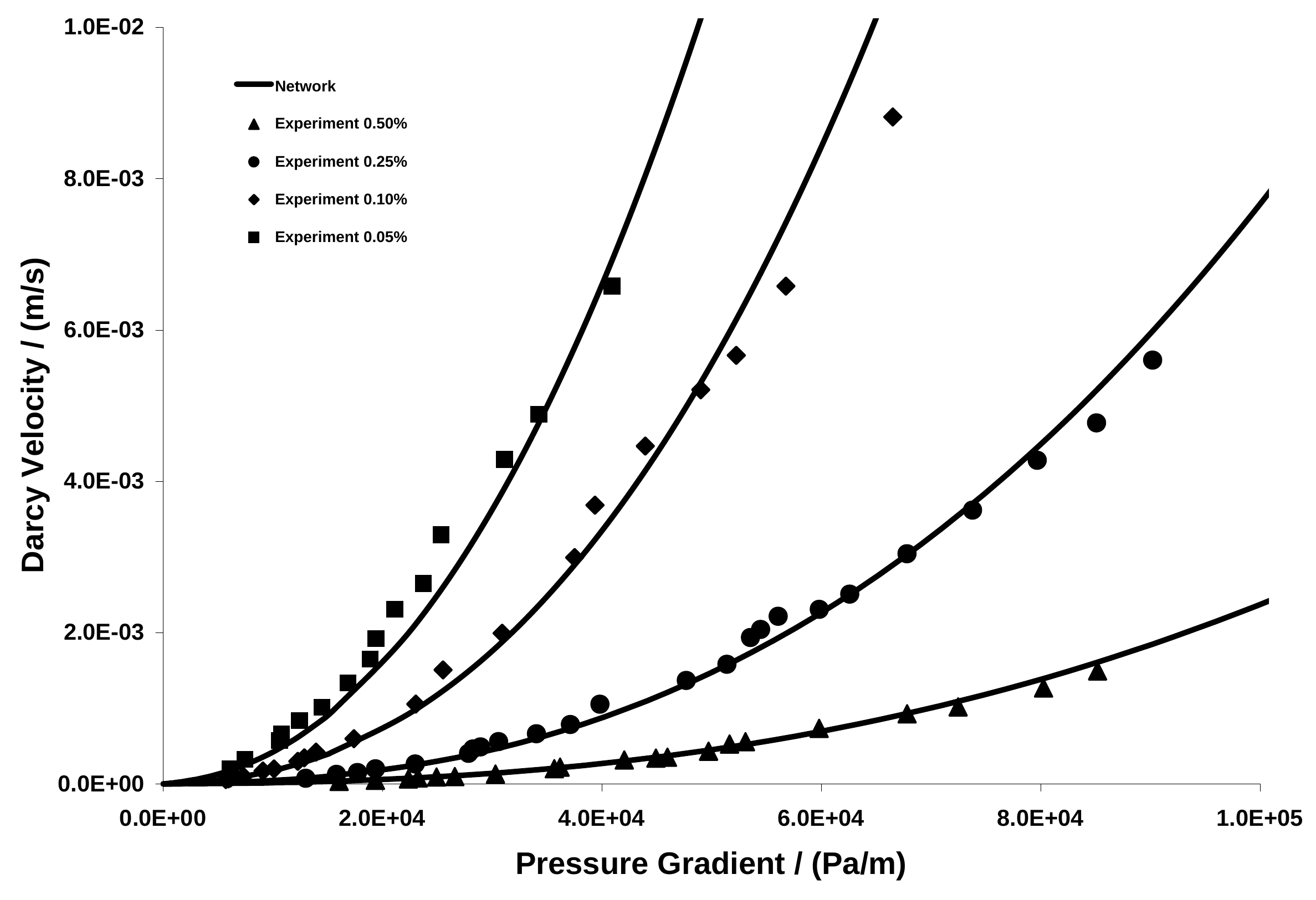}
  \caption[Park's \ELLIS\ experimental data sets for polyacrylamide solutions
  with 0.50\%, 0.25\%, 0.10\% and 0.05\% weight concentration flowing through a coarse packed bed of glass
  beads having $K=3413$\,Darcy and $\phi=0.42$ alongside the simulation results obtained with a
  scaled \sandp\ network having the same $K$ presented as $q$ vs. $|\nabla P|$]
  {Park's \ELLIS\ experimental data sets for polyacrylamide solutions with 0.50\%, 0.25\%, 0.10\%
  and 0.05\% weight concentration flowing through a coarse packed bed of glass
  beads having $K=3413$\,Darcy and $\phi=0.42$ alongside the simulation results obtained with a
  scaled \sandp\ network having the same $K$ presented as $q$ vs. $|\nabla P|$.}
  \label{ParkEllisQP}
\end{figure}

%%%%%%%%%%%%%%%%%%%%%%%%%%%%%%%%%%%%   Park Ellis v-q  %%%%%%%%%%%%%%%%%%%%%%%%%%%%%%%%%%%%%%%%

\begin{figure} [!h]
  \centering{}
  \includegraphics
  [scale=0.47]
  {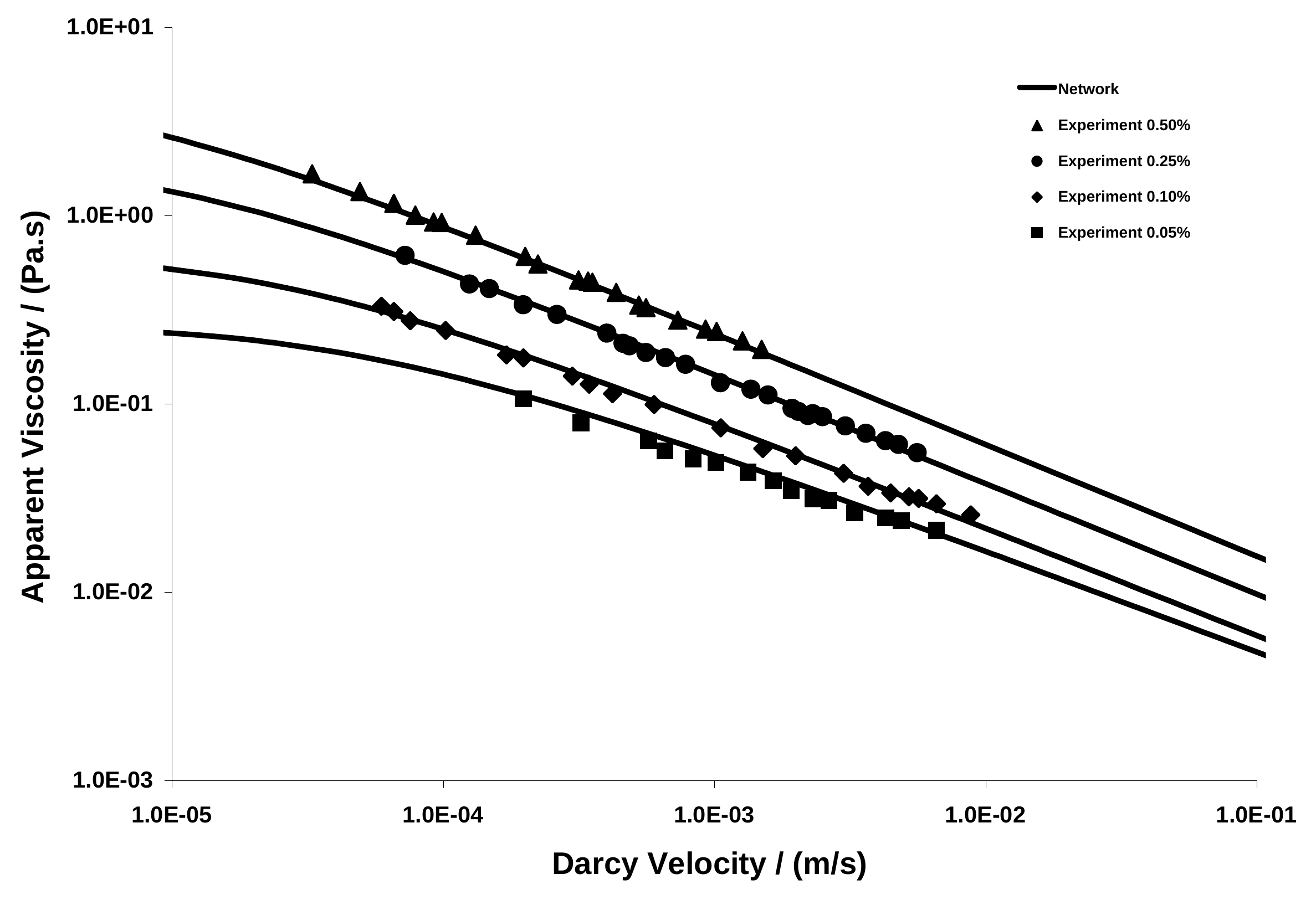}
  \caption[Park's \ELLIS\ experimental data sets for polyacrylamide solutions
  with 0.50\%, 0.25\%, 0.10\% and 0.05\% weight concentration flowing through a coarse packed bed of glass
  beads having $K=3413$\,Darcy and $\phi=0.42$ alongside the simulation results obtained with a
  scaled \sandp\ network having the same $K$ presented as $\aVis$ vs. $q$]
  {Park's \ELLIS\ experimental data sets for polyacrylamide solutions with 0.50\%, 0.25\%, 0.10\%
  and 0.05\% weight concentration flowing through a coarse packed bed of glass
  beads having $K=3413$\,Darcy and $\phi=0.42$ alongside the simulation results obtained with a
  scaled \sandp\ network having the same $K$ presented as $\aVis$ vs. $q$.}
  \label{ParkEllisVQ}
\end{figure}

%%%%%%%%%%%%%%%%%%%%%%%%%%%%%%%%%%%%%%%%%%%%%
\vspace{0cm}

\begin{table} [!h]
\centering
\begin{tabular}{|c|c|c|c|}
\hline
{\verb|   | \bf Solution \verb|   |} &   {\verb|   | $\lVis$ (Pa.s) \verb|   |} &    {\verb|    | $\eAlpha$ \verb|    |} & {\verb|   | $\hsS$ (Pa) \verb|  |} \\
\hline
   0.50\%  &    4.35213 &    2.4712 &     0.7185 \\
\hline
   0.25\% &    1.87862 &     2.4367 &     0.5310 \\
\hline
   0.10\% &    0.60870 &     2.3481 &     0.3920 \\
\hline
  0.05\%  &    0.26026 &     2.1902 &     0.3390 \\
\hline
\end{tabular}
\vspace{0.1cm} %
\caption[The bulk rheology of Park's \ELLIS\ experimental data]
{The bulk rheology of Park's \ELLIS\ experimental data.}
\vspace{1.0cm}
\label{parkEllisTable} %
\end{table}

%SSSSSSSSSSSSSSSSSSSSSSSSSSSSSSSSSSSSS
\subsection{Balhoff} \label{}
A complete data set for guar gum solution of 0.72\,\% concentration
with \ELLIS\ parameters $\lVis=2.672$\,Pa.s, $\eAlpha=3.46$ and
$\hsS=9.01$\,Pa flowing through a packed bed of glass beads having
$K=4.19 \times 10^{-9}$\,m$^{2}$ and $\phi=0.38$ was investigated.
The bulk rheology was given by Balhoff in his dissertation
\cite{balhoffthesis} and the \insitu\ experimental data was obtained
from Balhoff by private communication.

\vspace{0.2cm}

The simulation results with the experimental data points are
presented in Figure (\ref{BalhoffEllisQP}) as Darcy velocity versus
pressure gradient, and Figure (\ref{BalhoffEllisVQ}) as apparent
viscosity versus Darcy velocity. As seen, the agreement is very
good. Again, a discrepancy is observed in the high flow rate region
because of the absence of a high-shear \NEW\ plateau in the \ELLIS\
model.

%%%%%%%%%%%%%%%%%%%%%%%%%%%%%%%%%%%   Balhoff Ellis q-P & v-q  %%%%%%%%%%%%%%%%%%%%%%%%%%%%%%%%%%%
%
\begin{figure} [!h]
  \centering{}
  \includegraphics
  [scale=0.5]
  {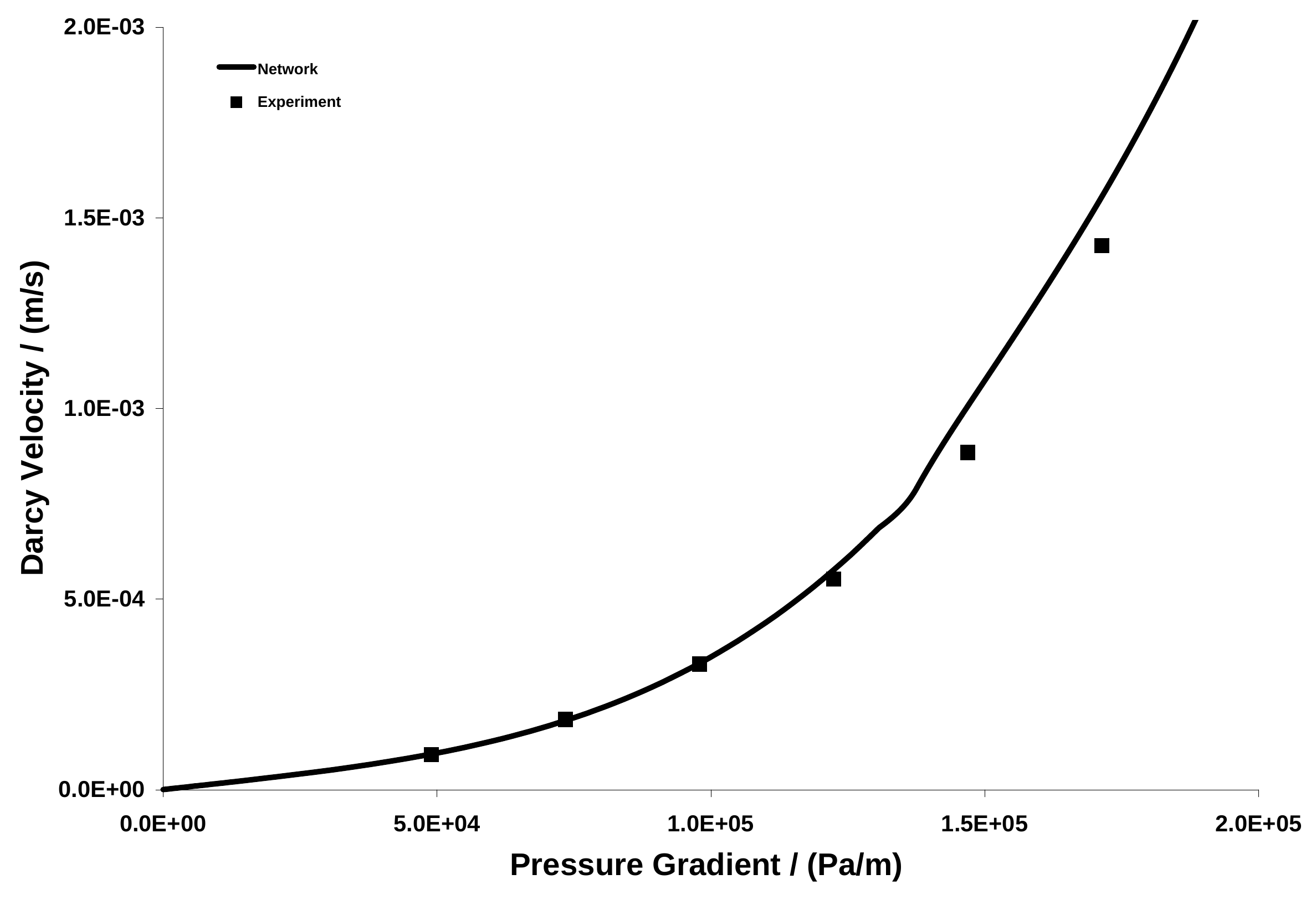}
  \caption[Balhoff's \ELLIS\ experimental data set for guar gum solution with 0.72\% concentration flowing
  through a packed bed of glass beads having $K=4.19 \times 10^{-9}$\,m$^{2}$ and $\phi=0.38$
  alongside the simulation results obtained with a scaled \sandp\ network having the same $K$ presented as $q$ vs. $|\nabla P|$]
  {Balhoff's \ELLIS\ experimental data set for guar gum solution with 0.72\% concentration flowing
  through a packed bed of glass beads having $K=4.19 \times 10^{-9}$\,m$^{2}$ and $\phi=0.38$
  alongside the simulation results obtained with a scaled \sandp\ network having the same $K$ presented as $q$ vs. $|\nabla P|$.}
  \label{BalhoffEllisQP}
\end{figure}
\vspace{0.5cm}
\begin{figure} [!h]
  \centering{}
  \includegraphics
  [scale=0.5]
  {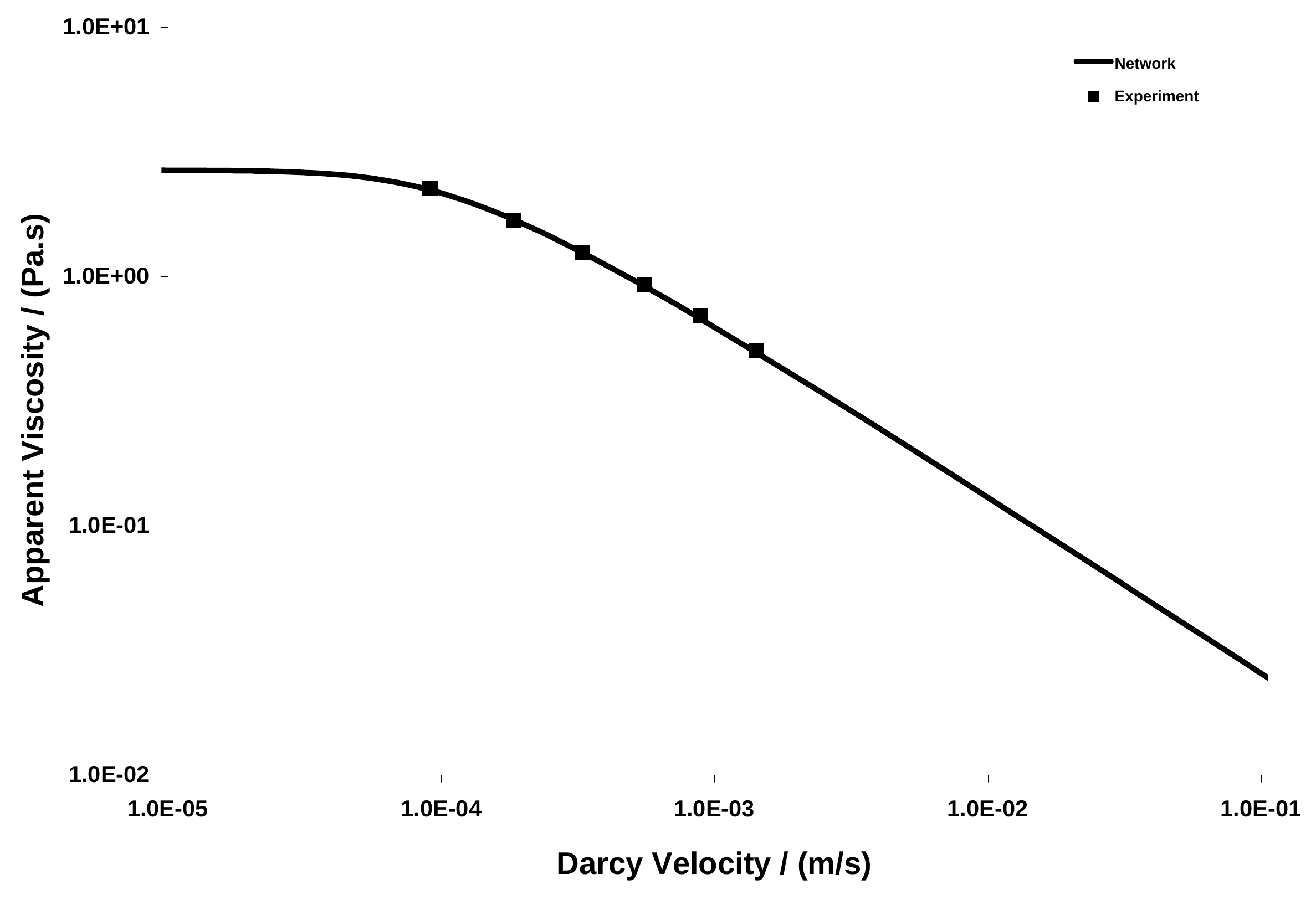}
  \caption[Balhoff's \ELLIS\ experimental data set for guar gum solution with 0.72\% concentration flowing
  through a packed bed of glass beads having $K=4.19 \times 10^{-9}$\,m$^{2}$ and $\phi=0.38$
  alongside the simulation results obtained with a scaled \sandp\ network having the same $K$ presented as $\aVis$ vs. $q$]
  {Balhoff's \ELLIS\ experimental data set for guar gum solution with 0.72\% concentration flowing
  through a packed bed of glass beads having $K=4.19 \times 10^{-9}$\,m$^{2}$ and $\phi=0.38$
  alongside the simulation results obtained with a scaled \sandp\ network having the same $K$ presented as $\aVis$ vs. $q$.}
  \label{BalhoffEllisVQ}
\end{figure}

\newpage

%XXXXXXXXXXXXXXXXXXXXXXXXXXXXXXXXXXXXXXXXXXXXXXXXXXXXXXXXXXXXXXXXXXXXXX
\section{\HB\ Model} \label{}
Three complete collections of experimental data found in the
literature on \HB\ fluid were investigated. Limited agreement with
the network model predictions was obtained.

%SSSSSSSSSSSSSSSSSSSSSSSSSSSSSSSSSSSSS
\subsection{Park} \label{ParkHB}
In this collection \cite{parkthesis}, eight complete data sets are presented. The fluid is an
aqueous solution of Polymethylcellulose (PMC) with two different molecular weights, PMC 25 and
PMC 400, each with concentration of 0.3\% and 0.5\% weight. For each of the four solutions,
two packed beds of spherical uniform-in-size glass beads, coarse and fine, were used.

\vspace{0.2cm}

The \insitu\ experimental data, alongside the fluids' bulk rheology
and the properties of the porous media, were tabulated in Park's
thesis. However, the permeability of the two beds, which is needed
to scale our network, is missing. To overcome this difficulty,
\Darcy's law was applied to the \NEW\ flow results of the fine bed,
as presented in Table M-1 in Park's thesis \cite{parkthesis}, to
extract the permeability of this bed from the slope of the best fit
line to the data points. In his dissertation, Park presented the
permeability relation of Equation (\ref{Kformula}) where $C^{''}$ is
a dimensionless constant to be determined from experiment and have
been assigned in the literature a value between 150 and 180
depending on the author. To find the permeability of the coarse bed,
we obtained a value for $C^{''}$ from Equation (\ref{Kformula}) by
substituting the permeability of the fine bed, with the other
relevant parameters. This value ($\simeq164$) was then substituted
in the formula with the relevant parameters to obtain the
permeability of the coarse bed.

\vspace{0.2cm}

We used our \nNEW\ code with two scaled \sandp\ networks and the
bulk rheology presented in Table (\ref{parkHerschelTable}) to
simulate the flow. The simulation results with the corresponding
experimental data sets are presented in Figures (\ref{ParkHBCoarse})
and (\ref{ParkHBFine}). As seen, the predictions are poor. One
possible reason is the high \shThin\ nature of the solutions, with
$n$ between 0.57 and 0.66. This can produce a large discrepancy even
for a small error in $n$. The failure to predict the threshold yield
pressure is also noticeable. Retention and other similar phenomena
may be ruled out as a possible cause for the higher experimental
threshold yield pressure by the fact that the solutions, according
to Park, were filtered to avoid gel formation.

%%%%%%%%%%%%%%%%%%%%%%%%%%%%%%%   Park \HB\  %%%%%%%%%%%%%%%%%%%%%%%%%%%%%%%%%%%%%%%%%
\newpage
\begin{figure}[!h]
  \centering{}
  \includegraphics
  [scale=0.5]
  {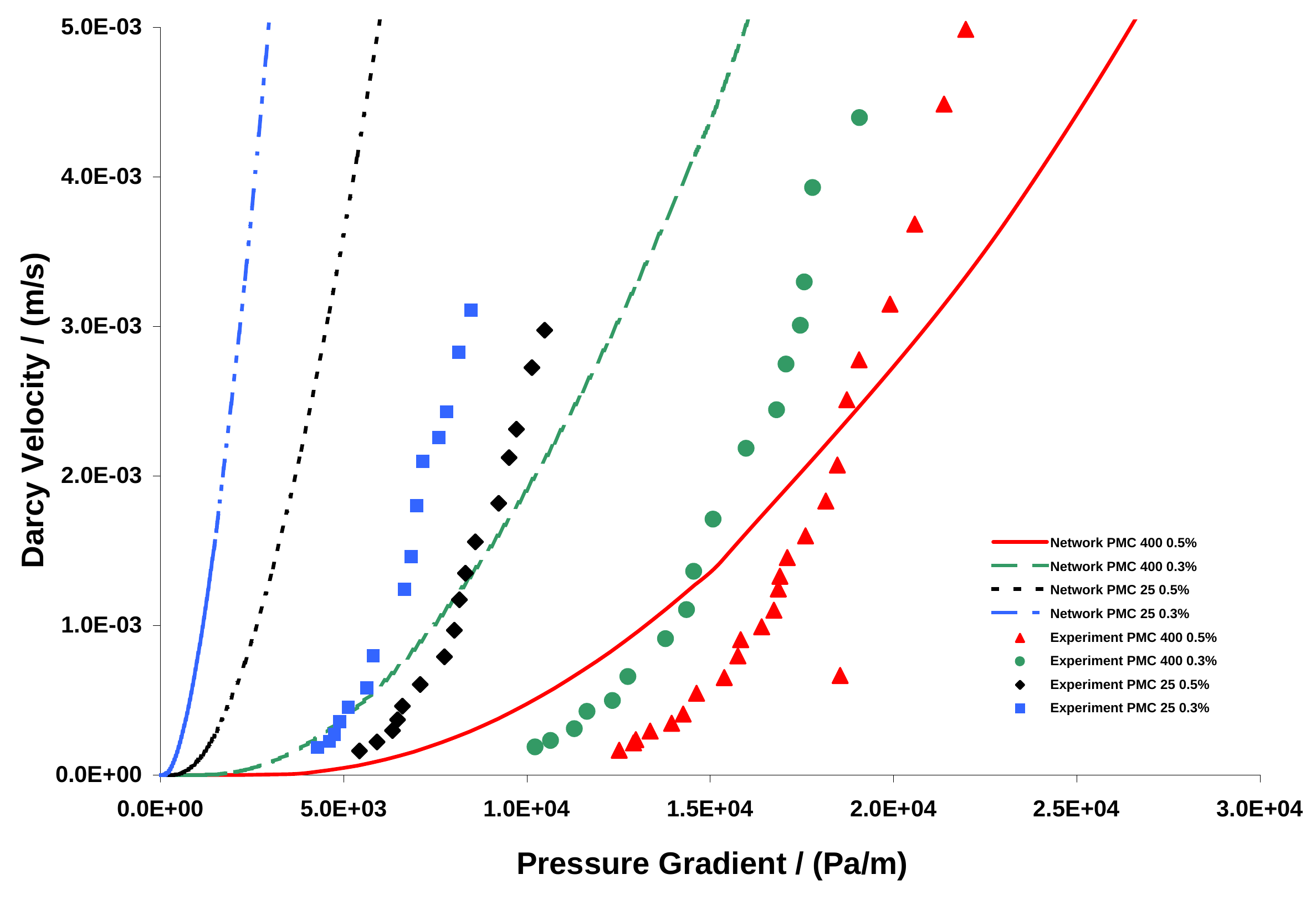}
  \caption[Park's \HB\ experimental data group for aqueous solutions of PMC 400 and PMC 25
  with 0.5\% and 0.3\% weight concentration flowing through a coarse packed bed of glass
  beads having $K=3413$\,Darcy and $\phi=0.42$ alongside the simulation results obtained with a
  scaled \sandp\ network having same $K$]
  {Park's \HB\ experimental data group for aqueous solutions of PMC 400 and PMC 25
  with 0.5\% and 0.3\% weight concentration flowing through a coarse packed bed of glass
  beads having $K=3413$\,Darcy and $\phi=0.42$ alongside the simulation results obtained with a
  scaled \sandp\ network having same $K$.}
  \label{ParkHBCoarse}
\end{figure}
\vspace{0.5cm}
\begin{figure}[!h]
  \centering{}
  \includegraphics
  [scale=0.5]
  {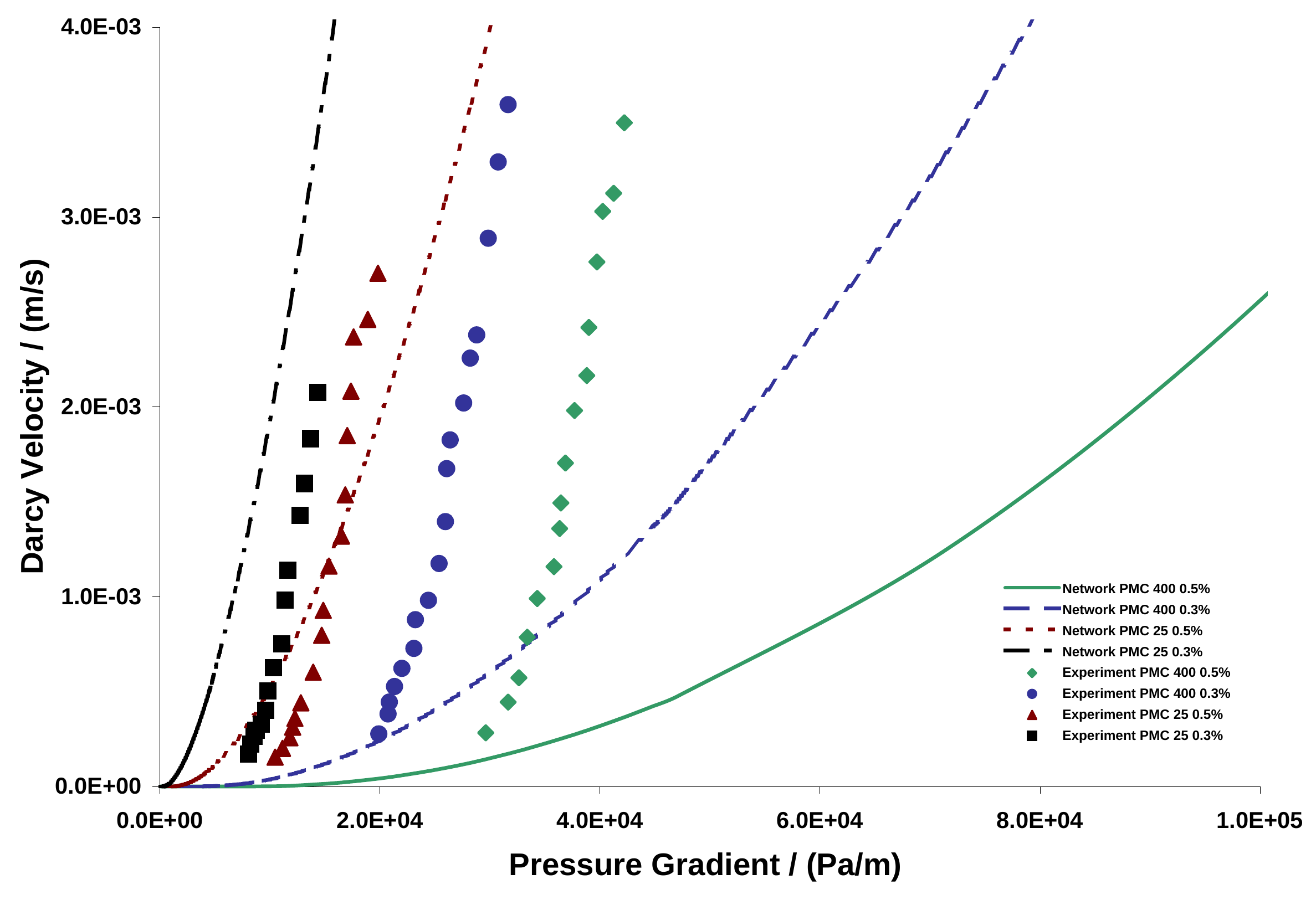}
  \caption[Park's \HB\ experimental data group for aqueous solutions of PMC 400 and PMC 25
  with 0.5\% and 0.3\% weight concentration flowing through a fine packed bed of glass
  beads having $K=366$\,Darcy and $\phi=0.39$ alongside the simulation results obtained with a
  scaled \sandp\ network having same $K$]
  {Park's \HB\ experimental data group for aqueous solutions of PMC 400 and PMC 25
  with 0.5\% and 0.3\% weight concentration flowing through a fine packed bed of glass
  beads having $K=366$\,Darcy and $\phi=0.39$ alongside the simulation results obtained with a
  scaled \sandp\ network having same $K$.}
  \label{ParkHBFine}
\end{figure}

%SSSSSSSSSSSSSSSSSSSSSSSSSSSSSSSSSSSSS
\subsection{Al-Fariss and Pinder} \label{}
In this collection \cite{alfariss1}, there are sixteen complete sets
of data for waxy oils with the bulk and \insitu\ rheologies. The
porous media consist of two packed beds of sand having different
dimensions, porosity, permeability and grain size. We used the bulk
rheology given by the authors and extracted the \insitu\ rheology
from the relevant graphs. The bulk rheology and bed properties are
given in Table (\ref{alfarissRheologyTable}).

\vspace{0.2cm}

The \nNEW\ code was used with two scaled \sandp\ networks and the
bulk rheology to simulate the flow. A sample of the simulation
results, with the corresponding \insitu\ experimental data points,
is shown in Figures (\ref{AlfarissHBWax25}) and
(\ref{AlfarissHBCrude}). Analyzing the experimental and network
results reveals that while the network behavior is consistent,
considering the underlying bulk rheology, the experimental data
exhibit an inconsistent pattern. This is evident when looking at the
\insitu\ behavior as a function of the bulk rheology which, in turn,
is a function of temperature.

\vspace{0.2cm}

The sixteen data sets are divided into four groups. In each group
the fluid and the porous medium are the same but the fluid
temperature varies. On analyzing the \insitu\ data, one can discover
that there is no obvious correlation between the fluid properties
and its temperature. An example is the 4.0\% wax in Clarus B group
where an increase in temperature from 14\,$^{\circ}$C to
16\,$^{\circ}$C results in a drop in the flow at high pressures
rather than rise, opposite to what is expected from the general
trend of the experimental data and the fact that the viscosity
usually decreases on increasing the temperature, as the bulk
rheology tells. One possibility is that in some cases the wax-oil
mix may not be homogenous, so other physical phenomena such as wax
precipitation took place. Such complex phenomena are accounted for
neither in our model nor in the Al-Fariss and Pinder model. This
might be inferred from the more consistent experimental results for
the waxy crude oil.

\vspace{0.2cm}

It is noteworthy that the Al-Fariss and Pinder model failed to cope
with these irregularities. As a result they arbitrarily modified the
model parameters on case-by-case basis to fit the experimental data.

%%%%%%%%%%%%%%%%%%%%%%%%%%%%%%%   Al-Fariss \HB\   %%%%%%%%%%%%%%%%%%%%%%%%%%%%%%%%%%%%%%%%%
\newpage
\begin{figure}[!h]
  \centering{}
  \includegraphics
  [scale=0.5]
  %[scale=0.5,width=0.9\textwidth,height=0.55\textwidth, trim = 0 0 0 0]
  {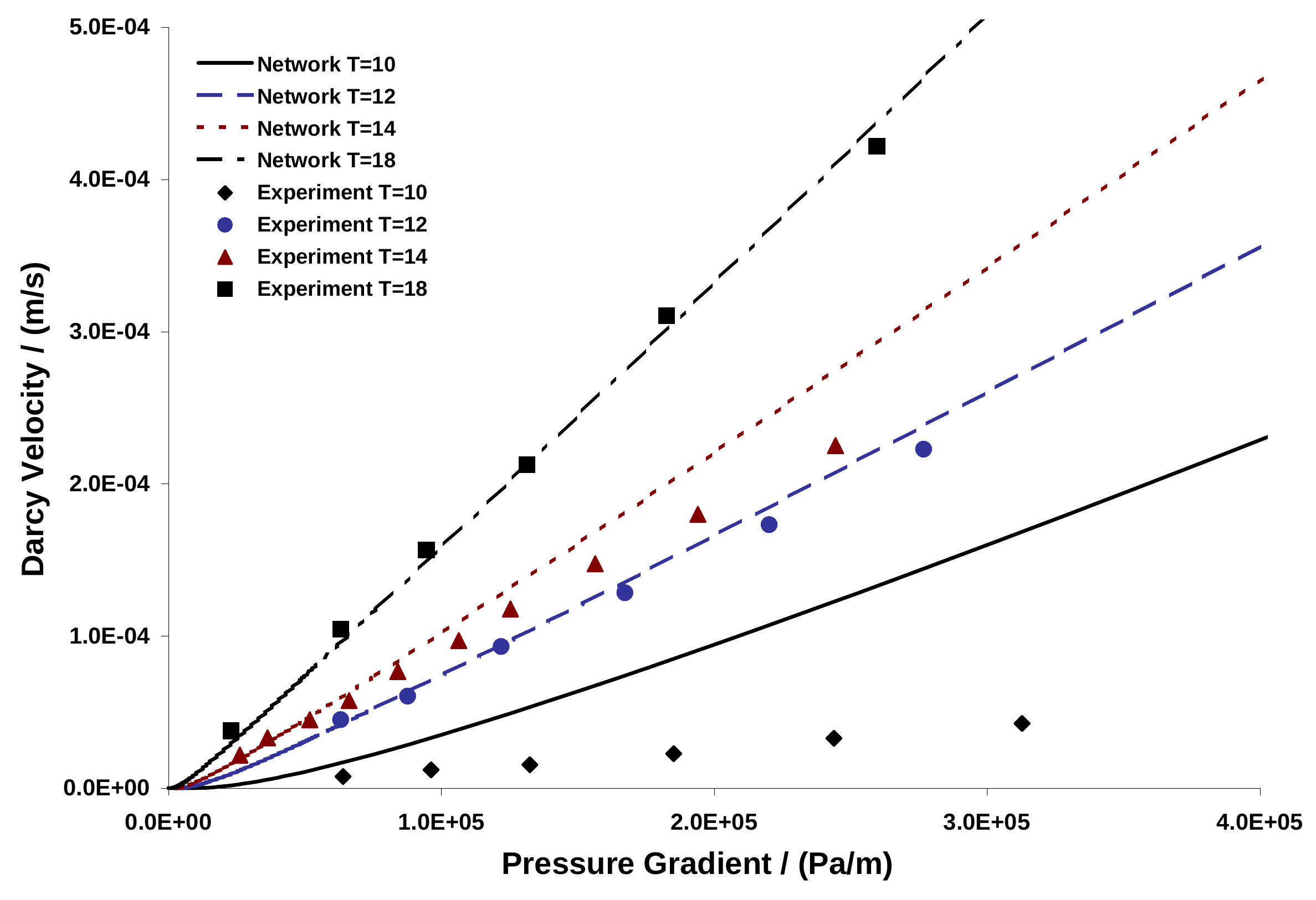}
  \caption[Al-Fariss and Pinder's \HB\ experimental data group for 2.5\% wax in Clarus B oil flowing
  through a column of sand having $K=315$\,Darcy and $\phi=0.36$ alongside the simulation results
  obtained with a scaled \sandp\ network having the same $K$ and $\phi$. The temperatures, T, are in
  $^{\circ}$C]
  {Al-Fariss and Pinder's \HB\ experimental data group for 2.5\% wax in Clarus B oil flowing
  through a column of sand having $K=315$\,Darcy and $\phi=0.36$ alongside the simulation results
  obtained with a scaled \sandp\ network having the same $K$ and $\phi$. The temperatures, T, are in
  $^{\circ}$C.}
  \label{AlfarissHBWax25}
\end{figure}
\vspace{0.5cm}
\begin{figure}[!h]
  \centering{}
  \includegraphics
  [scale=0.5]
  %[scale=0.5,width=0.9\textwidth,height=0.55\textwidth, trim = 0 0 0 0]
  {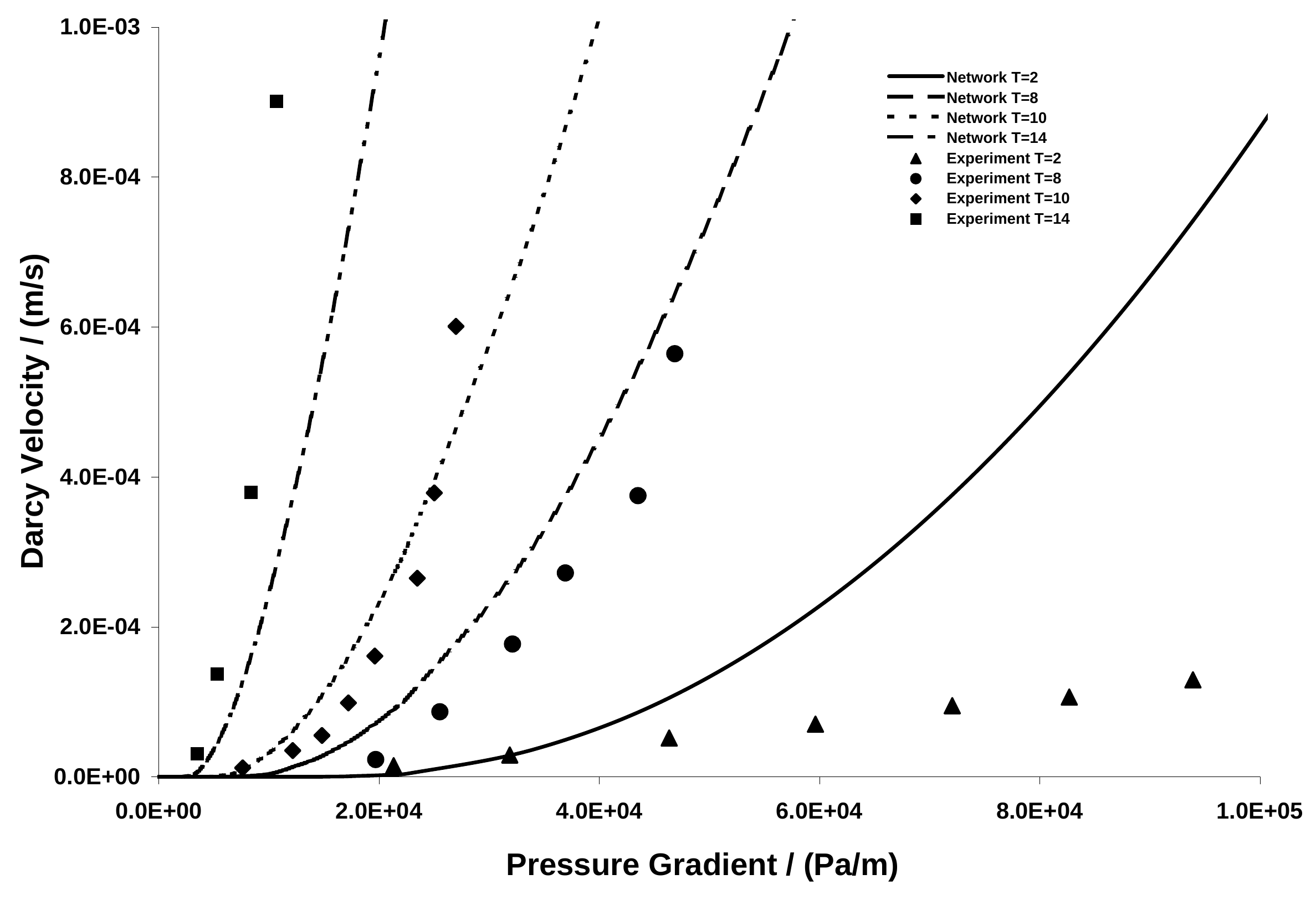}
  \caption[Al-Fariss and Pinder's \HB\ experimental data group for waxy crude oil flowing
  through a column of sand having $K=1580$\,Darcy and $\phi=0.44$ alongside the simulation results
  obtained with a scaled \sandp\ network having the same $K$. The temperatures, T, are in
  $^{\circ}$C]
  {Al-Fariss and Pinder's \HB\ experimental data group for waxy crude oil flowing
  through a column of sand having $K=1580$\,Darcy and $\phi=0.44$ alongside the simulation results
  obtained with a scaled \sandp\ network having the same $K$. The temperatures, T, are in
  $^{\circ}$C.}
  \label{AlfarissHBCrude}
\end{figure}

%SSSSSSSSSSSSSSSSSSSSSSSSSSSSSSSSSSSSS
\subsection{Chase and Dachavijit} \label{}
In this collection \cite{chase1}, there are ten complete data sets
for \BING\ aqueous solutions of Carbopol 941 with concentration
varying between 0.15 and 1.3 mass percent. The porous medium is a
packed column of spherical glass beads having a narrow size
distribution. The bulk rheology, which is extracted from a digitized
image and given in Table (\ref{chaseRheologyTable}), represents the
actual experimental data points rather than
the least square fitting suggested by the authors.%

\vspace{0.2cm}

Our \nNEW\ code was used to simulate the flow of a \BING\ fluid with
the extracted bulk rheology through a scaled \sandp\ network. The
scaling factor was chosen to have a permeability that produces a
best fit to the most \NEW\-like data set, excluding the first data
set with the lowest concentration due to a very large relative error
and a lack of fit to the trend line.

\vspace{0.2cm}

The simulation results, with the corresponding \insitu\ experimental
data sets extracted from digitized images of the relevant graphs,
are presented in Figures (\ref{ChaseHB1}) and (\ref{ChaseHB2}). The
fit is good in most cases. The experimental data in some cases shows
irregularities which may suggest large experimental errors or other
physical phenomena, such as retention, taking place. This erratic
behavior cannot fit a consistent pattern.

%%%%%%%%%%%%%%%%%%%%%%%%%%%%%%%   Chase \HB\   %%%%%%%%%%%%%%%%%%%%%%%%%%%%%%%%%%%%%%%%%
\newpage
\begin{figure}[!h]
  \centering{}
  \includegraphics
  [scale=0.5]
  {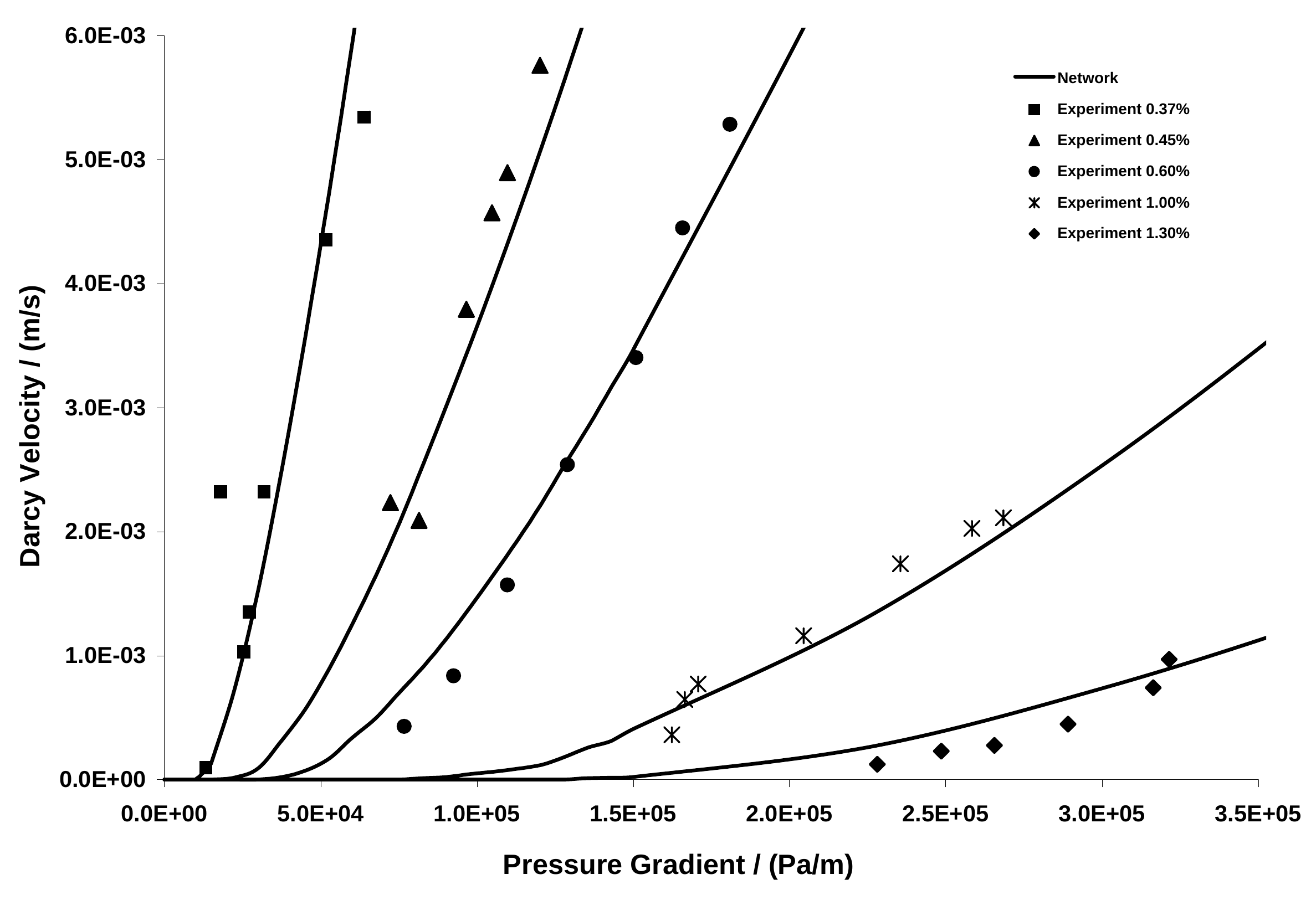}
  \caption[Network simulation results with the corresponding experimental data points
  of Chase and Dachavijit for a \BING\ aqueous solution of Carbopol 941 with various concentrations
  (0.37\%, 0.45\%, 0.60\%, 1.00\% and 1.30\%) flowing through a packed column of glass beads]
  {Network simulation results with the corresponding experimental data points
  of Chase and Dachavijit for a \BING\ aqueous solution of Carbopol 941 with various concentrations
  (0.37\%, 0.45\%, 0.60\%, 1.00\% and 1.30\%) flowing through a packed column of glass beads.}
  \label{ChaseHB1}
\end{figure}
\vspace{0.5cm}
\begin{figure}[!h]
  \centering{}
  \includegraphics
  [scale=0.5]
  {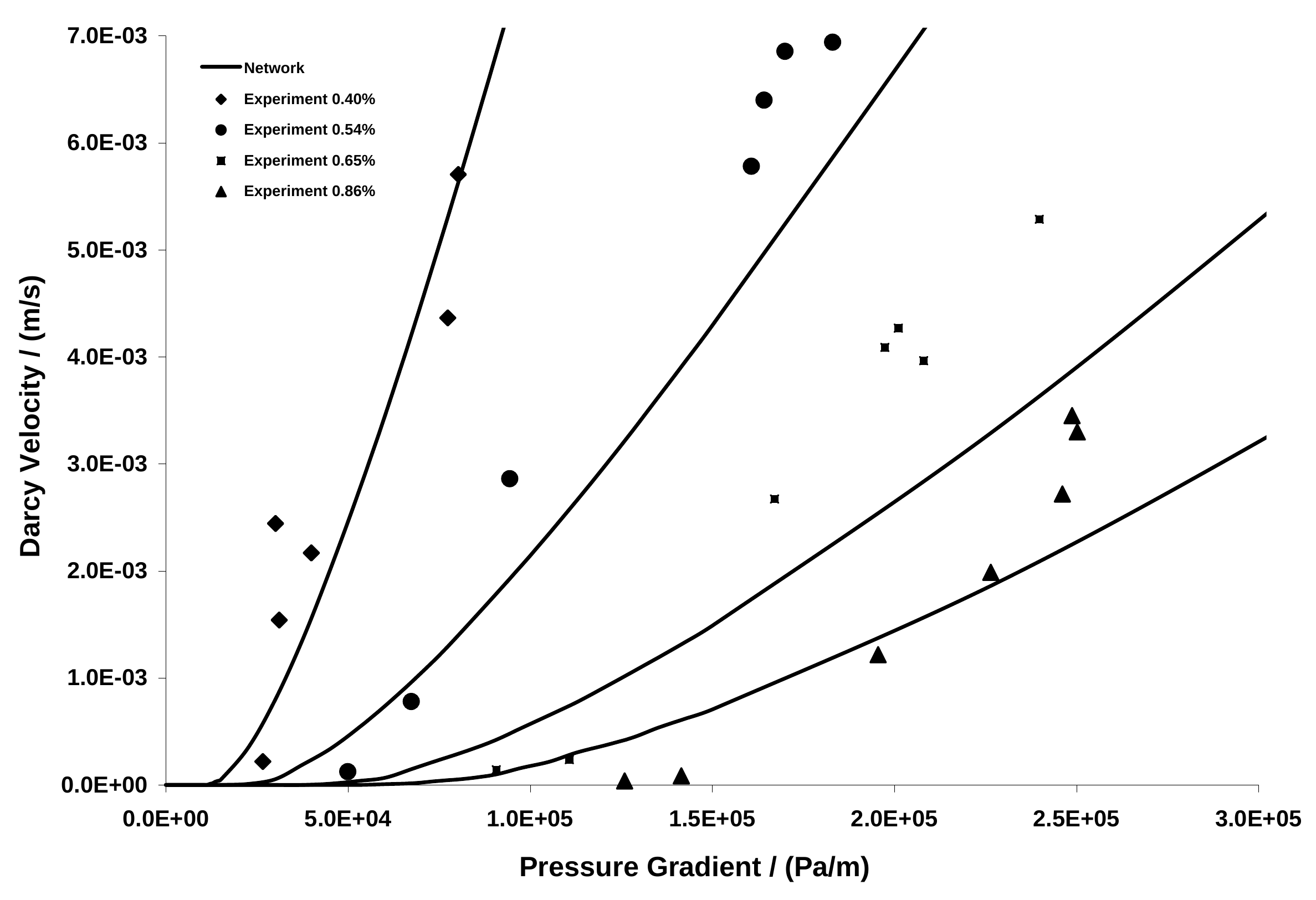}
  \caption[Network simulation results with the corresponding experimental data points
  of Chase and Dachavijit for a \BING\ aqueous solution of Carbopol 941 with various concentrations
  (0.40\%, 0.54\%, 0.65\% and 0.86\%) flowing through a packed column of glass beads]
  {Network simulation results with the corresponding experimental data points
  of Chase and Dachavijit for a \BING\ aqueous solution of Carbopol 941 with various concentrations
  (0.40\%, 0.54\%, 0.65\% and 0.86\%) flowing through a packed column of glass beads.}
  \label{ChaseHB2}
\end{figure}

%%%%%%%%%%%%%%%%%%%%%%%%%%%%%%%%%%%%%%%%%%%%%

\newpage

\hspace{1.0cm} \\

\begin{table}  [!h]
\centering
\begin{tabular}{|c|c|c|c|}
\hline
{\verb|     | \bf Solution \verb|     |} &   {\verb|   | $C$ (Pa.s$^{n}$) \verb|   |} &    {\verb|    | $n$ \verb|    |} & {\verb|   | $\ysS$ (Pa) \verb|   |} \\
\hline
0.50\% PMC 400  &      0.116 &       0.57 &      0.535 \\
\hline
0.30\% PMC 400  &      0.059 &       0.61 &      0.250 \\
\hline
0.50\% PMC 25  &      0.021 &       0.63 &      0.072 \\
\hline
0.30\% PMC 25  &      0.009 &       0.66 &      0.018 \\
\hline
\end{tabular}
\caption[The bulk rheology of Park's \HB\ experimental data] %
{The bulk rheology of Park's \HB\ experimental data.} %
\label{parkHerschelTable}
\end{table}

%%%%%%%%%%%%%%%%%%%%%%%%%%%%%%%%%%%%%%%%%%%%%

\vspace{2.0cm}

\begin{table} [!h]
\centering
\begin{tabular}{|c|c|c|c|c|c|c|}
\hline
                  \multicolumn{ 5}{|c}{{\bf Fluid Properties}} & \multicolumn{ 2}{|c|}{{\bf Bed Properties}} \\
\hline
{ Wax (\%)} & { T ($^{\circ}$C) } & { $C$ (Pa.s$^{n}$)} &    {\verb|  | $n$ \verb|  |} & { $\ysS$ (Pa)} & {\verb| | $K$ (m$^{2}$) \verb| |} &  {\verb|  | $\phi$ \verb|  |} \\
\hline
       2.5 &         10 &      0.675 &       0.89 &      0.605 &   3.15E-10 &       0.36 \\
\hline
       2.5 &         12 &      0.383 &       0.96 &      0.231 &   3.15E-10 &       0.36 \\
\hline
       2.5 &         14 &      0.300 &       0.96 &      0.142 &   3.15E-10 &       0.36 \\
\hline
       2.5 &         18 &      0.201 &       0.97 &      0.071 &   3.15E-10 &       0.36 \\
\hline
       4.0 &         12 &      1.222 &       0.77 &      3.362 &   3.15E-10 &       0.36 \\
\hline
       4.0 &         14 &      0.335 &       0.97 &      3.150 &   3.15E-10 &       0.36 \\
\hline
       4.0 &         16 &      0.461 &       0.88 &      1.636 &   3.15E-10 &       0.36 \\
\hline
       4.0 &         18 &      0.436 &       0.85 &      0.480 &   3.15E-10 &       0.36 \\
\hline
       4.0 &         20 &      0.285 &       0.90 &      0.196 &   3.15E-10 &       0.36 \\
\hline
       5.0 &         16 &      0.463 &       0.87 &      3.575 &   3.15E-10 &       0.36 \\
\hline
       5.0 &         18 &      0.568 &       0.80 &      2.650 &   3.15E-10 &       0.36 \\
\hline
       5.0 &         20 &      0.302 &       0.90 &      1.921 &   3.15E-10 &       0.36 \\
\hline
     Crude &          2 &      0.673 &       0.54 &      2.106 &   1.58E-09 &       0.44 \\
\hline
     Crude &          8 &      0.278 &       0.61 &      0.943 &   1.58E-09 &       0.44 \\
\hline
     Crude &         10 &      0.127 &       0.70 &      0.676 &   1.58E-09 &       0.44 \\
\hline
     Crude &         14 &      0.041 &       0.81 &      0.356 &   1.58E-09 &       0.44 \\
\hline
\end{tabular}
\caption[The bulk rheology and bed properties for the \HB\ experimental data of Al-Fariss and Pinder] %
{The bulk rheology and bed properties for the \HB\ experimental data of Al-Fariss and Pinder.} %
\label{alfarissRheologyTable}
\end{table}

%%%%%%%%%%%%%%%%%%%%%%%%%%%%%%%   Chase bulk rheology  %%%%%%%%%%%%%%%%%%%%%%%%%%%%%%%%%%%%%%%%%

\vspace{1.0cm}

\begin{table} [!h]
\centering
\begin{tabular}{|c|c|c|}
\hline
{\bf Concentration (\%)} & {\bf \verb|   | C (Pa.s) \verb|   |} & {\bf \verb|   | $\ysS$ (Pa) \verb|   |} \\
\hline
      0.15 &      0.003 &       0.08 \\
\hline
      0.37 &      0.017 &       2.06 \\
\hline
      0.40 &      0.027 &       2.39 \\
\hline
      0.45 &      0.038 &       4.41 \\
\hline
      0.54 &      0.066 &       4.37 \\
\hline
      0.60 &      0.057 &       7.09 \\
\hline
      0.65 &      0.108 &       8.70 \\
\hline
      0.86 &      0.136 &      12.67 \\
\hline
      1.00 &      0.128 &      17.33 \\
\hline
      1.30 &      0.215 &      28.46 \\
\hline
\end{tabular}
\caption[The bulk rheology of Chase and Dachavijit experimental data for a \BING\ fluid ($n=1.0$)]
        {The bulk rheology of Chase and Dachavijit experimental data for a \BING\ fluid ($n=1.0$).}
        \vspace{1.0cm}
\label{chaseRheologyTable} %
\end{table}

%XXXXXXXXXXXXXXXXXXXXXXXXXXXXXXXXXXXXXXXXXXXXXXXXXXXXXXXXXXXXXXXXXXXXXX
\section{Assessing \ELLIS\ and \HB\ Results} \label{}
The experimental validation of \ELLIS\ model is generally good. What
is remarkable is that this simple model was able to predict several
experimental data sets without introducing any arbitrary factors.
All we did is to use the bulk rheology and the bed properties as an
input to the \nNEW\ code. In fact, these results were obtained with
a scaled \sandp\ network which in most cases is not an ideal
representation of the porous medium used. Also remarkable is the use
of an analytical expression for the volumetric flow rate instead of
an empirical one as done by Lopez \cite{lopezthesis, lopez1} in the
case of the \CARREAU\ model.

\vspace{0.4cm}

Regarding the \HB\ model, the experimental validation falls short of
expectation. Nevertheless, even coming this close to the
experimental data is a success, being aware of the level of
sophistication of the model and the complexities of the
three-dimensional networks, and considering the fact that the
physics so far is relatively simple. Incorporating more physics in
the model and using better void space description in the form of
more realistic networks can improve the results.

\vspace{0.2cm}

However, it should be remarked that the complexity of the \yields\
phenomenon may be behind this relative failure of the \HB\ model
implementation. The flow of \yields\ fluids in porous media is
apparently too complex to describe by a simple rheological model
such as \HB, too problematic to investigate by primitive
experimental techniques and too difficult to model by the available
tools of pore-scale modeling at the current level of sophistication.
This impression is supported by the fact that much better results
are obtained for non-\yields\ fluids, like \ELLIS\, using the same
tools and techniques. One major limitation of our network model with
regard to the \yields\ fluids is that we use analytical expressions
for cylindrical tubes based on the concept of equivalent radius
$R_{eq}$ which we presented earlier. This is far from the reality as
the void space is highly complex in shape. The result is that the
yield condition for circular tube becomes invalid approximation to
the actual yield condition for the pore space. The simplistic nature
of the yield condition in porous media is highlighted by the fact
that in almost all cases of disagreement between the network and the
experimental results the network produced a lower yield value. Other
limitations and difficulties associated with the \yields\ fluids in
general, and hence affect our model, will be discussed in the next
section.

\vspace{0.2cm}

In summary, \yields\ fluid results are extremely sensitive to how
the fluid is characterized, how the void space is described and how
the yield process is modeled. In the absence of a comprehensive and
precise incorporation of all these factors in the modeling
procedure, pore scale modeling of \yields\ fluids in porous media
remains an inaccurate approximation that may not produce
quantitatively sensible predictions.

\def\baselinestretch{1}
\chapter{\YieldS\ Analysis} \label{Yield}
\def\baselinestretch{1.66}
\Yields\ or viscoplastic fluids can sustain shear stresses, that is
a certain amount of stress must be exceeded before the flow starts.
So an ideal \yields\ fluid is a solid before yield and a fluid
after. Accordingly, the viscosity of the substance changes from an
infinite to a finite value. However, the physical situation suggests
that it is more realistic to regard a \yields\ substance as a fluid
whose viscosity as a function of applied stress has a discontinuity
as it drops sharply from a very high value on exceeding a critical
\yields.

\vspace{0.2cm}

There are many controversies and unsettled issues in the \nNEW\
literature about \yields\ phenomenon and \yields\ fluids. In fact,
even the concept of a \yields\ has received much recent criticism,
with evidence presented to suggest that most materials weakly yield
or creep near zero strain rate. The supporting argument is that any
material will flow provided that one waits long enough. These
conceptual difficulties are supported by practical and experimental
complications. For example, the value of the \yields\ for a
particular fluid is difficult to measure consistently and it may
vary by more than one order of magnitude depending on the
measurement technique \cite{birdbook, carreaubook, Barnes1999,
BalmforthC2001}.

\vspace{0.2cm}

Several constitutive equations to describe liquids with \yields\ are
in use; the most popular ones are \BING, Casson and \HB. Some have
suggested that the \yields\ values obtained via such models should
be considered model parameters and not real material properties.

\vspace{0.2cm}

There are several difficulties in working with the \yields\ fluids
and validating the experimental data. One difficulty is that the
\yields\ value is usually obtained by extrapolating a plot of shear
stress to zero shear rate \cite{carreaubook, ParkHB1973, alfariss1,
balhoffthesis}. This extrapolation can result in a variety of values
for \yields, depending on the distance from the shear stress axis
experimentally accessible by the instrument used. The vast majority
of \yields\ data reported results from such extrapolations, making
most values in the literature instrument-dependent
\cite{carreaubook}. Another method used to measure \yields\ is by
lowering the shear rate until the shear stress approaches a
constant. This may be identified as the dynamic \yields\
\cite{larsonbook1999}. The results obtained using these methods may
not agree with the static \yields\ measured directly without
disturbing the microstructure during the measurement. The latter
seems more relevant for the flow initiation under gradual increase
in pressure gradient as in the case of flow in porous media.
Consequently, the accuracy of the predictions made using flow
simulation models in conjunction with such experimental data is
limited.

\vspace{0.2cm}

Another difficulty is that while in the case of pipe flow the
\yields\ value is a property of the fluid, in the case of flow in
porous media there are strong indications that in a number of
situations it may depend on both the fluid and the porous medium
itself \cite{vradis1, bearbook}. One possible explanation is that
the \yields\ value may depend on the size and shape of the pore
space when the polymer molecules become comparable in size to the
void space. The implicit assumption that the \yields\ value at pore
scale is the same as the yield value at bulk may not be self
evident. This makes the predictions of the network models based on
analytical solution to the flow in a single tube combined with the
bulk rheology less accurate. When the duct size is small, as it is
usually the case in porous media, flow of macromolecule solutions
normally displays deviations from predictions based on corresponding
viscometric data \cite{LiuM1998}. Consequently, the concept of
equivalent radius $R_{eq}$, which is used in the network modeling,
though is completely appropriate for the \NEW\ flow and reasonably
appropriate for the \nNEW\ with no \yields, seems inappropriate for
\yields\ fluids as the yield depends on the actual shape of the void
space rather than the equivalent radius and flow conductance.

\vspace{0.2cm}

In porous media, threshold yield pressure is expected to be directly
proportional to the \yields\ of the fluid and inversely proportional
to the porosity and permeability of the media \cite{WangHC2006}. A
shortcoming of using continuum models, such as an extended \Darcy's
law, to study the flow of \yields\ fluids in porous media is that
these models are unable to correctly describe the network behavior
at transition where the network is partly flowing. The reason is
that according to these models the network is either fully blocked
or fully flowing whereas in reality the network smoothly yields. For
instance, these models predict for \BING\ fluids a linear
relationship between Darcy velocity and pressure gradient with an
intercept at threshold yield gradient whereas our network model and
that of others \cite{balhoff1}, supported by experimental evidence,
predict a nonlinear behavior at transition stage \cite {vradis1}.
Some authors have concluded that in a certain range the macroscopic
flow rate of \BING\ plastic in a network depends quadratically on
the departure of the applied pressure difference from its minimum
value \cite{chen3}. Our network simulation results confirm this
quadratic correlation. Samples of this behavior for the \sandp\ and
Berea networks are given in Figures (\ref{SPquadratic}) and
(\ref{Bquadratic}). As can be seen, the match between the network
points and the least-square fitting quadratic curve is almost
perfect in the case of the \sandp. For the \Berea\ network the
agreement is less obvious. A possible cause is the inhomogeneity of
the \Berea\ network which makes the flow less regular.

%\newpage

%%%%%%%%%%%%%%%%%%%%%%%%%%%%%%%%%%%

\begin{figure}[!t]
  \centering{}
  \includegraphics
  [scale=0.53]
  {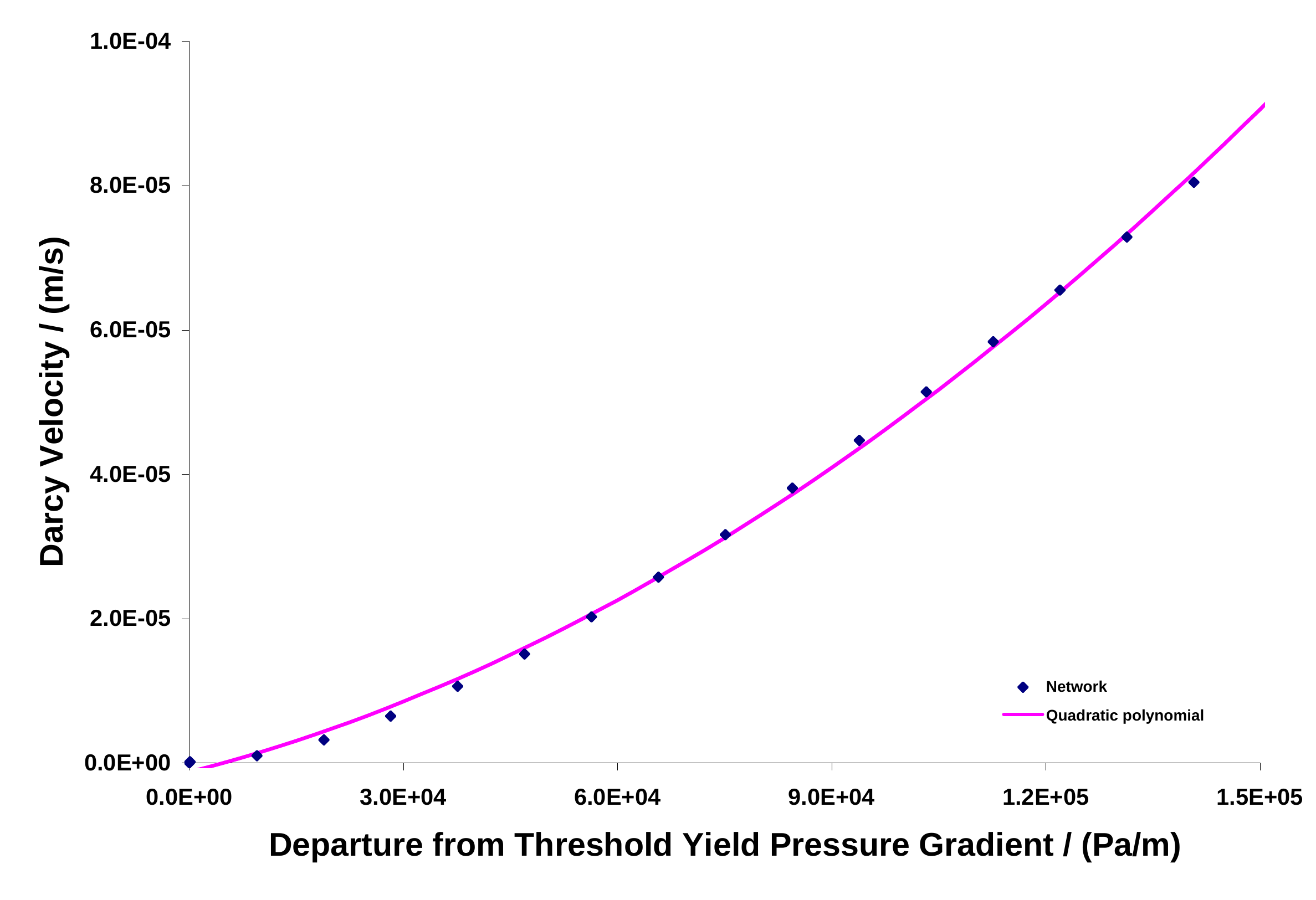}
  \caption[Comparison between the \sandp\ network simulation results for a \BING\ fluid
  with $\tau_{o}=1.0$Pa and the least-square fitting quadratic
  $y=2.39\times10^{-15}\,x^{2}+2.53\times10^{-10}\,x-1.26\times10^{-6}$]
  {Comparison between the \sandp\ network simulation results for a \BING\ fluid
  with $\tau_{o}=1.0$Pa and the least-square fitting quadratic
  $y=2.39\times10^{-15}\,x^{2}+2.53\times10^{-10}\,x-1.26\times10^{-6}$.}
  \label{SPquadratic}
\end{figure}

\vspace{0.5cm}

\begin{figure}[!h]
  \centering{}
  \includegraphics
  [scale=0.53]
  {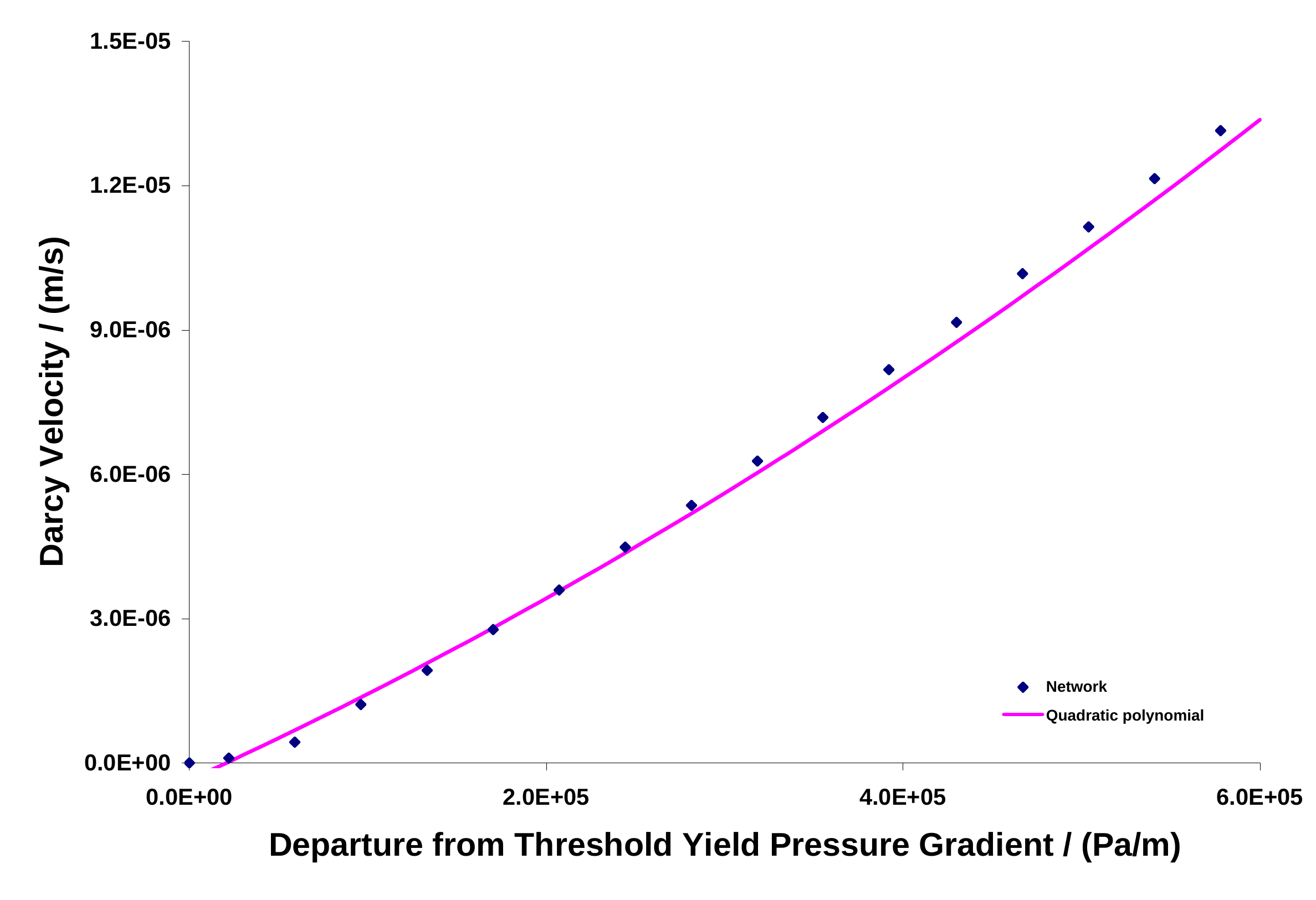}
  \caption[Comparison between the \Berea\ network simulation results for a \BING\ fluid
  with $\tau_{o}=1.0$Pa and the least-square fitting quadratic
  $y=1.00\times10^{-17}\,x^{2}+1.69\times10^{-11}\,x-3.54\times10^{-7}$]
  {Comparison between the \Berea\ network simulation results for a \BING\ fluid
  with $\tau_{o}=1.0$Pa and the least-square fitting quadratic
  $y=1.00\times10^{-17}\,x^{2}+1.69\times10^{-11}\,x-3.54\times10^{-7}$.}
  \label{Bquadratic}
\end{figure}

%SSSSSSSSSSSSSSSSSSSSSSSSSSSSSSSSSSSSS
\section{Predicting the Yield Pressure of a Network} \label{}
Predicting the threshold yield pressure of a \yields\ fluid in
porous media in its simplest form may be regarded as a special case
of the more general problem of finding the threshold conduction path
in disordered media consisting of elements with randomly distributed
thresholds. This problem was analyzed by Roux and Hansen \cite
{kharabaf1, roux1} in the context of studying the conduction of an
electric network of diodes by considering two different cases, one
in which the path is directed (no backtracking) and one in which it
is not. They suggested that the minimum overall threshold potential
difference across the network is akin to a \percolation\ threshold
and studied its dependence on the lattice size.

\vspace{0.2cm}

Kharabaf and Yortsos \cite {kharabaf1} noticed that a firm
connection of the lattice-threshold problem to \percolation\ appears
to be lacking and the relation of the Minimum Threshold Path (MTP)
to the minimum path of \percolation, if it indeed exists, is not
self-evident. They presented a new algorithm, \InvPM\ (IPM), for the
construction of the MTP, based on which its properties can be
studied. This algorithm will be closely examined later on.

\vspace{0.2cm}

In a series of studies on generation and mobilization of foam in
porous media, Rossen \etal\  \cite {rossen1, rossen2} analyzed the
threshold yield pressure using \percolation\ theory concepts and
suggested a simple \percolation\ model. In this model, the
\percolation\ cluster is first found, then the MTP was approximated
as a subset of this cluster that samples
those bonds with the smallest individual thresholds \cite {chen3}.%

\vspace{0.2cm}

Chen \etal\ \cite {chen3} elucidated the \InvPM\ method of Kharabaf.
They further extended this method to incorporate dynamic effects due
to the viscous friction following the onset of mobilization.

\vspace{0.2cm}

It is tempting to consider the \yields\ fluid behavior in porous
media as a \percolation\ phenomenon to be analyzed by the classical
\percolation\ theory. However, three reasons, at least, may suggest
otherwise:
\begin{enumerate}

    \item The conventional \percolation\ models can be applied only if the conducting elements are
    homogeneous, i.e. it is assumed in these models that the intrinsic property of all
    elements in the network are equal. However, this assumption cannot be justified for most kinds of
    media where the elements vary in their conduction and yield properties.
    Therefore, to apply \percolation\ principles, a generalization to the conventional
    \percolation\ theory is needed as suggested by Selyakov and Kadet \cite {selyakovbook}.

    \item The network elements cannot yield independently as a spanning path bridging the inlet
    to the outlet is a necessary condition for yield. This contradicts the \percolation\ theory assumption
    that the elements conduct independently.

    \item The pure \percolation\ approach ignores the dynamic aspects of the pressure field, that is a
    stable pressure configuration is a necessary condition which may not coincide with the
    simple \percolation\ requirement. The \percolation\ condition, as required by the classic
    \percolation\ theory, determines the stage at which the network starts flowing according to the intrinsic
    properties of its elements as an ensemble of conducting bonds regardless of the dynamic
    aspects introduced by the external pressure field.

\end{enumerate}

In this thesis, two approaches to predict the network threshold
yield pressure are presented and carefully examined: the \InvPM\ of
Kharabaf, and the \PathMP\ which is a novel approach that we
propose. Both approaches are implemented in our three-dimensional
network model of \nNEW\ code and the results are extracted. An
extensive analysis is conducted to compare the prediction and
performance of these approaches and relate their results to the
network threshold yield pressure as obtained from flow simulation.

%SSSSSSSSSSSSSSSSSSSSSSSSSSSSSSSSSSSSS
\subsection{\InvPM\ (IPM)} \label{}
The IPM is a way for finding the inlet-to-outlet path that minimizes
the sum of the values of a property assigned to the individual
elements of the network, and hence finding this minimum. For a
\yields\ fluid, this reduces to finding the inlet-to-outlet path
that minimizes the yield pressure. The yield pressure of this path
is taken as the network threshold yield pressure. An algorithm to
find the threshold yield pressure according to IPM is outlined
below:
\begin{enumerate}

    \item Initially, the nodes on the inlet are considered to be sources and the nodes on the
    outlet and    inside are targets. The inlet nodes are assigned a pressure value of 0.0.
    According to the IPM, a source cannot be a target and vice versa, i.e. they are disjoint
    sets and remain so in all stages.

    \item Starting from the source nodes, a step forward is made in which the yield front advances
    one bond from a single source node. The condition for choosing this step is that the sum of
    the source pressure plus the yield pressure of the bond connecting the source to the target node
    is the minimum of all similar sums from the source nodes to the possible target nodes.
    This sum is assigned to the target node.

    \item This target node loses its status as a target and obtains the status of a source.

    \item The last two steps are repeated until the outlet is reached, i.e. when the target is
    an outlet node. The pressure assigned to this target node is regarded as the yield
    pressure of the network.

\end{enumerate}

\vspace{0.2cm}

This algorithm was implemented in our \nNEW\ flow simulation code as
outlined below:
\begin{enumerate}

    \item Two boolean vectors and one double vector each of size $N$, where $N$ is the number of
    the network nodes, are used. The boolean vectors are used to store the status of the
    nodes, one for the sources and one for the targets. The double vector is used for storing
    the yield pressure assigned to the nodes. The vectors are initialized as explained already.

    \item A loop over all sources is started in which the sum of the pressure value of the
    source node plus the threshold yield pressure of the bond connecting the source node to
    one of the possible target nodes is computed. This is applied to all targets connected
    to all sources.

    \item The minimum of these sums is assigned to the respective target node. This target
    node is added to the source list and removed from the target list.

    \item The last two steps are repeated until the respective target node is verified as an outlet
    node. The pressure assigned to the target node in the last loop is the network threshold
    yield pressure.

\end{enumerate}
A flowchart of the IPM algorithm is given in Figure (\ref{IPMFlowchart}).

\vspace{0.2cm}

As implemented, the memory requirement for a network with $N$ nodes
is 10$N$ bytes which is a trivial cost even for a large network.
However, the CPU time for a medium-size network is considerable and
expected to rise sharply with increasing size of the network. A
sample of the IPM data is presented in Table (\ref{ipmPmpSPTable})
and Table (\ref{ipmPmpBereaTable}) for the sand pack and Berea
networks respectively, and in Table (\ref{ipmPmpCubicSPTable}) and
Table (\ref{ipmPmpCubicBereaTable}) for their cubic equivalents
which are described in Section (\ref{RandomCubicComp}).

%SSSSSSSSSSSSSSSSSSSSSSSSSSSSSSSSSSSSS
\subsection{\PathMP\ (PMP)} \label{}
This is a novel approach that we developed. It is based on a similar
assumption to that upon which the IPM is based, that is the network
threshold yield pressure is the minimum sum of the threshold yield
pressures of the individual elements of all possible paths from the
inlet to the outlet. However, it is computationally different and is
more efficient than the IPM in terms of the CPU time and memory.

\vspace{0.2cm}

According to the PMP, to find the threshold yield pressure of a network:
\begin{enumerate}

    \item All possible paths of serially-connected bonds from inlet to outlet are
    inspected. We impose a condition on the spanning path that there is
    no flow component opposite to the pressure gradient across the network in any part
    of the path, i.e. backtracking is not allowed.

    \item For each path, the threshold yield pressure of each bond is computed and the sum
    of these pressures is found.

    \item The network threshold yield pressure is taken as the minimum of these sums.

\end{enumerate}

\vspace{0.2cm}

Despite its conceptual simplicity, the computational aspects of this
approach for a three-dimensional topologically-complex network is
formidable even for a relatively small network. It is highly
demanding in terms of memory and CPU time. However, the algorithm
was implemented in an efficient way. In fact, the algorithm was
implemented in three different forms producing identical results but
varying in their memory and CPU time requirements. The form outlined
below is the slowest one in terms of CPU time but the most efficient
in terms of memory:
\begin{enumerate}

    \item A double vector of size $N$, where $N$ is the number of the network nodes, is used to store the
    yield pressure assigned to the nodes. The nodes on the inlet are initialized with zero pressure
    while the remaining nodes are initialized with infinite pressure (very high value).

    \item A source is an inlet or an inside node with a finite pressure while a target is a node
    connected to a source through a single bond with the condition that the spatial coordinate
    of the target in the direction of the pressure gradient across the network is higher than
    the spatial coordinate of the source.

    \item Looping over all sources, the threshold yield pressure of each bond
    connecting a source to a target is computed and the sum of this yield pressure plus the
    pressure of the source node is found. The minimum of this sum and the current yield pressure
    of the target node is assigned to the target node.

    \item The loop in the last step is repeated until a stable state is reached when no
    target node changes its pressure during looping over all sources.

    \item A loop over all outlet nodes is used to find the minimum pressure assigned to the
    outlet nodes. This pressure is taken as the network threshold yield pressure.

\end{enumerate}
The PMP algorithm is graphically presented in a flow chart in Figure
(\ref{PMPFlowchart}).

\vspace{0.2cm}

For a network with $N$ nodes the memory requirement is 8$N$ bytes
which is still insignificant even for a large network. The CPU time
is trivial even for a relatively large network and is expected to
rise almost linearly with the size of the network. A sample of the
PMP data is presented in Tables (\ref{ipmPmpSPTable}),
(\ref{ipmPmpBereaTable}), (\ref{ipmPmpCubicSPTable}) and
(\ref{ipmPmpCubicBereaTable}).

%SSSSSSSSSSSSSSSSSSSSSSSSSSSSSSSSSSSSS
\subsection{Analyzing IPM and PMP} \label{}
As implemented in our \nNEW\ code, the two methods are very
efficient in terms of memory requirement especially the PMP. The PMP
is also very efficient in terms of CPU time. This is not the case
with the IPM which is time-demanding compared to the PMP, as can be
seen in Tables (\ref{ipmPmpSPTable}), (\ref{ipmPmpBereaTable}),
(\ref{ipmPmpCubicSPTable}) and (\ref{ipmPmpCubicBereaTable}). So,
the PMP is superior to the IPM in performance.

\vspace{0.2cm}

The two methods are used to investigate the threshold yield pressure
of various slices of the \sandp\ and \Berea\ networks and their
cubic equivalent with different location and width, including the
whole width, to obtain different networks with various
characteristics. A sample of the results is presented in Tables
(\ref{ipmPmpSPTable}), (\ref{ipmPmpBereaTable}),
(\ref{ipmPmpCubicSPTable}) and (\ref{ipmPmpCubicBereaTable}).

\vspace{0.2cm}

It is noticeable that in most cases the IPM and PMP predict much
lower threshold yield pressures than the network as obtained from
solving the pressure field across the network in the flow
simulation, as described in Section (\ref{ModelingFlowPorous}),
particularly for the \Berea\ network. The explanation is that both
algorithms are based on the assumption that the threshold yield
pressure of serially-connected bonds is the sum of their yield
pressures. This assumption can be challenged by the fact that the
pressure gradient across the ensemble should reach the threshold
yield gradient for the bottleneck of the ensemble if yield should
occur. Moreover the IPM and PMP disregard the global pressure field
which can communicate with the internal nodes of the
serially-connected ensemble as it is part of a network and not an
isolated collection of bonds. It is not obvious that the IPM and PMP
condition should agree with the requirement of a stable and
mathematically consistent pressure filed as defined in Section
(\ref{ModelingFlowPorous}). Such an agreement should be regarded as
an extremely unlikely coincident. The static nature of the IPM and
PMP is behind this failure, that is their predictions are based on
the intrinsic properties of the network irrespective of the dynamic
aspects introduced by the pressure field. Good predictions can be
made only if the network is considered as a dynamic entity.

\vspace{0.2cm}

Several techniques were employed to investigate the divergence
between the IPM and PMP predictions and the solver results. One of
these techniques is to inspect the yield path at threshold when
using the solver to find the pressure field. The results were
conclusive; that is at the network yield point all elements in the
path have already reached their yield pressure. In particular, there
is a single element, a bottleneck, that is exactly at its threshold
yield pressure. This confirms the reliability of the network model
as it is functioning in line with the design. Beyond this, no
significance should be attached to this as a novel science.

\vspace{0.2cm}

One way of assessing the results is to estimate the average throat
size very roughly when flow starts in a \yields\ fluid by applying
the \percolation\ theory, as discussed by Rossen and Mamun
\cite{rossen2}, then comparing this to the average throat size as
found from the pressure solver and from the IPM and PMP. However, we
did not follow this line of investigation because we do not believe
that network yield is a \percolation\ phenomenon. Furthermore, the
\nNEW\ code in its current state has no \percolation\ capability.

\vspace{0.2cm}

In most cases of the random networks the predictions of the IPM and
PMP agree. However, when they disagree the PMP gives a higher value
of yield pressure. The reason is that backtracking is allowed in IPM
but not in PMP. When the actual path of minimum sum has a backward
component, which is not allowed by the PMP, the alternative path of
next minimum sum with no backtracking is more restrictive and hence
has a higher yield pressure value. This explanation is supported by
the predictions of the regular networks where the two methods
produce identical results as backtracking is less likely to occur in
a cubic network since it involves a more tortuous and restrictive
path with more than one connecting bond.

%%%%%%%%%%%%%%%%%%%%%%%%%%%%%%%   IPM flowchart   %%%%%%%%%%%%%%%%%%%%%%%%%%%%%%%%%%%%%%%%%
\newpage
\hspace{0.0cm}    \vspace{0.0cm}
\begin{figure}[!h]
  \centering{}
  \includegraphics
  [scale=0.90, trim = 0 -20 0 20]
  {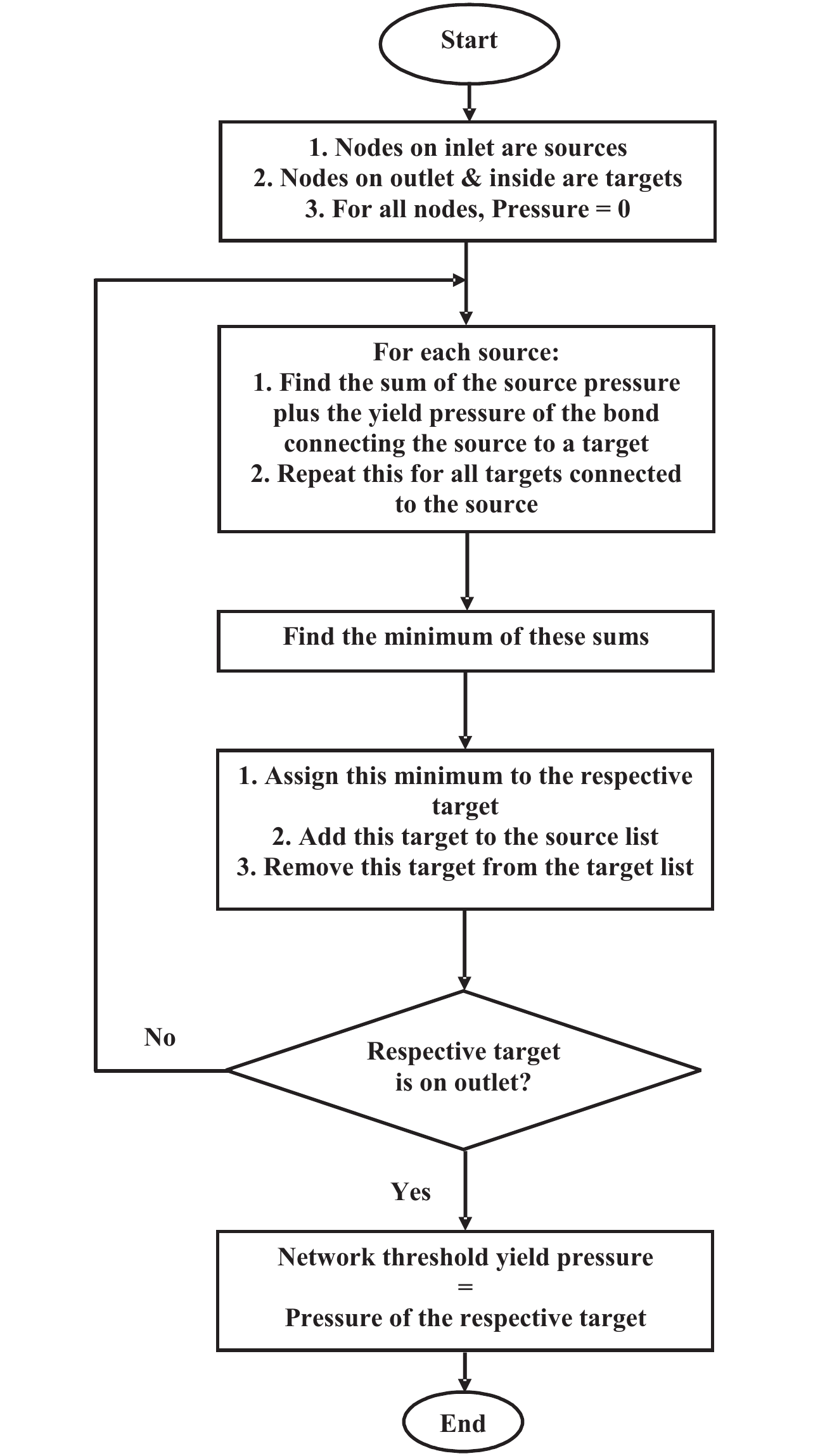}
  \caption[Flowchart of the \InvPM\ (IPM) algorithm]
  {Flowchart of the \InvPM\ (IPM) algorithm.}
  \label{IPMFlowchart}
\end{figure}

%%%%%%%%%%%%%%%%%%%%%%%%%%%%%%%   PMP flowchart   %%%%%%%%%%%%%%%%%%%%%%%%%%%%%%%%%%%%%%%%%
\newpage
\hspace{0.0cm}    \vspace{0.0cm}
\begin{figure}[!h]
  \centering{}
  \includegraphics
  [scale=0.9, trim = 0 -5 0 20]
  {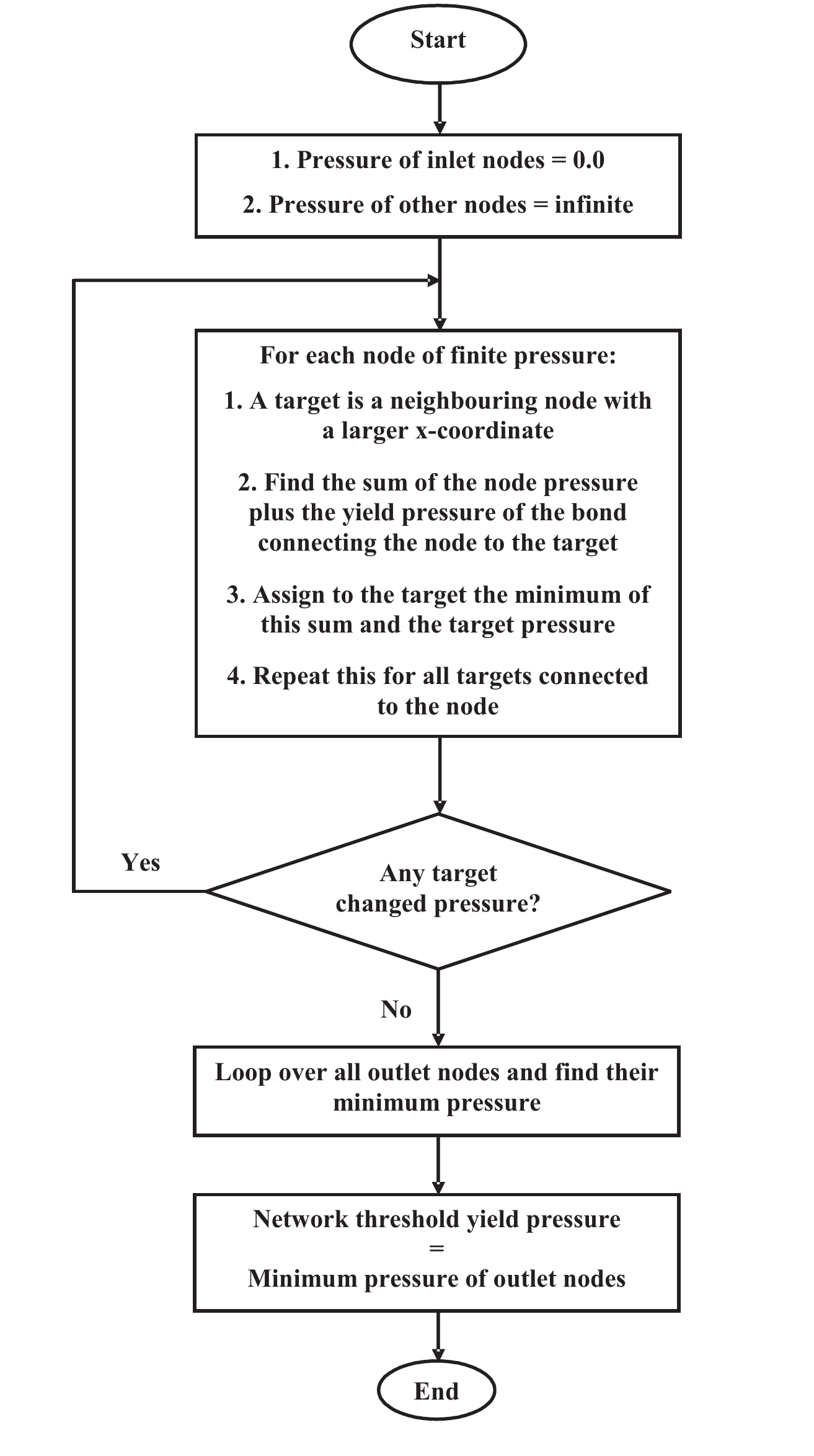}
  \caption[Flowchart of the \PathMP\ (PMP) algorithm]
  {Flowchart of the \PathMP\ (PMP) algorithm.}
  \label{PMPFlowchart}
\end{figure}

%%%%%%%%%%%%%%%%%%%%%%%%%%%%%%%   SP IPM-PMP table   %%%%%%%%%%%%%%%%%%%%%%%%%%%%%%%%%%%%%%%%%
\textheight = 745pt         %
\newpage

\begin{table} [h]
\centering %
\caption[Comparison between IPM and PMP for various slices of \sandp] %
{Comparison between IPM and PMP for various slices of the \sandp.} %
\label{ipmPmpSPTable} %
\vspace{0.5cm} %

\begin{tabular}{|c|c|c|c|c|c|c|c|c|c|}
\hline
           & \multicolumn{ 2}{c|}{{\bf Boundary}} & \multicolumn{ 3}{c|}{{\bf P$_{y}$ (Pa)}} & \multicolumn{ 2}{c|}{{\bf Iterations}} & \multicolumn{ 2}{c|}{{\bf Time (s)}} \\
\hline
 {\bf No.} & {\bf $x_{_{l}}$} & {\bf $x_{_{u}}$} & {\bf Network} &  {\bf IPM} &  {\bf PMP} &  {\bf IPM} &  {\bf PMP} &  {\bf IPM} &  {\bf PMP} \\
\hline
         1 &        0.0 &        1.0 &      80.94 &      53.81 &      54.92 &       2774 &         14 &       2.77 &       0.17 \\
\hline
         2 &        0.0 &        0.9 &      71.25 &      49.85 &      51.13 &       2542 &         11 &       2.47 &       0.13 \\
\hline
         3 &        0.0 &        0.8 &      61.14 &      43.96 &      44.08 &       2219 &         10 &       2.09 &       0.11 \\
\hline
         4 &        0.0 &        0.7 &      56.34 &      38.47 &      38.74 &       1917 &          9 &       1.75 &       0.08 \\
\hline
         5 &        0.0 &        0.6 &      51.76 &      32.93 &      33.77 &       1607 &         10 &       1.42 &       0.08 \\
\hline
         6 &        0.0 &        0.5 &      29.06 &      21.52 &      21.52 &       1048 &          9 &       0.84 &       0.06 \\
\hline
         7 &        0.0 &        0.4 &      24.24 &      18.45 &      18.45 &        897 &          9 &       0.69 &       0.05 \\
\hline
         8 &        0.0 &        0.3 &      18.45 &      12.97 &      12.97 &        622 &          7 &       0.44 &       0.03 \\
\hline
         9 &        0.1 &        1.0 &      69.75 &      47.67 &      48.78 &       2323 &         12 &       2.41 &       0.16 \\
\hline
        10 &        0.1 &        0.9 &      60.24 &      43.71 &      44.70 &       2084 &          9 &       2.14 &       0.09 \\
\hline
        11 &        0.1 &        0.8 &      49.53 &      37.82 &      37.94 &       1756 &          9 &       1.75 &       0.09 \\
\hline
        12 &        0.1 &        0.7 &      43.68 &      32.27 &      32.27 &       1471 &          9 &       1.42 &       0.09 \\
\hline
        13 &        0.1 &        0.6 &      39.72 &      26.64 &      26.64 &       1176 &         10 &       1.08 &       0.08 \\
\hline
        14 &        0.1 &        0.5 &      18.82 &      15.39 &      15.39 &        602 &          9 &       0.52 &       0.06 \\
\hline
        15 &        0.1 &        0.4 &      14.40 &      12.32 &      12.32 &        448 &          8 &       0.38 &       0.03 \\
\hline
        16 &        0.2 &        1.0 &      53.81 &      43.26 &      43.86 &       2060 &         12 &       2.08 &       0.14 \\
\hline
        17 &        0.2 &        0.9 &      46.44 &      39.30 &      40.19 &       1843 &         10 &       1.86 &       0.09 \\
\hline
        18 &        0.2 &        0.8 &      37.13 &      30.27 &      30.27 &       1372 &          9 &       1.33 &       0.08 \\
\hline
        19 &        0.2 &        0.7 &      37.84 &      24.28 &      24.28 &       1044 &          9 &       0.94 &       0.06 \\
\hline
        20 &        0.2 &        0.6 &      29.32 &      19.96 &      19.96 &        811 &          8 &       0.69 &       0.05 \\
\hline
        21 &        0.2 &        0.5 &      21.26 &      10.98 &      10.98 &        416 &          7 &       0.33 &       0.03 \\
\hline
        22 &        0.3 &        1.0 &      53.19 &      37.85 &      37.85 &       1727 &         11 &       1.75 &       0.09 \\
\hline
        23 &        0.3 &        0.9 &      45.64 &      33.56 &      33.56 &       1508 &          9 &       1.50 &       0.08 \\
\hline
        24 &        0.3 &        0.8 &      37.36 &      23.42 &      23.42 &       1006 &          7 &       0.95 &       0.05 \\
\hline
        25 &        0.3 &        0.7 &      28.43 &      17.43 &      17.43 &        672 &          7 &       0.58 &       0.05 \\
\hline
        26 &        0.3 &        0.6 &      19.09 &      10.94 &      10.94 &        355 &          7 &       0.27 &       0.03 \\
\hline
        27 &        0.4 &        1.0 &      47.23 &      30.46 &      30.96 &       1378 &         11 &       1.42 &       0.09 \\
\hline
        28 &        0.4 &        0.9 &      28.35 &      24.44 &      24.44 &       1062 &          8 &       1.08 &       0.06 \\
\hline
        29 &        0.4 &        0.8 &      24.28 &      15.82 &      15.82 &        623 &          8 &       0.59 &       0.05 \\
\hline
        30 &        0.4 &        0.7 &      10.00 &       9.83 &       9.83 &        301 &          6 &       0.24 &       0.03 \\
\hline
        31 &        0.5 &        1.0 &      36.25 &      24.68 &      24.68 &       1119 &          8 &       1.08 &       0.05 \\
\hline
        32 &        0.5 &        0.9 &      28.20 &      19.25 &      19.25 &        821 &          7 &       0.77 &       0.03 \\
\hline
        33 &        0.5 &        0.8 &      20.85 &      13.71 &      13.71 &        528 &          7 &       0.47 &       0.03 \\
\hline
\end{tabular}
\end{table}

\vspace* {0.0cm}    %
{\tiny \noindent
 {\begin{spacing}{1.1} \noindent
 Note: this data is for a \BING\ fluid ($n=1.0$) with $\ysS=1.0$Pa. The PMP is implemented in three forms.
 The ``Iterations'' and ``time'' data is for the slowest but the least memory demanding form. The
 machine used is a Pentium 3.4\,GHz processor workstation.
 \end{spacing}}
} %
\textheight = 23cm

%%%%%%%%%%%%%%%%%%%%%%%%%%%%%%%   Berea IPM-PMP table   %%%%%%%%%%%%%%%%%%%%%%%%%%%%%%%%%%%%%%%%%
\textheight = 745pt         %
\newpage

\begin{table} [h]
\centering %
\caption[Comparison between IPM and PMP for various slices of \Berea] %
{Comparison between IPM and PMP for various slices of \Berea.} %
\label{ipmPmpBereaTable} %
\vspace{0.5cm} %
\begin{tabular}{|c|c|c|c|c|c|c|c|c|c|}
\hline
           & \multicolumn{ 2}{c|}{{\bf Boundary}} & \multicolumn{ 3}{c|}{{\bf P$_{y}$ (Pa)}} & \multicolumn{ 2}{c|}{{\bf Iterations}} & \multicolumn{ 2}{c|}{{\bf Time (s)}} \\
\hline
 {\bf No.} & {\bf $x_{_{l}}$} & {\bf $x_{_{u}}$} & {\bf Network} &  {\bf IPM} &  {\bf PMP} &  {\bf IPM} &  {\bf PMP} &  {\bf IPM} &  {\bf PMP} \\
\hline
         1 &        0.0 &        1.0 &     331.49 &     121.68 &     121.68 &       8729 &         29 &      20.31 &       1.11 \\
\hline
         2 &        0.0 &        0.9 &     251.20 &     102.98 &     104.46 &       7242 &         26 &      17.50 &       0.89 \\
\hline
         3 &        0.0 &        0.8 &     235.15 &      84.55 &      87.51 &       5759 &         24 &      13.02 &       0.72 \\
\hline
         4 &        0.0 &        0.7 &     148.47 &      62.32 &      62.32 &       3968 &         21 &       8.39 &       0.53 \\
\hline
         5 &        0.0 &        0.6 &     115.80 &      43.24 &      43.24 &       2311 &         18 &       3.94 &       0.38 \\
\hline
         6 &        0.0 &        0.5 &      59.49 &      27.34 &      27.34 &       1146 &         17 &       1.25 &       0.30 \\
\hline
         7 &        0.0 &        0.4 &      23.13 &       8.14 &       8.14 &        361 &         18 &       0.31 &       0.25 \\
\hline
         8 &        0.0 &        0.3 &       3.92 &       3.72 &       3.72 &        265 &         12 &       0.14 &       0.11 \\
\hline
         9 &        0.1 &        1.0 &     283.88 &     119.38 &     119.38 &       8402 &         31 &      19.56 &       1.13 \\
\hline
        10 &        0.1 &        0.9 &     208.97 &     100.67 &     102.15 &       7002 &         26 &      16.14 &       0.81 \\
\hline
        11 &        0.1 &        0.8 &     189.73 &      82.24 &      84.23 &       5575 &         22 &      12.49 &       0.61 \\
\hline
        12 &        0.1 &        0.7 &     128.55 &      60.02 &      60.02 &       3876 &         20 &       8.02 &       0.45 \\
\hline
        13 &        0.1 &        0.6 &      98.56 &      40.93 &      40.93 &       2417 &         16 &       4.64 &       0.30 \\
\hline
        14 &        0.1 &        0.5 &      37.81 &      25.04 &      25.04 &       1270 &         15 &       2.05 &       0.24 \\
\hline
        15 &        0.1 &        0.4 &      28.52 &      22.02 &      22.02 &       1073 &         21 &       1.66 &       0.36 \\
\hline
        16 &        0.2 &        1.0 &     295.77 &     110.33 &     115.60 &       7424 &         27 &      21.77 &       0.88 \\
\hline
        17 &        0.2 &        0.9 &     222.84 &      94.58 &      95.48 &       6228 &         22 &      13.67 &       0.61 \\
\hline
        18 &        0.2 &        0.8 &     191.51 &      72.05 &      72.05 &       4589 &         20 &       9.77 &       0.47 \\
\hline
        19 &        0.2 &        0.7 &     147.06 &      58.60 &      58.60 &       3605 &         18 &       7.34 &       0.34 \\
\hline
        20 &        0.2 &        0.6 &      82.05 &      39.52 &      39.52 &       2237 &         15 &       4.09 &       0.23 \\
\hline
        21 &        0.2 &        0.5 &      29.57 &      23.62 &      23.62 &       1150 &         11 &       1.80 &       0.13 \\
\hline
        22 &        0.3 &        1.0 &     295.95 &      96.73 &     102.00 &       6223 &         21 &      13.19 &       0.61 \\
\hline
        23 &        0.3 &        0.9 &     219.45 &      80.97 &      81.88 &       5076 &         21 &      10.49 &       0.52 \\
\hline
        24 &        0.3 &        0.8 &     171.16 &      58.45 &      58.45 &       3460 &         17 &       6.77 &       0.34 \\
\hline
        25 &        0.3 &        0.7 &     126.21 &      48.02 &      48.02 &       2723 &         15 &       5.05 &       0.23 \\
\hline
        26 &        0.3 &        0.6 &      56.09 &      34.52 &      34.52 &       1856 &         12 &       3.09 &       0.16 \\
\hline
        27 &        0.4 &        1.0 &     214.73 &      79.42 &      84.69 &       4923 &         22 &       9.92 &       0.56 \\
\hline
        28 &        0.4 &        0.9 &     149.53 &      63.67 &      64.58 &       3787 &         17 &       7.50 &       0.36 \\
\hline
        29 &        0.4 &        0.8 &     106.15 &      41.14 &      41.14 &       2227 &         14 &       3.99 &       0.25 \\
\hline
        30 &        0.4 &        0.7 &      60.82 &      30.71 &      30.71 &       1530 &         13 &       2.45 &       0.17 \\
\hline
        31 &        0.5 &        1.0 &     125.88 &      67.61 &      67.61 &       4211 &         19 &       8.69 &       0.38 \\
\hline
        32 &        0.5 &        0.9 &      94.17 &      54.47 &      54.49 &       3267 &         15 &       6.53 &       0.24 \\
\hline
        33 &        0.5 &        0.8 &      83.00 &      31.94 &      31.94 &       1711 &         12 &       3.13 &       0.14 \\
\hline
\end{tabular}
\end{table}

\vspace* {0.0cm}    %
{\tiny \noindent
 {\begin{spacing}{1.1} \noindent
 Note: this data is for a \BING\ fluid ($n=1.0$) with $\ysS=1.0$Pa. The PMP is implemented in three forms.
 The ``Iterations'' and ``time'' data is for the slowest but the least memory demanding form. The
 machine used is a Pentium 3.4\,GHz processor workstation.
 \end{spacing}}
} %
\textheight = 23cm

%%%%%%%%%%%%%%%%%%%%%%%%%%%%%%%   Cubic SP IPM-PMP table   %%%%%%%%%%%%%%%%%%%%%%%%%%%%%%%%%%%%%%%%%
\textheight = 745pt         %
\newpage

\begin{table} [h]
\centering %
\caption[Comparison between IPM and PMP for various slices of cubic network with similar properties to \sandp] %
{Comparison between IPM and PMP for various slices of a cubic network with similar properties to the \sandp.} %
\label{ipmPmpCubicSPTable} %
\vspace{0.5cm} %

\begin{tabular}{|c|c|c|c|c|c|c|c|c|c|}
\hline
           & \multicolumn{ 2}{c|}{{\bf Boundary}} & \multicolumn{ 3}{c|}{{\bf P$_{y}$ (Pa)}} & \multicolumn{ 2}{c|}{{\bf Iterations}} & \multicolumn{ 2}{c|}{{\bf Time (s)}} \\
\hline
 {\bf No.} & {\bf $x_{_{l}}$} & {\bf $x_{_{u}}$} & {\bf Network} &  {\bf IPM} &  {\bf PMP} &  {\bf IPM} &  {\bf PMP} &  {\bf IPM} &  {\bf PMP} \\
\hline
         1 &        0.0 &        1.0 &     108.13 &      64.73 &      64.73 &       2860 &          9 &       1.39 &       0.14 \\
\hline
         2 &        0.0 &        0.9 &      81.72 &      53.94 &      53.94 &       2381 &          7 &       1.16 &       0.11 \\
\hline
         3 &        0.0 &        0.8 &      81.58 &      46.06 &      46.06 &       2006 &          7 &       0.97 &       0.09 \\
\hline
         4 &        0.0 &        0.7 &      62.99 &      39.05 &      39.05 &       1678 &          6 &       0.81 &       0.06 \\
\hline
         5 &        0.0 &        0.6 &      53.14 &      32.90 &      32.90 &       1401 &          6 &       0.67 &       0.05 \\
\hline
         6 &        0.0 &        0.5 &      40.20 &      27.50 &      27.50 &       1147 &          5 &       0.53 &       0.05 \\
\hline
         7 &        0.0 &        0.4 &      31.97 &      22.22 &      22.22 &        912 &          5 &       0.42 &       0.03 \\
\hline
         8 &        0.0 &        0.3 &      28.70 &      18.32 &      18.32 &        745 &          5 &       0.34 &       0.03 \\
\hline
         9 &        0.1 &        1.0 &      86.28 &      55.75 &      55.75 &       2422 &          9 &       1.19 &       0.14 \\
\hline
        10 &        0.1 &        0.9 &      63.74 &      44.95 &      44.95 &       1946 &          7 &       0.97 &       0.09 \\
\hline
        11 &        0.1 &        0.8 &      62.52 &      37.07 &      37.07 &       1568 &          7 &       0.77 &       0.08 \\
\hline
        12 &        0.1 &        0.7 &      42.70 &      30.07 &      30.07 &       1247 &          5 &       0.59 &       0.05 \\
\hline
        13 &        0.1 &        0.6 &      34.64 &      23.92 &      23.92 &        979 &          5 &       0.47 &       0.05 \\
\hline
        14 &        0.1 &        0.5 &      27.78 &      18.52 &      18.52 &        732 &          5 &       0.34 &       0.03 \\
\hline
        15 &        0.1 &        0.4 &      19.81 &      13.24 &      13.24 &        496 &          4 &       0.23 &       0.02 \\
\hline
        16 &        0.2 &        1.0 &      90.10 &      48.19 &      48.19 &       2200 &          5 &       1.08 &       0.08 \\
\hline
        17 &        0.2 &        0.9 &      73.98 &      39.92 &      39.92 &       1834 &          5 &       0.88 &       0.05 \\
\hline
        18 &        0.2 &        0.8 &      61.36 &      32.21 &      32.21 &       1475 &          5 &       0.72 &       0.05 \\
\hline
        19 &        0.2 &        0.7 &      39.98 &      25.40 &      25.40 &       1158 &          5 &       0.55 &       0.05 \\
\hline
        20 &        0.2 &        0.6 &      30.63 &      19.24 &      19.24 &        878 &          5 &       0.41 &       0.05 \\
\hline
        21 &        0.2 &        0.5 &      21.62 &      13.84 &      13.84 &        650 &          4 &       0.30 &       0.02 \\
\hline
        22 &        0.3 &        1.0 &      47.09 &      44.33 &      44.33 &       1879 &          5 &       0.91 &       0.06 \\
\hline
        23 &        0.3 &        0.9 &      37.88 &      35.73 &      35.73 &       1484 &          5 &       0.72 &       0.05 \\
\hline
        24 &        0.3 &        0.8 &      30.84 &      25.88 &      25.88 &       1045 &          5 &       0.50 &       0.03 \\
\hline
        25 &        0.3 &        0.7 &      24.86 &      20.62 &      20.62 &        794 &          5 &       0.38 &       0.03 \\
\hline
        26 &        0.3 &        0.6 &      17.87 &      14.65 &      14.65 &        528 &          4 &       0.25 &       0.02 \\
\hline
        27 &        0.4 &        1.0 &      45.45 &      37.15 &      37.15 &       1657 &          5 &       0.81 &       0.05 \\
\hline
        28 &        0.4 &        0.9 &      36.02 &      29.30 &      29.30 &       1302 &          5 &       0.63 &       0.03 \\
\hline
        29 &        0.4 &        0.8 &      26.66 &      20.42 &      20.42 &        914 &          5 &       0.44 &       0.03 \\
\hline
        30 &        0.4 &        0.7 &      19.90 &      15.16 &      15.16 &        675 &          4 &       0.33 &       0.02 \\
\hline
        31 &        0.5 &        1.0 &      52.31 &      30.35 &      30.35 &       1392 &          6 &       0.67 &       0.05 \\
\hline
        32 &        0.5 &        0.9 &      37.67 &      22.08 &      22.08 &       1016 &          6 &       0.47 &       0.05 \\
\hline
        33 &        0.5 &        0.8 &      21.99 &      16.33 &      16.33 &        764 &          5 &       0.36 &       0.03 \\
\hline
\end{tabular}
\end{table}

\vspace* {0.0cm}    %
{\tiny \noindent
 {\begin{spacing}{1.1} \noindent
 Note: this data is for a \BING\ fluid ($n=1.0$) with $\ysS=1.0$Pa. The PMP is implemented in three forms.
 The ``Iterations'' and ``time'' data is for the slowest but the least memory demanding form. The
 machine used is a dual core 2.4\,GHz processor workstation.
 \end{spacing}}
} %
\textheight = 23cm

%%%%%%%%%%%%%%%%%%%%%%%%%%%%%%%   Cubic Berea IPM-PMP table   %%%%%%%%%%%%%%%%%%%%%%%%%%%%%%%%%%%%%%%%%
\textheight = 745pt         %
\newpage

\begin{table} [h]
\centering %
\caption[Comparison between IPM and PMP for various slices of cubic network with similar properties to \Berea] %
{Comparison between IPM and PMP for various slices of a cubic network with similar properties to \Berea.} %
\label{ipmPmpCubicBereaTable} %
\vspace{0.5cm} %

\begin{tabular}{|c|c|c|c|c|c|c|c|c|c|}
\hline
           & \multicolumn{ 2}{c|}{{\bf Boundary}} & \multicolumn{ 3}{c|}{{\bf P$_{y}$ (Pa)}} & \multicolumn{ 2}{c|}{{\bf Iterations}} & \multicolumn{ 2}{c|}{{\bf Time (s)}} \\
\hline
 {\bf No.} & {\bf $x_{_{l}}$} & {\bf $x_{_{u}}$} & {\bf Network} &  {\bf IPM} &  {\bf PMP} &  {\bf IPM} &  {\bf PMP} &  {\bf IPM} &  {\bf PMP} \\
\hline
         1 &        0.0 &        1.0 &     128.46 &     100.13 &     100.13 &      10366 &         10 &      10.03 &       0.47 \\
\hline
         2 &        0.0 &        0.9 &     113.82 &      89.34 &      89.34 &       9188 &          9 &       8.89 &       0.38 \\
\hline
         3 &        0.0 &        0.8 &     100.77 &      81.07 &      81.07 &       8348 &          9 &       8.08 &       0.34 \\
\hline
         4 &        0.0 &        0.7 &      87.36 &      69.89 &      69.89 &       7167 &          8 &       6.91 &       0.28 \\
\hline
         5 &        0.0 &        0.6 &      75.21 &      58.59 &      58.59 &       5963 &          9 &       5.80 &       0.28 \\
\hline
         6 &        0.0 &        0.5 &      61.25 &      47.31 &      47.31 &       4747 &          7 &       4.45 &       0.17 \\
\hline
         7 &        0.0 &        0.4 &      50.22 &      37.11 &      37.11 &       3703 &          7 &       3.49 &       0.14 \\
\hline
         8 &        0.0 &        0.3 &      36.63 &      25.10 &      25.10 &       2429 &          6 &       2.25 &       0.09 \\
\hline
         9 &        0.1 &        1.0 &     111.99 &      88.24 &      88.24 &       8958 &         10 &       8.64 &       0.42 \\
\hline
        10 &        0.1 &        0.9 &      98.19 &      77.45 &      77.45 &       7762 &         10 &       7.38 &       0.38 \\
\hline
        11 &        0.1 &        0.8 &      86.46 &      65.83 &      65.83 &       6543 &          9 &       6.24 &       0.30 \\
\hline
        12 &        0.1 &        0.7 &      71.83 &      54.06 &      54.06 &       5309 &          8 &       5.00 &       0.23 \\
\hline
        13 &        0.1 &        0.6 &      58.74 &      45.49 &      45.49 &       4405 &          9 &       4.16 &       0.22 \\
\hline
        14 &        0.1 &        0.5 &      46.35 &      34.40 &      34.40 &       3216 &          7 &       3.03 &       0.13 \\
\hline
        15 &        0.1 &        0.4 &      34.56 &      24.20 &      24.20 &       2166 &          7 &       1.97 &       0.09 \\
\hline
        16 &        0.2 &        1.0 &     112.66 &      76.32 &      76.32 &       7742 &          9 &       7.34 &       0.36 \\
\hline
        17 &        0.2 &        0.9 &      96.83 &      65.57 &      65.57 &       6577 &          9 &       6.17 &       0.30 \\
\hline
        18 &        0.2 &        0.8 &      82.38 &      56.60 &      56.60 &       5615 &          8 &       5.27 &       0.24 \\
\hline
        19 &        0.2 &        0.7 &      68.62 &      44.65 &      44.65 &       4369 &          7 &       4.06 &       0.17 \\
\hline
        20 &        0.2 &        0.6 &      51.78 &      35.40 &      35.40 &       3398 &          8 &       3.13 &       0.17 \\
\hline
        21 &        0.2 &        0.5 &      36.82 &      25.56 &      25.56 &       2358 &          6 &       2.11 &       0.08 \\
\hline
        22 &        0.3 &        1.0 &      98.13 &      63.93 &      63.93 &       6576 &          8 &       6.25 &       0.28 \\
\hline
        23 &        0.3 &        0.9 &      82.63 &      53.18 &      53.18 &       5403 &          8 &       5.06 &       0.23 \\
\hline
        24 &        0.3 &        0.8 &      69.09 &      45.16 &      45.16 &       4557 &          8 &       4.27 &       0.19 \\
\hline
        25 &        0.3 &        0.7 &      53.13 &      34.73 &      34.73 &       3445 &          6 &       3.20 &       0.13 \\
\hline
        26 &        0.3 &        0.6 &      39.52 &      25.51 &      25.51 &       2503 &          7 &       2.28 &       0.11 \\
\hline
        27 &        0.4 &        1.0 &     138.82 &      52.26 &      52.26 &       5468 &          9 &       5.16 &       0.28 \\
\hline
        28 &        0.4 &        0.9 &     107.95 &      41.47 &      41.47 &       4259 &          7 &       3.94 &       0.17 \\
\hline
        29 &        0.4 &        0.8 &      71.37 &      33.20 &      33.20 &       3412 &          7 &       3.16 &       0.14 \\
\hline
        30 &        0.4 &        0.7 &      35.52 &      22.02 &      22.02 &       2223 &          6 &       2.02 &       0.09 \\
\hline
        31 &        0.5 &        1.0 &      74.88 &      46.07 &      46.07 &       4847 &          8 &       4.59 &       0.20 \\
\hline
        32 &        0.5 &        0.9 &      56.75 &      35.28 &      35.28 &       3683 &          7 &       3.44 &       0.16 \\
\hline
        33 &        0.5 &        0.8 &      41.24 &      24.59 &      24.59 &       2551 &          7 &       2.34 &       0.11 \\
\hline
\end{tabular}
\end{table}

\vspace* {0.0cm}    %
{\tiny \noindent
 {\begin{spacing}{1.1} \noindent
 Note: this data is for a \BING\ fluid ($n=1.0$) with $\ysS=1.0$Pa. The PMP is implemented in three forms.
 The ``Iterations'' and ``time'' data is for the slowest but the least memory demanding form. The
 machine used is a dual core 2.4\,GHz processor workstation.
 \end{spacing}}
} %
\textheight = 23cm

\def\baselinestretch{1}
\chapter{\Vy} \label{Viscoelasticity}
\def\baselinestretch{1.66}
\Vc\ substances exhibit a dual nature of behavior by showing signs
of both viscous fluids and elastic solids. In its most simple form,
\vy\ can be modeled by combining \Newton's law for viscous fluids
(stress $\propto$ rate of strain) with \Hook's law for elastic
solids (stress $\propto$ strain), as given by the original \Maxwell\
model and extended by the Convected \Maxwell\ models for the
nonlinear \vc\ fluids. Although this idealization predicts several
basic \vc\ phenomena, it does so only qualitatively \cite{larsonbook1988}.%

\vspace{0.2cm}

The behavior of \vc\ fluids is drastically different from that of
\NEW\ and inelastic \nNEW\ fluids. This includes the presence of
normal stresses in shear flows, sensitivity to deformation type, and
memory effects such as stress relaxation and \timedep\ viscosity.
These features underlie the observed peculiar \vc\ phenomena such as
rod-climbing (\Weissenberg\ effect), die swell and open-channel
siphon \cite{Boger1987, larsonbook1988}. Most \vc\ fluids exhibit
\shThin\ and an elongational viscosity that is both strain and
extensional strain rate dependent, in contrast to \NEW\ fluids where
the elongational viscosity is constant \cite{Boger1987}.

\vspace{0.2cm}

The behavior of \vc\ fluids at any time is dependent on their recent
deformation history, that is they possess a fading memory of their
past. Indeed a material that has no memory cannot be elastic, since
it has no way of remembering its original shape. Consequently, an
ideal \vc\ fluid should behave as an elastic solid in sufficiently
rapid deformations and as a \NEW\ liquid in sufficiently slow
deformations. The justification is that the larger the strain rate,
the more strain is imposed on the sample within the memory span of
the fluid \cite{Boger1987, birdbook, larsonbook1988}.

\vspace{0.2cm}

Many materials are \vc\ but at different time scales that may not be
reached. Dependent on the time scale of the flow, \vc\ materials
mainly show viscous or elastic behavior. The particular response of
a sample in a given experiment depends on the time scale of the
experiment in relation to a natural time of the material. Thus, if
the experiment is relatively slow, the sample will appear to be
viscous rather than elastic, whereas, if the experiment is
relatively fast, it will appear to be elastic rather than viscous.
At intermediate time scales mixed \vc\ response is observed.
Therefore the concept of a natural time of a material is important
in characterizing the material as viscous or elastic. The ratio
between the material time scale and the time scale of the flow is
indicated by a non-dimensional number: the \Deborah\ or the
\Weissenberg\ number \cite{barnesbookHW1993, wapperomthesis}.

\vspace{0.2cm}

A common feature of \vc\ fluids is stress relaxation after a sudden
shearing displacement where stress overshoots to a maximum then
starts decreasing exponentially and eventually settles to a steady
state value. This phenomenon also takes place on cessation of steady
shear flow where stress decays over a finite measurable length of
time. This reveals that \vc\ fluids are able to store and release
energy in contrast to inelastic fluids which react instantaneously
to the imposed deformation \cite{birdbook, deiberthesis,
larsonbook1988}.

\vspace{0.2cm}

A defining characteristic of \vc\ materials associated with stress
relaxation is the relaxation time which may be defined as the time
required for the shear stress in a simple shear flow to return to
zero under constant strain condition. Hence for a \Hookean\ elastic
solid the relaxation time is infinite, while for a \NEW\ fluid the
relaxation of the stress is immediate and the relaxation time is
zero. Relaxation times which are infinite or zero are never realized
in reality as they correspond to the mathematical idealization of
\Hookean\ elastic solids and \NEW\ liquids. In practice, stress
relaxation after the imposition of constant strain condition takes
place over some finite non-zero time interval \cite{owensbook2002}.

\vspace{0.2cm}

The complexity of \vy\ is phenomenal and the subject is notorious
for being extremely difficult and challenging. The constitutive
equations for \vc\ fluids are much too complex to be treated in a
general manner. Further complications arise from the confusion
created by the presence of other phenomena such as wall effects and
polymer-wall interactions, and these appear to be system specific.
Therefore, it is doubtful that a general fluid model capable of
predicting all the flow responses of \vc\ system with enough
mathematical simplicity or tractability can emerge in the
foreseeable future \cite{Wissler1971, deiberthesis, ChhabraCM2001}.
Understandably, despite the huge amount of literature composed in
the last few decades on this subject, almost all these studies have
investigated very simple cases in which substantial simplifications
have been made using basic \vc\ models.

\vspace{0.2cm}

In the last few decades, a general consensus has emerged that in the
flow of \vc\ fluids through porous media elastic effects should
arise, though their precise nature is unknown or controversial. In
porous media, \vc\ effects can be important in certain cases. When
they are, the actual pressure gradient will exceed the purely
viscous gradient beyond a critical flow rate, as observed by several
investigators. The normal stresses of high molecular polymer
solutions can explain in part the high flow resistances encountered
during \vc\ flow through porous media. It is argued that the very
high normal stress differences and \TR\ ratios associated with
polymeric fluids will produce increasing values of apparent
viscosity when flow channels in the porous medium are of rapidly
changing cross section.

\vspace{0.2cm}

Important aspects of \nNEW\ flow in general and \vc\ flow in
particular through porous media are still presenting serious
challenge for modeling and quantification. There are intrinsic
difficulties of characterizing \nNEW\ effects in the flow of polymer
solutions and the complexities of the local geometry of the porous
medium which give rise to a complex and pore-space-dependent flow
field in which shear and extension coexist in various proportions
that cannot be quantified. Flows through porous media cannot be
classified as pure shear flows as the \convdiv\ passages impose a
predominantly extensional flow fields especially at high flow rates.
Moreover, the extension viscosity of many \nNEW\ fluids increases
dramatically with the extension rate. As a consequence, the
relationship between the pressure drop and flow rate very often do
not follow the observed \NEW\ and inelastic \nNEW\ trend. Further
complication arises from the fact that for complex fluids the stress
depends not only on whether the flow is a shearing, extensional, or
mixed type, but also on the whole history of the velocity gradient
\cite{MarshallM1967, DaubenM1967, PearsonT2002, PlogK, MendesN2002,
larsonbook1999}.

%XXXXXXXXXXXXXXXXXXXXXXXXXXXXXXXXXXXXXXXXXXXXXXXXXXXXXXXXXXXXXXXXXXXXXX
\section{Important Aspects for Flow in Porous Media} \label{ImportantAspects}
Strong experimental evidence indicates that the flow of \vc\ fluids
through packed beds can exhibit rapid increases in the pressure
drop, or an increase in the apparent viscosity, above that expected
for a comparable purely viscous fluid. This increase has been
attributed to the extensional nature of the flow field in the pores
caused by the successive expansions and contractions that a fluid
element experiences as it traverses the pore space. Even though the
flow field at pore level is not an ideal extensional flow due to the
presence of shear and rotation, the increase in flow resistance is
referred to as an extension thickening effect \cite{ThienK1987,
PlogK, DeiberS1981, PilitsisB1989}.

\vspace{0.2cm}

There are two major interrelated aspects that have strong impact on
the flow through porous media. These are extensional flow and
\convdiv\ geometry.

%XXXXXXXXXXXXXXXXXXXXXXXXXXXXXXXXXXXXXXXXXXXXXXXXXXXXXXXXXXXXXXXXXXXXXX
\subsection{Extensional Flow}\label{ExtensionalFlow}
One complexity in the flow in general and through porous media in
particular usually arises from the coexistence of shear and
extensional components; sometimes with the added complication of
inertia. Pure shear or elongational flow is very much the exception
in practical situations, especially in the flow through porous
media. By far the most common situation is for mixed flow to occur
where deformation rates have components parallel and perpendicular
to the principal flow direction. In such flows, the elongational
components may be associated with the \convdiv\ flow paths
\cite{barnesbookHW1993, sorbiebook}.

\vspace{0.2cm}

A general consensus has emerged recently that the flow through
packed beds has a substantial extensional component and typical
polymer solutions exhibit strain hardening in extension, which is
mainly responsible for the reported dramatic increases in pressure
drop. Thus in principle the shear viscosity alone is inadequate to
explain the observed excessive pressure gradients. One is therefore
interested to know the relative importance of elastic and viscous
effects or equivalently the relationship between normal and shear
stresses for different shear rates \cite{carreaubook, ChhabraCM2001,
deiberthesis}.

\vspace{0.2cm}

Elongational flow is fundamentally different from shear, the
material property characterizing the flow is not the viscosity, but
the elongational viscosity. The behavior of the extensional
viscosity function is very often qualitatively different from that
of the shear viscosity function. For example, highly elastic polymer
solutions that possess a viscosity which decreases monotonically in
shear often exhibit an extensional viscosity that increases
dramatically with strain rate. Thus, while the shear viscosity is
\shThin\, the extensional viscosity is extension thickening. A fluid
for which the extension viscosity increases with increasing
elongation rate is said to be tension-thickening, whilst, if it
decreases with increasing elongation rate it is said to be
tension-thinning \cite{larsonbook1999, barnesbookHW1993,
wapperomthesis}.

\vspace{0.2cm}

An extensional or elongational flow is one in which fluid elements
are subjected to extensions and compressions without being rotated
or sheared. The study of the extensional flow in general as a
relevant variable has only received attention in the last few
decades with the realization that extensional flow is of significant
relevance in many practical situations. Moreover, \nNEW\ elastic
liquids often exhibit dramatically different extensional flow
characteristics from \NEW\ liquids. Before that, rheology was
largely dominated by shear flow. The historical convention of
matching only shear flow data with theoretical predictions in
constitutive modeling may have to be rethought in those areas of
interest where there is a large extensional contribution.
Extensional flow experiments can be viewed as providing critical
tests for any proposed constitutive equations \cite{larsonbook1999,
deiberthesis, barnesbookHW1993, ChengH1984}.

\vspace{0.2cm}

The elongation or extension viscosity $\exVis$, also called \TR\
viscosity, is defined as the ratio of tensile stress and rate of
elongation under condition of steady flow when both these quantities
attain constant values. Mathematically, it is given by
\begin{equation}\label{ExtensionViscosity}
    \exVis = \frac{\sS_{E}}{\elongR}
\end{equation}
where $\sS_{E}$ ($= \sS_{11} - \sS_{22}$) is the normal stress
difference and $\elongR$ is the elongation rate. The stress
$\sS_{11}$ is in the direction of the elongation while $\sS_{22}$ is
in a direction perpendicular to the elongation \cite{lodgebook1964,
wapperomthesis, barnesbookHW1993}.

\vspace{0.2cm}

Polymeric fluids show a non-constant elongational viscosity in
steady and unsteady elongational flow. In general, it is a function
of the extensional strain rate, just as the shear viscosity is a
function of shear rate. However, the behavior of the extensional
viscosity function is frequently qualitatively different from that
of the shear viscosity  \cite{larsonbook1999, barnesbookHW1993}.

\vspace{0.2cm}

It is generally agreed that it is far more difficult to measure
extensional viscosity than shear viscosity. There is therefore a
gulf between the strong desire to measure extensional viscosity and
the likely expectation of its fulfilment \cite{barnesbookHW1993,
larsonbook1988}. A major difficulty that hinders the progress in
this field is that it is difficult to achieve steady elongational
flow and quantify it precisely. Despite the fact that many
techniques have been developed for measuring the elongational flow
properties, so far these techniques failed to produce consistent
outcome as they generate results which can differ by several orders
of magnitude. This indicates that these results are dependent on the
method and instrument of measurement. This is highlighted by the
view that the extensional viscometers provide measurements of an
extensional viscosity rather than the extensional viscosity. The
situation is made more complex by the fact that it is almost
impossible to generate a pure extensional flow since a shear
component is always present in real flow situations. This makes the
measurements doubtful and inconclusive \cite{birdbook, carreaubook,
ChhabraR1999, MendesN2002}.

\vspace{0.2cm}

For \NEW\ and purely viscous inelastic \nNEW\ fluids the
elongational viscosity is a constant that only depends on the type
of elongational deformation. Moreover, the viscosity measured in a
shearing flow can be used to predict the stress in other types of
deformation. For example, in a simple uniaxial elongational flow of
a \NEW\ fluid the following relationship is satisfied
\begin{equation}\label{exVisNewt}
    \exVis = 3 \lVis
\end{equation}

For \vc\ fluids the flow behavior is more complex and the
extensional viscosity, like the shear viscosity, depends on both the
strain rate and the time following the onset of straining. The
rheological behavior of a complex fluid in extension is often very
different from that in shear. Polymers usually have extremely high
extensional viscosities which can be orders of magnitude higher than
those expected on the basis of \NEW\ theory. Moreover, the
extensional viscosities of elastic polymer solutions can be
thousands of times greater than the shear viscosities. To measure
the departure of the ratio of extensional to shear viscosity from
its \NEW\ behavior, the rheologists have introduced what is known as
the \TR\ ratio usually defined as
\begin{equation}\label{Trouton}
    \Tr = \frac{\exVis(\elongR)}{\sVis(\sqrt{3}\elongR)}
\end{equation}

For elastic liquids, the \TR\ ratio is expected to be always greater
than or equal to 3, with the value 3 only attained at vanishingly
small strain rates. These liquids are known for having very high
\TR\ ratios which can be as high as l0$^4$. This behavior has to be
expected especially when the fluid combines \shThin\ with
tension-thickening. However, it should be remarked that even for the
fluids that show tension-thinning behavior the associated \TR\
ratios usually increase with strain rate and are still significantly
in excess of the inelastic value of three \cite{denysthesis,
owensbook2002, CrochetW1983, larsonbook1988, larsonbook1999,
ChengH1984, barnesbookHW1993}.

\vspace{0.2cm}

Figures (\ref{ShearViscosity}) and (\ref{ExtensionViscosity})
compare the elongational viscosity to the shear viscosity at
isothermal condition for a typical \vc\ fluid at various extension
and shear rates. As seen in Figure (\ref{ShearViscosity}), the shear
viscosity curve can be divided into three regions. For low-shear
rates, compared to the time scale of the fluid, the viscosity is
approximately constant. It then equals the so-called zero-shear-rate
viscosity. After this initial plateau the viscosity rapidly
decreases with increasing shear-rate. This behavior is \shThin.
However, there are some materials for which the reverse behavior is
observed, that is the viscosity increases with shear rate giving
rise to \shThik. For high shear rates the viscosity often
approximates a constant value again. The constant viscosity extremes
at low and high shear rates are known as the lower and upper \NEW\
plateau, respectively \cite{birdbook, larsonbook1988,
barnesbookHW1993, wapperomthesis}.

\vspace{0.2cm}

In Figure (\ref{ExtensionViscosity}) the typical behavior of the
elongational viscosity, $\exVis$, of a \vc\ fluid as a function of
elongation rate, $\elongR$, has been depicted. As seen, the
elongational viscosity curve can be divided into three regions. At
low elongation rates the elongational viscosity approaches a
constant value known as the zero-elongation-rate elongational
viscosity, which is three times the zero-shear-rate viscosity, just
as for a \NEW\ fluid. For somewhat larger elongation rates the
elongational viscosity increases with increasing elongation rate
until a maximum constant value is reached. If the elongation rate is
increased once more the elongational viscosity may decrease again. A
fluid for which $\exVis$ increases with increasing $\elongR$ is said
to be tension-thickening, whilst if $\exVis$ decreases with
increasing $\elongR$ it is said to be tension-thinning. It should be
remarked that two polymeric liquids which may have essentially the
same behavior in shear can show a different response in extension
\cite{wapperomthesis, owensbook2002, denysthesis, barnesbookHW1993,
larsonbook1999, birdbook}.

%%%%%%%%%%%%%%% Shear viscosity

\begin{figure}[!t]
  \centering{}
  \includegraphics
  [scale=0.6]
  {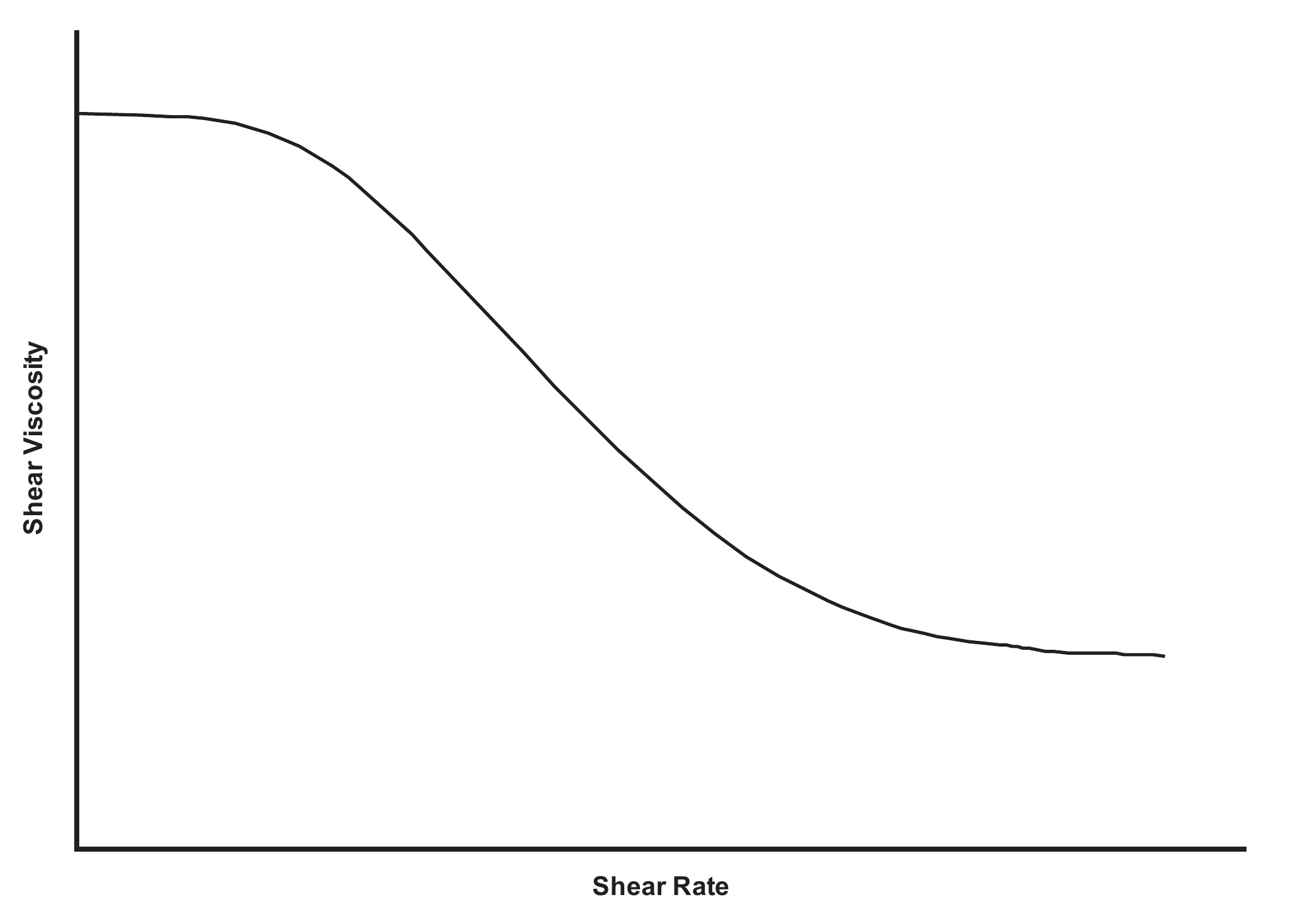}
  \caption[Typical behavior of shear viscosity $\sVis$ as a function of shear rate $\sR$ in shear flow on a log-log scale]
  {Typical behavior of shear viscosity $\sVis$ as a function of shear rate $\sR$ in shear flow on a log-log scale.}
  \label{ShearViscosity}
\end{figure}

%%%%%%%%%%%%%%% Extensional viscosity

\vspace{0.5cm}

\begin{figure}[!t]
  \centering{}
  \includegraphics
  [scale=0.6]
  {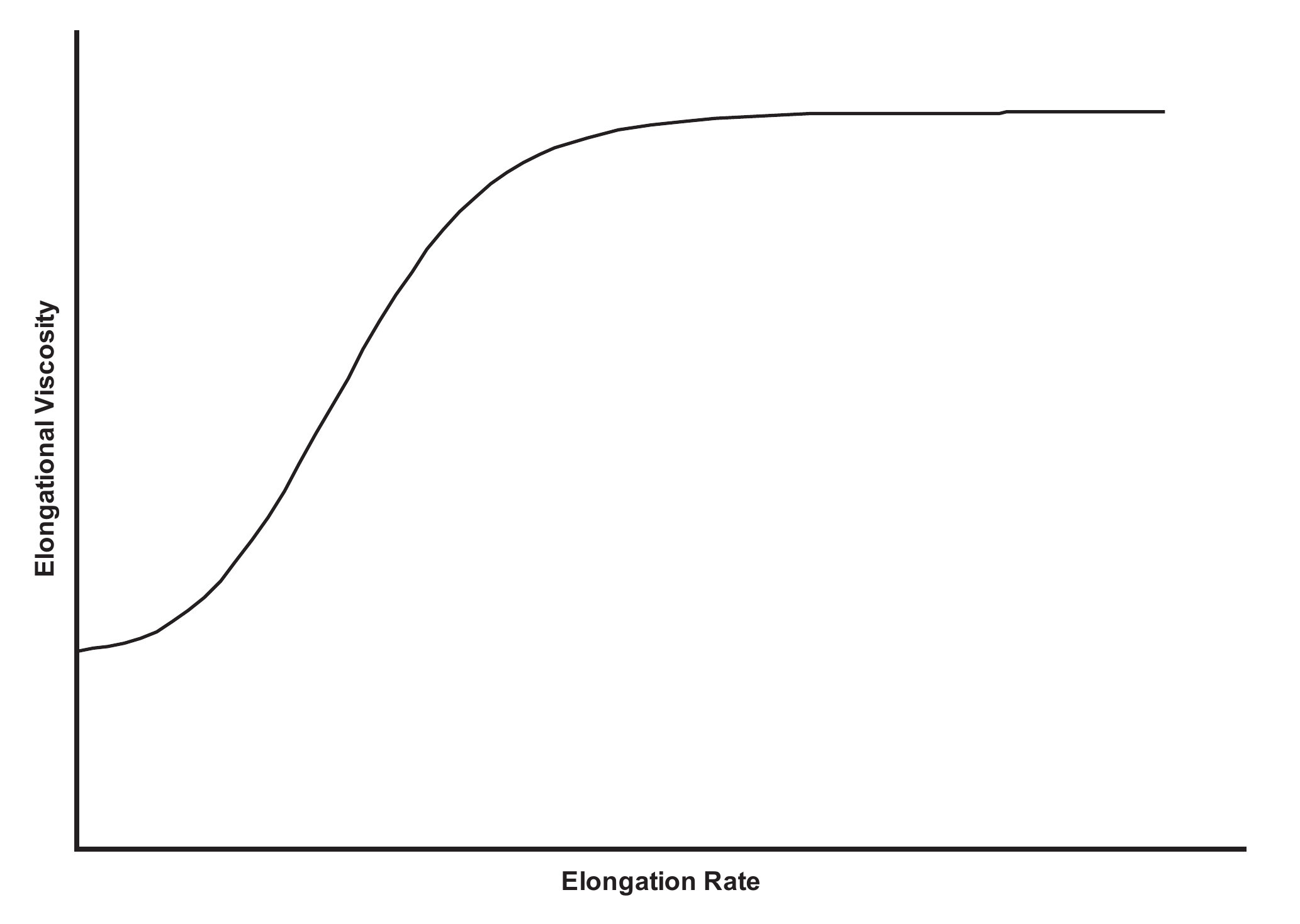}
  \caption[Typical behavior of elongational viscosity $\exVis$ as a function of elongation rate $\elongR$ in extensional flow on a log-log scale]
  {Typical behavior of elongational viscosity $\exVis$ as a function of elongation rate $\elongR$ in extensional flow on a log-log scale.}
  \label{ExtensionViscosity}
\end{figure}

%%%%%%%%%%%%%%%%

%XXXXXXXXXXXXXXXXXXXXXXXXXXXXXXXXXXXXXXXXXXXXXXXXXXXXXXXXXXXXXXXXXXXXXX
\subsection{\ConvDiv\ Geometry} \label{ConvergingDiverging}
An important aspect that characterizes the flow in porous media and
makes it distinct from bulk is the presence of \convdiv\ flow paths.
This geometric factor significantly affects the flow and accentuate
elastic responses. Several arguments have been presented in the
literature to explain the effect of \convdiv\ geometry on the flow
behavior.

\vspace{0.2cm}

One argument is that \vc\ flow in porous media differs from that of
\NEW\ flow, primarily because the \convdiv\ nature of flow in porous
media gives rise to normal stresses which are not solely
proportional to the local shear rate. As the elastic nature of the
fluid becomes more pronounced, an increase in flow resistance in
excess of that due to purely shearing forces occurs
\cite{TalwarK1992}. A second argument is that in a pure shear flow
the velocity gradient is perpendicular to the flow direction and the
axial velocity is only a function of the radial coordinate, while in
an elongational flow a velocity gradient in the flow direction
exists and the axial velocity is a function of the axial coordinate.
Any geometry which involves a diameter change will generate a flow
field with elongational characteristics. Therefore the flow field in
porous media is complex and involves both shear and elongational
components \cite{denysthesis}. A third argument suggests that
elastic effects are expected in the flow through porous media
because of the acceleration and deceleration of the fluid in the
interstices of the bed upon entering and leaving individual pores
\cite{GogartyLF1972, ParkHB1973}.

\vspace{0.2cm}

Despite this diversity, there is a general consensus that in porous
media the \convdiv\ nature of the flow paths brings out both the
extensional and shear properties of the fluid. The principal mode of
deformation to which a material element is subjected as the flow
converges into a constriction involves both a shearing of the
material element and a stretching or elongation in the direction of
flow, while in the diverging portion the flow involves both shearing
and compression. The actual channel geometry determines the ratio of
shearing to extensional contributions. In many realistic situations
involving \vc\ flows the extensional contribution is the most
important of the two modes. As porous media flow involves
elongational flow components, the coil-stretch phenomenon can also
take place. Consequently, a suitable model pore geometry is one
having converging and diverging sections which can reproduce the
elongational nature of the flow, a feature that is not exhibited by
straight capillary tubes \cite{MarshallM1967, deiberthesis,
denysthesis, DurstHI1987, PilitsisB1989}.

\vspace{0.2cm}

For long time, the straight capillary tube has been the conventional
model for porous media and packed beds. Despite the general success
of this model with the \NEW\ and inelastic \nNEW\ flow, its failure
with elastic flow was remarkable. To redress the flaws of this
model, the undulating tube and \convdiv\ channel were proposed in
order to include the elastic character of the flow. Various
corrugated tube models with different simple geometries have been
used as a first step to model the effect of \convdiv\ geometry on
the flow of \vc\ fluids in porous media (e.g. \cite{MarshallM1967,
DeiberS1981, SouvaliotisB1992}). Those geometries include conically
shaped sections, sinusoidal corrugation and abrupt expansions and
contractions. Similarly, a bundle of \convdiv\ tubes forms a better
model for a porous medium in \vc\ flow than the normally used bundle
of straight capillary tubes, as the presence of diameter variations
makes it possible to account for elongational contributions.

\vspace{0.2cm}

Many investigators have attempted to capture the role of the
successive \convdiv\ character of packed bed flow by numerically
solving the flow equations in conduits of periodically varying cross
sections. Different opinions on the success of these models can be
found in the literature. Some of these are presented in the
literature review \cite{denysthesis, deiberthesis, TalwarK1992,
ChhabraCM2001, PilitsisB1989, PodolsakTF1997} in Chapter
(\ref{Literature}). With regards to modeling \vc\ flow in regular or
random networks of \convdiv\ capillaries, very little work has been
done.

%XXXXXXXXXXXXXXXXXXXXXXXXXXXXXXXXXXXXXXXXXXXXXXXXXXXXXXXXXXXXXXXXXXXXXX
\section{\Vc\ Effects in Porous Media} \label{ViscoelasticEffects}
In packed bed flows, the main manifestation of \steadys\ \vc\
effects is the excess pressure drop or dilatancy behavior in
different flow regimes above what is accounted for by shear
viscosity of the flowing liquid. Qualitatively, this behavior has
been attributed to memory effects. Another explanation is that it is
due to extensional flow. However, both explanations are valid and
justified as long as the flow regime is considered. It should be
remarked that the geometry of the porous media must be taken into
account when considering elastic responses \cite{ChhabraCM2001,
GogartyLF1972}.

\vspace{0.2cm}

It is generally agreed that the flow of \vc\ fluids in packed beds
results in a greater pressure drop than that which can be ascribed
to the shear-rate-dependent viscosity. Fluids which exhibit
elasticity deviate from viscous flow above some critical velocity in
porous flow. At low flow rates, the pressure drop is determined
largely by shear viscosity, and the viscosity and elasticity are
approximately the same as in bulk. As the flow rate is gradually
increased, \vc\ effects begin to appear in various flow regimes.
Consequently, the \insitu\ rheological characteristics become
significantly different from those in bulk rheology as the
elasticity dramatically increases showing strong dilatant behavior
\cite{carreaubook, GogartyLF1972, ChhabraCM2001}.

\vspace{0.2cm}

Although experimental evidence for \vc\ effects is convincing, an
equally convincing theoretical analysis is not available. The
general argument is that when the fluid suffers a significant
deformation in a time comparable to the relaxation time of the
fluid, elastic effects become important. When the pressure drop is
plotted against a suitably defined \Deborah\ or \Weissenberg\
number, beyond a critical value of the \Deborah\ number, the
pressure drop increases rapidly \cite{Wissler1971, carreaubook}.

\vspace{0.2cm}

The complexity of the \vc\ flow in porous media is aggravated by the
possible occurrence of other non-\vc\ phenomena which have similar
effects as \vy. These phenomena include adsorption of the polymers
on the capillary walls, mechanical retention and partial pore
blockage. All these effects also lead to pressure drops well in
excess to that expected from the shear viscosity levels.
Consequently, the interpretation of many observed effects is
confused and controversial. Some authors may interpret some
observations in terms of partial pore blockage whereas others insist
on \nNEW\ effects including \vy\ as an explanation. However, none of
these have proved to be completely satisfactory. Yet, no one can
rule out the possibility of simultaneous occurrence of these elastic
and inelastic phenomena with further complication and confusion. A
disturbing fact is the observation made by several researchers, e.g.
Sadowski \cite{sadowskithesis}, that reproducible results could be
obtained only in constant flow rate experiments because at constant
pressure drop the velocity kept decreasing. This kind of observation
indicates deposition of polymer on the solid surface by one
mechanism or another, and cast doubt on some explanations which
involve elasticity. At constant flow rate the increased pressure
drop seems to provide the necessary force to keep a reproducible
portion of the flow channels open \cite{DaubenM1967, Dullien1975,
Savins1969, ChhabraCM2001}.

\vspace{0.2cm}

There are three principal \vc\ effects that have to be accounted for
in \vy\ investigation: \trans\ time-dependence, \steadys\
time-dependence and dilatancy at high flow rates.

%XXXXXXXXXXXXXXXXXXXXXXXXXXXXXXXXXXXXXXXXXXXXXXXXXXXXXXXXXXXXXXXXXXXXXX
\subsection{\Trans\ Time-Dependence} \label{TransientTimeDependence}
\Trans\ or \timedep\ \vc\ behavior has been observed in bulk on the
startup and cessation of processes involving the displacement of
\vc\ materials, and on a sudden change of rate or reversal in the
direction of deformation. During these \trans\ states, there are
frequently overshoots and undershoots in stress as a function of
time which scale with strain. However, if the fluid is strained at a
constant rate, these transients die out in the course of time, and
the stress approaches a \steadys\ value that depends only on the
strain rate. For example, under initial flow conditions stresses can
reach magnitudes which are substantially more important than their
\steadys\ values, whereas the relaxation on a sudden cessation of
strain can vary substantially in various circumstances
\cite{carreaubook, larsonbook1999, tannerbook2000, birdbook,
BautistaSPM1999}.%

\vspace{0.2cm}

\Trans\ responses are usually investigated in the context of bulk
rheology, despite the fact that there is no available theory that
can quantitatively predict this behavior. As a consequence of this
limitation to bulk, the literature of \vc\ flow in porous media is
almost entirely dedicated to the \steadys\ situation with hardly any
work on the \insitu\ \timedep\ \vc\ flows. However, \trans\ behavior
should be expected in similar situations in the flow through porous
media. One reason for this gap is the absence of a proper
theoretical structure and the experimental difficulties associated
with these flows \insitu. Another possible explanation is that the
\insitu\ \trans\ flows may have less important practical
applications.

%XXXXXXXXXXXXXXXXXXXXXXXXXXXXXXXXXXXXXXXXXXXXXXXXXXXXXXXXXXXXXXXXXXXXXX
\subsection{\SteadyS\ Time-Dependence} \label{SteadyTimeDependence}
By this, we mean the effects that may arise in the \steadys\ flow
due to time dependency, as the time-dependence characteristics of
the \vc\ material must have an impact on the \steadys\ behavior.
Therefore, elastic effects should be expected during \steadys\ flow
in porous media because of the time-dependence nature of the flow at
the pore level \cite{GogartyLF1972}.

\vspace{0.2cm}

Depending on the time scale of the flow, \vc\ materials may show
viscous or elastic behavior. The particular response in a given
process depends on the time scale of the process in relation to a
natural time of the material \cite{wapperomthesis,
barnesbookHW1993}. With regard to this time-scale dependency of \vc\
flow process at pore scale, it has been suggested that the fluid
relaxation time and the rate of elongation or contraction that
occurs as the fluid flows through a channel or pore with varying
cross-sectional area should be used to represent \vc\ behavior
\cite{MarshallM1967, hirasaki1974}.

\vspace{0.2cm}

Sadowski and Bird \cite{SadowskiB1965} analyzed the viscometric flow
problem in a long straight circular tube and argued that in such a
flow no \timedep\ elastic effects are expected. However, in a porous
medium elastic effects may occur. As the fluid moves through a
tortuous channel in the porous medium, it encounters a capriciously
changing cross-section. If the fluid relaxation time is small with
respect to the transit time through a contraction or expansion in a
tortuous channel, the fluid will readjust to the changing flow
conditions and no elastic effects would be observed. If, on the
other hand, the fluid relaxation time is large with respect to the
time to go through a contraction or expansion, then the fluid will
not accommodate and elastic effects will be observed in the form of
an extra pressure drop or an increase in the apparent viscosity.
Thus the concept of a ratio of characteristic times emerges as an
ordering parameter in \nNEW\ flow through porous media. This
indicates the importance of the ratio of the natural time of a fluid
to the duration time of a process, i.e. the residence time of the
fluid element in a process \cite{Savins1969}. It should be remarked
that this formulation relies on an approximate dual-nature ordering
scheme and is valid for some systems and flow regimes. More
elaborate formulation will be given in the next paragraph in
conjunction with the intermediate plateau phenomenon.

\vspace{0.2cm}

One of the \steadys\ \vc\ phenomena observed in the flow through
porous media and can be qualified as a \timedep\ effect, among other
possibilities such as retention, is the intermediate plateau at
medium flow rates as demonstrated in Figure (\ref{VERheology2}). A
possible explanation is that at low flow rates before the appearance
of the intermediate plateau the fluid behaves inelastically like any
typical \shThin\ fluid. This implies that in the low flow regime
\vc\ effects are negligible as the fluid is able to respond to its
local state of deformation essentially instantly, that is it does
not remember its earlier configurations or deformation rates and
hence behaves as a purely viscous fluid. As the flow rate increases
above a threshold, a point will be reached at which the solid-like
characteristics of \vc\ materials begin to appear in the form of an
increased pressure drop or increased apparent viscosity as the time
of process becomes comparable to the natural time of fluid, and
hence a plateau is observed. At higher flow rates the process time
is short compared to the natural time of the fluid and hence the
fluid has no time to react as the fluid is not an ideal elastic
solid which reacts instantaneously. Since the precess time is very
short, no overshoot will occur at the tube constriction as a
measurable finite time is needed for the overshoot to take place.
The result is that no increase in the pressure drop will be observed
in this flow regime and the normal \shThin\ behavior is resumed with
the eventual high flow rate plateau \cite{MarshallM1967,
LetelierSC2002}.

\vspace{0.2cm}

It is worth mentioning that Dauben and Menzie \cite{DaubenM1967}
have discussed a similar issue in conjunction with the effect of
\divconv\ geometry on the flow through porous media. They argued
that the process of expansion and contraction repeated many times
may account in part for the high pressure drops encountered during
\vc\ flow in porous media. The fluid relaxation time combined with
the flow velocity determines the response of the \vc\ fluid in the
suggested \divconv\ model. Accordingly, it is possible that if the
relaxation time is long enough or the fluid velocity high enough the
fluid can pass through the large opening before it has had time to
respond to a stress change imposed at the opening entrance, and
hence the fluid would behave more as a viscous and less like a \vc.

%XXXXXXXXXXXXXXXXXXXXXXXXXXXXXXXXXXXXXXXXXXXXXXXXXXXXXXXXXXXXXXXXXXXXXX
\subsection{Dilatancy at High Flow Rates} \label{DilatancyatHighFlowRates}
The third distinctive feature of \vc\ flow is the dilatant behavior
in the form of excess pressure losses at high flow rates as
schematically depicted in Figure (\ref{VERheology3}). Certain
polymeric solutions which exhibit \shThin\ behavior in a viscometric
flow seem to exhibit a \shThik\ response under appropriate
conditions of flow through porous media. Under high flow conditions
through porous media, abnormal increases in flow resistance which
resemble a \shThik\ response have been observed in flow experiments
involving a variety of dilute to moderately concentrated solutions
of high molecular weight polymers \cite{Savins1969, sorbiebook}.

\vspace{0.2cm}

This phenomenon can be attributed to stretch-thickening behavior due
to the dominance of extensional over shear flow. At high flow rates
strong extensional flow effects do occur and the extensional
viscosity rises very steeply with increasing extension rate. As a
result, the non-shear terms become much larger than the shear terms.
For \NEW\ fluids, the extensional viscosity is just three times the
shear viscosity. However, for \vc\ fluids the shear and extensional
viscosities often behave oppositely, that is while the shear
viscosity is generally a decreasing function of the shear rate, the
extensional viscosity increases as the extension rate is increased.
The consequence is that the pressure drop will be governed by the
extension thickening behavior and the apparent viscosity rises
sharply. Other possibilities such as physical retention are less
likely to take place at these high flow rates \cite{DurstHI1987,
MendesN2002}.

\vspace{0.2cm}

Finally a question may arise that why dilatancy at high flow rates
occurs in some situations while intermediate plateau occurs in
others? It seems that the combined effect of the fluid and porous
media properties and the nature of the process is behind this
difference.

%XXXXXXXXXXXXXXXXXXXXXXXXXXXXXXXXXXXXXXXXXXXXXXXXXXXXXXXXXXXXXXXXXXXXXX
\section{\BauMan\ Model} \label{BautistaManero}
This is a relatively simple model that combines the \OldB\
constitutive equation for \vy\ (\ref{OBM1}) and the \FRED's kinetic
equation for flow-induced structural changes usually associated with
\thixotropy. The model requires six parameters that have physical
significance and can be estimated from rheological measurements.
These parameters are the low and high shear rate viscosities, the
elastic modulus, the relaxation time, and two other constants
describing the build up and break down of viscosity.

\vspace{0.2cm}

The kinetic equation of \FRED\ that accounts for the destruction and
construction of structure is given by
\begin{equation}\label{Fredrickson}
    \D \Vis t = \frac{\Vis}{\rxTimF} \left( 1 - \frac{\Vis}{\lVis} \right)
                + \kF \Vis \left( 1 - \frac{\Vis}{\hVis} \right) \sTen : \rsTen
\end{equation}
where  $\Vis$ is the \nNEW\ viscosity, $t$ is the time of
deformation, $\rxTimF$ is the relaxation time upon the cessation of
steady flow, $\lVis$ and $\hVis$ are the viscosities at zero and
infinite shear rates respectively, $\kF$ is a parameter that is
related to a critical stress value below which the material exhibits
primary creep, $\sTen$ is the stress tensor and $\rsTen$ is the rate
of strain tensor. In this model, $\rxTimF$ is a structural
relaxation time, whereas $\kF$ is a kinetic constant for structure
break down. The elastic modulus $\Go$ is related to these parameters
by $\Go = \Vis/\rxTimF$ \cite{BautistaSPM1999, BautistaSLPM2000,
ManeroBSP2002, TardyA2005}.

\vspace{0.2cm}

\BauMan\ model was originally proposed for the rheology of worm-like
micellar solutions which usually have an upper \NEW\ plateau, and
show strong signs of \shThin. The model, which incorporates \shThin,
elasticity and \thixotropy, can be used to describe the complex
rheological behavior of \vc\ systems that also exhibit \thixotropy\
and \rheopexy\ under shear flow. The model predicts creep behavior,
stress relaxation and the presence of \thixotropic\ loops when the
sample is subjected to \trans\ stress cycles. The \BauMan\ model has
also been found to fit steady shear, oscillatory and \trans\
measurements of \vc\ solutions \cite{BautistaSLPM2000, TardyA2005,
BautistaSPM1999, ManeroBSP2002}.

%XXXXXXXXXXXXXXXXXXXXXXXXXXXXXXXXXXXXXXXXXXXXXXXXXXXXXXXXXXXXXXXXXXXXXX
\subsection{\Tardy\ Algorithm}\label{TardyAlgorithm}
This algorithm is proposed by \Tardy\ to compute the pressure
drop-flow rate relationship for the \steadys\ flow of a \BauMan\
fluid in simple capillary network models. The bulk rheology of the
fluid and the dimensions of the capillaries making up the network
are used as inputs to the models \cite{TardyA2005}. In the following
paragraphs we outline the basic components of this algorithm and the
logic behind it. This will be followed by some mathematical and
technical details related to the implementation of this algoritm in
our \nNEW\ code.

\vspace{0.2cm}

The flow in a single capillary can be described by the following
general relation
\begin{equation}\label{generalFlowPresRel}
    Q = G' \Delta P
\end{equation}
where $Q$ is the volumetric flow rate, $G'$ is the flow conductance
and $\Delta P$ is the pressure drop. For a particular capillary and
a specific fluid, $G'$ is given by
\begin{eqnarray}
  G' &=& G'(\mu) = \textrm{constant}                     \verb|      |   \textrm{\NEW\ Fluid} \nonumber \\
  G' &=& G'(\mu, \Delta P)            \verb|            |      \textrm{Purely viscous \nNEW\ Fluid} \nonumber \\
  G' &=& G'(\mu, \Delta P, t)         \verb|           |      \textrm{Fluid with memory}
\end{eqnarray}

For a network of capillaries, a set of equations representing the
capillaries and satisfying mass conservation should be solved
simultaneously to produce a consistent pressure field, as presented
in Chapter (\ref{Modeling}). For \NEW\ fluid, a single iteration is
needed to solve the pressure field since the conductance is known in
advance as the viscosity is constant. For purely viscous \nNEW\
fluid, we start with an initial guess for the viscosity, as it is
unknown and pressure-dependent, and solve the pressure field
iteratively updating the viscosity after each iteration cycle until
convergence is reached. For memory fluids, the dependence on time
must be taken into account when solving the pressure field
iteratively. Apparently, there is no general strategy to deal with
such situation. However, for the \steadys\ flow of memory fluids a
sensible approach is to start with an initial guess for the flow
rate and iterate, considering the effect of the local pressure and
viscosity variation due to \convdiv\ geometry, until convergence is
achieved. This approach is adopted by \Tardy\ to find the flow of a
\BauMan\ fluid in a simple capillary network model. The \Tardy\
algorithm proceeds as follows \cite{TardyA2005}
\begin{itemize}

    \item For fluids without memory the capillary is unambiguously defined by its
    radius and length. For fluids with memory, where going from one section to
    another with different radius is important,
    the capillary should be modeled with contraction to account for
    the effect of \convdiv\ geometry on the flow. The reason is that
    the effects of fluid memory take place on going through a radius change,
    as this change induces a change in shear rate with a consequent
    \thixotropic\ break-down or build-up of viscosity or elastic characteristic times competition.
    Examples of the \convdiv\ geometries are given in Figure (\ref{ConvDivGeom}).

    \item Each capillary is discretized in the flow direction and a discretized form
    of the flow equations is used assuming a prior knowledge of the stress and viscosity
    at the inlet of the network.

    \item Starting with an initial guess for the flow rate and using iterative technique,
    the pressure drop as a function of the flow rate is found for each capillary.

    \item Finally, the pressure field for the whole network is found iteratively
    until convergence is achieved. Once the pressure field is found the flow rate through each
    capillary in the network can be computed and the total flow rate
    through the network can be determined by summing and averaging the
    flow through the inlet and outlet capillaries.

\end{itemize}

A modified version of the \Tardy\ algorithm was implemented in our
\nNEW\ code. In the following, we outline the main steps of this
algorithm and its implementation.
\begin{itemize}

    \item From a one-dimensional \steadys\ \FRED\ equation in which the
    partial time derivative is written in the form $\Dp{}{t} = V
    \Dp{}{x}$ where the fluid velocity $V$ is given by $Q / \pi r^{2}$ and
    the average shear rate in the tube with radius $r$ is given by $Q / \pi r^{3}$, the
    following equation is obtained

    \begin{equation}\label{simplifiedFred}
        V \D \Vis x = \frac{\Vis}{\rxTimF} \left( \frac{\lVis - \Vis}{\lVis} \right)
        + \kF \Vis \left( \frac{\hVis - \Vis}{\hVis} \right) \sTenC \rsTenC
    \end{equation}

    \item Similarly, another simplified equation is obtained from
    the \OldB\ model

    \begin{equation}\label{simplifiedOldB}
        \sTenC + \frac{V \Vis}{\Go} \D \sTenC x = 2 \Vis \rsTenC
    \end{equation}

    \item The \convdiv\ feature of the capillary is implemented in the
    form of a parabolic profile as outlined in Appendix \ref{AppConvDiv}.

    \item Each capillary in the network is discretized into $m$
    slices each with width $\delta x = L/m$ where $L$ is the capillary length.

    \item From Equation (\ref{simplifiedOldB}), applying the
    simplified assumption $\D \sTenC x = \frac{(\sTenC_{2} - \sTenC_{1})}{\delta
    x}$ where the subscripts $1$ and $2$ stand for the inlet and
    outlet of the slice respectively, we obtain an expression for
    $\sTenC_{2}$ in terms of $\sTenC_{1}$ and $\Vis_{2}$

    \begin{equation}\label{tau2}
        \sTenC_{2} = \frac{2 \Vis_{2} \rsTenC + \frac{V \Vis_{2} \sTenC_{1}}{\Go \delta x}}
        {1 + \frac{V \Vis_{2}}{\Go \delta x}}
    \end{equation}

    \item From Equations (\ref{simplifiedFred}) and (\ref{tau2}), applying $\D \Vis x = \frac{(\Vis_{2} - \Vis_{1})}{\delta x}$,
    a third order polynomial in $\Vis_{2}$ is obtained. The coefficients of the four terms
    of this polynomial are
    \begin{eqnarray} \label{mu2}
      \Vis_{2}^{3}: \hspace{1.0cm} - \frac{V}{\rxTimF \lVis \Go \delta x} - \frac{2 \kF \rsTenC^{2}}{\hVis} - \frac{\kF \rsTenC \sTenC_{1} V}{\hVis \Go \delta x} \hspace{2.4cm} \nonumber \\
      \Vis_{2}^{2}: \hspace{1.0cm} - \frac{1}{\rxTimF \lVis} - \frac{V^{2}}{\Go (\delta x)^{2}} + \frac{V}{\rxTimF \Go \delta x} + 2 \kF \rsTenC^{2} + \frac{\kF \rsTenC \sTenC_{1} V}{\Go \delta x} \nonumber \\
      \Vis_{2}^{1}: \hspace{1.0cm} - \frac{V}{\delta x} + \frac{1}{\rxTimF} + \frac{V^{2} \Vis_{1}}{\Go (\delta x)^{2}} \hspace{4.2cm} \nonumber \\
      \Vis_{2}^{0}: \hspace{1.0cm} \frac{V \Vis_{1}}{\delta x} \hspace{7.1cm}
    \end{eqnarray}

    \item The algorithm starts by assuming a \NEW\ flow in a network
    of straight capillaries. Accordingly, the volumetric flow rate $Q$,
    and consequently $V$ and $\rsTenC$, for each capillary are obtained.

    \item Starting from the inlet of the network where the
    viscosity and the stress are assumed having known values of
    $\lVis$ and $R \Delta P / 2 L$ for each capillary respectively,
    the computing of the \nNEW\ flow in a \convdiv\
    geometry takes place in each capillary independently by
    calculating $\Vis_{2}$ and $\sTenC_{2}$ slice by slice,
    where the values from the previous slice are used for
    $\Vis_{1}$ and $\sTenC_{1}$ of the current slice.

    \item For the capillaries which are not at the inlet of the
    network, the initial values of the viscosity and stress at the
    inlet of the capillary are found by computing the $Q$-weighted
    average from all the capillaries that feed into the pore which is at the
    inlet of the corresponding capillary.

    \item To find $\Vis_{2}$ of a slice, ``rtbis'' which is a
    bisection method from the Numerical Recipes \cite{NumericalRecipes} is
    used. To eliminate possible non-physical roots, the interval for
    the accepted root is set between zero and $3\lVis$ with $\lVis$
    used in the case of failure. These conditions are logical as
    long as the slice is reasonably thin and the flow and fluid are
    physically viable. In the case of a convergence failure, error
    messages are issued to inform the user. No failure has been detected during
    the many runs of this algorithm. Moreover, extensive sample inspection
    of the $\Vis_{2}$ values has been carried out and proved to be sensible and realistic.

    \item The value found for the $\Vis_{2}$ is used to find
    $\sTenC_{2}$ which is needed as an input to the next slice.

    \item Averaging the value of $\Vis_{1}$ and $\Vis_{2}$, the
    viscosity for the slice is found and used with \POIS\ law to
    find the pressure drop across the slice.

    \item The total pressure drop across the whole capillary is computed by
    summing up the pressure drops across its individual slices. This total is
    used with \POIS\ law to find the effective viscosity for the
    capillary as a whole.

    \item Knowing the effective viscosities for all capillaries of
    the network, the pressure field is solved iteratively using the Algebraic
    Multi-Grid (AMG) solver, and hence the total volumetric flow
    rate from the network and the apparent viscosity are found.

\end{itemize}

%XXXXXXXXXXXXXXXXXXXXXXXXXXXXXXXXXXXXXXXXXXXXXXXXXXXXXXXXXXXXXXXXXXXXXX
\subsection{Initial Results of the Modified \Tardy\ Algorithm}\label{ResultsTardyAlgorithm}
The modified \Tardy\ algorithm was tested and assessed. Various
qualitative aspects were verified and proved to be correct. Some
general results and conclusions are outlined below with sample
graphs and data for the \sandp\ and \Berea\ networks using a
calculation box with $x_{_{l}}$=0.5 and $x_{_{u}}$=0.95. We would
like to remark that despite our effort to use typical values for the
parameters and variables, in some cases we were forced, for
demonstration purposes, to use eccentric values to accentuate the
features of interest. In Table (\ref{wormlikeMicellar}) we present
some values for the \BauMan\ model parameters as obtained from the
wormlike micellar system studied by Anderson and co-workers
\cite{AndersonPS2006} to give an idea of the parameter ranges in
real fluids. The system is a solution of a surfactant concentrate [a
mixture of the cationic surfactant erucyl
bis(hydroxyethyl)methylammonium chloride (EHAC) and 2-propanol in a
3:2 weight ratio] in an aqueous solution of potassium chloride.

\vspace{0.2cm}

We wish also to insist that this initial investigation is part of
the process of testing and debugging the algorithm. Therefore, it
should not be regarded as comprehensive or conclusive as we believe
that this is a huge field which requires another PhD for proper
investigation. We make these initial results and conclusions
available to other researchers for further investigation.

%%%%%%%%%%%%%%%%%%%%%%%%%%%%%%%%%%%%%%%%%%

\begin{table} [h]
\centering %
\caption[Some values of the wormlike micellar system studied by
Anderson and co-workers, which is a solution of surfactant
concentrate (a mixture of EHAC and 2-propanol) in an
aqueous solution of potassium chloride] %
{Some values of the wormlike micellar system studied by Anderson and
co-workers \cite{AndersonPS2006} which is a solution of surfactant
concentrate (a mixture of EHAC and 2-propanol) in an
aqueous solution of potassium chloride.} %
\label{wormlikeMicellar} %
\vspace{0.5cm} %
\begin{tabular}{|l|l|}
\hline

Parameter   \verb|  |  &           Value \verb|  | \\

\hline

$\Go$ (Pa)              &          1 - 10 \\

$\lVis$ (Pa.s)          &          115 - 125 \\

$\hVis$ (Pa.s)          &          0.00125 - 0.00135 \\

$\rxTimF$ (s)           &          1 - 30 \\

$\kF$ (Pa$^{-1}$)       &          10$^{^{-3}}$ - 10$^{^{-6}}$ \\

\hline
\end{tabular}
\end{table}

%%%%%%%%%%%%%%%%%%%%%%%%%%%%%%%%%%%%%%%%%%

%XXXXXXXXXXXXXXXXXXXXXXXXXXXXXXXXXXXXXXXXXXXXXXXXXXXXXXXXXXXXXXXXXXXXXX
\subsubsection{Convergence-Divergence}\label{}
A dilatant effect due to the \convdiv\ feature has been detected
relative to a network of straight capillaries. Because the
conductance of the capillaries is reduced by tightening the radius
at the middle, an increase in the apparent viscosity is to be
expected. As the corrugation feature of the tubes is exacerbated by
narrowing the radius at the middle, the dilatant behavior is
intensified. The natural explanation is that the increase in
apparent viscosity should be proportionate to the magnitude of
tightening. This feature is presented in Figures (\ref{TardySP1})
and (\ref{TardyB1}) for the \sandp\ and \Berea\ networks
respectively on a log-log scale.

\begin{figure}[!t]
  \centering{}
  \includegraphics
  [scale=0.5]
  {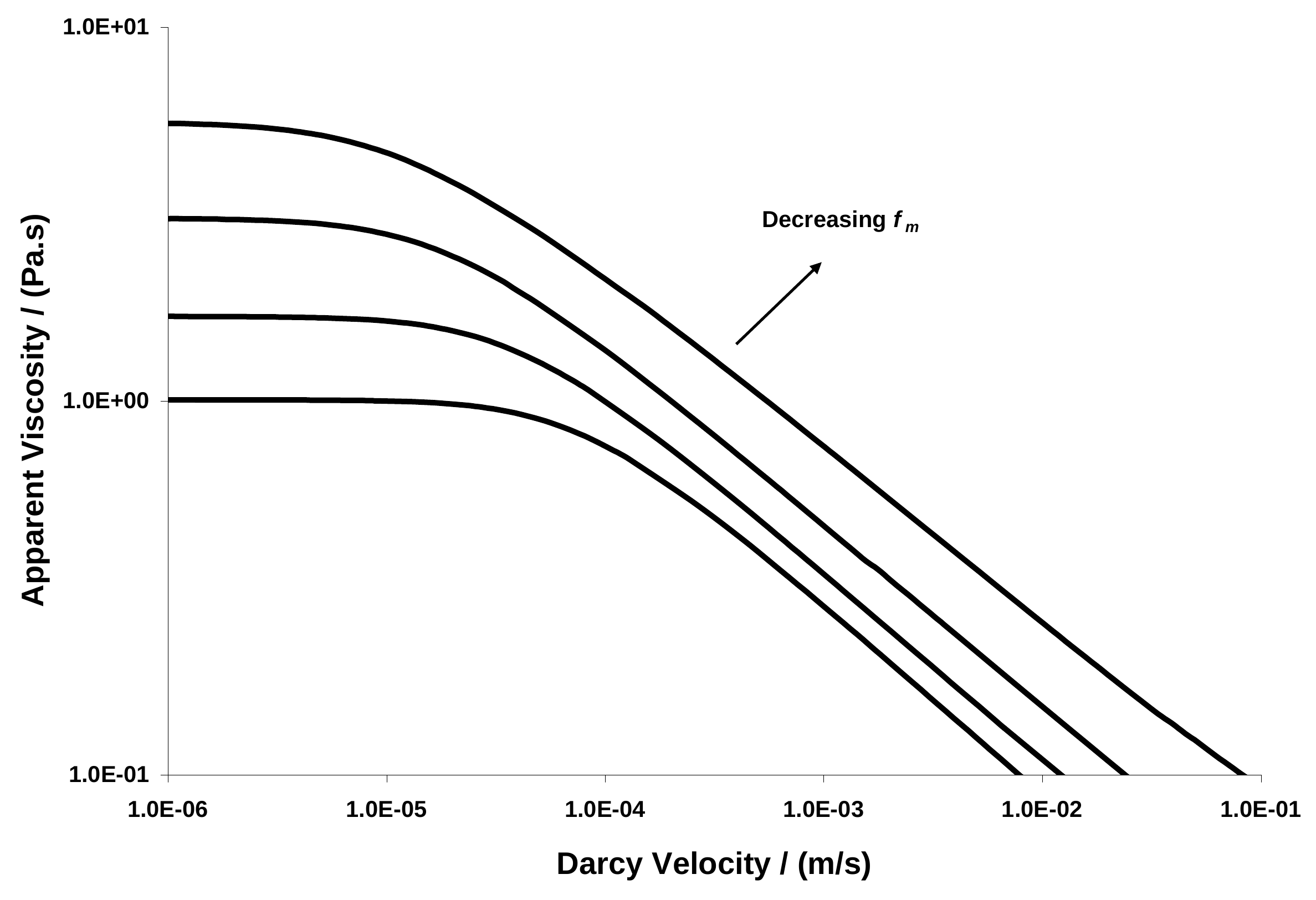}
  \caption[The \Tardy\ algorithm \sandp\ results for $\Go$=0.1\,Pa, $\hVis$=0.001\,Pa.s, $\lVis$=1.0\,Pa.s, $\rxTimF$=1.0\,s,
            $\kF$=10$^{-5}$\,Pa$^{-1}$, $\fe$=1.0, $m$=10 slices, with varying $\fm$ (1.0, 0.8, 0.6 and 0.4)]
  {The \Tardy\ algorithm \sandp\ results for $\Go$=0.1\,Pa, $\hVis$=0.001\,Pa.s, $\lVis$=1.0\,Pa.s, $\rxTimF$=1.0\,s,
            $\kF$=10$^{-5}$\,Pa$^{-1}$, $\fe$=1.0, $m$=10 slices, with varying $\fm$ (1.0, 0.8, 0.6 and 0.4).}
  \label{TardySP1}
\end{figure}
\begin{figure}[!h]
  \centering{}
  \includegraphics
  [scale=0.5]
  {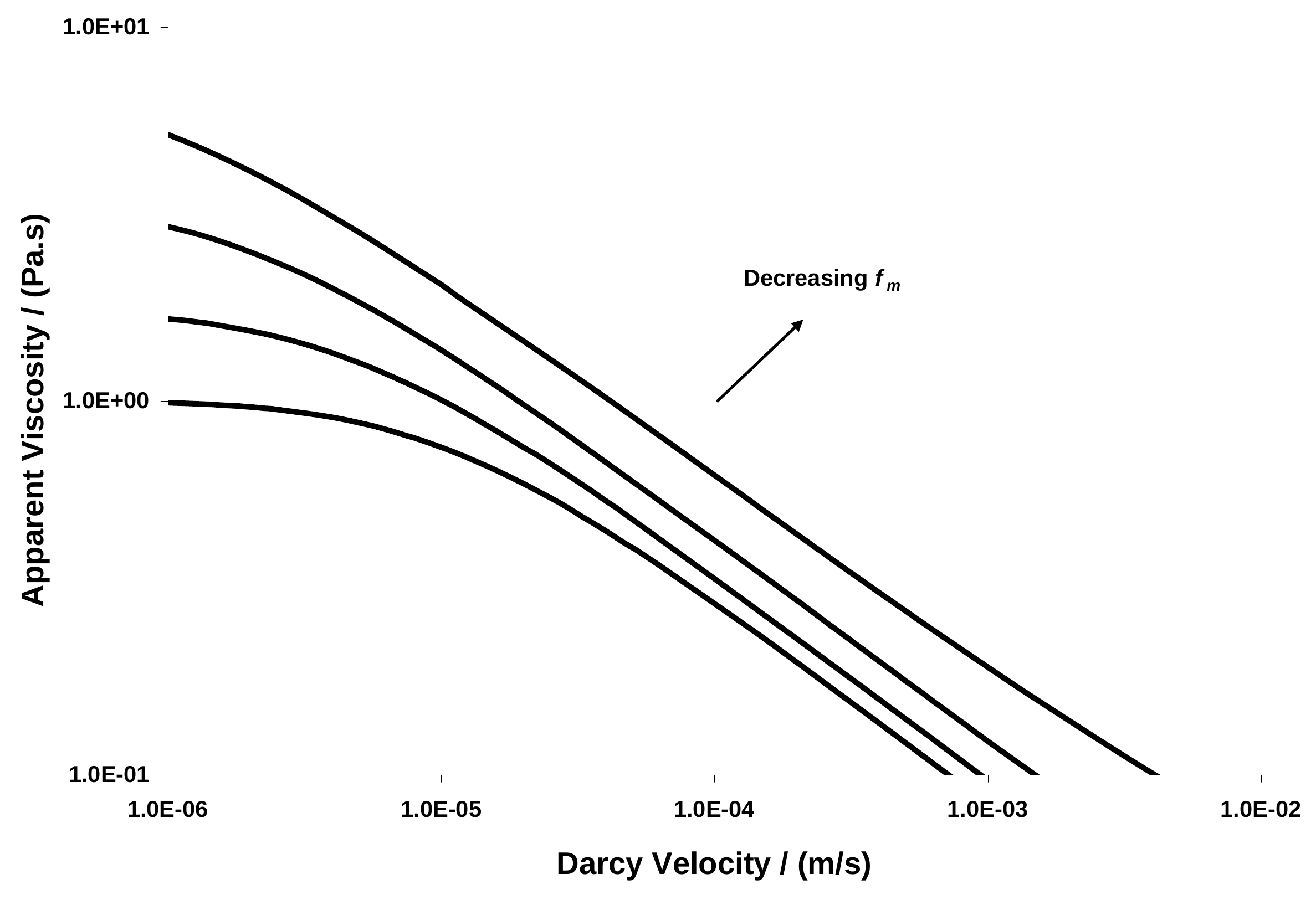}
  \caption[The \Tardy\ algorithm \Berea\ results for $\Go$=0.1\,Pa, $\hVis$=0.001\,Pa.s, $\lVis$=1.0\,Pa.s, $\rxTimF$=1.0\,s,
            $\kF$=10$^{-5}$\,Pa$^{-1}$, $\fe$=1.0, $m$=10 slices, with varying $\fm$ (1.0, 0.8, 0.6 and 0.4)]
  {The \Tardy\ algorithm \Berea\ results for $\Go$=0.1\,Pa, $\hVis$=0.001\,Pa.s, $\lVis$=1.0\,Pa.s, $\rxTimF$=1.0\,s,
            $\kF$=10$^{-5}$\,Pa$^{-1}$, $\fe$=1.0, $m$=10 slices, with varying $\fm$ (1.0, 0.8, 0.6 and 0.4).}
  \label{TardyB1}
\end{figure}

%XXXXXXXXXXXXXXXXXXXXXXXXXXXXXXXXXXXXXXXXXXXXXXXXXXXXXXXXXXXXXXXXXXXXXX
\subsubsection{\DivConv}\label{}
The investigation of the effect of \divconv\ geometry by expanding
the radius of the capillaries at the middle revealed a thinning
effect relative to the straight and \convdiv\ geometries, as seen in
Figures (\ref{TardySP2}) and (\ref{TardyB2}) for the \sandp\ and
\Berea\ networks respectively on a log-log scale. This is due to the
increase in the conductance of the capillaries by enlarging the
radius at the middle.

\begin{figure}[!t]
  \centering{}
  \includegraphics
  [scale=0.5]
  {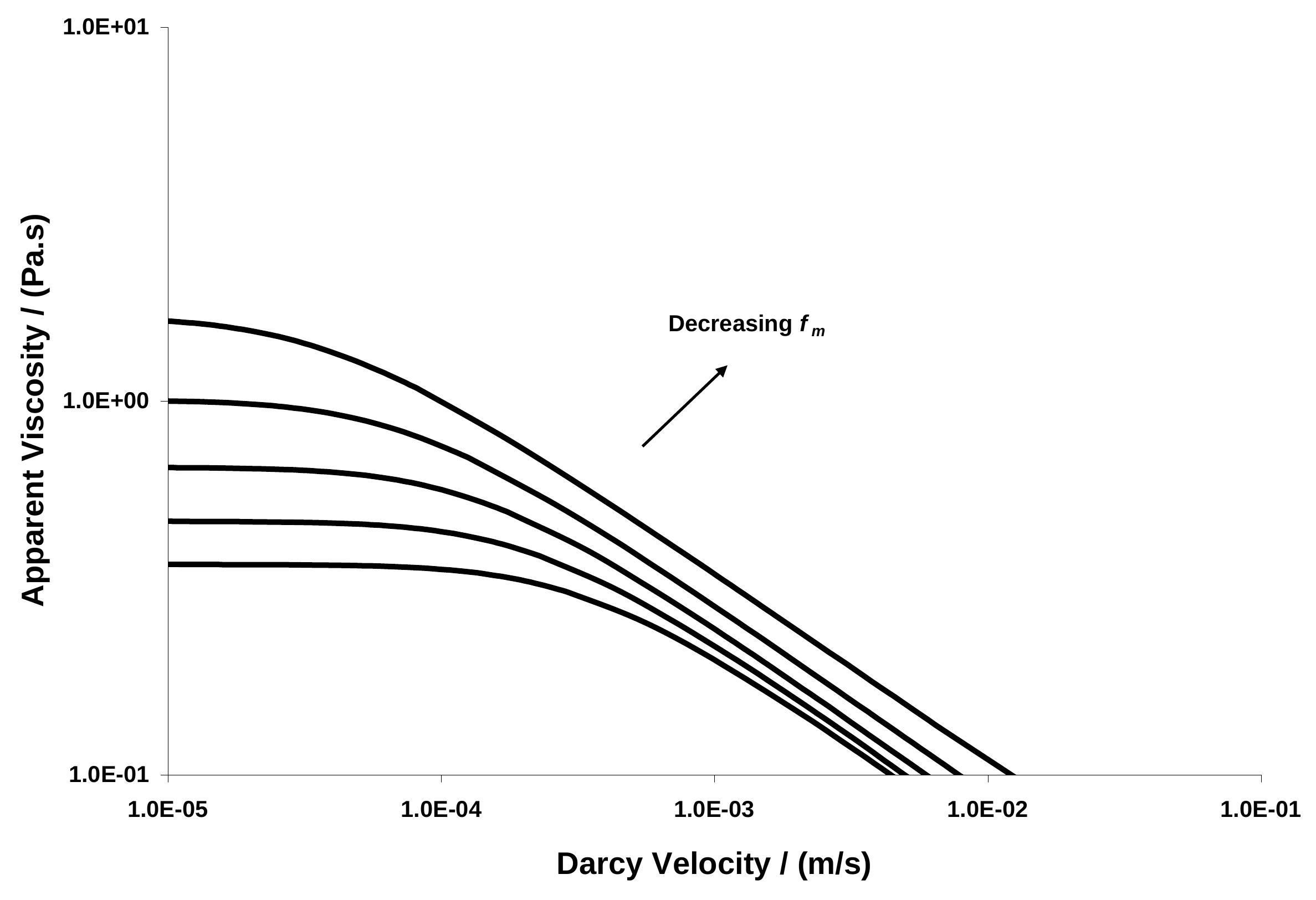}
  \caption[The \Tardy\ algorithm \sandp\ results for $\Go$=0.1\,Pa, $\hVis$=0.001\,Pa.s, $\lVis$=1.0\,Pa.s, $\rxTimF$=1.0\,s,
            $\kF$=10$^{-5}$\,Pa$^{-1}$, $\fe$=1.0, $m$=10 slices, with varying $\fm$ (0.8, 1.0, 1.2, 1.4 and 1.6)]
  {The \Tardy\ algorithm \sandp\ results for $\Go$=0.1\,Pa, $\hVis$=0.001\,Pa.s, $\lVis$=1.0\,Pa.s, $\rxTimF$=1.0\,s,
            $\kF$=10$^{-5}$\,Pa$^{-1}$, $\fe$=1.0, $m$=10 slices, with varying $\fm$ (0.8, 1.0, 1.2, 1.4 and 1.6).}
  \label{TardySP2}
\end{figure}
\begin{figure}[!h]
  \centering{}
  \includegraphics
  [scale=0.5]
  {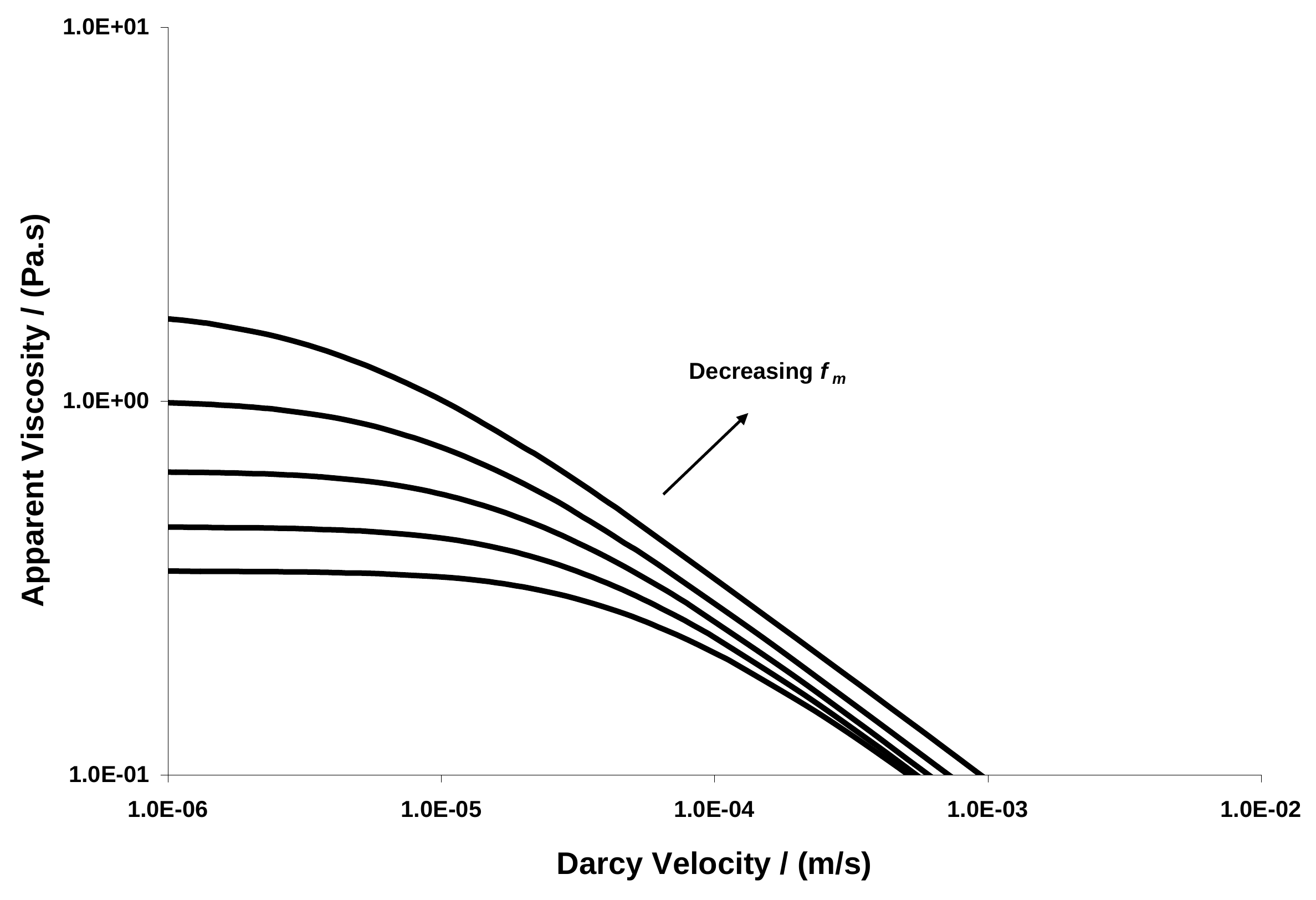}
  \caption[The \Tardy\ algorithm \Berea\ results for $\Go$=0.1\,Pa, $\hVis$=0.001\,Pa.s, $\lVis$=1.0\,Pa.s, $\rxTimF$=1.0\,s,
            $\kF$=10$^{-5}$\,Pa$^{-1}$, $\fe$=1.0, $m$=10 slices, with varying $\fm$ (0.8, 1.0, 1.2, 1.4 and 1.6)]
  {The \Tardy\ algorithm \Berea\ results for $\Go$=0.1\,Pa, $\hVis$=0.001\,Pa.s, $\lVis$=1.0\,Pa.s, $\rxTimF$=1.0\,s,
            $\kF$=10$^{-5}$\,Pa$^{-1}$, $\fe$=1.0, $m$=10 slices, with varying $\fm$ (0.8, 1.0, 1.2, 1.4 and 1.6).}
  \label{TardyB2}
\end{figure}

%XXXXXXXXXXXXXXXXXXXXXXXXXXXXXXXXXXXXXXXXXXXXXXXXXXXXXXXXXXXXXXXXXXXXXX
\subsubsection{Number of Slices}\label{}
As the number of slices of the capillaries increases, the algorithm
converges to a stable and constant solution within acceptable
numerical errors. This indicates that the numerical aspects of the
algorithm are functioning correctly because the effect of
discretization  errors is expected to diminish by increasing the
number of slices and hence decreasing their width. A sample graph of
apparent viscosity versus number of slices for a typical data point
is presented in Figure (\ref{TardySP3}) for the \sandp\ and in
Figure (\ref{TardyB3}) for \Berea.

\begin{figure}[!t]
  \centering{}
  \includegraphics
  [scale=0.5]
  {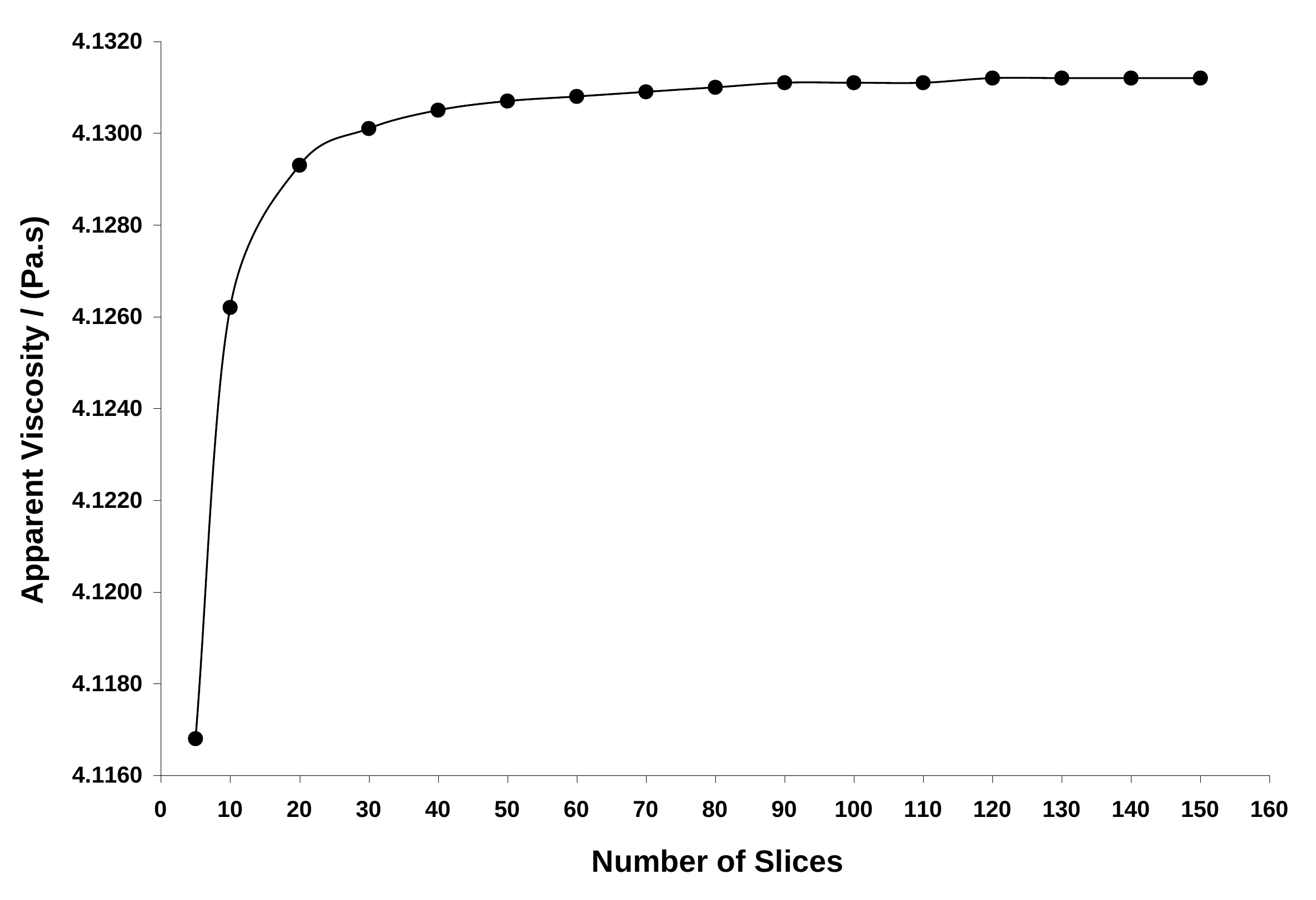}
  \caption[The \Tardy\ algorithm \sandp\ results for $\Go$=1.0\,Pa, $\hVis$=0.001\,Pa.s, $\lVis$=1.0\,Pa.s, $\rxTimF$=1.0\,s,
            $\kF$=10$^{-5}$\,Pa$^{-1}$, $\fe$=1.0, $\fm$=0.5, with varying number of slices for
            a typical data point ($\Delta P$=100\,Pa)]
  {The \Tardy\ algorithm \sandp\ results for $\Go$=1.0\,Pa, $\hVis$=0.001\,Pa.s, $\lVis$=1.0\,Pa.s, $\rxTimF$=1.0\,s,
            $\kF$=10$^{-5}$\,Pa$^{-1}$, $\fe$=1.0, $\fm$=0.5, with varying number of slices for
            a typical data point ($\Delta P$=100\,Pa).}
  \label{TardySP3}
\end{figure}
\begin{figure}[!h]
  \centering{}
  \includegraphics
  [scale=0.5]
  {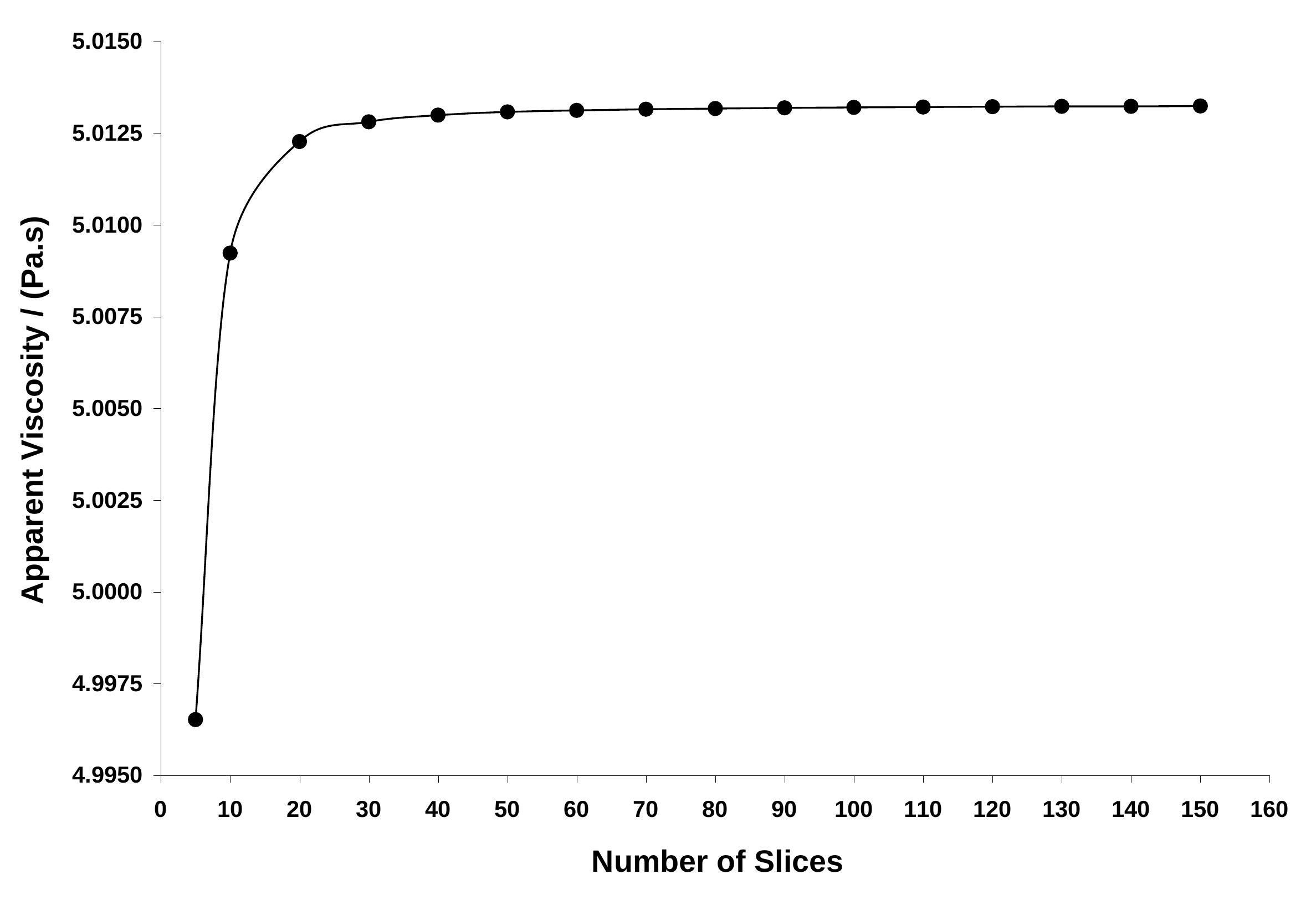}
  \caption[The \Tardy\ algorithm \Berea\ results for $\Go$=1.0\,Pa, $\hVis$=0.001\,Pa.s, $\lVis$=1.0\,Pa.s, $\rxTimF$=1.0\,s,
            $\kF$=10$^{-5}$\,Pa$^{-1}$, $\fe$=1.0, $\fm$=0.5, with varying number of slices for
            a typical data point ($\Delta P$=200\,Pa)]
  {The \Tardy\ algorithm \Berea\ results for $\Go$=1.0\,Pa, $\hVis$=0.001\,Pa.s, $\lVis$=1.0\,Pa.s, $\rxTimF$=1.0\,s,
            $\kF$=10$^{-5}$\,Pa$^{-1}$, $\fe$=1.0, $\fm$=0.5, with varying number of slices for
            a typical data point ($\Delta P$=200\,Pa).}
  \label{TardyB3}
\end{figure}

%XXXXXXXXXXXXXXXXXXXXXXXXXXXXXXXXXXXXXXXXXXXXXXXXXXXXXXXXXXXXXXXXXXXXXX
\subsubsection{\Boger\ Fluid}\label{}
A \Boger\ fluid behavior was observed when setting $\lVis = \hVis$.
However, the apparent viscosity increased as the \convdiv\ feature
is intensified. This feature is demonstrated in Figure
(\ref{TardySP4}) for the \sandp\ and in Figure (\ref{TardyB4}) for
\Berea\ on a log-log scale. Since \Boger\ fluid is a limiting and
obvious case, this behavior indicates that the model, as
implemented, is well-behaved. The viscosity increase is a natural
\vc\ response to the radius tightening at the middle, as discussed
earlier.

\begin{figure}[!t]
  \centering{}
  \includegraphics
  [scale=0.5]
  {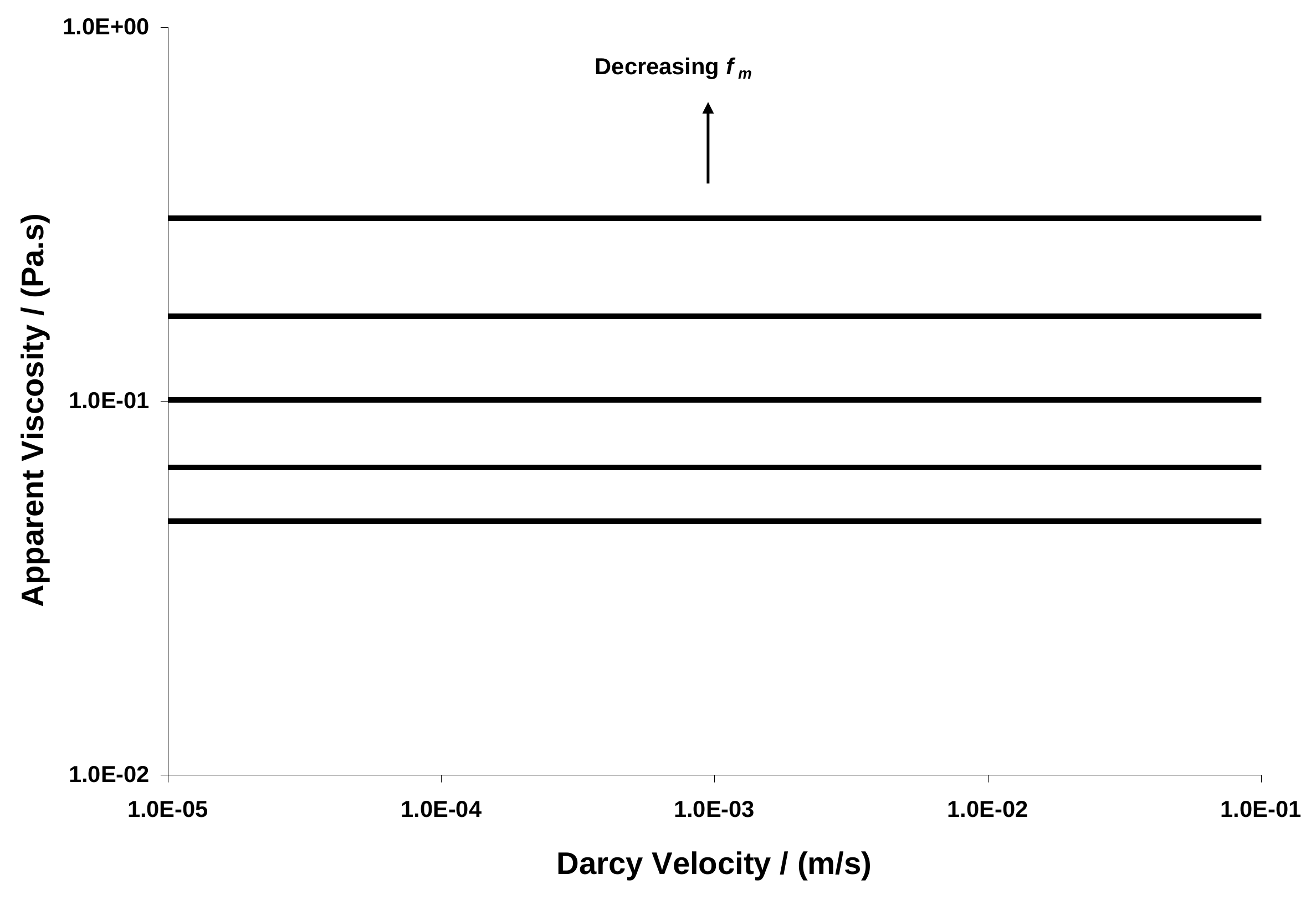}
  \caption[The \Tardy\ algorithm \sandp\ results for $\Go$=1.0\,Pa, $\hVis$=$\lVis$=0.1\,Pa.s, $\rxTimF$=1.0\,s,
            $\kF$=10$^{-5}$\,Pa$^{-1}$, $\fe$=1.0, $m$=10 slices, with varying $\fm$ (0.6, 0.8, 1.0, 1.2 and 1.4)]
  {The \Tardy\ algorithm \sandp\ results for $\Go$=1.0\,Pa, $\hVis$=$\lVis$=0.1\,Pa.s, $\rxTimF$=1.0\,s,
            $\kF$=10$^{-5}$\,Pa$^{-1}$, $\fe$=1.0, $m$=10 slices, with varying $\fm$ (0.6, 0.8, 1.0, 1.2 and 1.4).}
  \label{TardySP4}
\end{figure}
\begin{figure}[!h]
  \centering{}
  \includegraphics
  [scale=0.5]
  {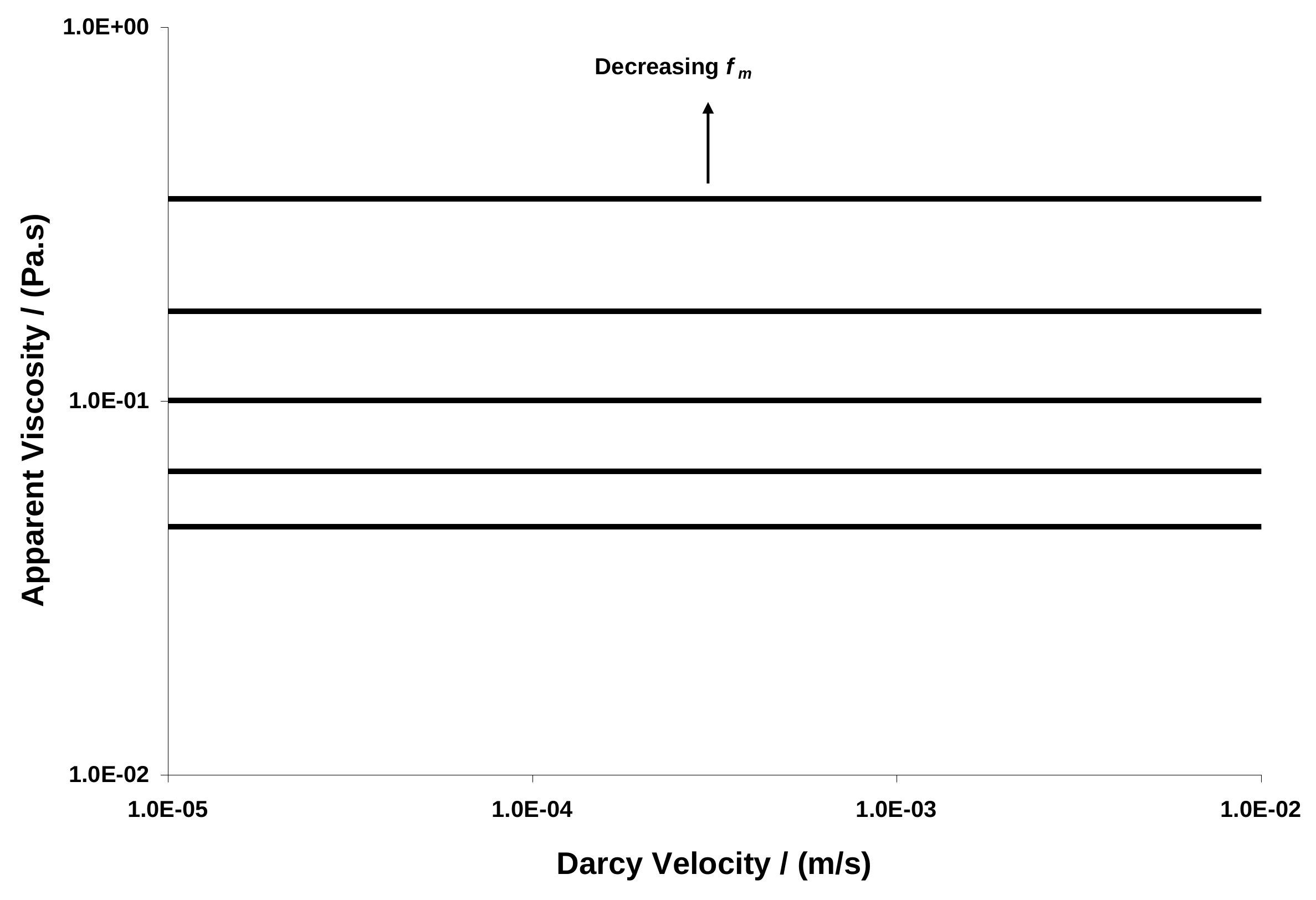}
  \caption[The \Tardy\ algorithm \Berea\ results for $\Go$=1.0\,Pa, $\hVis$=$\lVis$=0.1\,Pa.s, $\rxTimF$=1.0\,s,
            $\kF$=10$^{-5}$\,Pa$^{-1}$, $\fe$=1.0, $m$=10 slices, with varying $\fm$ (0.6, 0.8, 1.0, 1.2 and 1.4)]
  {The \Tardy\ algorithm \Berea\ results for $\Go$=1.0\,Pa, $\hVis$=$\lVis$=0.1\,Pa.s, $\rxTimF$=1.0\,s,
            $\kF$=10$^{-5}$\,Pa$^{-1}$, $\fe$=1.0, $m$=10 slices, with varying $\fm$ (0.6, 0.8, 1.0, 1.2 and 1.4).}
  \label{TardyB4}
\end{figure}

%XXXXXXXXXXXXXXXXXXXXXXXXXXXXXXXXXXXXXXXXXXXXXXXXXXXXXXXXXXXXXXXXXXXXXX
\subsubsection{Elastic Modulus}\label{}
The effect of the elastic modulus $\Go$ was investigated for
\shThin\ fluids, i.e. $\lVis > \hVis$, by varying this parameter
over several orders of magnitude while holding the others constant.
It was observed that by increasing $\Go$, the high-shear apparent
viscosities were increased while the low-shear viscosities remained
constant and have not been affected. However, the increase at
high-shear rates has almost reached a saturation point where beyond
some limit the apparent viscosities converged to certain values
despite a large increase in $\Go$. A sample of the results for this
investigation is presented in Figure (\ref{TardySP5}) for the
\sandp\ and in Figure (\ref{TardyB5}) for \Berea\ on a log-log
scale. The stability at low-shear rates is to be expected because at
low flow rate regimes near the lower \NEW\ plateau the \nNEW\
effects due to elastic modulus are negligible. As the elastic
modulus increases, an increase in apparent viscosity due to elastic
effects contributed by elastic modulus occurs. Eventually, a
saturation will be reached when the contribution of this factor is
controlled by other dominant factors and mechanisms from the porous
medium and flow regime.

\begin{figure}[!t]
  \centering{}
  \includegraphics
  [scale=0.5]
  {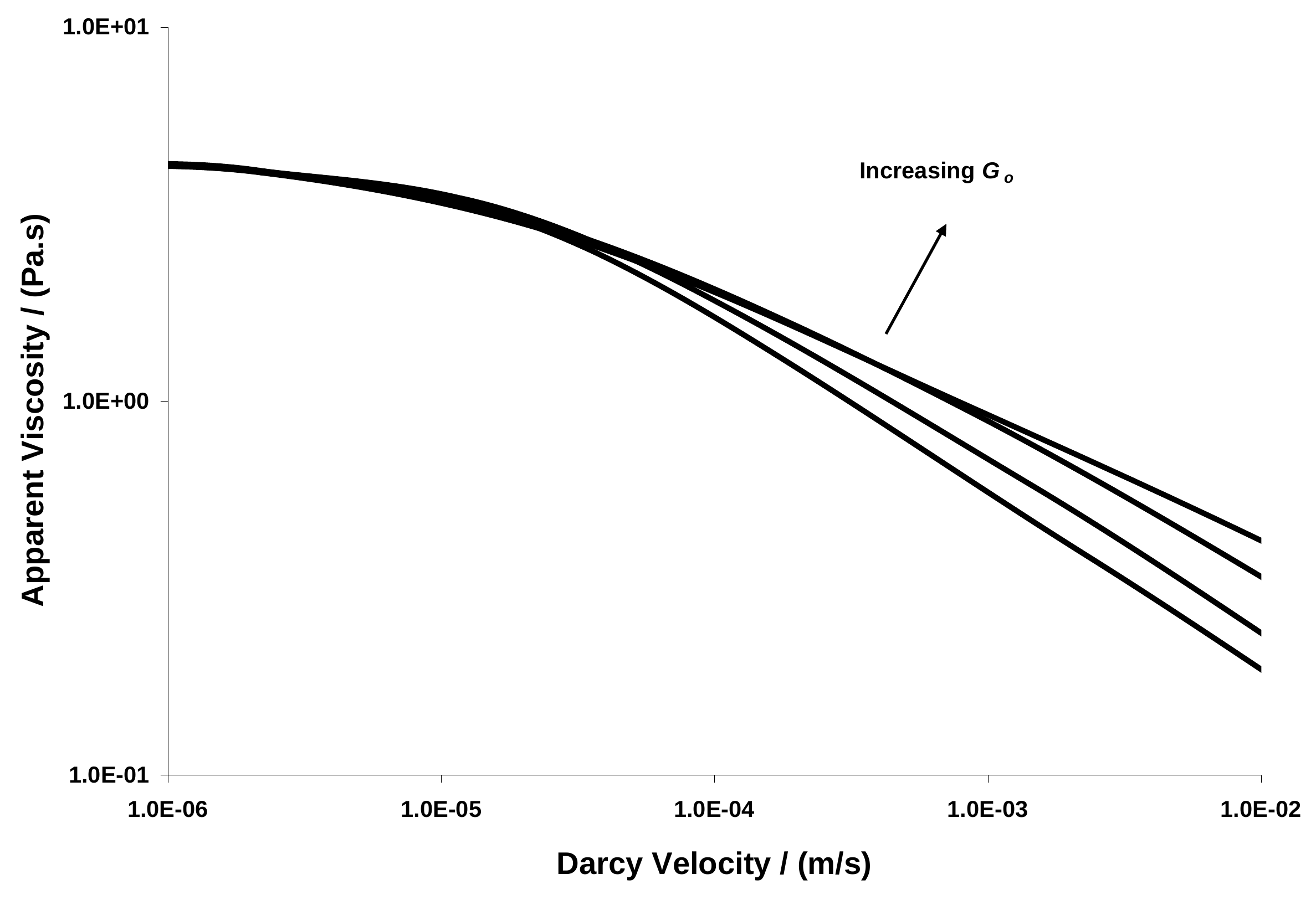}
  \caption[The \Tardy\ algorithm \sandp\ results for $\hVis$=0.001\,Pa.s, $\lVis$=1.0\,Pa.s, $\rxTimF$=1.0\,s,
            $\kF$=10$^{-5}$\,Pa$^{-1}$, $\fe$=1.0, $\fm$=0.5, $m$=10 slices, with varying $\Go$ (0.1, 1.0, 10 and 100\,Pa)]
  {The \Tardy\ algorithm \sandp\ results for $\hVis$=0.001\,Pa.s, $\lVis$=1.0\,Pa.s, $\rxTimF$=1.0\,s,
            $\kF$=10$^{-5}$\,Pa$^{-1}$, $\fe$=1.0, $\fm$=0.5, $m$=10 slices, with varying $\Go$ (0.1, 1.0, 10 and 100\,Pa).}
  \label{TardySP5}
\end{figure}
\begin{figure}[!h]
  \centering{}
  \includegraphics
  [scale=0.5]
  {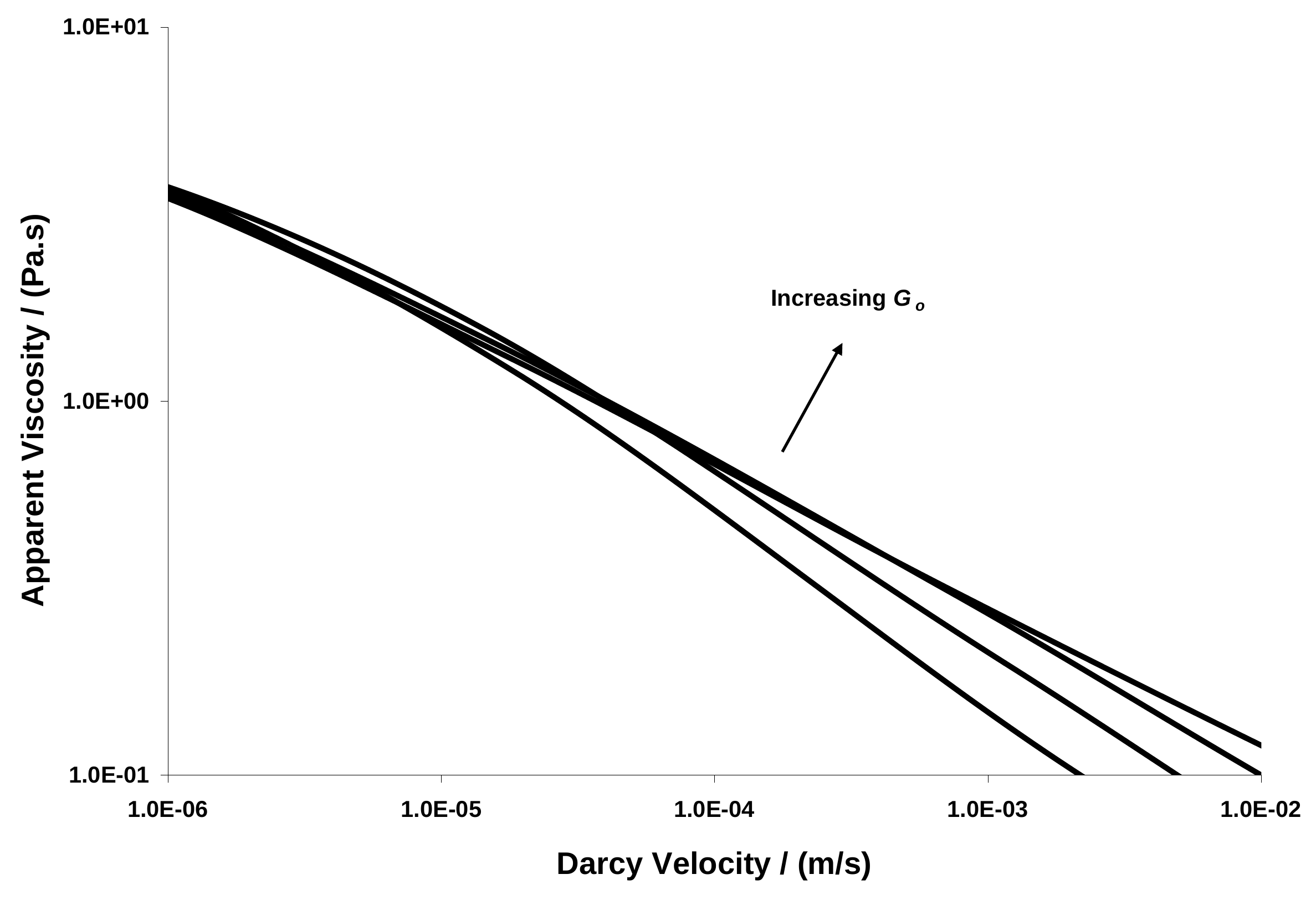}
  \caption[The \Tardy\ algorithm \Berea\ results for $\hVis$=0.001\,Pa.s, $\lVis$=1.0\,Pa.s, $\rxTimF$=1.0\,s,
            $\kF$=10$^{-5}$\,Pa$^{-1}$, $\fe$=1.0, $\fm$=0.5, $m$=10 slices, with varying $\Go$ (0.1, 1.0, 10 and 100\,Pa)]
  {The \Tardy\ algorithm \Berea\ results for $\hVis$=0.001\,Pa.s, $\lVis$=1.0\,Pa.s, $\rxTimF$=1.0\,s,
            $\kF$=10$^{-5}$\,Pa$^{-1}$, $\fe$=1.0, $\fm$=0.5, $m$=10 slices, with varying $\Go$ (0.1, 1.0, 10 and 100\,Pa).}
  \label{TardyB5}
\end{figure}

%XXXXXXXXXXXXXXXXXXXXXXXXXXXXXXXXXXXXXXXXXXXXXXXXXXXXXXXXXXXXXXXXXXXXXX
\subsubsection{Shear-Thickening}\label{}
A slight \shThik\ effect was observed when setting $\lVis < \hVis$
while holding the other parameters constant. The effect of
increasing $\Go$ on the apparent viscosities at high-shear rates was
similar to the effect observed for the \shThin\ fluids though it was
at a smaller scale. However, the low-shear viscosities are not
affected, as in the case of \shThin\ fluids. A convergence for the
apparent viscosities at high-shear rates for large values of $\Go$
was also observed as for \shThin. A sample of the results obtained
in this investigation is presented in Figures (\ref{TardySP6}) and
(\ref{TardyB6}) for the \sandp\ and \Berea\ networks respectively on
a log-log scale. It should be remarked that as the \BauMan\ is
originally a \shThin\ model, it should not be expected to
fully-predict \shThik\ phenomenon. However, the observed behavior is
a good sign for our model because as we push the algorithm beyond
its limit, the results are still qualitatively reasonable.

\begin{figure}[!t]
  \centering{}
  \includegraphics
  [scale=0.5]
  {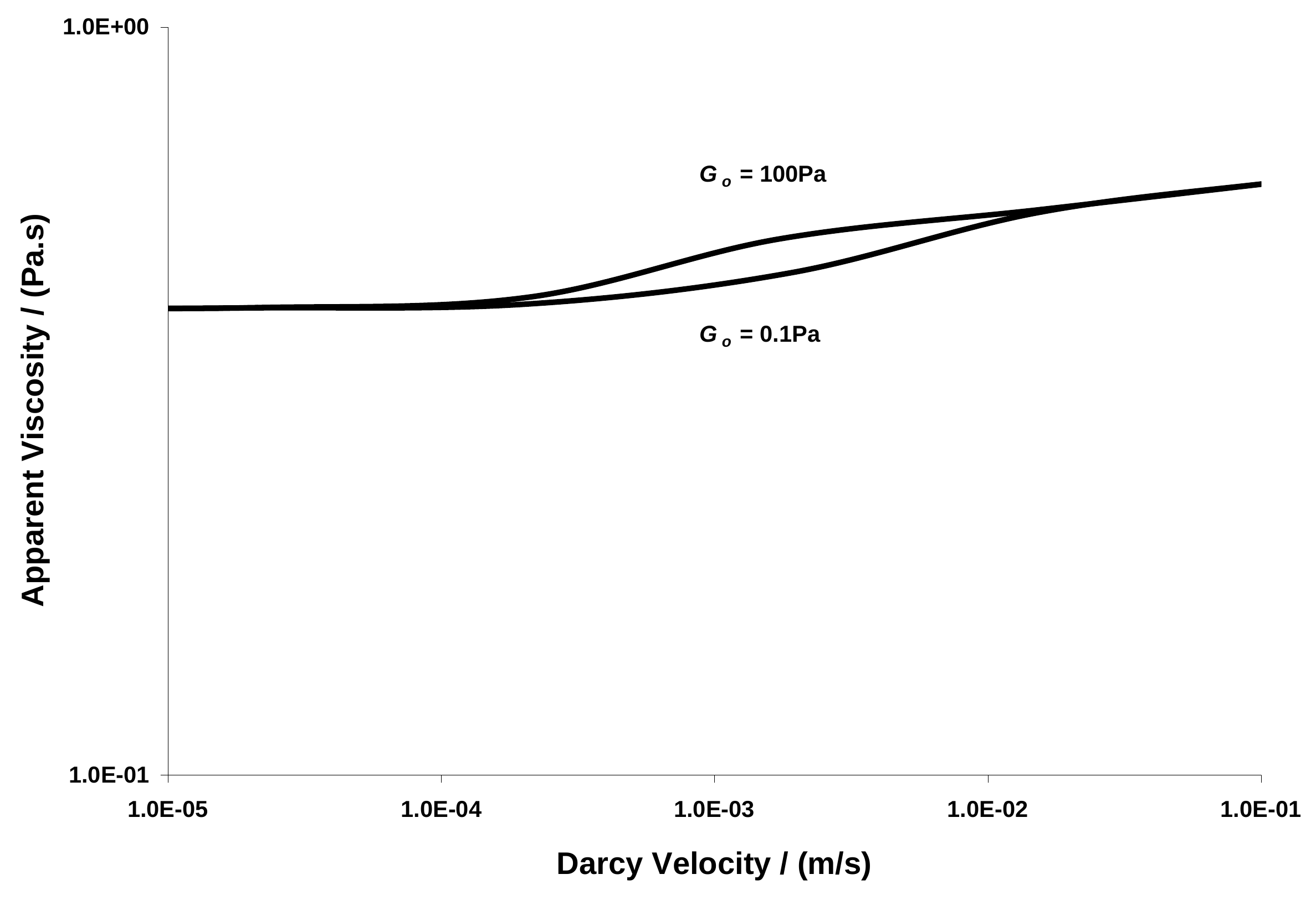}
  \caption[The \Tardy\ algorithm \sandp\ results for $\hVis$=10.0\,Pa.s, $\lVis$=0.1\,Pa.s, $\rxTimF$=1.0\,s,
            $\kF$=10$^{-4}$\,Pa$^{-1}$, $\fe$=1.0, $\fm$=0.5, $m$=10 slices, with varying $\Go$ (0.1 and 100\,Pa)]
  {The \Tardy\ algorithm \sandp\ results for $\hVis$=10.0\,Pa.s, $\lVis$=0.1\,Pa.s, $\rxTimF$=1.0\,s,
            $\kF$=10$^{-4}$\,Pa$^{-1}$, $\fe$=1.0, $\fm$=0.5, $m$=10 slices, with varying $\Go$ (0.1 and 100\,Pa).}
  \label{TardySP6}
\end{figure}
\begin{figure}[!h]
  \centering{}
  \includegraphics
  [scale=0.5]
  {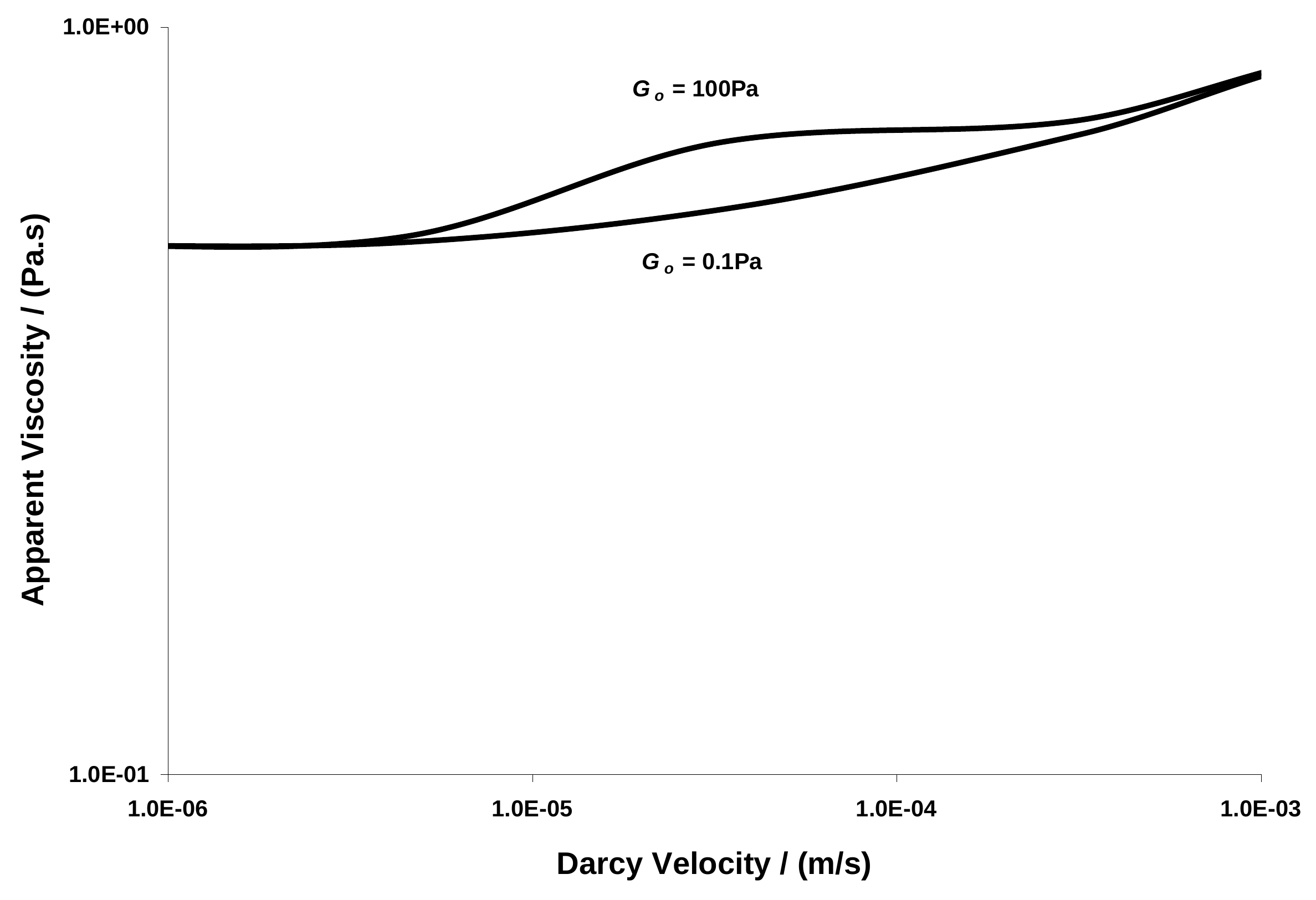}
  \caption[The \Tardy\ algorithm \Berea\ results for $\hVis$=10.0\,Pa.s, $\lVis$=0.1\,Pa.s, $\rxTimF$=1.0\,s,
            $\kF$=10$^{-4}$\,Pa$^{-1}$, $\fe$=1.0, $\fm$=0.5, $m$=10 slices, with varying $\Go$ (0.1 and 100\,Pa)]
  {The \Tardy\ algorithm \Berea\ results for $\hVis$=10.0\,Pa.s, $\lVis$=0.1\,Pa.s, $\rxTimF$=1.0\,s,
            $\kF$=10$^{-4}$\,Pa$^{-1}$, $\fe$=1.0, $\fm$=0.5, $m$=10 slices, with varying $\Go$ (0.1 and 100\,Pa).}
  \label{TardyB6}
\end{figure}

%XXXXXXXXXXXXXXXXXXXXXXXXXXXXXXXXXXXXXXXXXXXXXXXXXXXXXXXXXXXXXXXXXXXXXX
\subsubsection{Relaxation Time}\label{}
The effect of the structural relaxation time $\rxTimF$ was
investigated for \shThin\ fluids by varying this parameter over
several orders of magnitude while holding the others constant. It
was observed that by increasing the structural relaxation time, the
apparent viscosities were steadily decreased. However, the decrease
at high-shear rates has almost reached a saturation point where
beyond some limit the apparent viscosities converged to certain
values despite a large increase in $\rxTimF$. A sample of the
results is given in Figure (\ref{TardySP7}) for the \sandp\ and
Figure (\ref{TardyB7}) for \Berea\ on a log-log scale. As we
discussed earlier in this chapter, the effects of relaxation time
are expected to ease as the relaxation time increases beyond a limit
such that the effect of interaction with capillary constriction is
negligible. Despite the fact that this feature requires extensive
investigation for quantitative confirmation, the observed behavior
seems qualitatively reasonable considering the flow regimes and the
size of relaxation times of the sample results.

\begin{figure}[!t]
  \centering{}
  \includegraphics
  [scale=0.5]
  {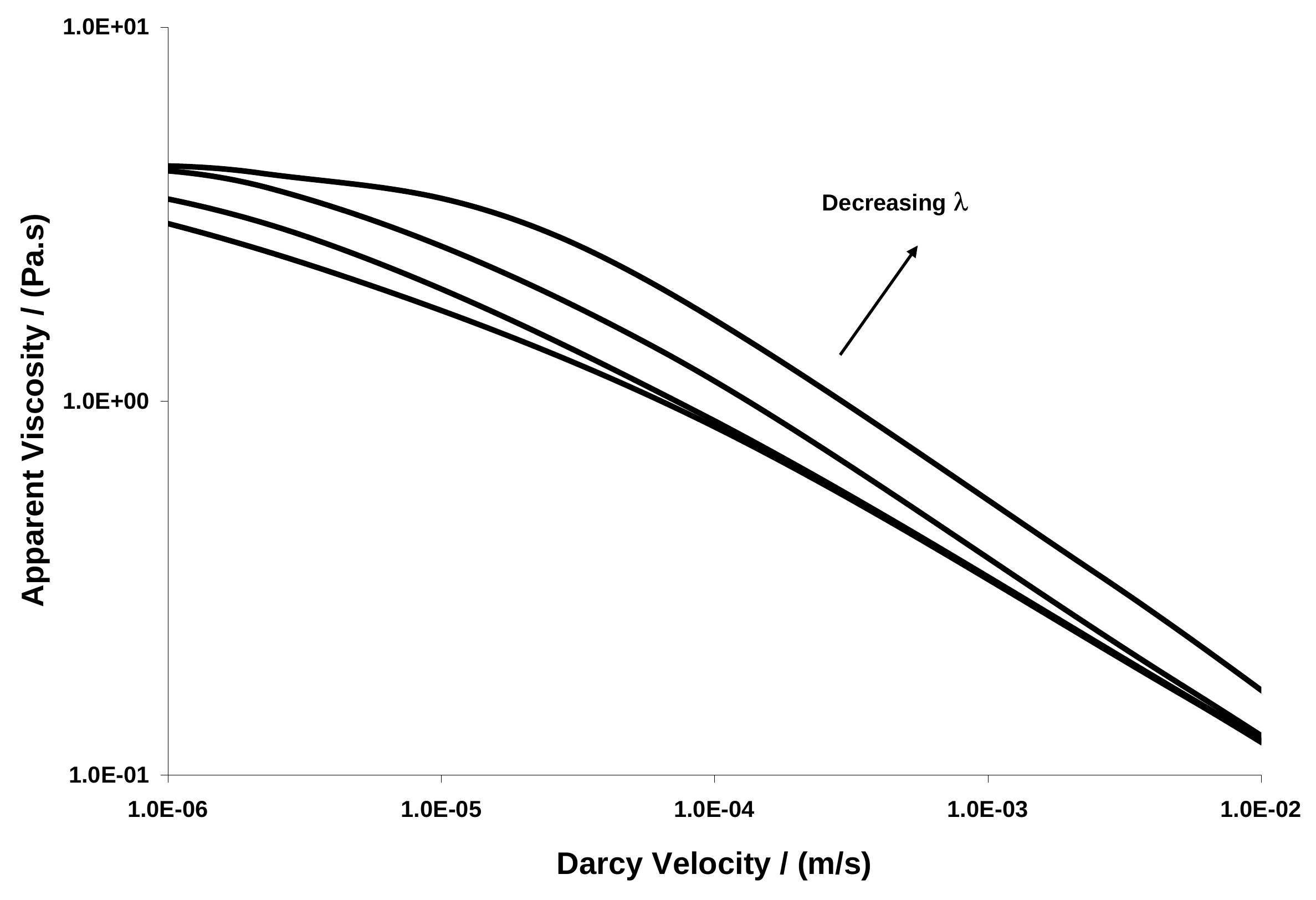}
  \caption[The \Tardy\ algorithm \sandp\ results for $\Go$=1.0\,Pa,  $\hVis$=0.001\,Pa.s, $\lVis$=1.0\,Pa.s,
            $\kF$=10$^{-5}$\,Pa$^{-1}$, $\fe$=1.0, $\fm$=0.5, $m$=10 slices, with varying $\rxTimF$ (0.1, 1.0, 10, and 100\,s)]
  {The \Tardy\ algorithm \sandp\ results for $\Go$=1.0\,Pa,  $\hVis$=0.001\,Pa.s, $\lVis$=1.0\,Pa.s,
            $\kF$=10$^{-5}$\,Pa$^{-1}$, $\fe$=1.0, $\fm$=0.5, $m$=10 slices, with varying $\rxTimF$ (0.1, 1.0, 10, and 100\,s).}
  \label{TardySP7}
\end{figure}
\begin{figure}[!h]
  \centering{}
  \includegraphics
  [scale=0.5]
  {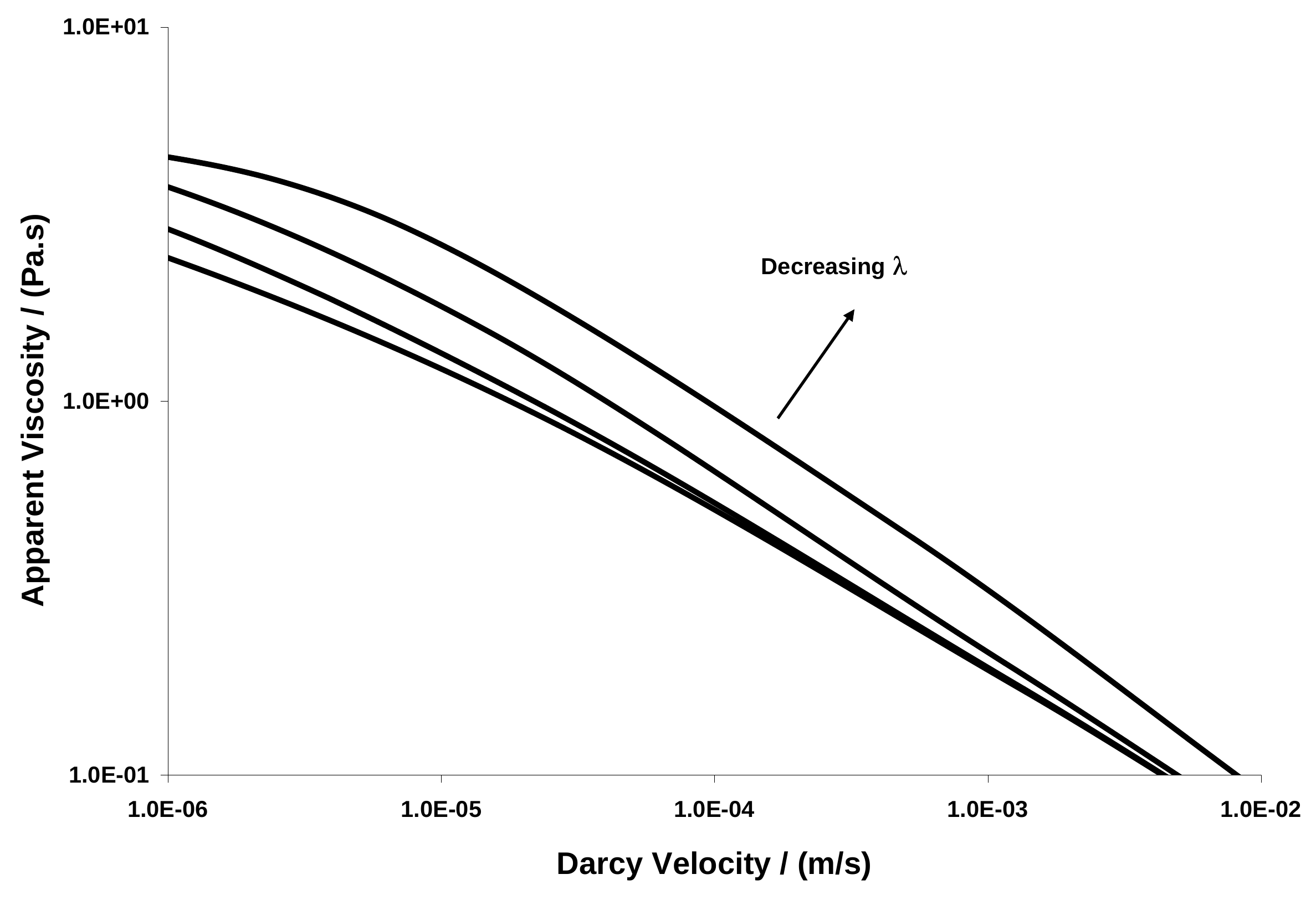}
  \caption[The \Tardy\ algorithm \Berea\ results for $\Go$=1.0\,Pa,  $\hVis$=0.001\,Pa.s, $\lVis$=1.0\,Pa.s,
            $\kF$=10$^{-5}$\,Pa$^{-1}$, $\fe$=1.0, $\fm$=0.5, $m$=10 slices, with varying $\rxTimF$ (0.1, 1.0, 10, and 100\,s)]
  {The \Tardy\ algorithm \Berea\ results for $\Go$=1.0\,Pa,  $\hVis$=0.001\,Pa.s, $\lVis$=1.0\,Pa.s,
            $\kF$=10$^{-5}$\,Pa$^{-1}$, $\fe$=1.0, $\fm$=0.5, $m$=10 slices, with varying $\rxTimF$ (0.1, 1.0, 10, and 100\,s).}
  \label{TardyB7}
\end{figure}

%XXXXXXXXXXXXXXXXXXXXXXXXXXXXXXXXXXXXXXXXXXXXXXXXXXXXXXXXXXXXXXXXXXXXXX
\subsubsection{Kinetic Parameter}\label{}
The effect of the kinetic parameter for structure break down $\kF$
was investigated for \shThin\ fluids by varying this parameter over
several orders of magnitude while holding the others constant. It
was observed that by increasing the kinetic parameter, the apparent
viscosities were generally decreased. However, in some cases the
low-shear viscosities has not been substantially affected. A sample
of the results is given in Figures (\ref{TardySP8}) and
(\ref{TardyB8}) for the \sandp\ and \Berea\ networks respectively on
a log-log scale. The decrease in apparent viscosity on increasing
the kinetic parameter is a natural response as the parameter
quantifies structure break down and hence reflects thinning
mechanisms. It is a general trend that the low flow rate regimes
near the lower \NEW\ plateau are usually less influenced by \nNEW\
effects. It will therefore come as no surprise that in some cases
the low-shear viscosities experienced minor changes.

\begin{figure}[!t]
  \centering{}
  \includegraphics
  [scale=0.5]
  {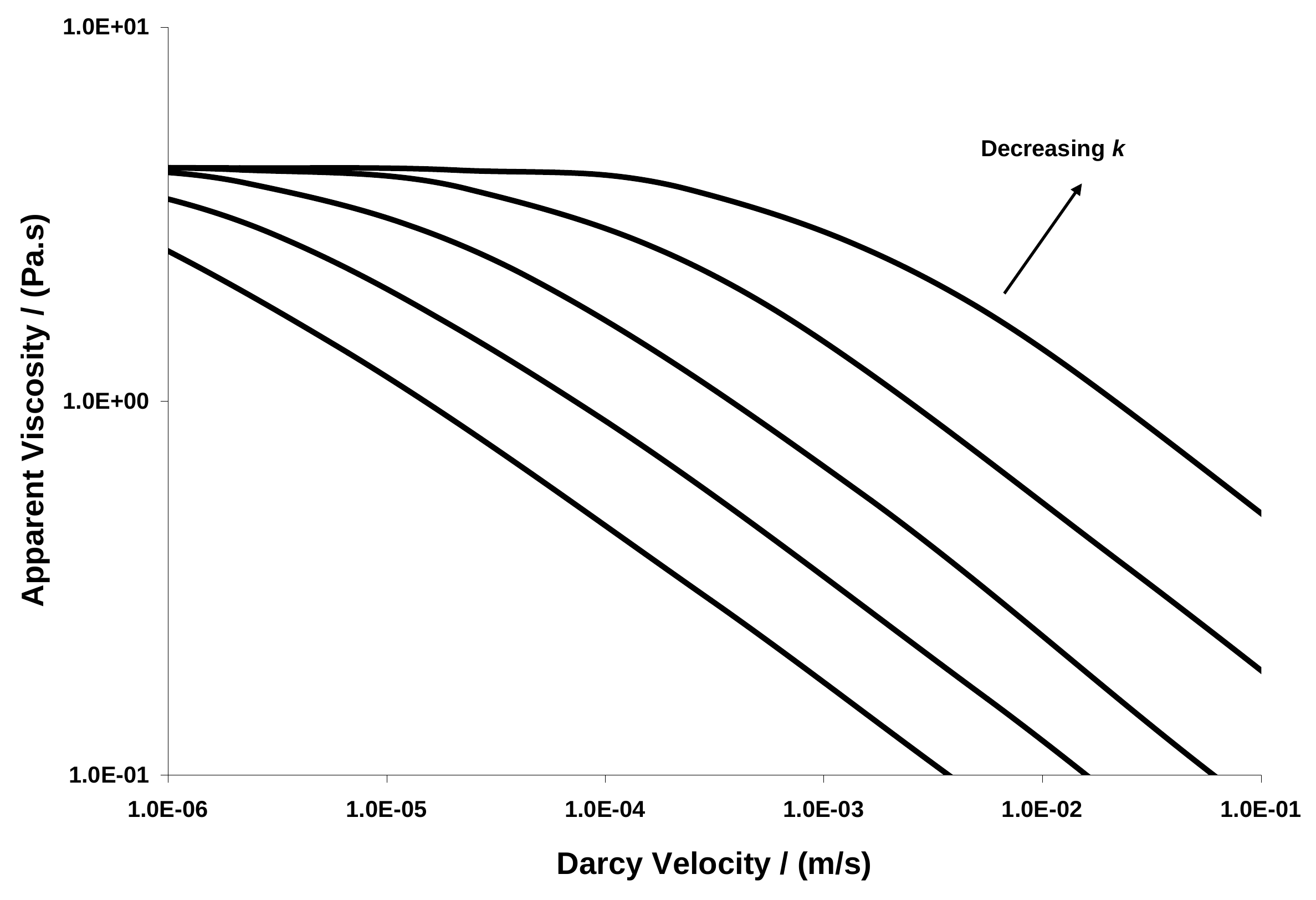}
  \caption[The \Tardy\ algorithm \sandp\ results for $\Go$=1.0\,Pa,  $\hVis$=0.001\,Pa.s, $\lVis$=1.0\,Pa.s,
            $\rxTimF$=10\,s, $\fe$=1.0, $\fm$=0.5, $m$=10 slices, with varying $\kF$ ($10^{-3}$,
            $10^{-4}$, $10^{-5}$, $10^{-6}$ and $10^{-7}$\,Pa$^{-1}$)]
  {The \Tardy\ algorithm \sandp\ results for $\Go$=1.0\,Pa,  $\hVis$=0.001\,Pa.s, $\lVis$=1.0\,Pa.s,
            $\rxTimF$=10\,s, $\fe$=1.0, $\fm$=0.5, $m$=10 slices, with varying $\kF$ ($10^{-3}$,
            $10^{-4}$, $10^{-5}$, $10^{-6}$ and $10^{-7}$\,Pa$^{-1}$).}
  \label{TardySP8}
\end{figure}
\begin{figure}[!h]
  \centering{}
  \includegraphics
  [scale=0.5]
  {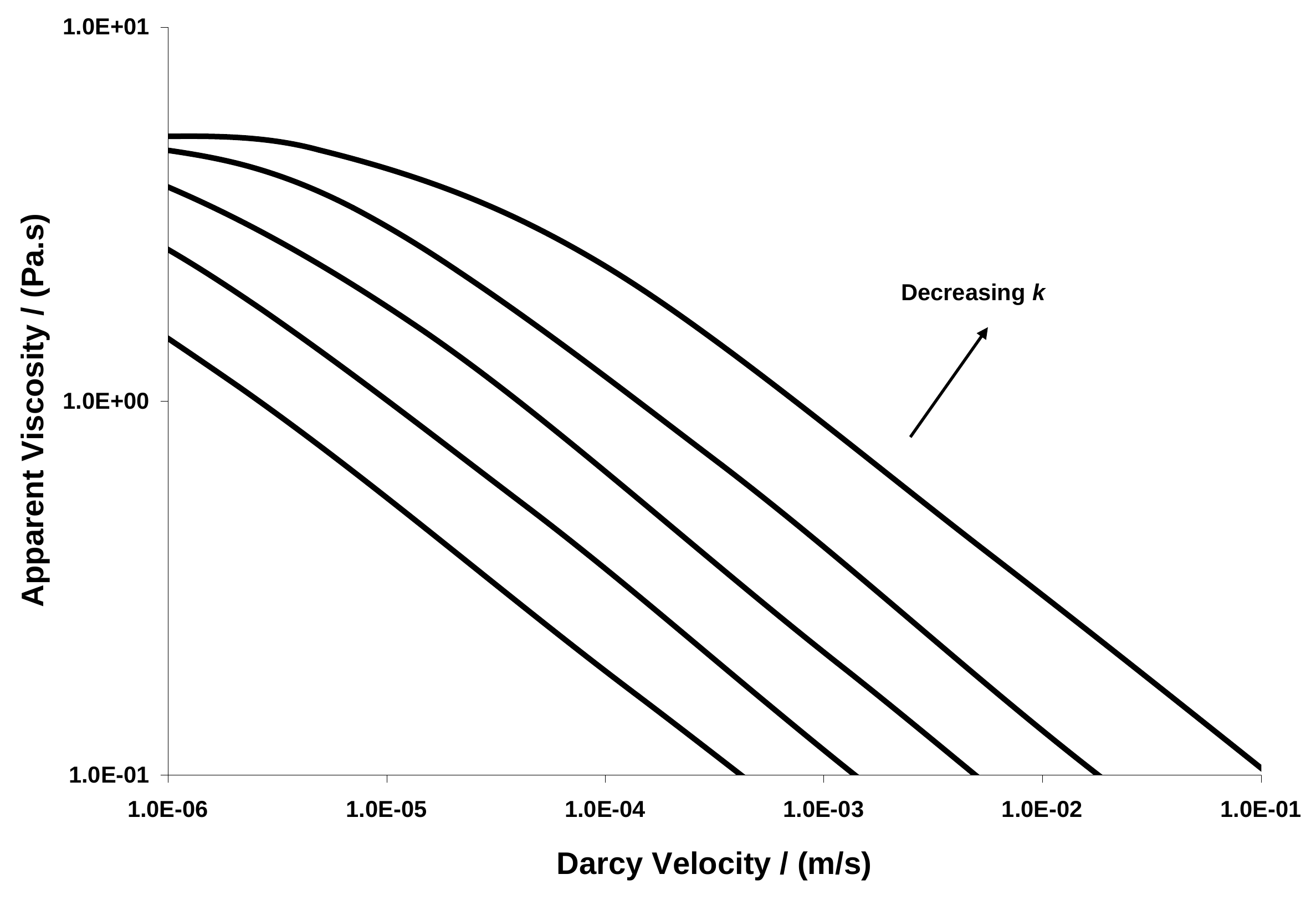}
  \caption[The \Tardy\ algorithm \Berea\ results for $\Go$=1.0\,Pa,  $\hVis$=0.001\,Pa.s, $\lVis$=1.0\,Pa.s,
            $\rxTimF$=10\,s, $\fe$=1.0, $\fm$=0.5, $m$=10 slices, with varying $\kF$ ($10^{-3}$,
            $10^{-4}$, $10^{-5}$, $10^{-6}$ and $10^{-7}$\,Pa$^{-1}$)]
  {The \Tardy\ algorithm \Berea\ results for $\Go$=1.0\,Pa,  $\hVis$=0.001\,Pa.s, $\lVis$=1.0\,Pa.s,
            $\rxTimF$=1.0\,s, $\fe$=1.0, $\fm$=0.5, $m$=10 slices, with varying $\kF$ ($10^{-3}$,
            $10^{-4}$, $10^{-5}$, $10^{-6}$ and $10^{-7}$\,Pa$^{-1}$).}
  \label{TardyB8}
\end{figure}

\def\baselinestretch{1}
\chapter{\Thixotropy\ and \Rheopexy} \label{Thixotropy}
\def\baselinestretch{1.66}
\Timedep\ fluids are defined to be those fluids whose viscosity
depends on the duration of flow under isothermal conditions. For
these fluids, a \timeind\ \steadys\ viscosity is eventually reached
in steady flow situation. The time effect is often reversible though
it may be partial. As a consequence, the trend of the viscosity
change is overturned in time when the stress is reduced. The
phenomenon is generally attributed to \timedep\ \thixotropic\
breakdown or \rheopectic\ buildup of some particulate structure
under relatively high stress followed by structural change in the
opposite direction for lower stress, though the exact mechanism may
not be certain. Furthermore, \thixotropic\ buildup and \rheopectic\
breakdown may also be a possibility \cite{carreaubook, Collyer1974,
EscudierP1996, DullaertM2005, Barnes1997}.

\vspace{0.2cm}

\Timedep\ fluids are generally divided into two main categories:
\thixotropic\ (work softening) whose viscosity gradually decreases
with time at a constant shear rate, and \rheopectic\ (work hardening
or anti-\thixotropy\ or negative \thixotropy) whose viscosity
increases under similar circumstances without an overshoot which is
a characteristic feature of \vy. However, it has been proposed that
\rheopexy\ and negative \thixotropy\ are different, and hence three
categories of \timedep\ fluids do exist \cite{owensbook2002,
carreaubook, ChengE1965}. It should be emphasized that in this
thesis we may rely on the context and use ``\thixotropy''
conveniently to indicate non-elastic time-dependence in general
where the meaning is obvious.

\vspace{0.2cm}

\Thixotropic\ fluids may be described as \shThin\ while the
\rheopectic\ as \shThik, in the sense that these effects take place
on shearing, though they are effects of time-dependence. However, it
has been suggested that \thixotropy\ invariably occurs in
circumstances where the liquid is \shThin\ (in the sense that
viscosity levels decrease with increasing shear rate, other things
being equal) and the anti-\thixotropy\ is usually associated with
\shThik. This may be behind the occasional confusion between
\thixotropy\ and \shThin\ \cite{Barnes1997, BalmforthC2001,
barnesbookHW1993, Mewis1979}.

\vspace{0.2cm}

A substantial number of complex fluids display time-dependence
phenomena, with \thixotropy\ being more commonplace and better
understood than \rheopexy. Various mathematical models have been
proposed to describe time-dependence behavior. These models include
microstructural, continuum mechanics, and structural kinetics models
\cite{Mewis1979, DullaertM2005}.

\vspace{0.2cm}

\Thixotropic\ and \rheopectic\ behaviors may be detected by the
hysteresis loop test, as well as by the more direct steady shear
test. In the loop test the substance is sheared cyclically and a
graph of stress versus shear rate is obtained. A \timedep\ fluid
should display a hysteresis loop the area of which is a measure of
the degree of \thixotropy\ or \rheopexy\ and may be used to quantify
time-dependent behavior \cite{Collyer1974, ChhabraR1999,
ChengE1965}.

\vspace{0.2cm}

In theory, time-dependence effects can arise from \thixotropic\
structural change or from \timedep\ \vy\ or from both effects
simultaneously. The existence of these two different types of
\timedep\ rheological behavior is generally recognized. Although it
is convenient to distinguish between these as separate types of
phenomena, real fluids can exhibit both types of rheological
behavior simultaneously. Several physical distinctions between \vc\
and \thixotropic\ time-dependence have been made. The important one
is that while the time-dependence of \vc\ fluids arises because the
response of stresses and strains in the fluid to changes in imposed
strains and stresses respectively is not instantaneous, in
\thixotropic\ fluids such response is instantaneous and the
\timedep\ behavior arises purely because of changes in the structure
of the fluid as a consequence of strain. Despite the fact that the
mathematical theory of \vy\ has been developed to an advanced level,
especially on the continuum mechanical front, relatively little work
has been done on \thixotropy\ and \rheopexy. One reason is the lack
of a comprehensive framework to describe the dynamics of
\thixotropy. This may partly explain why \thixotropy\ is rarely
incorporated in the constitutive equation when modeling the flow of
\nNEW\ fluids. The underlying assumption is that in these situations
the \thixotropic\ effects have a negligible impact on the resulting
flow field, and this allows great mathematical simplifications
\cite{JoyeP1971, ChengE1965, EscudierP1996, PearsonT2002,
PritchardP2006, BillinghamF1993}.
\begin{figure} [!t]
  \centering{}
  \includegraphics
  [scale=0.6]
  {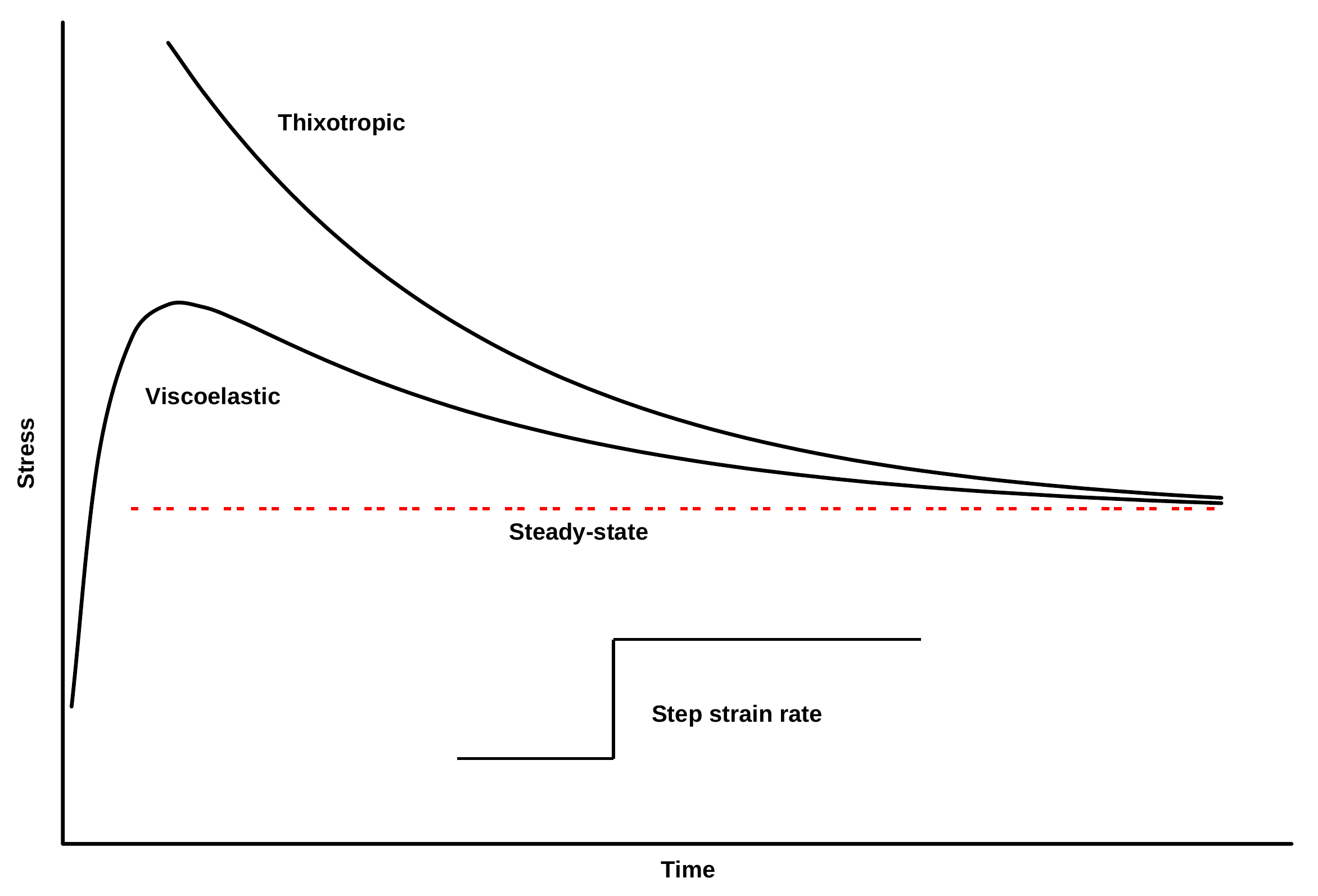}
  \caption[Comparison between time dependency in \thixotropic\ and \vc\ fluids following a step increase in strain rate]
  {Comparison between time dependency in \thixotropic\ and \vc\ fluids following a step increase in strain rate.}
  \label{ThixoVE}
\end{figure}

Several behavioral distinctions can be made to differentiate between
\vy\ and \thixotropy. These include the presence or absence of some
characteristic elastic features such as recoil and normal stresses.
However, these signs may be of limited use in some practical
situations involving complex fluids where these two phenomena
coexist. In Figure (\ref{ThixoVE}) the behavior of these two types
of fluids in response to step change in strain rate is compared.
Although both fluids show signs of dependency on the past history,
the graph suggests that inelastic \thixotropic\ fluids do not
possess a memory in the same sense as \vc\ materials. The behavioral
difference, such as the absence of elastic effects or the difference
in the characteristic time scale associated with these phenomena,
confirms this suggestion \cite{Barnes1997, carreaubook, JoyeP1971,
Mewis1979}.

\vspace{0.2cm}

\Thixotropy, like \vy, is a phenomenon which can appear in a large
number of systems. The time scale for \thixotropic\ changes is
measurable in various materials including important commercial and
biological products. However, the investigation of \thixotropy\ has
been hampered by several difficulties. Consequently, the suggested
\thixotropic\ models seem unable to present successful quantitative
description of \thixotropic\ behavior especially for the \trans\
state. In fact, even the most characteristic property of
\thixotropic\ fluids, i.e. the decay of viscosity or stress under
steady shear conditions, presents difficulties in modeling and
characterization \cite{Mewis1979, JoyeP1971, Godfrey1973}.

\vspace{0.2cm}

The lack of a comprehensive theoretical framework to describe
\thixotropy\ is matched by a severe shortage on the experimental
side. One reason is the difficulties confronted in measuring
\thixotropic\ systems with sufficient accuracy. The result is that
very few systematic data sets can be found in the literature and
this deficit hinders the progress in this field. Moreover, the
characterization of the data in the absence of an agreed-upon
mathematical structure may be questionable \cite{EscudierP1996,
DullaertM2005}.

%XXXXXXXXXXXXXXXXXXXXXXXXXXXXXXXXXXXXXXXXXXXXXXXXXXXXXXXXXXXXXXXXXXXXXX
\section{Simulating \TimeDep\ Flow in Porous Media} \label{SimulatingTDFPM}
There are three major cases of simulating the flow of \timedep\
fluids in porous media:

\begin{itemize}

    \item The flow of strongly strain-dependent fluid in a porous medium which
    is not very homogeneous. This is very difficult to model because
    it is difficult to track the fluid elements in the pores and throats and determine
    their deformation history. Moreover, the viscosity
    function may not be well defined due to the mixing of fluid elements
    with various deformation history in the individual pores and throats.

    \item The flow of strain-independent or weakly strain-dependent fluid
    through porous media in general. A possible strategy is to apply single \timedep\
    viscosity function to all pores and throats at each instant of time and hence
    simulating time development as a sequence of \NEW\ states.

    \item The flow of strongly strain-dependent fluid in a very
    homogeneous porous medium such that the fluid is subject to the same
    deformation in all network elements. The strategy for modeling this flow is
    to define an effective pore strain rate. Then using a very small time
    step the strain rate in the next instant of time can be found assuming constant strain rate.
    As the change in the strain rate is now known, a correction to the viscosity due to
    strain-dependency can be introduced.

    \vspace{0.1cm}

    There are two possible problems in this strategy. The first is edge
    effects in the case of injection from a reservoir since the
    deformation history of the fluid elements at the inlet is different
    from the deformation history of the fluid elements inside the
    network. The second is that considerable computing resources may be
    required if very small time steps are used.

\end{itemize}

In the current work, thixotropic behavior is modeled within the
\Tardy\ algorithm as the \BauMan\ fluid incorporates thixotropic as
well as \vc\ effects.

\def\baselinestretch{1}
\chapter{Conclusions and Future Work} \label{Conclusions}
\def\baselinestretch{1.66}
%
%SSSSSSSSSSSSSSSSSSSSSSSSSSSSSSSSSSSSSSSSSSSSSSSSSSSSSSSSSSSSSSSSSSSSSSSSSS
\section{Conclusions} \label{}
Here, we present some general results and conclusions that can be
drawn from this work
\begin{itemize}

\item The network model has been extended to account for different types of \nNEW\
rheology: \ELLIS\ and \HB\ models for \timeind, and \BauMan\ for
\vc. The basis of the implementation in the first case is an
analytical expression for the nonlinear relationship between flow
rate and pressure drop in a single cylindrical element. These
expressions are used in conjunction with an iterative technique to
find the total flow across the network. For the \BauMan\ \vc\ model,
the basis is a numerical algorithm originally suggested by \Tardy. A
modified version of this algorithm was used to investigate the
effect of \convdiv\ geometry on the \steadys\ \vc\ flow.

\item The general trends in behavior for shear-independent, \shThin\ and \shThik\
fluids with and without \yields\ for the \timeind\ category have
been examined. Compared to an equivalent uniform bundle of tubes
model, our networks yield at lower pressure gradients. Moreover,
some elements remain blocked even at very high pressure gradients.
The results also reveal that \shThin\ accentuates the heterogeneity
of the network, while \shThik\ makes the flow more uniform.

\item A comparison between the \sandp\ and \Berea\ networks and
a cubic network matching their characteristics has been made and
some useful conclusions have been drawn.

\item The network model successfully predicted several experimental data sets in the
literature on \ELLIS\ fluids. Less satisfactory results were
obtained for the \HB\ fluids.

\item The predictions were less satisfactory for \yields\ fluids of
\HB\ model. The reason is the inadequate representation of the pore
space structure with regard to the yield process, that is the yield
in a network depends on the actual shape of the void space rather
than the flow conductance of the pores and throats.  Other possible
reasons for this failure are experimental errors and the occurrence
of other physical effects which are not accounted for in our model,
such as precipitation and adsorption.

\item Several numerical algorithms related to \yields\ have been
implemented in our \nNEW\ code and a comparison has been made
between the two algorithms whose objective is to predict the
threshold yield pressure of a network. These algorithms include the
\InvPM\ (IPM) of Kharabaf and the \PathMP\ (PMP) which is a novel
and computationally more efficient technique than the IPM. The two
methods gave similar predictions for the random networks and
identical predictions for the cubic networks. An explanation was
given for the difference observed in some cases of random networks
as the algorithms are implemented with different assumptions on the
possibility of flow backtracking.

\item A comparison was also made between the IPM and PMP algorithms on one hand
and the network flow simulation results on the other. An explanation
was given for why these algorithms predict lower threshold yield
pressure than the network simulation results.

\item The implementation of the \BauMan\ model in the form of a
modified \Tardy\ algorithm for \steadys\ \vc\ flow has been examined
and assessed. The generic behavior indicates qualitatively correct
trends. A conclusion was reached that the current implementation is
a proper basis for investigating the \steadys\ \vc\ effects, and
possibly some \thixotropic\ effects, due to \convdiv\ geometries in
porous media. This implementation can also be used as a suitable
base for further future development.

\end{itemize}

%SSSSSSSSSSSSSSSSSSSSSSSSSSSSSSSSSSSSSSSSSSSSSSSSSSSSSSSSSSSSSSSSSSSSSSSSSS
\section{Recommendations for Future Work} \label{}
Our recommendations for the future work are

\begin{itemize}

\item Extending the current \nNEW\ model to account for other physical
phenomena such as adsorption and wall exclusion.

\item Implementing other \timeind\ fluid models.

\item Implementing two-phase flow of two \nNEW\ fluids.

\item Considering more complex void space description either by
using more elaborate networks or by superimposing more elaborate
details on the current networks.

\item Extending the analysis of the \yields\ fluids and
investigating the effect of the actual shape of the void space on
the yield point instead of relating the yield to the conductance of
cylindrically-shaped capillaries.

\item Examining the effect of \convdiv\ geometry on
the \yields\ of the network. The current implementation of the
\convdiv\ capillaries in the \Tardy\ \vc\ algorithm will facilitate
this task.

\item Elaborating the modified \Tardy\ algorithm and thoroughly investigating its
\vc\ and \thixotropic\ predictions in quantitative terms.

\item Developing and implementing other \trans\ and
\steadys\ \vc\ algorithms.

\item Developing and implementing \timedep\ algorithms in
the form of pure \thixotropic\ and \rheopectic\ models such as the
\SEM, though \thixotropic\ effects are integrated within the
\BauMan\ fluid which is already implemented in the \Tardy\
algorithm.

\end{itemize}

} %End of: \setlength{\parskip}

\phantomsection \addcontentsline{toc}{chapter}{\protect \numberline{} Bibliography} %
\bibliographystyle{unsrt}    % New

%\bibliography{Biblio}        % New

\begin{thebibliography}{100}

\bibitem{skellandbook}
A.H.P. Skelland.
\newblock {\em Non-Newtonian Flow and Heat Transfer}.
\newblock John Wiley and Sons Inc., 1967.

\bibitem{ChhabraR1999}
R.P. Chhabra;~J.F. Richardson.
\newblock {\em Non-Newtonian Flow in the Process Industries}.
\newblock Butterworth Heinemann Publishers, 1999.

\bibitem{owensbook2002}
R.G. Owens;~T.N. Phillips.
\newblock {\em Computational Rheology}.
\newblock Imperial College Press, 2002.

\bibitem{OsP2004}
R.G.M. van Os; T.N.~Phillips.
\newblock Spectral element methods for transient viscoelastic flow problems.
\newblock {\em Journal of Computational Physics}, 201(1):286--314, 2004.

\bibitem{barnesbookHW1993}
H.A. Barnes; J.E Hutton;~K. Walters.
\newblock {\em An introduction to rheology}.
\newblock Elsevier, 1993.

\bibitem{birdbook}
R.B. Bird; R.C. Armstrong;~O. Hassager.
\newblock {\em Dynamics of Polymeric Liquids}, volume~1.
\newblock John Wily \& Sons, second edition, 1987.

\bibitem{Collyer1974}
A.A. Collyer.
\newblock Time dependent fluids.
\newblock {\em Physics Education}, 9:38--44, 1974.

\bibitem{TardyA2005}
P.~Tardy;~V.J. Anderson.
\newblock Current modelling of flow through porous media.
\newblock {\em Private communication}, 2005.

\bibitem{SadowskiB1965}
T.J. Sadowski;~R.B. Bird.
\newblock Non-\textrm{N}ewtonian flow through porous media \textrm{I.
  T}heoretical.
\newblock {\em Transactions of the Society of Rheology}, 9(2):243--250, 1965.

\bibitem{Savins1969}
J.G. Savins.
\newblock Non-\textrm{N}ewtonian flow through porous media.
\newblock {\em Industrial and Engineering Chemistry}, 61(10):18--47, 1969.

\bibitem{carreaubook}
P.J. Carreau; D. De Kee;~R.P. Chhabra.
\newblock {\em Rheology of Polymeric Systems}.
\newblock Hanser Publishers, 1997.

\bibitem{LiuM1998}
S.~Liu;~J.H. Masliyah.
\newblock On non-\rm{N}ewtonian fluid flow in ducts and porous media -
  \rm{O}ptical rheometry in opposed jets and flow through porous media.
\newblock {\em Chemical Engineering Science}, 53(6):1175--1201, 1998.

\bibitem{larsonbook1999}
R.G. Larson.
\newblock {\em The Structure and Rheology of Complex Fluids}.
\newblock Oxford University Press, 1999.

\bibitem{Hulsen1996-2}
M.A. Hulsen.
\newblock {\em A discontinuous Galerkin method with splitting applied to
  visco-elastic flow}.
\newblock Faculty of Mechanical Engineering and Marine Technology, Delft
  University of Technology, 1986.

\bibitem{Keunings2004}
R.~Keunings.
\newblock Micro-macro methods for the multi-scale simulation of viscoelastic
  flow using molecular models of kinetic theory.
\newblock {\em Rheology Reviews}, pages 67--98, 2004.

\bibitem{Denn1990}
M.M. Denn.
\newblock Issues in viscoelastic fluid mechanics.
\newblock {\em Annual Review of Fluid Mechanics}, 22:13--34, 1990.

\bibitem{deiberthesis}
J.A. Deiber.
\newblock {\em Modeling the flow of Newtonian and viscoelastic fluids through
  porous media}.
\newblock PhD thesis, Princeton University, 1978.

\bibitem{larsonbook1988}
R.G. Larson.
\newblock {\em Constitutive Equations for Polymer Melts and Solutions}.
\newblock Butterworth Publishers, 1988.

\bibitem{kronjagerthesis}
V.J. Kronj\"{a}ger.
\newblock {\em Numerical studies of viscoelastic shear turbulence}.
\newblock Thesis, University of Marburg, 2001.

\bibitem{MenaML1987}
B.~Mena; O. Manero;~L.G. Leal.
\newblock The influence of rheological properties on the slow flow past
  spheres.
\newblock {\em Journal of Non-Newtonian Fluid Mechanics}, 26(2):247--275, 1987.

\bibitem{maxwell1}
J.C. Maxwell.
\newblock On the dynamical theory of gases.
\newblock {\em Philosophical Transactions of the Royal Society of London},
  157:49--88, 1867.

\bibitem{Jeffreysbook}
H.~Jeffreys.
\newblock {\em The Earth Its Origin, History and Physical Constitution}.
\newblock Cambridge University Press, second edition, 1929.

\bibitem{MardonesG1990}
J.~Mart\'{\i}nez-Mardones;~C. P\'{e}rez-Garc\'{\i}a.
\newblock Linear instability in viscoelastic fluid convection.
\newblock {\em Journal of Physics: Condensed Matter}, 2(5):1281--1290, 1990.

\bibitem{white1}
J.L. White;~A.B. Metzner.
\newblock Thermodynamic and heat transport considerations for viscoelastic
  fluids.
\newblock {\em Chemical Engineering Science}, 20:1055--1062, 1965.

\bibitem{tannerbook2000}
R.I. Tanner.
\newblock {\em Engineering Rheology}.
\newblock Oxford University Press, 2nd edition, 2000.

\bibitem{BalmforthC2001}
N.J. Balmforth;~R.V. Craster.
\newblock {\em Geophysical Aspects of Non-Newtonian Fluid Mechanics}.
\newblock Springer, 2001.

\bibitem{Boger1987}
D.V. Boger.
\newblock Viscoelastic flows through contractions.
\newblock {\em Annual Review of Fluid Mechanics}, 19:157--182, 1987.

\bibitem{Godfrey1973}
J.C. Godfrey.
\newblock Steady shear measurement of thixotropic fluid properties.
\newblock {\em Rheologica Acta}, 12(4):540--545, 1973.

\bibitem{Barnes1997}
H.A. Barnes.
\newblock Thixotropy -- a review.
\newblock {\em Journal of Non-Newtonian Fluid Mechanics}, 70(1):1--33, 1997.

\bibitem{Sadowski1965}
T.J. Sadowski.
\newblock Non-\textrm{N}ewtonian flow through porous media \textrm{II.
  E}xperimental.
\newblock {\em Transactions of the Society of Rheology}, 9(2):251--271, 1965.

\bibitem{sadowskithesis}
T.J. Sadowski.
\newblock {\em Non-Newtonian flow through porous media}.
\newblock PhD thesis, University of Wisconsin, 1963.

\bibitem{parkthesis}
H.C. Park.
\newblock {\em The flow of non-Newtonian fluids through porous media}.
\newblock PhD thesis, Michigan State University, 1972.

\bibitem{ParkHB1973}
H.C. Park; M.C. Hawley;~R.F. Blanks.
\newblock The flow of non-\textrm{N}ewtonian solutions through packed beds.
\newblock {\em SPE 4722}, 1973.

\bibitem{balhoffthesis}
M.T. Balhoff.
\newblock {\em Modeling the flow of non-Newtonian fluids in packed beds at the
  pore scale}.
\newblock PhD thesis, Louisiana State University, 2005.

\bibitem{balhoff2}
M.T. Balhoff;~K.E. Thompson.
\newblock A macroscopic model for shear-thinning flow in packed beds based on
  network modeling.
\newblock {\em Chemical Engineering Science}, 61(2006):698--719, 2005.

\bibitem{pascal1}
H.~Pascal.
\newblock Nonsteady flow through porous media in the presence of a threshold
  gradient.
\newblock {\em Acta Mechanica}, 39:207--224, 1981.

\bibitem{alfariss1}
T.F. Al-Fariss;~K.L. Pinder.
\newblock Flow of a shear-thinning liquid with yield stress through porous
  media.
\newblock {\em SPE 13840}, 1984.

\bibitem{alfariss2}
T.F. Al-Fariss.
\newblock A new correlation for non-\textrm{N}ewtonian flow through porous
  media.
\newblock {\em Computers and Chemical Engineering}, 13(4/5):475--482, 1989.

\bibitem{wu1}
Y.S. Wu; K. Pruess;~P.A. Witherspoon.
\newblock Flow and displacement of \textrm{B}ingham non-\textrm{N}ewtonian
  fluids in porous media.
\newblock {\em SPE Reservoir Engineering, SPE 20051}, pages 369--376, 1992.

\bibitem{chaplain1}
V.~Chaplain; P. Mills; G. Guiffant;~P. Cerasi.
\newblock Model for the flow of a yield fluid through a porous medium.
\newblock {\em Journal de Physique II}, 2:2145--2158, 1992.

\bibitem{saffman1}
P.G. Saffman.
\newblock A theory of dispersion in a porous medium.
\newblock {\em Journal of Fluid Mechanics}, 6:321--349, 1959.

\bibitem{vradis1}
G.C. Vradis;~A.L. Protopapas.
\newblock Macroscopic conductivities for flow of \textrm{B}ingham plastics in
  porous media.
\newblock {\em Journal of Hydraulic Engineering}, 119(1):95--108, 1993.

\bibitem{chase1}
G.G. Chase;~P. Dachavijit.
\newblock Incompressible cake filtration of a yield stress fluid.
\newblock {\em Separation Science and Technology}, 38(4):745--766, 2003.

\bibitem{kuzhir1}
P.~Kuzhir; G. Bossis; V. Bashtovoi;~O. Volkova.
\newblock Flow of magnetorheological fluid through porous media.
\newblock {\em European Journal of Mechanics B/Fluids}, 22:331--343, 2003.

\bibitem{balhoff1}
M.T. Balhoff;~K.E. Thompson.
\newblock Modeling the steady flow of yield-stress fluids in packed beds.
\newblock {\em AIChE Journal}, 50(12):3034--3048, 2004.

\bibitem{sorbiebook}
K.S. Sorbie.
\newblock {\em Polymer-Improved Oil Recovery}.
\newblock Blakie and Son Ltd, 1991.

\bibitem{christopher1965}
R.~Christopher;~S. Middleman.
\newblock Power-law flow through a packed tube.
\newblock {\em Industrial \& Engineering Chemistry Fundamentals},
  4(4):422--426, 1965.

\bibitem{gaitonde1966}
N.~Gaitonde;~S. Middleman.
\newblock Flow of viscoelastic fluids through porous media.
\newblock {\em SPE Symposium on Mechanics of Rheologically Complex Fluids,
  15-16 December, Houston, Texas, SPE 1685}, 1966.

\bibitem{MarshallM1967}
R.J. Marshall;~A.B. Metzner.
\newblock Flow of viscoelastic fluids through porous media.
\newblock {\em Industrial \& Engineering Chemistry Fundamentals},
  6(3):393--400, 1967.

\bibitem{Wissler1971}
E.H. Wissler.
\newblock Viscoelastic effects in the flow of non-\textrm{N}ewtonian fluids
  through a porous medium.
\newblock {\em Industrial \& Engineering Chemistry Fundamentals},
  10(3):411--417, 1971.

\bibitem{TalwarK1992}
K.K. Talwar;~B. Khomami.
\newblock Application of higher order finite element methods to viscoelastic
  flow in porous media.
\newblock {\em Journal of Rheology}, 36(7):1377--1416, 1992.

\bibitem{GogartyLF1972}
W.B. Gogarty; G.L. Levy;~V.G. Fox.
\newblock Viscoelastic effects in polymer flow through porous media.
\newblock {\em SPE 47th Annual Fall Meeting, 8-11 October, San Antonio, Texas,
  SPE 4025}, 1972.

\bibitem{hirasaki1974}
G.J. Hirasaki;~G.A. Pope.
\newblock Analysis of factors influencing mobility and adsorption in the flow
  of polymer solution through porous media.
\newblock {\em SPE 47th Annual Fall Meeting, 8-11 October, San Antonio, Texas,
  SPE 4026}, 1972.

\bibitem{DeiberS1981}
J.A. Deiber;~W.R. Schowalter.
\newblock Modeling the flow of viscoelastic fluids through porous media.
\newblock {\em AIChE Journal}, 27(6):912--920, 1981.

\bibitem{DurstHI1987}
F.~Durst; R. Haas;~W. Interthal.
\newblock The nature of flows through porous media.
\newblock {\em Journal of Non-Newtonian Fluid Mechanics}, 22(2):169--189, 1987.

\bibitem{ChmielewskiPJ1990}
C.~Chmielewski; C.A. Petty;~K. Jayaraman.
\newblock Crossflow of elastic liquids through arrays of cylinders.
\newblock {\em Journal of Non-Newtonian Fluid Mechanics}, 35(2-3):309--325,
  1990.

\bibitem{ChmielewskiJ1992}
C.~Chmielewski;~K. Jayaraman.
\newblock The effect of polymer extensibility on crossflow of polymer solutions
  through cylinder arrays.
\newblock {\em Journal of Rheology}, 36(6):1105--1126, 1992.

\bibitem{ChmielewskiJ1993}
C.~Chmielewski;~K. Jayaraman.
\newblock Elastic instability in crossflow of polymer solutions through
  periodic arrays of cylinders.
\newblock {\em Journal of Non-Newtonian Fluid Mechanics}, 48(3):285--301, 1993.

\bibitem{PilitsisSB1991}
S.~Pilitsis; A. Souvaliotis;~A.N. Beris.
\newblock Viscoelastic flow in a periodically constricted tube: \textrm{T}he
  combined effect of inertia, shear thinning, and elasticity.
\newblock {\em Journal of Rheology}, 35(4):605--646, 1991.

\bibitem{souvaliotis1996}
A.~Souvaliotis;~A.N. Beris.
\newblock Spectral collocation/domain decomposition method for viscoelastic
  flow simulations in model porous geometries.
\newblock {\em Computer Methods in Applied Mechanics and Engineering},
  129(1):9--28, 1996.

\bibitem{HuaS1998}
C.C. Hua;~J.D. Schieber.
\newblock Viscoelastic flow through fibrous media using the \rm{CONNFFESSIT}
  approach.
\newblock {\em Journal of Rheology}, 42(3):477--491, 1998.

\bibitem{garrouch1999}
A.A. Garrouch.
\newblock A viscoelastic model for polymer flow in reservoir rocks.
\newblock {\em SPE Asia Pacific Oil and Gas Conference and Exhibition, 20-22
  April, Jakarta, Indonesia, SPE 54379}, 1999.

\bibitem{koshiba2000}
T.~Koshiba; N. Mori; K. Nakamura;~S. Sugiyama.
\newblock Measurement of pressure loss and observation of the flow field in
  viscoelastic flow through an undulating channel.
\newblock {\em Journal of Rheology}, 44(1):65--78, 2000.

\bibitem{KhuzhayorovAR2000}
B.~Khuzhayorov; J.-L. Auriault;~P. Royer.
\newblock Derivation of macroscopic filtration law for transient linear
  viscoelastic fluid flow in porous media.
\newblock {\em International Journal of Engineering Science}, 38(5):487--504,
  2000.

\bibitem{HuifenXDQX2001}
X.~Huifen; Y. XiangAn; W. Dexi; L. Qun;~Z. Xuebin.
\newblock Prediction of \textrm{IPR} curve of oil wells in visco-elastic
  polymer solution flooding reservoirs.
\newblock {\em SPE Asia Pacific Improved Oil Recovery Conference, 6-9 October,
  Kuala Lumpur, Malaysia, SPE 72122}, 2001.

\bibitem{MendesN2002}
P.R.S. Mendes;~M.F. Naccache.
\newblock A constitutive equation for extensional-thickening fluids flowing
  through porous media.
\newblock {\em Proceedings of 2002 ASME International Mechanical Engineering
  Congress \& Exposition, November 17-22, New Orleans, Louisiana, USA}, 2002.

\bibitem{DolejsCSD2002}
V.~Dolej$\check{\rm{s}}$; J. Cakl;~B. $\check{\rm{S}}$i$\check{\rm{s}}$ka;
  P.~Dole$\check{\rm{c}}$ek.
\newblock Creeping flow of viscoelastic fluid through fixed beds of particles.
\newblock {\em Chemical Engineering and Processing}, 41(2):173--178, 2002.

\bibitem{MasulehP2004}
S.H. Momeni-Masuleh;~T.N. Phillips.
\newblock Viscoelastic flow in an undulating tube using spectral methods.
\newblock {\em Computers \& fluids}, 33(8):1075--1095, 2004.

\bibitem{ChengE1965}
D.C. Cheng;~F. Evans.
\newblock Phenomenological characterization of the rheological behaviour of
  inelastic reversible thixotropic and antithixotropic fluids.
\newblock {\em British Journal of Applied Physics}, 16(11):1599--1617, 1965.

\bibitem{PearsonT2002}
J.R.A. Pearson;~P.M.J. Tardy.
\newblock Models for flow of non-\rm{N}ewtonian and complex fluids through
  porous media.
\newblock {\em Journal of Non-Newtonian Fluid Mechanics}, 102:447--473, 2002.

\bibitem{PritchardP2006}
D.~Pritchard;~J.R.A. Pearson.
\newblock Viscous fingering of a thixotropic fluid in a porous medium or a
  narrow fracture.
\newblock {\em Journal of Non-Newtonian Fluid Mechanics}, 135(2-3):117--127,
  2006.

\bibitem{WangHC2006}
S.~Wang; Y. Huang;~F. Civan.
\newblock Experimental and theoretical investigation of the \rm{Z}aoyuan field
  heavy oil flow through porous media.
\newblock {\em Journal of Petroleum Science and Engineering}, 50(2):83--101,
  2006.

\bibitem{Hulsen1990}
M.A. Hulsen.
\newblock {\em A numerical method for solving steady 2D and axisymmetrical
  viscoelastic flow problems with an application to inertia effects in
  contraction flows}.
\newblock Faculty of Mechanical Engineering, Delft University of Technology,
  1990.

\bibitem{Hulsen1986}
M.A. Hulsen.
\newblock {\em Analysis of the equations for viscoelastic flow: Type, boundary
  conditions, and discontinuities}.
\newblock Faculty of Mechanical Engineering, Delft University of Technology,
  1986.

\bibitem{Keunings2003}
R.~Keunings.
\newblock Finite element methods for integral viscoelastic fluids.
\newblock {\em Rheology Reviews}, pages 167--195, 2003.

\bibitem{blunt2}
M.J. Blunt; M.D. Jackson; M. Piri;~P.H. Valvatne.
\newblock Detailed physics, predictive capabilities and macroscopic
  consequences for pore-network models of multiphase flow.
\newblock {\em Advances in Water Resources}, 25:1069--1089, 2002.

\bibitem{blunt1}
M.J. Blunt.
\newblock Flow in porous media - pore-network models and multiphase flow.
\newblock {\em Colloid and Interface Science}, 6(3):197--207, 2001.

\bibitem{valvatnethesis}
P.H. Valvatne.
\newblock {\em Predictive pore-scale modelling of multiphase flow}.
\newblock PhD thesis, Imperial College London, 2004.

\bibitem{valvatne1}
P.H. Valvatne;~M.J. Blunt.
\newblock Predictive pore-scale modeling of two-phase flow in mixed wet media.
\newblock {\em Water Resources Research}, 40(W07406), 2004.

\bibitem{lopezthesis}
X.~Lopez.
\newblock {\em Pore-scale modelling of non-Newtonian flow}.
\newblock PhD thesis, Imperial College London, 2004.

\bibitem{lopez1}
X.~Lopez; P.H. Valvatne;~M.J. Blunt.
\newblock Predictive network modeling of single-phase non-\textrm{N}ewtonian
  flow in porous media.
\newblock {\em Journal of Colloid and Interface Science}, 264:256--265, 2003.

\bibitem{LopezB2004}
X.~Lopez;~M.J. Blunt.
\newblock Predicting the impact of non-\textrm{N}ewtonian rheology on relative
  permeability using pore-scale modeling.
\newblock {\em SPE Annual Technical Conference and Exhibition, 26—29 September,
  Houston, Texas, SPE 89981}, 2004.

\bibitem{OrenBA1997}
P.E. {\O}ren; S. Bakke;~O.J. Amtzen.
\newblock Extending predictive capabilities to network models.
\newblock {\em SPE Annual Technical Conference and Exhibition, San Antonio,
  Texas}, (SPE 38880), 1997.

\bibitem{OrenB2003}
P.E. {\O}ren;~S. Bakke.
\newblock Reconstruction of berea sandstone and pore-scale modelling of
  wettability effects.
\newblock {\em Journal of Petroleum Science and Engineering}, 39:177--199,
  2003.

\bibitem{rugebook}
J.W. Ruge;~K. St\"{u}ben.
\newblock {\em Multigrid Methods: Chapter 4 (Algebraic Multigrid), Frontiers in
  Applied Mathematics}.
\newblock SIAM, 1987.

\bibitem{Barnes1995}
H.A. Barnes.
\newblock A review of the slip (wall depletion) of polymer solutions, emulsions
  and particle suspensions in viscometers: its cause, character, and cure.
\newblock {\em Journal of Non-Newtonian Fluid Mechanics}, 56(3):221--251, 1995.

\bibitem{FattalK2005}
R.~Fattal;~R. Kupferman.
\newblock Time-dependent simulation of viscoelastic flows at high
  \rm{W}eissenberg number using the log-conformation representation.
\newblock {\em Journal of Non-Newtonian Fluid Mechanics}, 126(1):23--37, 2005.

\bibitem{MoresiDM2003}
L.N. Moresi; F. Dufour;~H. M\"{u}hlhaus.
\newblock A lagrangian integration point finite element method for large
  deformation modeling of viscoelastic geomaterials.
\newblock {\em Journal Of Computational Physics}, 184(2):476--497, 2003.

\bibitem{denysthesis}
K.F.J. Denys.
\newblock {\em Flow of polymer solutions through porous media}.
\newblock PhD thesis, Delft University of Technology, 2003.

\bibitem{PilitsisB1989}
S.~Pilitsis;~A.N. Beris.
\newblock Calculations of steady-state viscoelastic flow in an undulating tube.
\newblock {\em Journal of Non-Newtonian Fluid Mechanics}, 31(3):231--287, 1989.

\bibitem{ZaitounK1986}
A.~Zaitoun;~N. Kohler.
\newblock Two-phase flow through porous media: Effect of an adsorbed polymer
  layer.
\newblock {\em SPE Annual Technical Conference and Exhibition, 2-5 October,
  Houston, Texas, SPE 18085}, 1988.

\bibitem{Barnes1999}
H.A. Barnes.
\newblock The yield stress---a review or `$\pi\alpha\nu\tau\alpha \
  \rho\varepsilon\iota$'---everything flows?
\newblock {\em Journal of Non-Newtonian Fluid Mechanics}, 81(1):133--178, 1999.

\bibitem{bearbook}
J.~Bear.
\newblock {\em Dynamics of Fluids in Porous Media}.
\newblock American Elsevier, 1972.

\bibitem{chen3}
M.~Chen; W.R. Rossen;~Y.C. Yortsos.
\newblock The flow and displacement in porous media of fluids with yield
  stress.
\newblock {\em Chemical Engineering Science}, 60:4183--4202, 2005.

\bibitem{kharabaf1}
H.~Kharabaf;~Y.C. Yortsos.
\newblock Invasion percolation with memory.
\newblock {\em Physical Review}, 55(6):7177--7191, 1996.

\bibitem{roux1}
S.~Roux;~A. Hansen.
\newblock A new algorithm to extract the backbone in a random resistor network.
\newblock {\em Journal of Physics A}, 20:L1281--Ll285, 1987.

\bibitem{rossen1}
W.R. Rossen;~P.A. Gauglitz.
\newblock Percolation theory of creation and mobilization of foam in porous
  media.
\newblock {\em A.I.Ch.E. Journal}, 36:1176--1188, 1990.

\bibitem{rossen2}
W.R. Rossen;~C.K. Mamun.
\newblock Minimal path for transport in networks.
\newblock {\em Physical Review B}, 47:11815 – 11825, 1993.

\bibitem{selyakovbook}
V.I. Selyakov;~V.V. Kadet.
\newblock {\em Percolation Models for Transport in Porous Media with
  Applications to Reservoir Engineering}.
\newblock Kluwer Academic Publishers, 1996.

\bibitem{wapperomthesis}
P.~Wapperom.
\newblock {\em Nonisothermal flows of viscoelastic fluids Thermodynamics,
  analysis and numerical simulation}.
\newblock PhD thesis, Delft University of Technology, 1996.

\bibitem{ChhabraCM2001}
R.P. Chhabra; J. Comiti;~I. Macha\v{c}.
\newblock Flow of non-\rm{N}ewtonian fluids in fixed and fluidised beds.
\newblock {\em Chemical Engineering Science}, 56(1):1--27, 2001.

\bibitem{DaubenM1967}
D.L. Dauben;~D.E. Menzie.
\newblock Flow of polymer solutions through porous media.
\newblock {\em Journal of Petroleum Technology, SPE 1688-PA}, pages 1065--1073,
  1967.

\bibitem{PlogK}
J.P. Plog;~W.M. Kulicke.
\newblock Flow behaviour of dilute hydrocolloid solutions through porous media.
  \rm{I}nfluence of molecular weight, concentration and solvent.
\newblock {\em Institute of Technical and Macromolecular Chemistry, University
  of Hamburg}.

\bibitem{ThienK1987}
N.~Phan-Thien;~M.M.K. Khan.
\newblock Flow of an \rm{O}ldroyd-type fluid through a sinusoidally corrugated
  tube.
\newblock {\em Journal of Non-Newtonian Fluid Mechanics}, 24:203--220, 1987.

\bibitem{ChengH1984}
D.C.H. Cheng;~N.I. Heywood.
\newblock Flow in pipes. \rm{I}. \rm{F}low of homogeneous fluids.
\newblock {\em Physics in Technology}, 15(5):244--251, 1984.

\bibitem{lodgebook1964}
A.S. Lodge.
\newblock {\em Elastic Liquids An Introductory Vector Treatment of
  Finite-strain Polymer Rheology}.
\newblock Academic Press, 1964.

\bibitem{CrochetW1983}
M.J. Crochet;~K. Walters.
\newblock Numerical methods in non-\rm{N}ewtonian fluid mechanics.
\newblock {\em Annual Review of Fluid Mechanics}, 15(241-260), 1983.

\bibitem{SouvaliotisB1992}
A.~Souvaliotis;~A.N. Beris.
\newblock Applications of domain decomposition spectral collocation methods in
  viscoelastic flows through model porous media.
\newblock {\em Journal of Rheology}, 36(7):1417--1453, 1992.

\bibitem{PodolsakTF1997}
A.K. Podolsak; C. Tiu;~T.N. Fang.
\newblock Flow of non-\rm{N}ewtonian fluids through tubes with abrupt
  expansions and contractions (square wave tubes).
\newblock {\em Journal of Non-Newtonian Fluid Mechanics}, 71(1):25--39, 1997.

\bibitem{Dullien1975}
F.A.L. Dullien.
\newblock Single phase flow through porous media and pore structure.
\newblock {\em The Chemical Engineering Journal}, 10(1):1--34, 1975.

\bibitem{BautistaSPM1999}
F.~Bautista;~J.M. de~Santos; J.E. Puig; O.~Manero.
\newblock Understanding thixotropic and antithixotropic behavior of
  viscoelastic micellar solutions and liquid crystalline dispersions. \rm{I}.
  \rm{T}he model.
\newblock {\em Journal of Non-Newtonian Fluid Mechanics}, 80(2):93--113, 1999.

\bibitem{LetelierSC2002}
M.F. Letelier; D.A. Siginer;~C. Caceres.
\newblock Pulsating flow of viscoelastic fluids in straight tubes of arbitrary
  cross-section-\rm{P}art \rm{I}: longitudinal field.
\newblock {\em International Journal of Non-Linear Mechanics}, 37(2):369--393,
  2002.

\bibitem{BautistaSLPM2000}
F.~Bautista; J.F.A. Soltero; J.H. P\'{e}rez-L\'{o}pez; J.E. Puig;~O. Manero.
\newblock On the shear banding flow of elongated micellar solutions.
\newblock {\em Journal of Non-Newtonian Fluid Mechanics}, 94(1):57--66, 2000.

\bibitem{ManeroBSP2002}
O.~Manero; F. Bautista; J.F.A. Soltero;~J.E. Puig.
\newblock Dynamics of worm-like micelles: the \rm{C}ox-\rm{M}erz rule.
\newblock {\em Journal of Non-Newtonian Fluid Mechanics}, 106(1):1--15, 2002.

\bibitem{NumericalRecipes}
W.H. Press; S.A. Teukolsky; W.T. Vetterling;~B.P. Flannery.
\newblock {\em Numerical Recipes in C++ The Art of Scientific Computing}.
\newblock Cambridge University Press, 2nd edition, 2002.

\bibitem{AndersonPS2006}
V.J. Anderson; J.R.A. Pearson;~J.D. Sherwood.
\newblock Oscillation superimposed on steady shearing: \rm{M}easurements and
  predictions for wormlike micellar solutions.
\newblock {\em Journal of Rheology}, 50(5):771--796, 2006.

\bibitem{EscudierP1996}
M.P. Escudier;~F. Presti.
\newblock Pipe flow of a thixotropic liquid.
\newblock {\em Journal of Non-Newtonian Fluid Mechanics}, 62(2):291--306, 1996.

\bibitem{DullaertM2005}
K.~Dullaert;~J. Mewis.
\newblock Thixotropy: Build-up and breakdown curves during flow.
\newblock {\em Journal of Rheology}, 49(6):1213--1230, 2005.

\bibitem{Mewis1979}
J.~Mewis.
\newblock Thixotropy - a general review.
\newblock {\em Journal of Non-Newtonian Fluid Mechanics}, 6(1):1--20, 1979.

\bibitem{JoyeP1971}
D.D. Joye;~G.W. Poehlein.
\newblock Characteristics of thixotropic behavior.
\newblock {\em Journal of Rheology}, 15(1):51--61, 1971.

\bibitem{BillinghamF1993}
J.~Billingham;~J.W.J. Ferguson.
\newblock Laminar, unidirectional flow of a thixotropic fluid in a circular
  pipe.
\newblock {\em Journal of Non-Newtonian Fluid Mechanics}, 47(21-55), 1993.

\bibitem{whitebook}
F.M. White.
\newblock {\em Viscous Fluid Flow}.
\newblock McGraw Hill Inc., second edition, 1991.

\bibitem{boger1977}
D.V. Boger.
\newblock A highly elastic constant-viscosity fluid.
\newblock {\em Journal of Non-Newtonian Fluid Mechanics}, 3:87--91, 1977/1978.

\end{thebibliography}

%%%%%%%%%%%%%%%%%%%%%%%%%%%%%%%%%%%%%%%%%%%%%%%%%%%%%%%%%%%%%%%%%%%%%%%%%%%%%%%%%%%%%%%%%%%%%%%%%%%%%%%%%%%%

%%%%%%%%%%%%%%%%%%%%%%%%%%%%%%%%%%%%%%%%%%%%%%%%%%%%%%%%%%%%%%%%%%%%%%%%%%%%%%%%%%%%%%%%%%%%%%%%%%%%%%%%%%%%

%\phantomsection \addcontentsline{toc}{chapter}{\protect \numberline{} Appendices:}

\appendix

\def\baselinestretch{1}
\chapter[Flow Rate of \ELLIS\ Fluid in Cylindrical Tube]{} \label{AppEllisQ}

\begin{spacing}{2}
{\LARGE \bf The Flow Rate of an \ELLIS\ Fluid in a Cylindrical Tube}
\end{spacing}

\vspace{1.0cm}

\begin{figure} [!h]
  \centering{}
  \includegraphics
  [scale=0.5]
  {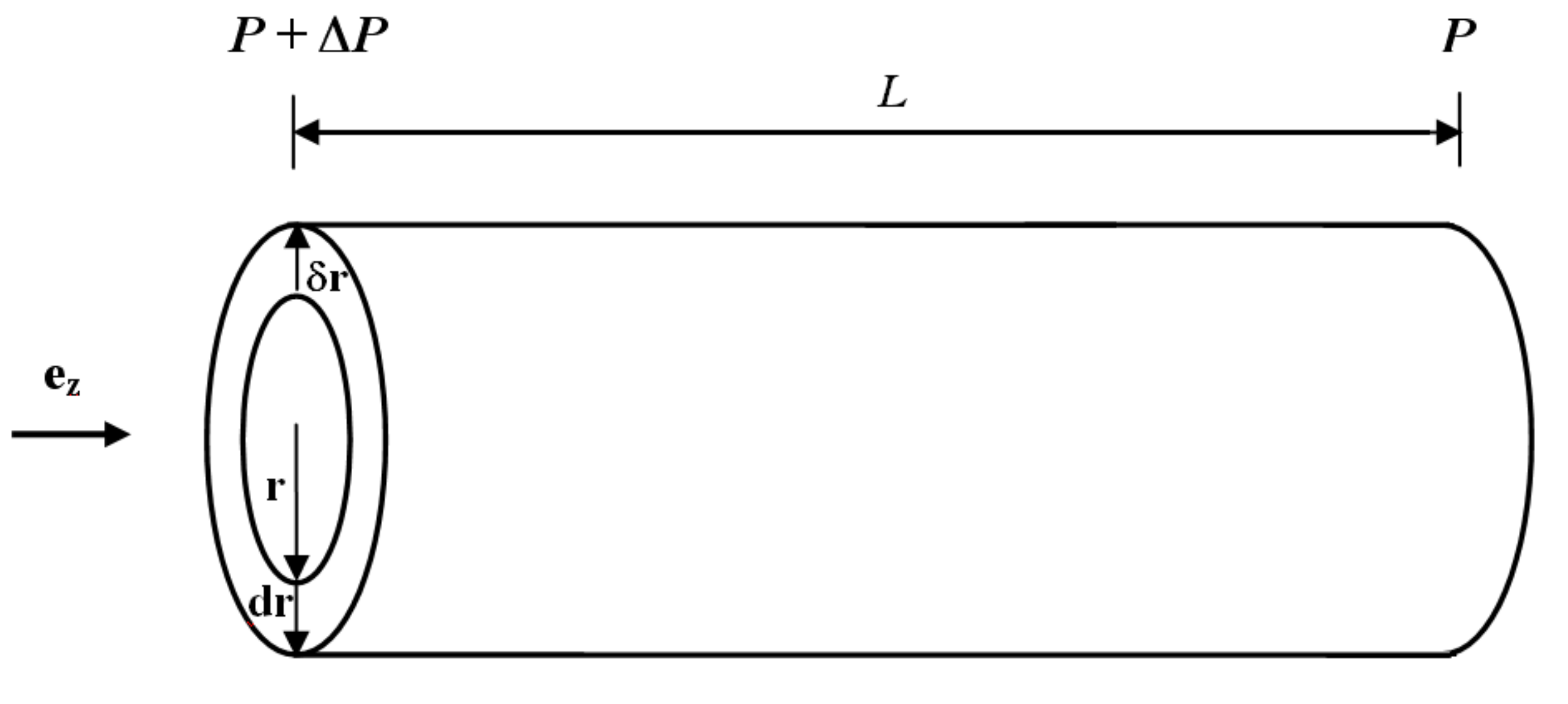}
  \caption[Schematic diagram of a cylindrical annulus used to derive analytical expressions for the
  volumetric flow rate of \ELLIS\ and \HB\ fluids]
  {Schematic diagram of a cylindrical annulus used to derive analytical expressions for the
  volumetric flow rate of \ELLIS\ and \HB\ fluids.}
  \label{Annulus}
\end{figure}

\def\baselinestretch{1.66}
\noindent Here, we derive an analytical expression for the
volumetric flow rate in a cylindrical duct, with inner radius $R$
and length $L$, assuming \ELLIS\ flow. We apply the well-known
general result, sometimes called the \Weissenberg-\Rabinowitsch\
equation or \Rabinowitsch-\Mooney\ equation \cite{birdbook,
LiuM1998} which relates the flow rate $Q$ and the shear stress at
the tube wall $\wsS$ for laminar flow of \timeind\ fluids in a
cylindrical tube. This equation, which can be derived considering
the volumetric flow rate through the differential annulus between
$r$ and $r + dr$ as shown in Figure (\ref{Annulus}), is given by
\cite{skellandbook, birdbook, carreaubook}
\begin{equation}\label{A.1}
    \frac{Q}{\pi R^{3}} = \frac{1}{\sS^{3}_{w}} \int_{0}^{\wsS} \sS^{2}\sR \ \mathrm{d} \sS
\end{equation}

\vspace{0.1cm}

The derivation is based on the definition of the viscosity as the
stress over strain rate. For \ELLIS\ fluid, the viscosity is given
by \cite{SadowskiB1965, Savins1969, birdbook, carreaubook}
\begin{equation}
    \Vis = \frac{\lVis}{1+ \left(\frac{\sS}{\hsS} \right)^{\eAlpha - 1}}
\end{equation}
where $\lVis$ is the low-shear viscosity, $\sS$ is the shear stress,
$\hsS$ is the shear stress at which $\Vis=\lVis/2$ and $\eAlpha$ is
an indicial parameter in the \ELLIS\ model.

\vspace{0.2cm}

From this expression we obtain a formula for the shear rate
\begin{equation}
    \sR
    = \frac{\sS}{\Vis}
    = \frac{\sS}{\lVis} \left[ 1 + \left( \frac{\sS}{\hsS}\right)^{\eAlpha-1}\right]
\end{equation}

\vspace{0.1cm}

On substituting this into Equation (\ref{A.1}), integrating and
simplifying we obtain
\begin{eqnarray}
    Q
    & = & \frac{\pi R^{3} \wsS}{4 \lVis}
    \left[ 1 + \frac{4}{\eAlpha + 3} \left( \frac{\wsS}{\hsS}\right)^{\eAlpha-1}
    \right] \nonumber \\
    & = & \frac{\pi R^{4} \Delta P}{8 L \lVis}
    \left[ 1 + \frac{4}{\eAlpha + 3} \left( \frac{R \Delta P}{2 L \hsS}\right)^{\eAlpha-1} \right]
\end{eqnarray}
where $\wsS = \Delta P R / 2 L$ and $\Delta P$ is the pressure drop
across the tube.

\vspace{2.0cm}

\def\baselinestretch{1}
\chapter[Flow Rate of \HB\ Fluid in Cylindrical Tube]{} \label{AppHerQ}
\def\baselinestretch{1.66}

\begin{spacing}{2}
{\LARGE \bf The Flow Rate of a \HB\ Fluid in a Cylindrical Tube}
\end{spacing}

\vspace{1.0cm}

\def\baselinestretch{1.66}

\noindent Here, we present two methods for deriving an analytical
expression for the volumetric flow rate in a cylindrical duct, with
inner radius $R$ and length $L$, assuming \HB\ flow.

%XXXXXXXXXXXXXXXXXXXXXXXXXXXXXXXXXXXXXXXXXXXXXXXXXXXXXXXXXXXXXXXXXXXXXX
\section{First Method} \label{}
In this method, Equation (\ref{A.1}) is used \cite{skellandbook} as
in the case of \ELLIS\ fluid. For a \HB\ fluid

\begin{equation}\label{A.2}
    \sR = \left(\frac{\sS}{C} - \frac{\ysS}{C} \right)^{\frac{1}{n}}
\end{equation}

\vspace{0.2cm}

On substituting (\ref{A.2}) into (\ref{A.1}), integrating and
simplifying we obtain
\begin{equation}\label{A.3}
    Q = \frac{8\pi}{C^\frac{1}{n}}\left(\frac{L}{\Delta P}\right)^{3}
    \left(\wsS - \ysS\right)^{1+\frac{1}{n}}\left[\frac{\left(\wsS - \ysS\right)^{2}}{3+1/n}+
    \frac{2\ysS \left(\wsS - \ysS\right)}{2+1/n} + \frac{\ysS^{2}}{1+1/n}\right]
\end{equation}

%XXXXXXXXXXXXXXXXXXXXXXXXXXXXXXXXXXXXXXXXXXXXXXXXXXXXXXXXXXXXXXXXXXXXXX
\section{Second Method} \label{}
In this method we use cylindrical polar coordinate system where the
tube axis coincides with the coordinate $z$-axis and the flow is in
the positive $z$-direction. \Newton's second law [$m${\bf a} =
$\sum_{i}${\bf F}$_{i}$] is applied to a cylindrical annulus of
fluid with inner radius $r$, outer radius $r$+d$r$ and length $L$,
as illustrated in Figure (\ref{Annulus}). Assuming negligible body
forces, we have two surface forces; one is due to the pressure
gradient and the other is due to the shear stress, that is
\begin{equation}\label{A.4}
   (2\pi r \delta rL \rho \mathrm{a}){\bf{e}}_{z} \cong (2\pi r \delta r\Delta P){\bf{e}}_{z} -
   [2\pi(r+\delta r)(\sS+\delta\sS)L-2\pi r\sS L]{\bf{e}}_{z}
\end{equation}
Dividing by $2\pi \delta rL$, simplifying and taking the scalar
form, we obtain
\begin{equation}\label{A.5}
 \rho \mathrm{a} \cong \frac{\Delta P}{L}-(\frac{\sS}{r}+\frac{\delta \sS}{\delta r}+\frac{\delta
 \sS}{r})
\end{equation}
Taking the limit when $\delta r\rightarrow 0$, and hence $\delta \sS
\rightarrow 0$, we obtain the exact relation
\begin{equation}\label{A.6}
 \rho \mathrm{a} = \frac{\Delta P}{L}-(\frac{\sS}{r}+\frac{\mathrm{d} \sS}{\mathrm{d} r})
\end{equation}

\vspace{0.2cm}

For steady flow, a [$=\frac{\mathrm{d}u_{z}}{\mathrm{d}t}$] is zero,
thus
\begin{equation}\label{A.7}
 \frac{\mathrm{d} \sS}{\mathrm{d} r} + \frac{\sS}{r}= \frac{\Delta P}{L}
\end{equation}
On integrating this relation we obtain
\begin{equation}\label{A.8}
 \sS = \frac{\Delta Pr}{2L}+\frac{A}{r}
\end{equation}
where $A$ is the constant of integration.

\vspace{0.2cm}

At $r=0$, $\sS$ is finite, so $A=0$. From this we obtain
\begin{equation}\label{A.9}
 \sS = \frac{\Delta Pr}{2L}
\end{equation}
From the definition of the shear rate, $\sR =
\frac{\mathrm{d}u_{z}}{\mathrm{d}r}$, we obtain
\begin{equation}\label{A.10}
 u_{z} = \int \sR \mathrm{d}r
\end{equation}
For \HB\ fluid
\begin{equation}\label{A.11}
    \sR = \left(\frac{\sS}{C} - \frac{\ysS}{C} \right)^{\frac{1}{n}}
\end{equation}
Substituting for $\sS$ from (\ref{A.9}) into (\ref{A.11}) we obtain
\begin{equation}\label{A.12}
    \sR = \left(\frac{\Delta P}{2LC}r - \frac{\ysS}{C} \right)^{\frac{1}{n}}
\end{equation}
Inserting (\ref{A.12}) into (\ref{A.10}), integrating and applying the second boundary
condition, i.e $u_{z}(r$=$R)=0$, we obtain \\
\begin{equation}\label{A.13}
    u_{z}(r) = \frac{n}{1+n}\left( \frac{2LC}{\Delta P} \right)
    \left[ \left( \frac{\Delta P}{2LC}R-\frac{\ysS}{C}\right)^{1+\frac{1}{n}}
    - \left( \frac{\Delta P}{2LC}r-\frac{\ysS}{C}\right)^{1+\frac{1}{n}} \right]
\end{equation}

\vspace{0.2cm}

In fact, Equation (\ref{A.10}) is valid only for $\sR > 0$, and this
condition requires that at any point with finite shear rate, $\sS >
\ysS$, as can be seen from (\ref{A.11}). Equation (\ref{A.9}) then
implies that the region with $r \leq r_{o} = \frac{2L\ysS}{\Delta
P}$ is a shear-free zone. From the continuity of the velocity field,
we conclude that
\begin{equation}\label{A.14}
    u_{z}(0\leq r \leq r_{o}) = \frac{n}{1+n}\left( \frac{2LC}{\Delta P} \right)
    \left( \frac{\Delta P}{2LC}R-\frac{\ysS}{C}\right)^{1+\frac{1}{n}}
\end{equation}
The volumetric flow rate is given by
\begin{equation}\label{A.15}
    Q = \int_{0}^{R}2\pi ru_{z} \mathrm{d}r
\end{equation}
that is
\begin{equation}\label{A.16}
    Q = \int_{0}^{r_{o}}2\pi ru_{z}(0\leq r \leq r_{o}) \mathrm{d}r
    + \int_{r_{o}}^{R}2\pi ru_{z}(r > r_{o}) \mathrm{d}r
\end{equation}
On inserting (\ref{A.14}) into the first term of (\ref{A.16}) and
(\ref{A.13}) into the second term of (\ref{A.16}), integrating and
simplifying we obtain
\begin{equation}\label{A.17}
    Q = \frac{8\pi}{C^\frac{1}{n}}\left(\frac{L}{\Delta P}\right)^{3}
    \frac{\left(\wsS -\ysS\right)^{1+\frac{1}{n}}}{(1+1/n)}\left[
    \frac{2(\wsS-\ysS)^{2}}{(3+1/n)(2+1/n)}
    +\frac{2\wsS(\ysS-\wsS)}{(2+1/n)}
    +\wsS^{2}\right]
    \hspace{-0.1cm}
\end{equation}

\vspace{0.2cm}

Although (\ref{A.3}) and (\ref{A.17}) look different, they are
mathematically identical, as each one can be obtained from the other
by algebraic manipulation.

\vspace{0.5cm}

There are three important special cases for \HB\ fluid. These, with
the volumetric flow rate expressions, are
\begin{enumerate}

    \item \NEW \\
    \begin{equation}\label{A.18}
      Q = \frac{\pi R^{4}\Delta P}{8LC}
    \end{equation}

    \item \Powlaw\ \\
    \begin{equation}\label{A.19}
      Q = \frac{\pi R^{4}\Delta P^{1/n}}{8LC^{1/n}} \left( \frac{4n}{3n+1} \right)
      \left( \frac{2L}{R} \right)^{1-1/n}
    \end{equation}

    \item \BING\ plastic \\
    \begin{equation}\label{A.20}
      Q = \frac{\pi R^{4}\Delta P}{8LC} \left[ \frac{1}{3}\left( \frac{\ysS}{\wsS}\right)^{4}
      - \frac{4}{3} \left( \frac{\ysS}{\wsS} \right) + 1 \right]
    \end{equation}

\end{enumerate}

Each one of these expressions can be derived directly by these two
methods. Moreover, they can be obtained by substituting the relevant
conditions in the flow rate expression for \HB\ fluid. This serves
as a consistency check.

\def\baselinestretch{1}
\chapter[Yield Condition for Single Tube]{} \label{AppTubeY}
\def\baselinestretch{1.66}

\begin{spacing}{2}
{\LARGE \bf Yield Condition for a Single Tube}
\end{spacing}

\vspace{1.0cm}

\def\baselinestretch{1.66}
\noindent The volumetric flow rate of a \HB\ fluid in a cylindrical
tube is given by Equation (\ref{A.3}). For a fluid with \yields, the
flow occurs {\em iff} $Q>0$. Assuming $\ysS$, $R$, $L$, $C$, $\Delta
P$, $n$ $>0$, it is straightforward to show that the condition $Q>0$
is satisfied {\em iff}\hspace{0.2cm} $\left(\wsS - \ysS\right) > 0$,
that is
\begin{equation}
    \wsS = \frac{\Delta PR}{2L} > \ysS
\end{equation}
Hence, the threshold pressure gradient, above which the flow starts, is
\begin{equation}
    |\nabla P_{th}| = {\frac{2\ysS}{R}}
\end{equation}

\vspace{0.1cm}

Alternatively, the flow occurs when the shear stress at wall exceeds
the \yields, i.e.
\begin{equation}
    \wsS > \ysS
\end{equation}
which leads to the same condition. This second argument is less obvious but more general than
the first.

\def\baselinestretch{1}
\chapter[Radius of Bundle of Tubes to Compare with Network]{} \label{AppTubeRad}

\begin{spacing}{2}
{\LARGE \bf The Radius of the Bundle of Tubes to Compare with the
Networks}
\end{spacing}

\vspace{1.0cm}

\def\baselinestretch{1.66}
\noindent In Chapter (\ref{NetTube}) a comparison was made between
two networks representing \sandp\ and \Berea\ porous media and a
bundle of capillaries of uniform radius model. The networks and the
bundle are assumed to have the same porosity and Darcy velocity
under \NEW\ flow condition.

\vspace{0.2cm}

Here, we derive an expression for the radius of the tubes in the
bundle. We equate \POIS\ for a single tube in the bundle divided by
$d^{2}$, where $d$ is the side of the square enclosing the tube
cross section in the bundle's matrix, to the Darcy velocity for the
network, that is
\begin{equation}\label{D.1}
    q = \frac{\pi R^{4}\Delta P}{8\Vis Ld^{2}} = \frac{K \Delta P}{\Vis L}
\end{equation}

Thus
\begin{equation}\label{D.2}
    \frac{\pi R^{4}}{8d^{2}} = K
\end{equation}

\vspace{0.2cm}

The porosity of a bundle of $N$ tubes is
\begin{equation}\label{D.3}
    \frac{NL\pi R^{2}}{NLd^{2}} = \frac{\pi R^{2}}{d^{2}}
\end{equation}

On equating this to the network porosity, $\phi$, substituting into
(\ref{D.2}) and rearranging we obtain
\begin{equation}\label{D.4}
    R = \sqrt{\frac{8K}{\phi}}
\end{equation}

\def\baselinestretch{1}
\chapter[Terminology for Flow and \Vy]{} \label{AppVE}
{ %start \renewcommand
\renewcommand{\thefootnote}{\fnsymbol{footnote}}

\begin{spacing}{2}
{\LARGE \bf Terminology of Flow and \Vy\footnote{In preparing this
Appendix, we consulted most of our references on \vy. The main ones
are \cite{skellandbook, OsP2004, birdbook, tannerbook2000,
whitebook, larsonbook1988, barnesbookHW1993, larsonbook1999,
wapperomthesis, boger1977}.} }
\end{spacing}

\vspace{1.0cm}

\def\baselinestretch{1.66}
\noindent A tensor is an array of numbers which transform according
to certain rules under coordinate change. In a three-dimensional
space, a tensor of order $n$ has $3^{n}$ components which may be
specified with reference to a given coordinate system. Accordingly,
a scalar is a zero-order tensor and a vector is a first-order
tensor.

\vspace{0.2cm}

A stress is defined as a force per unit area. Because both force and area are vectors,
considering the orientation of the normal to the area, stress has two directions associated
with it instead of one as with vectors. This indicates that stress should be represented by a
second-order tensor, given in Cartesian coordinate system by
\begin{equation}\label{strTens}
    \sTen = \left(
               \begin{array}{ccc}
                 \sTenC_{xx} & \sTenC_{xy} & \sTenC_{xz} \\
                 \sTenC_{yx} & \sTenC_{yy} & \sTenC_{yz} \\
                 \sTenC_{zx} & \sTenC_{zy} & \sTenC_{zz} \\
               \end{array}
             \right)
\end{equation}
where $\sTenC_{ij}$ is the stress in the $j$-direction on a surface
normal to the $i$-axis. A shear stress is a force that a flowing
liquid exerts on a surface, per unit area of that surface, in the
direction parallel to the flow. A normal stress is a force per unit
area acting normal or perpendicular to a surface. The components
with identical subscripts represent normal stresses while the others
represent shear stresses. Thus $\sTenC_{xx}$ is a normal stress
acting in the $x$-direction on a face normal to $x$-direction, while
$\sTenC_{yx}$ is a shear stress acting in the $x$-direction on a
surface normal to the $y$-direction, positive when material at
greater $y$ exerts a shear in the positive $x$-direction on material
at lesser $y$. Normal stresses are conventionally positive when
tensile. The stress tensor is symmetric that is $\sTenC_{ij} =
\sTenC_{ji}$ where $i$ and $j$ represent $x$, $y$ or $z$. This
symmetry is required by angular momentum considerations to satisfy
equilibrium of moments about the three axes of any fluid element.
This means that the state of stress at a point is determined by six,
rather than nine, independent stress components.

\vspace{0.2cm}

\Vc\ fluids show normal stress differences in steady shear flows.
The first normal stress difference $\fNSD$ is defined as
\begin{equation}\label{fNSD}
    \fNSD = \sTenC_{11} - \sTenC_{22}
\end{equation}
where $\sS_{11}$ is the normal stress component acting in the
direction of flow and $\sS_{22}$ is the normal stress in the
gradient direction. The second normal stress difference $\sNSD$ is
defined as
\begin{equation}\label{sNSD}
    \sNSD = \sTenC_{22} - \sTenC_{33}
\end{equation}
where $\sS_{33}$ is the normal stress component in the indifferent
direction. The magnitude of $\sNSD$ is in general much smaller than
$\fNSD$. For some \vc\ fluids, $\sNSD$ may be virtually zero. Often
not the first normal stress difference $\fNSD$ is given, but a
related quantity: the first normal stress coefficient. This
coefficient is defined by $\Psi_{1} = \frac{\fNSD}{\sR^{2}}$ and
decreases with increasing shear rate. Different conventions about
the sign of the normal stress differences do exist. However, in
general $\fNSD$ and $\sNSD$ show opposite signs. \Vc\ solutions with
almost the same viscosities may have very different values of normal
stress differences. The presence of normal stress differences is a
strong indication of \vy, though some associate these properties
with non-\vc\ fluids.

\vspace{0.2cm}

The total stress tensor, also called Cauchy stress tensor, is
usually divided into two parts: hydrostatic stress tensor and extra
stress tensor. The former represents the hydrostatic pressure while
the latter represents the shear and extensional stresses caused by
the flow. In equilibrium the pressure reduces to the hydrostatic
pressure and the extra stress tensor $\sTen$ vanishes. The extra
stress tensor is determined by the deformation history and has to be
specified by the constitutive equation of the particular fluid.

\vspace{0.2cm}

The rate-of-strain or rate-of-deformation tensor is a symmetric
second-order tensor which gives the components, both extensional and
shear, of the strain rate. In Cartesian coordinate system it is
given by:
\begin{equation}\label{rateStrTens}
    \rsTen = \left(
               \begin{array}{ccc}
                 \rsTenC_{xx} & \rsTenC_{xy} & \rsTenC_{xz} \\
                 \rsTenC_{yx} & \rsTenC_{yy} & \rsTenC_{yz} \\
                 \rsTenC_{zx} & \rsTenC_{zy} & \rsTenC_{zz} \\
               \end{array}
             \right)
\end{equation}
where $\rsTenC_{xx}$, $\rsTenC_{yy}$ and $\rsTenC_{zz}$ are the
extensional components while the others are the shear components.
These components are given by:
\begin{eqnarray} \label{rsTenComp}
  \hspace{-2.0cm}
  \rsTenC_{xx} = 2 \frac{\partial \vC_{x}}{\partial x}  \verb|       |
  \rsTenC_{yy} = 2 \frac{\partial \vC_{y}}{\partial y}  \verb|       |
  \rsTenC_{zz} = 2 \frac{\partial \vC_{z}}{\partial z}  \verb|       |
  \hspace{-4.0cm}
  \nonumber \\
  \rsTenC_{xy} = \rsTenC_{yx} =
  \frac{\partial \vC_{x}}{\partial y} + \frac{\partial \vC_{y}}{\partial x}
  \nonumber \\
  \rsTenC_{yz} = \rsTenC_{zy} =
  \frac{\partial \vC_{y}}{\partial z} + \frac{\partial \vC_{z}}{\partial y}
  \nonumber \\
  \rsTenC_{xz} = \rsTenC_{zx} =
  \frac{\partial \vC_{x}}{\partial z} + \frac{\partial \vC_{z}}{\partial x}
\end{eqnarray}
where $\vC_{x}$, $\vC_{y}$ and $\vC_{z}$ are the velocity components in the respective
directions $x$, $y$ and $z$.

\vspace{0.2cm}

The stress tensor is related to the rate-of-strain tensor by the
constitutive or rheological equation of the fluid which takes a
differential or integral form. The rate-of-strain tensor $\rsTen$ is
related to the fluid velocity vector $\fVel$, which describes the
steepness of velocity variation as one moves from point to point in
any direction in the flow at a given instant in time, by the
relation
\begin{equation}\label{rsTen-fVel}
    \rsTen = \nabla \fVel + (\nabla \fVel)^{\textrm{T}}
\end{equation}
where $(.)^{\textrm{T}}$ is the tensor transpose and $\nabla \fVel$
is the fluid velocity gradient tensor defined by
\begin{equation}\label{fVelGradTen}
    \nabla \fVel =
    \left(
    \begin{array}{ccc}
      \frac{\partial \vC_{x}}{\partial x} &
      \frac{\partial \vC_{x}}{\partial y} &
      \frac{\partial \vC_{x}}{\partial z}
      \\
      \frac{\partial \vC_{y}}{\partial x} &
      \frac{\partial \vC_{y}}{\partial y} &
      \frac{\partial \vC_{y}}{\partial z}
      \\
      \frac{\partial \vC_{z}}{\partial x} &
      \frac{\partial \vC_{z}}{\partial y} &
      \frac{\partial \vC_{z}}{\partial z}
    \end{array}
    \right)
\end{equation}
with $\fVel = (\vC_{x}, \vC_{y}, \vC_{z})$. It should be remarked
that the sign and index conventions used in the definitions of these
tensors are not universal.

\vspace{0.2cm}

A fluid possesses \vy\ if it is capable of storing elastic energy. A
sign of this is that stresses within the fluid persist after the
deformation has ceased. The duration of time over which appreciable
stresses persist after cessation of deformation gives an estimate of
what is called the relaxation time. The relaxation and retardation
times, $\rxTim$ and $\rdTim$ respectively, are important physical
properties of \vc\ fluids because they characterize whether \vy\ is
likely to be important within an experimental or observational
timescale. They usually have the physical significance that if the
motion is suddenly stopped the stress decays as $e^{-t/\rxTim}$, and
if the stress is removed the rate of strain decays as
$e^{-t/\rdTim}$.

\vspace{0.2cm}

For viscous flow, the \Rey\ number $Re$ is a dimensionless number
defined as the ratio of the inertial to viscous forces and is given
by
\begin{equation}\label{Reynolds}
    Re = \frac{\rho l_{c} v_{c}}{\mu}
\end{equation}
where $\rho$ is the mass density of the fluid, $l_{c}$ is a
characteristic length of the flow system, $v_{c}$ is a
characteristic speed of the flow and $\mu$ is the viscosity of the
fluid.

\vspace{0.2cm}

For \vc\ fluids the key dimensionless group is the \Deborah\ number
which is a measure of the elasticity of the fluid. This number may
be interpreted as the ratio of the magnitude of the elastic forces
to that of the viscous forces. It is defined as the ratio of a
characteristic time of the fluid $\fTim_{c}$ to a characteristic
time of the flow system $t_{c}$
\begin{equation}\label{Deborah}
    De = \frac{\fTim_{c}}{t_{c}}
\end{equation}
The \Deborah\ number is zero for a \NEW\ fluid and is infinite for a
\Hookean\ elastic solid. High \Deborah\ numbers correspond to
elastic behavior and low \Deborah\ numbers to viscous behavior. As
the characteristic times are process-dependent, materials may not
have a single \Deborah\ number associated with them.

\vspace{0.2cm}

Another dimensionless number which measures the elasticity of the
fluid is the \Weissenberg\ number $We$. It is defined as the product
of a characteristic time of the fluid $\fTim_{c}$ and a
characteristic strain rate $\sR_{c}$ in the flow
\begin{equation}\label{Weissenberg}
    We = \fTim_{c} \sR_{c}
\end{equation}

Other definitions to \Deborah\ and \Weissenberg\ numbers are also in
common use and some even do not differentiate between the two
numbers. The characteristic time of the fluid may be taken as the
largest time constant describing the molecular motions, or a time
constant in a constitutive equation. The characteristic time for the
flow can be the time interval during which a typical fluid element
experiences a significant sequence of kinematic events or it is
taken to be the duration of an experiment or experimental
observation. Many variations in defining and quantifying these
characteristics do exit, and this makes \Deborah\ and \Weissenberg\
numbers not very well defined and measured properties and hence the
interpretation of the experiments in this context may not be totally
objective.

\vspace{0.2cm}

The \Boger\ fluids are constant-viscosity purely elastic \nNEW\
fluids. They played important role in the recent development of the
theory of fluid elasticity as they allow dissociation between
elastic and viscous effects. \Boger\ realized that the complication
of variable viscosity effects can be avoided by employing test
liquids which consist of low concentrations of flexible high
molecular weight polymers in very viscous solvents, and these
solutions are nowadays called \Boger\ fluids.

} %end \renewcommand{\thefootnote}{\fnsymbol{footnote}}

\def\baselinestretch{1}
\chapter[\ConvDiv\ Geometry and Tube Discretization]{} \label{AppConvDiv}

\begin{spacing}{2}
{\LARGE \bf \ConvDiv\ Geometry and Tube Discretization}
\end{spacing}

\vspace{1.0cm}

\def\baselinestretch{1.66}
\begin{figure}[!t]
  \centering{}
  \includegraphics
  [scale=0.35]
  {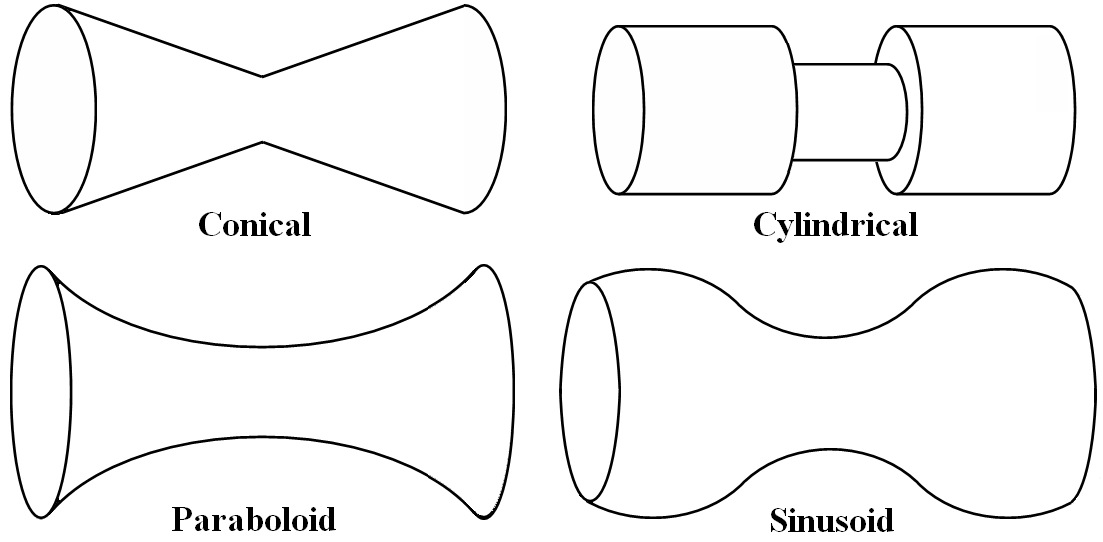}
  \caption[Examples of corrugated capillaries which can be used to model \convdiv\ geometry in porous media]
  {Examples of corrugated capillaries which can be used to model \convdiv\ geometry in porous media.}
  \label{ConvDivGeom}
\end{figure}

\begin{figure}[!t]
  \centering{}
  \includegraphics
  [scale=0.55]
  {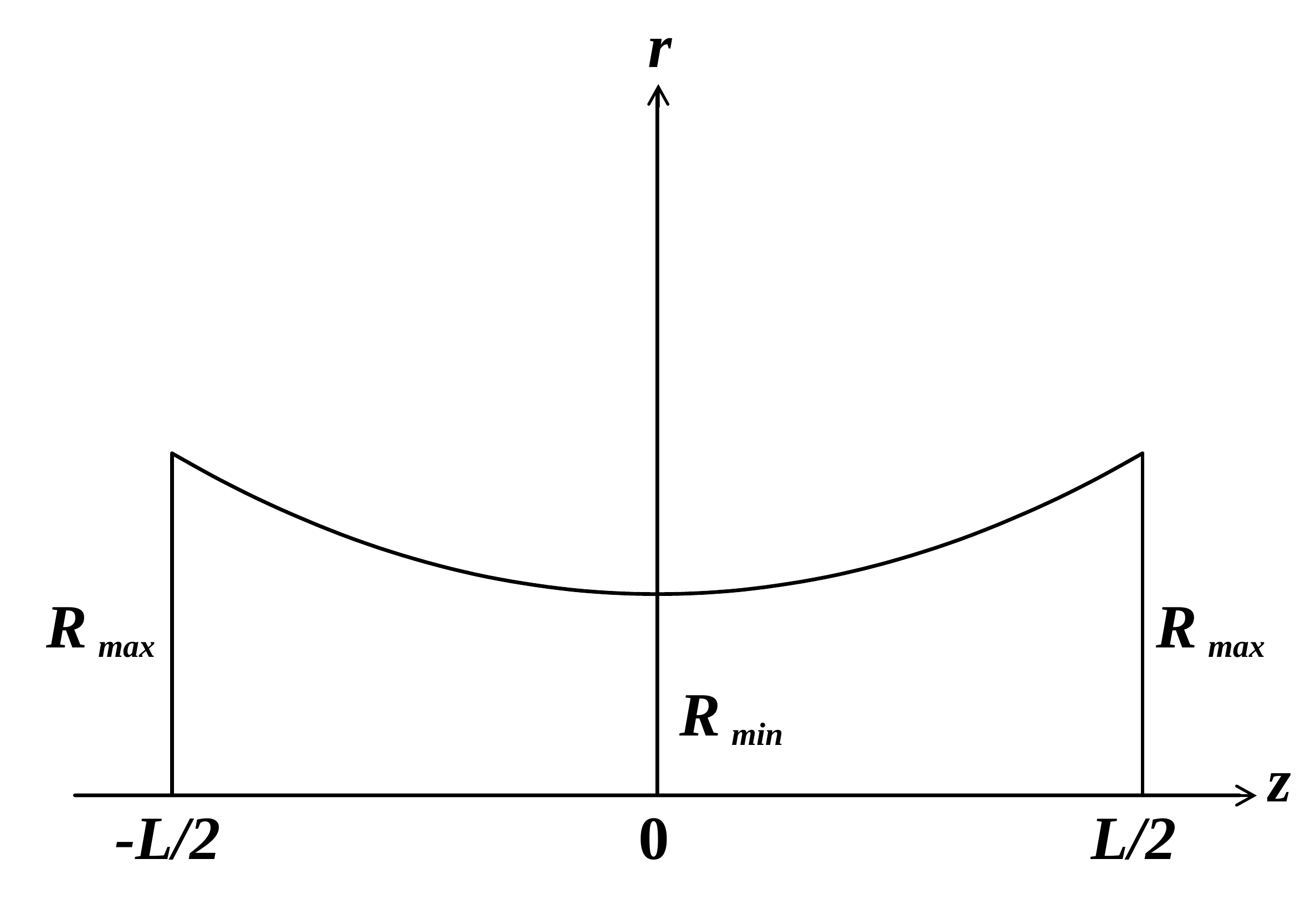}
  \caption[Radius variation in the axial direction for a corrugated paraboloid capillary using a cylindrical coordinate system]
  {Radius variation in the axial direction for a corrugated paraboloid capillary using a cylindrical coordinate system.}
  \label{ParaboloidTube}
\end{figure}

\noindent There are many simplified \convdiv\ geometries for
corrugated capillaries which can be used to model the flow of \vc\
fluids in porous media; some suggestions are presented in Figure
(\ref{ConvDivGeom}). In this study we adopted the paraboloid and
hence developed simple formulae to find the radius as a function of
the distance along the tube axis, assuming cylindrical coordinate
system, as shown in Figure (\ref{ParaboloidTube}).

\vspace{0.2cm}

For the paraboloid depicted in Figure (\ref{ParaboloidTube}), we
have
\begin{equation}\label{paraboloidRad}
    r(z) = a z^{2} + b z + c
\end{equation}

Since
\begin{equation}\label{paraboloidPoints}
    r(-L/2) = r(L/2) = R_{max}     \verb|       |     \&     \verb|       |     r(0) = R_{min}
\end{equation}
the paraboloid is uniquely identified by these three points. On
substituting and solving these equations simultaneously, we obtain
\begin{equation}\label{paraboloidCoef}
    a = \left(\frac{2}{L}\right)^{2} (R_{max} - R_{min})  \verb|       |  b = 0
    \verb|       |  \&     \verb|       |  c = R_{min}
\end{equation}
that is
\begin{equation}\label{paraboloidFinal}
    r(z) = \left(\frac{2}{L}\right)^{2} (R_{max} - R_{min}) z^{2} + R_{min}
\end{equation}

In the modified \Tardy\ algorithm for \steadys\ \vc\ flow which we
implemented in our \nNEW\ code, when a capillary is discretized into
$m$ slices the radius $r(z)$ is sampled at $m$ $z$-points given by
\begin{equation}\label{radiusPoints}
    z = -\frac{L}{2} + k \frac{L}{m}
    \verb|       | (k = 1,...,m)
\end{equation}

More complex polynomial and sinusoidal \convdiv\ profiles can also
be modeled using this approach.

\def\baselinestretch{1}
\chapter[Sand Pack and Berea Networks]{} \label{AppNetProp}

\begin{spacing}{2}
{\LARGE \bf The \SandP\ and \Berea\ Networks}
\end{spacing}

\vspace{1.0cm}

\def\baselinestretch{1.66}
\noindent The \sandp\ and \Berea\ networks used in this study are
those of Statoil ({\O}ren and coworkers \cite{OrenBA1997,
OrenB2003}). They are constructed from voxel images generated by
simulating the geological processes by which the porous medium was
formed. The physical and statistical properties of the networks are
presented in Tables (\ref{SPPropertiesTable}) and
(\ref{BereaPropertiesTable}).

\vspace{0.2cm}

The \Berea\ network has more complex and less homogeneous structure
than the \sandp. A sign of this is the bimodal nature of the \Berea\
pore and throat size distributions. The details can be found in
Lopez \cite{lopezthesis}. This fact may be disputed by the broader
distributions of the \sandp\ indicated by broader ranges and larger
standard deviations as seen in Tables (\ref{SPPropertiesTable}) and
(\ref{BereaPropertiesTable}). However, the reason for this is the
larger averages of the \sandp\ properties. By normalizing the
standard deviations to their corresponding averages, the \sandp\
will have narrower distributions for some of the most important
features that control the flow and determine the physical properties
of the network. These include the size distribution of inscribed
radius of pores and throats. Furthermore, our judgement, which is
shared by Lopez \cite{lopezthesis}, is based on more thorough
inspection to the behavior of the two networks in practical
situations.

\vspace{0.2cm}

Despite the fact that the \sandp\ network seems to have some curious
properties as it includes some eccentric elements such as
super-connected pores and very long throats, we believe that these
peculiarities have very little impact on the network behavior in
general. The reason is that these elements are very few and their
influence is diluted in a network of several thousands of normal
pores and throats, especially for the fluids with no \yields. As for
\yields\ fluids, the effect of these extreme elements may not be
negligible if they occurr to aggregate in a favorable combination.
However, it seems that this is not the case. A graphic demonstration
of this fact is displayed in Figure (\ref{SPVisualYieldStages})
where the number of elements spanning the network (8\,-10) when flow
starts is comparable to the lattice size of the network
(15$\times$15$\times$15). This rules out the possibility of a
correlated streak of unusually large pores and throats. Our
conviction is that 8\,-10 moderately-large throats spanning a
network of this size is entirely normal and do not involve eccentric
elements. It is totally natural for \yields\ flow to start with the
elements whose size is higher than the average.

\vspace{0.2cm}

It is noteworthy that the results of yield-free fluids are almost
independent of the calculation box as long as the pertinent bulk
properties are retained and provided that the width of the slice is
sufficiently large to be representative of the network and mask the
pore scale fluctuations. This is in sharp contrast with the \yields\
fluids where the results are highly dependent on the location of the
calculation box as the \yields\ of a network is highly dependent on
the nature of its elements and their topology and geometry. These
observations are confirmed by a large number of simulation runs; and
as for \yields\ fluids can be easily seen in Tables
(\ref{ipmPmpSPTable}-\ref{ipmPmpCubicBereaTable}).

%%%%%%%%%%%%%%%%%%%%%%%%%%%%%%%   Sand pack properties table   %%%%%%%%%%%%%%%%%%%%%%%%%%%%%%%%%%%%%%%%%
\newpage

\begin{table} []
\centering %
\caption[Physical and statistical properties of the \sandp\ network] %
{Physical and statistical properties of the \sandp\ network.} %
\label{SPPropertiesTable} %
\vspace{0.5cm} %
\begin{tabular}{|l|c|c|c|c|}
\hline
                          \multicolumn{ 5}{|c|}{{\bf General}} \\
\hline
{\bf Size} &     \multicolumn{ 4}{c|}{2.5mm $\times$ 2.5mm $\times$ 2.5mm} \\
\hline
{\bf Total number of elements} &                      \multicolumn{ 4}{c|}{13490} \\
\hline
{\bf Absolute permeability ($10^{-12}$m$^2$)} &                   \multicolumn{ 4}{c|}{93.18} \\
\hline
{\bf Formation factor} &                       \multicolumn{ 4}{c|}{2.58} \\
\hline
{\bf Porosity} &                      \multicolumn{ 4}{c|}{0.338} \\
\hline
{\bf Clay bound porosity} &                      \multicolumn{ 4}{c|}{0.00} \\
\hline
{\bf No. of connections to inlet} &                        \multicolumn{ 4}{c|}{195} \\
\hline
{\bf No. of connections to outlet} &                        \multicolumn{ 4}{c|}{158} \\
\hline
{\bf Triangular shaped elements (\%)} &                       \multicolumn{ 4}{c|}{94.69} \\
\hline
{\bf Square shaped elements (\%)} &                        \multicolumn{ 4}{c|}{1.56} \\
\hline
{\bf Circular shaped elements (\%)} &                        \multicolumn{ 4}{c|}{3.75} \\
\hline
{\bf Physically isolated elements (\%)} &                       \multicolumn{ 4}{c|}{0.00} \\
\hline
                            \multicolumn{ 5}{|c|}{{\bf Pores}} \\
\hline
{\bf Total number} &                      \multicolumn{ 4}{c|}{3567} \\
\hline
{\bf No. of triangular shaped} &                      \multicolumn{ 4}{c|}{3497} \\
\hline
{\bf No. of square shaped} &                        \multicolumn{ 4}{c|}{70} \\
\hline
{\bf No. of circular shaped} &                         \multicolumn{ 4}{c|}{0} \\
\hline
{\bf No. of physically isolated} &                          \multicolumn{ 4}{c|}{0} \\
\hline
    {\bf } & {\bf Min.} & {\bf Max.} & {\bf Ave.} & {\bf St. Dev.} \\
\hline
{\bf Connection number} &          1 &         99 &      5.465 &      4.117 \\
\hline
{\bf Center to center length ($10^{-6}$m)} &       5.00 &     2066.62 &     197.46 &      176.24 \\
\hline
{\bf Inscribed radius ($10^{-6}$m)} &       2.36 &      98.29 &      39.05 &       15.78 \\
\hline
{\bf Shape factor} &     0.0146 &     0.0625 &     0.0363 &     0.0081 \\
\hline
{\bf Volume ($10^{-15}$m$^3$)} &       0.13 &   73757.26 &     1062.42 &     2798.79 \\
\hline
{\bf Clay volume ($10^{-15}$m$^3$)} &       0.00 &    0.00 &      0.00 &     0.00 \\
\hline
                          \multicolumn{ 5}{|c|}{{\bf Throats}} \\
\hline
{\bf Total number} &                      \multicolumn{ 4}{c|}{9923} \\
\hline
{\bf No. of triangular shaped} &                      \multicolumn{ 4}{c|}{9277} \\
\hline
{\bf No. of square shaped} &                       \multicolumn{ 4}{c|}{140} \\
\hline
{\bf No. of circular shaped} &                        \multicolumn{ 4}{c|}{506} \\
\hline
{\bf No. of physically isolated} &                          \multicolumn{ 4}{c|}{0} \\
\hline
    {\bf } & {\bf Min.} & {\bf Max.} & {\bf Ave.} & {\bf St. Dev.} \\
\hline
{\bf Length ($10^{-6}$m)} &       1.67 &      153.13 &      30.61 &       16.31 \\
\hline
{\bf Inscribed radius ($10^{-6}$m)} &       0.50 &      85.57 &      23.71 &       11.19 \\
\hline
{\bf Shape factor} &     0.0067 &     0.0795 &     0.0332 &     0.0137 \\
\hline
{\bf Volume ($10^{-15}$m$^3$)} &       0.13 &     2688.18 &      155.93 &      151.18 \\
\hline
{\bf Clay volume ($10^{-15}$m$^3$)} &       0.00 &     0.00 &      0.00 &      0.00 \\
\hline
\end{tabular}
{\tiny
 {\begin{spacing}{1.1}
 \begin{flushleft}
 No. = Number \verb|    | Min. = Minimum \verb|    | Max. = Maximum \verb|    | Ave. = Average \verb|    | St. Dev. =
 Standard Deviation. \vspace{0.3cm} \\
 Note: The size-dependent properties, such as absolute permeability, are for the original
 network with no scaling. All data are for the complete network, i.e. for a calculation box
 with a lower boundary $x_{_{l}}=0.0$ and an upper boundary $x_{_{u}}=1.0$.
 \end{flushleft}
 \end{spacing}}
} %
\end{table}

%%%%%%%%%%%%%%%%%%%%%%%%%%%%%%%   Berea properties table   %%%%%%%%%%%%%%%%%%%%%%%%%%%%%%%%%%%%%%%%%
\newpage

\begin{table} [h]
\centering %
\caption[Physical and statistical properties of the \Berea\ network] %
{Physical and statistical properties of the \Berea\ network.} %
\label{BereaPropertiesTable} %
\vspace{0.5cm} %
\begin{tabular}{|l|c|c|c|c|}
\hline
                          \multicolumn{ 5}{|c|}{{\bf General}} \\
\hline
{\bf Size} &     \multicolumn{ 4}{c|}{3.0mm $\times$ 3.0mm $\times$ 3.0mm} \\
\hline
{\bf Total number of elements} &                      \multicolumn{ 4}{c|}{38495} \\
\hline
{\bf Absolute permeability ($10^{-12}$m$^2$)} &                   \multicolumn{ 4}{c|}{2.63} \\
\hline
{\bf Formation factor} &                       \multicolumn{ 4}{c|}{14.33} \\
\hline
{\bf Porosity} &                      \multicolumn{ 4}{c|}{0.183} \\
\hline
{\bf Clay bound porosity} &                      \multicolumn{ 4}{c|}{0.057} \\
\hline
{\bf No. of connections to inlet} &                        \multicolumn{ 4}{c|}{254} \\
\hline
{\bf No. of connections to outlet} &                        \multicolumn{ 4}{c|}{267} \\
\hline
{\bf Triangular shaped elements (\%)} &                       \multicolumn{ 4}{c|}{92.26} \\
\hline
{\bf Square shaped elements (\%)} &                        \multicolumn{ 4}{c|}{6.51} \\
\hline
{\bf Circular shaped elements (\%)} &                        \multicolumn{ 4}{c|}{1.23} \\
\hline
{\bf Physically isolated elements (\%)} &                       \multicolumn{ 4}{c|}{0.02} \\
\hline
                            \multicolumn{ 5}{|c|}{{\bf Pores}} \\
\hline
{\bf Total number} &                      \multicolumn{ 4}{c|}{12349} \\
\hline
{\bf No. of triangular shaped} &                      \multicolumn{ 4}{c|}{11794} \\
\hline
{\bf No. of square shaped} &                        \multicolumn{ 4}{c|}{534} \\
\hline
{\bf No. of circular shaped} &                         \multicolumn{ 4}{c|}{21} \\
\hline
{\bf No. of physically isolated} &                          \multicolumn{ 4}{c|}{6} \\
\hline
    {\bf } & {\bf Min.} & {\bf Max.} & {\bf Ave.} & {\bf St. Dev.} \\
\hline
{\bf Connection number} &          1 &         19 &      4.192 &      1.497 \\
\hline
{\bf Center to center length ($10^{-6}$m)} &       4.28 &     401.78 &     115.89 &      48.81 \\
\hline
{\bf Inscribed radius ($10^{-6}$m)} &       3.62 &      73.54 &      19.17 &       8.47 \\
\hline
{\bf Shape factor} &     0.0113 &     0.0795 &     0.0332 &     0.0097 \\
\hline
{\bf Volume ($10^{-15}$m$^3$)} &       1.56 &   10953.60 &     300.55 &     498.53 \\
\hline
{\bf Clay volume ($10^{-15}$m$^3$)} &       0.00 &    8885.67 &      88.30 &     353.42 \\
\hline
                          \multicolumn{ 5}{|c|}{{\bf Throats}} \\
\hline
{\bf Total number} &                      \multicolumn{ 4}{c|}{26146} \\
\hline
{\bf No. of triangular shaped} &                      \multicolumn{ 4}{c|}{23722} \\
\hline
{\bf No. of square shaped} &                       \multicolumn{ 4}{c|}{1972} \\
\hline
{\bf No. of circular shaped} &                        \multicolumn{ 4}{c|}{452} \\
\hline
{\bf No. of physically isolated} &                          \multicolumn{ 4}{c|}{3} \\
\hline
    {\bf } & {\bf Min.} & {\bf Max.} & {\bf Ave.} & {\bf St. Dev.} \\
\hline
{\bf Length ($10^{-6}$m)} &       1.43 &      78.94 &      13.67 &       5.32 \\
\hline
{\bf Inscribed radius ($10^{-6}$m)} &       0.90 &      56.85 &      10.97 &       7.03 \\
\hline
{\bf Shape factor} &     0.0022 &     0.0795 &     0.0346 &     0.0139 \\
\hline
{\bf Volume ($10^{-15}$m$^3$)} &       0.49 &     766.42 &      48.07 &      43.27 \\
\hline
{\bf Clay volume ($10^{-15}$m$^3$)} &       0.00 &     761.35 &      17.66 &      34.89 \\
\hline
\end{tabular}
{\tiny
 {\begin{spacing}{1.1}
 \begin{flushleft}
 No. = Number \verb|    | Min. = Minimum \verb|    | Max. = Maximum \verb|    | Ave. = Average \verb|    | St. Dev. =
 Standard Deviation. \vspace{0.3cm} \\
 Note: The size-dependent properties, such as absolute permeability, are for the original
 network with no scaling. All data are for the complete network, i.e. for a calculation box
 with a lower boundary $x_{_{l}}=0.0$ and an upper boundary $x_{_{u}}=1.0$.
 \end{flushleft}
 \end{spacing}}
} %
\end{table}

\def\baselinestretch{1}
\chapter[Manual of \NNEW\ Code]{} \label{AppManual}

\begin{spacing}{2}
{\LARGE \bf The Manual of the \NNEW\ Code}
\end{spacing}
\vspace{1.0cm}

\def\baselinestretch{1.66}
\definecolor{shadecolor}{rgb}{0.9,0.8,1} %
\noindent The computer code which is developed during this study is
based on the \nNEW\ code by Xavier Lopez \cite{lopezthesis, lopez1}.
The latter was constructed from an early version of the \NEW\ code
by Per Valvatne \cite{valvatnethesis, valvatne1}. The \nNEW\ code in
its current state simulates single-phase flow only, as the code is
tested and debugged for this case only. However, all features of the
two-phase flow which are inherited from the parent code are left
intact and can be easily activated for future development. Beside
the \CARREAU\ model inherited from the original code of Lopez, the
current code can simulate the flow of \ELLIS\ and \HB\ fluids.
Several algorithms related to \yields\ and a modified version of the
\Tardy\ algorithm to simulate \steadys\ \vc\ flow using a \BauMan\
model are also implemented. The code can be downloaded from this
URL:
\href{http://www3.imperial.ac.uk/earthscienceandengineering/research/perm/porescalemodelling/software}
{http://www3.imperial.ac.uk/earthscienceandengineering/research/perm/
\\ porescalemodelling/software}.

\vspace{0.2cm}

The program uses a keyword based input file ``0.inputFile.dat''. All
keywords are optional with no order required. If keywords are
omitted, default values will be used. Comments in the data file are
indicated by ``\%'', resulting in the rest of the line being
discarded. All data should be on a single line following the keyword
(possibly separated by comment lines).
\vspace{0.2cm}

The general flowing sequence of the program is as follows:
\begin{itemize}

    \item The program starts by reading the input and network data files followed by
    creating the network.

    \item For fluids with \yields, the program executes IPM (\InvPM)
    and PMP (\PathMP) algorithms to predict the threshold yield pressure of the
    network. This is followed by an iterative simulation algorithm to find the network's actual
    threshold yield pressure to the required number of decimal places.

    \item Single-phase flow with \NEW\ fluid is simulated and the fluid-related network
    properties, such as absolute permeability, are obtained.

    \item Single-phase flow with \NEW\ and \nNEW\ fluids is simulated over a pressure line.

\end{itemize}

In all stages, informative messages are issued about the program
flow and the data obtained. The program also creates several output
data files. These are:
\begin{enumerate}

    \item The main output data file which contains the data in the input file and
    the screen output. This will be discussed under the ``TITLE'' keyword.

    \item ``01.FNProp'' file which contains data on the network and fluid properties.

    \item ``1.Pressure'' file which contains the pressure points.

    \item ``2.Gradient'' file which contains the pressure gradients corresponding to the
    pressure points in the previous file.

    \item ``3.FlowRate'' file which contains the corresponding volumetric flow rates for the
    single-phase \nNEW\ flow.

    \item ``4.DarcyVelocity'' file which contains the corresponding Darcy velocities.

    \item ``5.Viscosity'' file which contains the corresponding apparent viscosities.

    \item ``6.AvrRadFlow'' file which contains the average radius of the flowing throats of the
    network at each pressure point. This is mainly useful in the case of a \yields\ fluid.

    \item ``7.PerCentFlow'' file which contains the percentage of the flowing throats at each
    pressure point. This is mainly useful in the case of a \yields\ fluid.

\end{enumerate}

The reason for keeping the data separated in different files is to
facilitate flexible post-processing by spreadsheets or other
programs. Two samples of spreadsheets used to process the data are
included. The spreadsheets are designed to read the data files
directly on launching the application. What the user needs is to
``Enable automatic refresh'' after redefining the path to the
required data files using ``Edit Text Import'' option. The relevant
cells to do this path redefinition are blue-colored.

\vspace{0.2cm}

It should be remarked that all the data in the files are valid data
obtained from converged-to pressure points. If for some reason the
program failed to converge at a pressure point, all the data related
to that point will be ignored and not saved to the relevant files to
keep the integrity of the data intact.

\vspace{0.2cm}

All physical data in the input file are in SI unit system.

{\noindent \bf \Large ------------------------------------------------------------------} \\
%%%%%%%%%%%%%%%%%%%%%%%%%%%%%%%%%%%%%%%%%%%%%%%%%%%%%%%%%%%%%%%%%%%%%%%%%%%%%%%%%%%%%%%%%%%%%%%%%
\noindent
{\bf \large TITLE} \vspace{0.4cm} \\
The program generates a file named ``{\em title}.prt'' containing
the date and time of simulation, the data in the input file and the
screen output. Moreover, the user can also define the directory to
which all the output files will be saved.
\begin{enumerate}

    \item Title of the ``.prt'' file.

    \item The directory to which all output files should be written.

\end{enumerate}
If this keyword is omitted, the default is ``0.Results\verb|      |./''. %

\begin{shaded} \hspace{-0.7cm}
{\normalsize \color{blue} TITLE}\\
\% Title of .prt file\verb|             |Directory of output files\\
\verb|     |0.Results\verb|                    |../Results/
\end{shaded}
{\noindent \bf \Large ------------------------------------------------------------------} \\
%%%%%%%%%%%%%%%%%%%%%%%%%%%%%%%%%%%%%%%%%%%%%%%%%%%%%%%%%%%%%%%%%%%%%%%%%%%%%%%%%%%%%%%%%%%%%%%%%
\noindent
{\bf \large NETWORK} \vspace{0.4cm} \\
This keyword specifies the files containing the network data. The
data is located in four ASCII files, with a common prefix specified
after this keyword. These files are: ``{\em
filename}\_\,node1.dat'', ``{\em filename}\_\,node2.dat'', ``{\em
filename}\_\,link1.dat'' and ``{\em filename}\_\,link2.dat". If
these data files are not located in the same directory as the
program, the filename should be preceded by its path relative to the
program. The structure of the network data files will be explained
in Appendix \ref{AppNetFiles}: ``The Structure of the Network Data
Files''. If this keyword is omitted, the default value is ``SP''
with no relative path.

\begin{shaded} \hspace{-0.7cm}
{\normalsize \color{blue} NETWORK}\\
\% Directory and prefix of the network data files \\
\verb|      |../Networks/SandPack/SP
\end{shaded}
{\noindent \bf \Large ------------------------------------------------------------------} \\
%%%%%%%%%%%%%%%%%%%%%%%%%%%%%%%%%%%%%%%%%%%%%%%%%%%%%%%%%%%%%%%%%%%%%%%%%%%%%%%%%%%%%%%%%%%%%%%%%
\noindent
{\bf \large PRS\_\,LINE} \vspace{0.4cm} \\
i.e. pressure line. This keyword defines the pressure line to be
scanned in the single-phase flow simulation. There are two options:
\begin{itemize}

    \item Reading the pressure line from a separate file. In this case an entry of zero
    followed by the relative path and the name of the pressure line data file are required.
    The pressure line file must contain the same data as in the second option.

    \item Reading the pressure line from the input file. In this case the first entry should
    be the number of pressure points plus one, the last entry is the outlet pressure (in Pa)
    and the entries in-between are the inlet pressure points (in Pa).

\end{itemize}
If this keyword is omitted, the default is a 15-point pressure line:
``0.1 0.5 1 5 10 50 100 500 1000 5000 10000 50000 100000 500000
1000000'' with an outlet pressure of zero. Some pressure line files
are stored in the ``PressureLines'' directory.

\begin{shaded} \hspace{-0.7cm}
{\normalsize \color{blue} PRS\_\,LINE}\\
\% First: No. of pressure points plus 1. Last: outlet pressure(Pa). Between: inlet \\
\% pressure points(Pa). If first entry is 0, data is read from ``PressureLine.dat'' file \\
\verb|  |8\verb|       |1\verb|  |10\verb|  |100\verb|  |1000\verb|  |10000\verb|  |100000
\verb| |1000000\verb|       |0
\end{shaded}
{\noindent \bf \Large ------------------------------------------------------------------} \\
%%%%%%%%%%%%%%%%%%%%%%%%%%%%%%%%%%%%%%%%%%%%%%%%%%%%%%%%%%%%%%%%%%%%%%%%%%%%%%%%%%%%%%%%%%%%%%%%%
\noindent
{\bf \large CALC\_\,BOX} \vspace{0.4cm} \\
i.e. calculation box. In the single-phase flow simulation, it may be
desired to use a slice of the network instead of the whole network.
Some reasons are reducing the CPU time needed to solve the pressure
field by reducing the network size, and using a network with
slightly different properties. By this keyword, the user has the
option to control the upper and lower boundaries of the network
across which the pressure drop is applied and the pressure field is
solved. The lower and upper boundaries, $x_{_{l}}$ and $x_{_{u}}$,
are dimensionless numbers between 0.0 (corresponding to the inlet
face) and 1.0 (corresponding to the outlet face) with $x_{_{l}} <
x_{_{u}}$. It should be remarked that the slice width must be
sufficiently large so the slice remains representative of the
network and the pore scale fluctuations are smoothed out.
\begin{enumerate}

    \item Relative position of the lower boundary, $x_{_{l}}$ (e.g. 0.2).

    \item Relative position of the upper boundary, $x_{_{u}}$ (e.g. 0.8).

\end{enumerate}

If this keyword is omitted, the whole network will be used in the
calculations, i.e. $x_{_{l}}=0.0$ and $x_{_{u}}=1.0$.

\begin{shaded} \hspace{-0.7cm}
{\normalsize \color{blue} CALC\_\,BOX}\\
\% Lower boundary, $x_{_{l}}$\verb|    |Upper boundary, $x_{_{u}}$ \\
\verb|         |0.0\verb|                  |1.0
\end{shaded}
{\noindent \bf \Large ------------------------------------------------------------------} \\
%%%%%%%%%%%%%%%%%%%%%%%%%%%%%%%%%%%%%%%%%%%%%%%%%%%%%%%%%%%%%%%%%%%%%%%%%%%%%%%%%%%%%%%%%%%%%%%%%
\noindent
{\bf \large NEWTONIAN} \vspace{0.4cm} \\
This keyword determines the viscosity of the \NEW\ fluid.
\begin{enumerate}
    \item Viscosity of the \NEW\ fluid (Pa.s).
\end{enumerate}

If this keyword is omitted, the default value is ``0.001''.

\begin{shaded} \hspace{-0.7cm}
{\normalsize \color{blue} NEWTONIAN}\\
\% Viscosity (Pa.s) \\%
\verb|    |0.025
\end{shaded}
{\noindent \bf \Large ------------------------------------------------------------------} \\
%%%%%%%%%%%%%%%%%%%%%%%%%%%%%%%%%%%%%%%%%%%%%%%%%%%%%%%%%%%%%%%%%%%%%%%%%%%%%%%%%%%%%%%%%%%%%%%%%
\noindent
{\bf \large NON\_\,NEWTONIAN} \vspace{0.4cm} \\
This keyword specifies the rheological model of the \nNEW\ fluid and
its parameters. If this keyword is omitted, the default is \NEW\
with a viscosity of ``0.001'' Pa.s. The \nNEW\ rheological models
which are implemented in this code are:

\vspace{1.5cm}

%%%%%%%%%%%%%%%%%%%%%%%%%%%%%%%%%%%%%%%%%%%%%%%%%%%%%%
\noindent {\bf \large 1. \CARREAU\ Model}

\vspace{0.2cm}

\noindent This model is given by

\begin{equation}\label{carreau}
    \Vis = \hVis + \frac{\lVis - \hVis}
    {\left[1+(\frac{\sR}{\crsR})^{2}\right]^{\frac{1-n}{2}}}
\end{equation}
where $\Vis$ is the fluid viscosity, $\hVis$ is the viscosity at
infinite shear, $\lVis$ is the viscosity at zero shear, $\sR$ is the
shear rate, $\crsR$ is the critical shear rate and $n$ is the flow
behavior index. The critical shear rate is given by

\begin{equation}\label{gammacr}
    \crsR=\left(\frac{\lVis}{C}\right)^{\frac{1}{n-1}}
\end{equation}
where $C$ is the consistency factor of the equivalent \shThin\ fluid
in the \powlaw\ formulation.

\vspace{0.2cm}

Although \CARREAU\ formulation models \shThin\ fluids with no
\yields, the code can simulate \CARREAU\ fluid with \yields. The
entries for \CARREAU\ model are

\begin{enumerate}

    \item ``C'' or ``c'' to identify the \CARREAU\ model.

    \item Consistency factor $C$ (Pa.s$^{n}$).

    \item Flow behavior index $n$ (dimensionless).

    \item Viscosity at infinite shear $\hVis$ (Pa.s).

    \item Viscosity at zero shear $\lVis$ (Pa.s).

    \item \Yields\ $\ysS$ (Pa). To use the original model with no \yields, the \yields\
    should be set to zero.

\end{enumerate}

\vspace{0.5cm}

%%%%%%%%%%%%%%%%%%%%%%%%%%%%%%%%%%%%%%%%%%%%%%%%%%%%%%
\noindent {\bf \large 2. \ELLIS\ Model}

\vspace{0.2cm}

\noindent This model is given by

\begin{equation}\label{ellis}
    \Vis=\frac{\lVis}{1+\left|\frac{\sS}{\hsS}\right|^{\eAlpha-1}}
\end{equation}
where $\Vis$ is the fluid viscosity, $\lVis$ is the viscosity at
zero shear, $\sS$ is the shear stress, $\hsS$ is the shear stress at
which $\Vis=\lVis/2$ and $\eAlpha$ is a dimensionless indicial
parameter related to the slope in the \powlaw\ region.

\vspace{0.2cm}

Although \ELLIS\ formulation models \shThin\ fluids with no \yields,
it is possible to simulate the model with \yields. The entries for
\ELLIS\ model are

\begin{enumerate}

    \item ``E'' or ``e'' to identify the \ELLIS\ model.

    \item Viscosity at zero shear $\lVis$ (Pa.s).

    \item Indicial parameter $\eAlpha$ (dimensionless).

    \item Shear stress at half initial viscosity $\hsS$ (Pa).

    \item \Yields\ $\ysS$ (Pa). To run the model with no \yields, the \yields\
    should be set to zero.

\end{enumerate}

\vspace{0.5cm}

%%%%%%%%%%%%%%%%%%%%%%%%%%%%%%%%%%%%%%%%%%%%%%%%%%%%%%
\noindent {\bf \large 3. \HB\ Model}

\vspace{0.2cm}

\noindent This model is given by

\begin{equation}\label{}
    \sS = \ysS + C\sR^{n}
\end{equation}
where $\sS$ is the shear stress, $\ysS$ is the \yields, $C$ is the
consistency factor, $\sR$ is the shear rate and $n$ is the flow
behavior index.

\vspace{0.2cm}

For \HB\ model the entries are

\begin{enumerate}

    \item ``H'' or ``h'' to identify the \HB\ model.

    \item Consistency factor $C$ (Pa.s$^{n}$).

    \item Flow behavior index $n$ (dimensionless).

    \item \Yields\ $\ysS$ (Pa).

\end{enumerate}

\vspace{0.5cm}

%%%%%%%%%%%%%%%%%%%%%%%%%%%%%%%%%%%%%%%%%%%%%%%%%%%%%%
\noindent {\bf \large 4. \Vy\ Model}

\vspace{0.2cm}

\noindent This is the modified \Tardy\ algorithm based on the
\BauMan\ model to simulate \steadys\ \vc\ flow, as described in
section (\ref{BautistaManero}).

\vspace{0.2cm}

The required entries are

\begin{enumerate}

    \item ``V'' or ``v'' to identify the \vc\ model.

    \item Elastic modulus $\Go$ (Pa).

    \item Retardation time $\rdTim$ (s). This parameter is not used
    in the current \steadys\ algorithm. However, it is included
    for possible future development.

    \item Viscosity at infinite shear rate $\hVis$ (Pa.s).

    \item Viscosity at zero shear rate $\lVis$ (Pa.s).

    \item Structural relaxation time $\rxTimF$ (s).

    \item Kinetic parameter for structure break down in \FRED\ model
    $\kF$ (Pa$^{-1}$).

    \item Multiplicative scale factor to obtain the parabolic tube radius at
    the entry from the equivalent radius of the straight tube $\fe$ (dimensionless).

    \item Multiplicative scale factor to obtain the parabolic tube radius at
    the middle from the equivalent radius of the straight tube $\fm$
    (dimensionless). For \convdiv\ geometry $\fe > \fm$,
    whereas for \divconv\ geometry $\fe < \fm$.

    \item The number of slices the corrugated tube is
    divided to during the step-by-step calculation of the pressure
    drop $m$ (dimensionless). A rough estimation of the value for
    this parameter to converge to a stable solution is about 10-20,
    although this may depend on other factors especially for the
    extreme cases.

\end{enumerate}

\begin{shaded} \hspace{-0.5cm}
{\normalsize \color{blue} NON\_\,NEWTONIAN}\\
\noindent
\begin{tabular}{lccccc}
\% Carreau (\rm{C}):  \verb|  | &  $C$(Pa.s$^{n}$)  & \verb|  | $n$       & \verb| | $\hVis$(Pa.s)     & \verb| | $\lVis$(Pa.s)  & \verb| | $\ysS$(Pa) \\
\% \ELLIS\ (\rm{E}):    \verb|  | &  $\lVis$(Pa.s)    & \verb|  | $\eAlpha$ & \verb| | $\hsS$(Pa)        &  \verb| |               & \verb| | $\ysS$(Pa) \\
\% Herschel (\rm{H}): \verb|  | &  $C$(Pa.s$^{n}$)  & \verb|  | $n$       & \verb| |                   &  \verb| |               & \verb| | $\ysS$(Pa) \\
\verb|     |H         \verb|  | &         0.1       & \verb|  | 0.85      & \verb| |                   &   \verb| |              &  \verb| |       1.0 \\
\end{tabular}
\end{shaded}
{\noindent \bf \Large ------------------------------------------------------------------} \\
%%%%%%%%%%%%%%%%%%%%%%%%%%%%%%%%%%%%%%%%%%%%%%%%%%%%%%%%%%%%%%%%%%%%%%%%%%%%%%%%%%%%%%%%%%%%%%%%%
\noindent
{\bf \large MAX\_\,ERROR} \vspace{0.4cm} \\
This keyword controls the error tolerance (i.e. the volumetric flow
rate relative error) for the \nNEW\ solver to converge. The
criterion for real convergence is that no matter how small the
tolerance is, the solver eventually converges given enough CPU time.
Therefore, to avoid accidental convergence, where the solutions of
two consecutive iterations comes within the error tolerance
coincidentally, the tolerance should be set to a sufficiently small
value so that the chance of this occurrence becomes negligible. As a
rough guide, it is recommended that the tolerance should not exceed
0.0001. Obviously reducing the tolerance usually results in an
increase in the number of iterations and hence CPU time.

\vspace{0.2cm}

Accidental convergence, if happened, can be detected by fluctuations
and glitches in the overall behavior resulting in non-smooth curves.
It is remarked that the convergence for \ELLIS\ and \HB\ models is
easily achieved even with extremely small tolerance, so the mostly
affected model by this parameter is \CARREAU\ where the convergence
may be seriously delayed if very small tolerance is chosen.

\begin{enumerate}
    \item Relative error tolerance in the volumetric flow rate (non-negative dimensionless real number).
\end{enumerate}

If this keyword is omitted, the default is ``$10^{-6}$''.

\begin{shaded} \hspace{-0.7cm}
{\normalsize \color{blue} MAX\_\,ERROR}\\
\% \NNEW\ solver convergence tolerance (recommended $<$ 0.0001). \\
\verb|  |1.0E-6
\end{shaded}
{\noindent \bf \Large ------------------------------------------------------------------} \\
%%%%%%%%%%%%%%%%%%%%%%%%%%%%%%%%%%%%%%%%%%%%%%%%%%%%%%%%%%%%%%%%%%%%%%%%%%%%%%%%%%%%%%%%%%%%%%%%%
\noindent
{\bf \large MAX\_\,ITERATIONS} \vspace{0.4cm} \\
This keyword determines the maximum number of iterations allowed
before abandoning the pressure point in the flow simulation. If this
number is reached before convergence, the pressure point will be
discarded and no data related to the point will be displayed on the
screen or saved to the output files.

\vspace{0.2cm}

As remarked earlier, the convergence for the \ELLIS\ and \HB\ models
is easily achieved. However, for \CARREAU\ model the convergence is
more difficult to attain. The main factors affecting \CARREAU\
convergence are the size of the error tolerance and the fluid
rheological properties. As the fluid becomes more \shThin\ by
decreasing the flow behavior index $n$, the convergence becomes
harder and the number of iterations needed to converge increases
sharply. This increase mainly occurs during the \shThin\ regime away
from the two \NEW\ plateaux. It is not unusual for the number of
iterations in these circumstances to reach or even exceed a hundred.
Therefore it is recommended, when running \CARREAU\ model, to set
the maximum number of iterations to an appropriately large value
(e.g. 150) if convergence should be achieved.

\begin{enumerate}
    \item Maximum number of \nNEW\ solver iterations before abandoning the attempt (positive integer).
\end{enumerate}

If this keyword is omitted, the default is ``200''. %

\begin{shaded} \hspace{-0.7cm}
{\normalsize \color{blue} MAX\_\,ITERATIONS}\\
\% Maximum number of iterations before stopping the solver (positive integer). \\
\verb|  |10
\end{shaded}
{\noindent \bf \Large ------------------------------------------------------------------} \\
%%%%%%%%%%%%%%%%%%%%%%%%%%%%%%%%%%%%%%%%%%%%%%%%%%%%%%%%%%%%%%%%%%%%%%%%%%%%%%%%%%%%%%%%%%%%%%%%%
\noindent
{\bf \large YIELD\_\,ALGORITHMS} \vspace{0.4cm} \\
The purpose of this keyword is to control the algorithms related to
\yields\ fluids. These algorithms are
\begin{itemize}

    \item \InvPM\ (IPM) algorithm to predict the network threshold yield
    pressure.

    \item \PathMP\ (PMP) algorithm to predict the network threshold yield
    pressure.

    \item Actual Threshold Pressure (ATP) algorithm to find the network threshold yield
    pressure from the solver.

\end{itemize}

\vspace{0.4cm}

The PMP was implemented in three different ways

\begin{itemize}

    \item PMP1 in which the memory requirement is minimized as it requires only $8N$ bytes of
    storage for a network with $N$ nodes. However, the CPU time is maximum.

    \item PMP2 which is a compromise between memory and CPU time requirements. The storage needed
    is $8N$+$M$ bytes for a network with $N$ nodes and $M$ bonds.

    \item PMP3 in which the CPU time is minimized with very large memory requirement of $8N$+$8M$
    bytes for a network with $N$ nodes and $M$ bonds.

\end{itemize}
These three variations of the PMP produce identical results.

\vspace{0.2cm}

The ATP is an iterative simulation algorithm which uses the solver
to find the network yield pressure to the required number of decimal
places. It requires two parameters: an additive-subtractive factor
used with its sub-multiples to step through the pressure line
seeking for the network yield pressure, and an integer identifying
the required number of decimal places for the threshold yield
pressure calculation.

\vspace{0.2cm}

Although the algorithm works for any positive real factor, to find
the yield pressure to the correct number of decimal places the
factor should be a power of 10 (e.g. 10 or 100). The size of the
factor does not matter although the time required to converge will
be affected. To minimize the CPU convergence time the size of the
factor should be chosen according to the network size and properties
and the value of the \yields. The details are lengthy and messy,
however for the \sandp\ and \Berea\ networks with fluid of a
\yields\ $\ysS$ between $1.0-10.0$\,Pa the recommended factor is
100.

\vspace{0.2cm}

Using non-positive integer for the number of decimal places means
reduction in accuracy, e.g. ``$-1$'' means finding the threshold
yield pressure to the nearest 10. This can be employed to  get an
initial rough estimate of the threshold yield pressure in a short
time and this may help in selecting a factor with an appropriate
size for speedy convergence.

\vspace{0.2cm}

The required entries for this keyword are

\begin{enumerate}

    \item Run IPM? (true ``T'' or false ``F'').

    \item Run PMP? (``1'' for PMP1. ``2'' for PMP2. ``3'' for PMP3. ``4'' for all PMPs. ``0''
    or any other character for none).

    \item Run ATP? (true ``T'' or false ``F'').

    \item Additive-subtractive factor for stepping through the pressure line while
    running the ATP  (multiple of 10).

    \item Number of decimal places for calculating the threshold yield pressure by ATP (integer).

\end{enumerate}

It should be remarked that none of these algorithms will be executed
unless the \nNEW\ phase has a \yields.

\vspace{0.2cm}

If this keyword is omitted, the default is ``F\verb|    |0\verb|    |F\verb|    |100\verb|    |1''.  %

\begin{shaded} \hspace{-0.7cm}
{\normalsize \color{blue} YIELD\_\,ALGORITHMS}\\
\% IPM?\verb|      |PMP?\verb|      |ATP?\verb|      |ATP factor\verb|      |ATP decimals \\ %
\verb|   |F\verb|          |0\verb|          |F\verb|           |100\verb|               |1 %
\end{shaded}
{\noindent \bf \Large ------------------------------------------------------------------} \\
%%%%%%%%%%%%%%%%%%%%%%%%%%%%%%%%%%%%%%%%%%%%%%%%%%%%%%%%%%%%%%%%%%%%%%%%%%%%%%%%%%%%%%%%%%%%%%%%%
%exception

\noindent
{\bf \large DRAW\_\,NET} \vspace{0.4cm} \\
This keyword enables the user to write script files to visualize the
network using Rhino program. If the fluid has no \yields, all the
throats and pores in the calculation box will be drawn once. If the
fluid has a \yields, the non-blocked elements in the calculation box
will be drawn at each pressure point. This helps in checking the
continuity of flow and having a graphic inspection when simulating
the flow of a \yields\ fluid. If the ATP algorithm is on, the
non-blocked elements at the threshold yield pressure will also be
drawn.

\begin{enumerate}
    \item Write Rhino script file(s)? (true ``T'' or false ``F'').
\end{enumerate}

If this keyword is omitted, the default is ``F''.

\begin{shaded} \hspace{-0.7cm}
{\normalsize \color{blue} DRAW\_\,NET}\\
\% Write Rhino script file(s)? \\
\verb|  |F
\end{shaded}
{\noindent \bf \Large ------------------------------------------------------------------} \\
%%%%%%%%%%%%%%%%%%%%%%%%%%%%%%%%%%%%%%%%%%%%%%%%%%%%%%%%%%%%%%%%%%%%%%%%%%%%%%%%%%%%%%%%%%%%%%%%%

%
          %include in the final

\def\baselinestretch{1}
\chapter[Structure of Network Data Files]{} \label{AppNetFiles}

\begin{spacing}{2}
{\LARGE \bf The Structure of the Network Data Files}
\end{spacing}

\vspace{0.5cm}

\def\baselinestretch{1.66}
\definecolor{shadecolor}{rgb}{0.9,0.8,1} %
\noindent The network data are stored in four ASCII files. The
format of these files is that of Statoil. The physical data are given in SI unit system. \\
{\noindent \bf \Large ------------------------------------------------------------------} \\
%%%%%%%%%%%%%%%%%%%%%%%%%%%%%%%%%%%%%%%%%%%%%%%%%%%%%%%%%%%%%%%%%%%%%%%%%%%%%%%%%%%%%%%%%%%%%%%%%
\noindent
{\bf \large Throat Data} \vspace{0.1cm} \\
The data for the throats are read from the link files. The structure of the link files is as
follows: \vspace{0.2cm} \\
{\large \bf \color{red} 1. {\em prefix}\,\_\,link1.dat file} \vspace{0.2cm}\\
The first line of the file contains a single entry that is the total
number of throats, say $N$, followed by $N$ data lines. Each of
these lines contains six data entries in the following
order: \vspace{0.2cm}\\
\indent 1. Throat index \\
\indent 2. Pore 1 index \\
\indent 3. Pore 2 index \\
\indent 4. Throat radius \\
\indent 5. Throat shape factor \hspace{0.6cm} \\
\indent 6. Throat total length (pore center to pore center)
\begin{shaded} \hspace{-0.7cm}
{\normalsize \color{blue} Example of {\em prefix}\,\_\,link1.dat file} \\
26146 \\
1\verb|   |-1\verb|   | 8\verb|   |0.349563E-04\verb|   |0.297308E-01\verb|   |0.160000E-03 \\
2\verb|   |-1\verb|   |53\verb|   |0.171065E-04\verb|   |0.442550E-01\verb|   |0.211076E-04 \\
3\verb|   |-1\verb|   |60\verb|   |0.198366E-04\verb|   |0.354972E-01\verb|   |0.300000E-04 \\
4\verb|   |-1\verb|   |68\verb|   |0.938142E-05\verb|   |0.323517E-01\verb|   |0.100000E-04
\end{shaded}
\noindent
{\large \bf \color{red} 2. {\em prefix}\,\_\,link2.dat file} \vspace{0.2cm}\\
For a network with $N$ throats, the file contains $N$ data lines.
Each line has eight data entries in the following order: \vspace{0.2cm}\\
\indent 1. Throat index \\
\indent 2. Pore 1 index \\
\indent 3. Pore 2 index \\
\indent 4. Length of pore 1 \\
\indent 5. Length of pore 2 \\
\indent 6. Length of throat \\
\indent 7. Throat volume \\
\indent 8. Throat clay volume
\begin{shaded} \hspace{-0.7cm}
{\normalsize \color{blue} Example of {\em prefix}\,\_\,link2.dat file} \\
\noindent {\scriptsize
22714\verb|  |10452\verb|  |10533\verb|  |0.178262E-04\verb|  |0.120716E-03\verb|  |0.239385E-04\verb|  |0.218282E-13\verb|  |0.137097E-14 \\
22715\verb|  |10452\verb|  |10612\verb|  |0.121673E-04\verb|  |0.747863E-04\verb|  |0.100000E-04\verb|  |0.266790E-13\verb|  |0.355565E-14 \\
22716\verb|  |10453\verb|  |10534\verb|  |0.100000E-04\verb|  |0.270040E-04\verb|  |0.139862E-04\verb|  |0.543278E-13\verb|  |0.863932E-14
}
\end{shaded}
{\noindent \bf \Large ------------------------------------------------------------------} \\
\noindent
{\bf \large Pore Data} \vspace{0.1cm} \\
The data for the pores are read from the node files. The structure of the node files is as
follows: \vspace{0.2cm} \\
{\large \bf \color{red} 1. {\em prefix}\,\_\,node1.dat file} \vspace{0.2cm}\\
The first line of the file contains four entries: the total number
of pores, the length ($x$-direction), width ($y$-direction) and
height ($z$-direction) of the network. For a network with $M$ pores,
the first line is followed by $M$ data lines each containing the
following data entries: \vspace{0.2cm}\\
\indent 1. Pore index \\
\indent 2. Pore $x$-coordinate \\
\indent 3. Pore $y$-coordinate \\
\indent 4. Pore $z$-coordinate \\
\indent 5. Pore connection number \\
\indent 6. For a pore with a connection number $i$ there are $2(i+1)$ entries as follows: \\
\indent \indent  A. The first $i$ entries are the connecting pores indices \\
\indent \indent  B. The $(i+1)$st entry is the pore ``inlet'' status (0 for false and 1 for true) \\
\indent \indent  C. The $(i+2)$nd entry is the pore ``outlet'' status (0 for false and 1 for true) \\
\indent \indent  D. The last $i$ entries are the connecting throats indices \vspace{0.2cm}\\
Note: the inlet/outlet pores are those pores which are connected to a throat whose other pore
is the inlet/outlet reservoir, i.e. the other pore has an index of -1/0. So if the $(i+1)$st
entry is 1, one of the connecting pores indices is -1, and if the $(i+2)$nd entry is 1, one of
the connecting pores indices is 0.
\begin{shaded} \hspace{-0.7cm}
{\normalsize \color{blue} Example of {\em prefix}\,\_\,node1.dat file} \\
\noindent {\footnotesize
12349\verb|  |0.300000E-02\verb|  |0.300000E-02\verb|  |0.300000E-02 \\
1\verb|  |0.350000E-03\verb|  |0.000000E+00\verb|  |0.700000E-04\verb|  |3\verb|  |796\verb|  |674\verb|  |2\verb|  |0\verb|  |0\verb|  |522\verb|  |523\verb|  |524 \\
2\verb|  |0.450000E-03\verb|  |0.500000E-04\verb|  |0.000000E+00\verb|  |3\verb|  |359\verb|  | 31\verb|  |1\verb|  |0\verb|  |0\verb|  |525\verb|  |526\verb|  |524 \\
3\verb|  |0.880000E-03\verb|  |0.100000E-04\verb|  |0.000000E+00\verb|  |1\verb|  |392\verb|   | 0\verb|  |0\verb|  |527
}
\end{shaded}
\noindent
{\large \bf \color{red} 2. {\em prefix}\,\_\,node2.dat file} \vspace{0.2cm}\\
For a network with $M$ pores, the file contains $M$ data lines. Each
line has five data
entries in the following order: \vspace{0.2cm}\\
\indent 1. Pore index \\
\indent 2. Pore volume \\
\indent 3. Pore radius \\
\indent 4. Pore shape factor \\
\indent 5. Pore clay volume
\begin{shaded} \hspace{-0.7cm}
{\normalsize \color{blue} Example of {\em prefix}\,\_\,node2.dat file} \\
\noindent
50\verb|  |0.373367E-13\verb|  |0.195781E-04\verb|  |0.336954E-01\verb|  |0.784623E-16 \\
51\verb|  |0.155569E-14\verb|  |0.821594E-05\verb|  |0.326262E-01\verb|  |0.471719E-16 \\
52\verb|  |0.171126E-13\verb|  |0.122472E-04\verb|  |0.329865E-01\verb|  |0.148506E-15
\end{shaded}
{\noindent \bf \Large ------------------------------------------------------------------} \\

%
        %include in the final (remove from final ?)

% XXXXXXXXXXXXXXXXXXXXXXXXXXXXXXXXX Index XXXXXXXXXXXXXXXXXXXXXXXXXXXXXXXXXX
\voffset = 0.0cm
\phantomsection \addcontentsline{toc}{chapter}{\protect \numberline{} Index} %
%\phantomsection \addcontentsline{toc}{section}{Index} %
\printindex

% XXXXXXXXXXXXXXXXXXXXXXXXXXXXXXXX Glossary XXXXXXXXXXXXXXXXXXXXXXXXXXXXXXXX
%\voffset = 0.0cm
%\phantomsection \addcontentsline{toc}{chapter}{\protect \numberline{} Index} %
%%\phantomsection \addcontentsline{toc}{section}{Index} %
%\printglossary

\end{document}